\documentclass[a4paper,12pt,twoside]{report}
\usepackage[left=2cm,right=2cm,top=2cm,bottom=3cm]{geometry}

\usepackage{graphicx}
\usepackage{verbatim}
\usepackage{latexsym}
\usepackage{mathchars}
\usepackage{amssymb}
\usepackage{amsmath}
\usepackage{setspace}
\usepackage{physics}
\usepackage{tikz}
\usetikzlibrary{positioning, decorations.pathmorphing, decorations.pathreplacing, decorations.shapes,patterns} 
\usepackage{pgfplots}
\usepackage{caption}
\usepackage{subcaption}
\usepackage{cite}
\usepackage{hyperref}
\usepackage{enumerate}

\input{blocked.sty}
\input{uhead.sty}
\input{boxit.sty}
\input{icthesis.sty}

\newcommand{\keybox}[1]{\color{red}\boxed{\color{black}#1}\color{black}} 

\newcommand{\dbar}{d\hspace*{-0.08em}\bar{}\hspace*{0.1em}}
\newcommand{\deltabar}{\delta\hspace*{-0.2em}\bar{}\hspace*{0.1em}}

\newcommand{\NN}{{\sf I\kern-0.14emN}}   
\newcommand{\ZZ}{{\sf Z\kern-0.45emZ}}   
\newcommand{\QQQ}{{\sf C\kern-0.48emQ}}   
\newcommand{\RR}{{\sf I\kern-0.14emR}}   

\newcommand{\lb}{\left(}
\newcommand{\rb}{\right)}
\newcommand{\lsb}{\left[}
\newcommand{\rsb}{\right]}

\newcommand{\normallinespacing}{\renewcommand{\baselinestretch}{1.5} \normalsize}

\newcommand{\narrowlinespacing}{\renewcommand{\baselinestretch}{1.0} \normalsize}

\newcommand{\syncc}{~\stackrel{\textstyle \rhd\kern-0.57em\lhd}{\scriptstyle L}~}

\allowdisplaybreaks
\begin{document}

\title{\LARGE {\bf A second-order stochastic effective theory for the long-distance behaviour of scalar fields in de Sitter spacetime}\\
 \vspace*{6mm}
}

\author{Archie Cable}
\submitdate{July 2023}

\normallinespacing
\maketitle

\preface
\addcontentsline{toc}{chapter}{Abstract}

\begin{abstract}

This thesis introduces an effective theory for the long-distance behaviour of scalar fields in de Sitter spacetime, known as the second-order stochastic theory, with the aim of computing scalar correlation functions that are useful in inflationary cosmology. The need for such a theory stems from the challenge standard perturbative quantum field theory calculations face when considering self-interacting scalar fields with a mass $m$. Focussing on quartic self-interactions, parameterised by the coupling $\lambda$, one finds that the perturbative expansion about the free field solution returns correlation functions that are of order $\frac{\lambda H^4}{m^4}$. Thus, the method breaks down beyond the limit $\lambda\ll m^4/H^4$ because the perturbative sum does not converge.

The stochastic theory was introduced as a non-perturbative method for computing these scalar correlation functions. Motivated by the existence of a de Sitter horizon, field modes can be separated into short and long wavelength modes. The expanding spacetime stretches these modes such that the long wavelengths are considered classical. The short wavelength modes contribute to the long-distance behaviour when they cross the horizon, which amounts to a statistical white noise contribution. The result is stochastic equations describing the behaviour of the long-distance modes. One can then apply the formalism of stochastic processes to obtain non-perturbative expressions for stochastic correlation functions. In much of the literature, one introduces a hard cut-off between long and short wavelength modes about the horizon to derive the stochastic equations. The approximation one must make here is $m\ll H$ and $\lambda\ll m^2/H^2$ such that the resulting equations are overdamped.

The second-order stochastic theory is introduced in this thesis to extend the regime of validity of the stochastic approach. Instead of using the cut-off procedure, we write stochastic equations that resemble the field equations of motion, leaving the stochastic parameters - the stochastic mass, quartic coupling and white noise contributions - as free parameters. We employ stochastic techniques to compute correlation functions and then use results from perturbative quantum field theory to fix the parameters such that they return physical quantities. In doing so, we improve upon the overdamped stochastic approach by extending the regime of validity to $m\lesssim H$ and $\lambda^2\ll m^4/H^4$ and by including ultraviolet effects from renormalisation in our stochastic theory. Additionally, we perform non-perturbative computations such that the second-order stochastic theory goes beyond the regime of validity of perturbative quantum field theory.

\end{abstract}
\cleardoublepage

\begin{declaration}
    \textbf{Statement of originality}
    
    I, Archie Cable, the author of this thesis, declare that the work produced here is original. It appears first either in this thesis or in previous publications [K1], [K2], [K3], which I have written. When I do use the work of others, I have referenced appropriately.

    Note that Figure \ref{fig:noise_plot} is taken from my first publication, Ref. [K1], Figures \ref{fig:fit_evals_v_trunc}, \ref{fig:eval_v_m2_OD_v_match_diff} and \ref{fig:eval_v_m2_QFT_v_OD_diff} are taken from Ref. [K2] while Figures \ref{fig:model_limitations_comparison}, \ref{fig:m20.1_match_v_OD_v_QFT}, \ref{fig:m20.1_match_renorm} and \ref{fig:m20.3_match_v_OD_v_QFT} are taken from Ref. [K3]. All other figures and tables are original to this thesis or referenced appropriately.

    \textbf{Copyright declaration}
    
    The copyright of this thesis rests with the author. Unless otherwise indicated, its contents are licensed under a Creative Commons Attribution-Non Commercial 4.0 International Licence (CC BY-NC).
    
    Under this licence, you may copy and redistribute the material in any medium or format. You may also create and distribute modified versions of the work. This is on the condition that: you credit the author and do not use it, or any derivative works, for a commercial purpose. 
    
    When reusing or sharing this work, ensure you make the licence terms clear to others by naming the licence and linking to the licence text. Where a work has been adapted, you should indicate that the work has been changed and describe those changes.
    
    Please seek permission from the copyright holder for uses of this work that are not included in this licence or permitted under UK Copyright Law.
\end{declaration}
\cleardoublepage

\addcontentsline{toc}{chapter}{Acknowledgements}

\begin{acknowledgements}

\begin{center}
\textit{It was the best of times, it was the worst of times.}
\end{center}
\begin{flushright}
\textit{-- Charles Dickens}
\end{flushright}

I would not count myself among the biggest of Charles Dickens' fans, but this quote aptly summarises my time as a PhD student. I have had some incredible experiences, travelled the globe and met some wonderful people, but there have certainly been some difficult challenges in a time punctuated by two years of COVID-19 lockdowns. There are so many people to thank for spurring me on through this PhD.

Firstly, I would like to thank my supervisor, Arttu Rajantie, for his support, feedback and wisdom that have helped guide me through my 4 years of PhD life. To my fellow year group, Aoibh, George and Victor, thanks for keeping me company along this perilous road, largely through the media of board games, climbing and beers! This sentiment extends to all my colleagues within the Theory Group, with whom I've shared countless lunches, seminars and Teams calls.

Beyond Imperial, I've been fortunate to meet some fantastic people who have shaped me as a physicist and a person. I'd like to thank my collaborators whom I've worked and continue to work with on projects beyond the scope of this thesis. To Ashley Wilkins, thanks for teaching me a thing or two about primordial black holes, and to Greg Kaplanek, I'm excited about our venture into `Opening Stochastic' and for guiding me on this path.  Across the pond in Maryland, thank you to Bei Lok Hu for his generous hospitality, pearls of wisdom and continued support from afar. Similar sentiments are extended to Anders Tranberg and Magdalena Eriksson, who were kind enough to host me in Stavanger. A special thank you to them for organising a week of sun in Norway! Closer to home, thank you to Gerasimos Rigopolous and David Wands, Andrew Gow and Joe Jackson for kindly hosting me at Newcastle and Portsmouth, and furthering my education.

I have also been very fortunate to be supported by an awesome array of family and friends; if I were to name them all, it would probably require an additional thesis. Special thanks must go to my parents, Lorna and Stuart, who have supported me from day one and have really, really tried to understand what my PhD work is about. At least we've moved on from ``he does black holes''... just. And I must thank my sister, Iona, who has been a constant support throughout my life. It would also be remiss of me to neglect my fellow COVID survivors, Ella and Ben, who managed to put up with me and, I would argue, thrive during the lockdowns. Merlin was the peak. 

Finally, for Fenella, I thank you for your love and support. You have been a constant throughout the rollercoaster ride of the last 4 years; truly, you are my light, even in the darkest of places. 
\end{acknowledgements}
\clearpage

\narrowlinespacing

\vspace*{4mm}

\textit{Wisdom has always been chasing you, but you've always been faster.}\\

\textit{\textit{-- Iroh}}

\normallinespacing

\body
\chapter{Introduction}

\section{A lay for the lay}

There are many non-experts - or lay persons - who have helped me produce this work, so I feel it would be remiss of me to launch immediately into technical detail without giving them anything to get their teeth into. Therefore, I begin with a summary of the motivations and achievements of this thesis, omitting any technical detail such that it can be read by a non-expert: \textit{a lay for the lay}, if you will. For the expert reader, this summary should be a good starting point, to get a flavour for my work before we dive into technical details in the bulk text.

When one looks at the night sky, pricked by millions upon millions of stars, it is natural to ponder philosophical questions. Where did we come from? Where are we going? Why are we here? It is a unique facet of the human psyche to ask such questions; more impressive still is the fact that these are not purely the musings of a curious mind but have become firmly embedded in the proof-driven world of science. \textit{Cosmology} - the study of the origin and evolution of the Universe - is focussed on answering these questions through observing patterns and correlations in the sky and relating them to mathematical models of our Universe. It is a remarkable subject where mathematicians and theoretical physicists sit in offices, pubs and restaurants, discussing and writing rows of equations to describe the mysteries of the Universe, while astronomers, astrophysicists and engineers develop groundbreaking technology to shine a light on the darkest secrets the cosmos has to offer. There is a give and take between the two groups, with mathematical theories guiding observation and observation constraining theories, all in the pursuit of answers to these fundamental questions about our existence.

I fall into the former of the two groups. I am a theoretical cosmologist, aiming to develop mathematical models to describe the Universe in its infancy. Holistically, I am searching for answers to the question ``where do we come from?'', though I will disappoint the reader now and state that the answer is not contained within this thesis (not even on pg. 42!). My study is of the \textit{inflationary epoch}, a time period immediately following the Big Bang - the phrase coined for the beginning of our Universe - before any structure was formed: before galaxies, stars, even before microscopic particles such as electrons. It is an example of a mathematical theory, initially developed to solve some problems with the cosmological model of our Universe's history\footnote{For a more in-depth discussion of the history of the early Universe, see Sec. \ref{sec:history_universe}, which is largely accessible to the lay person.}, that has now been backed by observational evidence. During cosmic inflation, the Universe underwent a period of accelerated expansion, where the distance between every point in space became larger and larger. Even though inflation lasted for an incredibly short time - a tiny fraction of a second - it has huge consequences for the present state of the Universe. In fact, if it weren't for inflation, there would be no galaxies, stars, planets nor, ultimately, us!

My work, although strongly motivated by inflation, is more focussed on the theoretical framework of the epoch, as opposed to doing ``actual cosmology''. The basic model of a Universe undergoing accelerated expansion is \textit{scalar quantum field theory in de Sitter spacetime}. Let me break this down. de Sitter spacetime\footnote{Don't worry too much about this seemingly sci-fi terminology. Spacetime is just the combination of 3-dimensional space with time!} is a theory of gravity that describes what the accelerated expansion of space actually looks like. For example, if I tell you that, at a given moment in time during inflation, two spatial points are separated by a given distance, you could use the mathematics of de Sitter spacetime to tell me the separation between the same two spatial points at some moment in the future, and you will find that it is significantly larger. Scalar field theory then describes the mechanism that drives such an accelerated expansion and, crucially, how it ends. In a gross oversimplification, the scalar fields (and there can be multiple!) tell us what particles exist during inflation. The one driving the expansion is known as \textit{the inflaton}, but there can be others such as \textit{spectators}, that don't actually do much during inflation but can have observable consequences in the present state of the Universe. Finally, quantum theory describes the microscopic behaviour of particles; it is probabilistic in nature, as opposed to the determinism of the classical world that we see around us. Indeed, quantum theory underpins everything in our Universe but, on large scales, much of the uncertainty associated with a probabilistic theory is erased\footnote{Think, for example, about 6-sided dice. If I roll 1 die, the result could be any number between 1 and 6. However, if I roll hundreds of thousands of dice all at once, one would find that roughly 1 in 6 would be 1, 1 in 6 would be 2, and so on. The probabilities have been averaged out. This is exactly what happens in Nature, but on an enormous scale.}. Inflation is a very unusual situation where it is essential to consider both quantum and gravitational theories in tandem. Scalar quantum field theory in de Sitter spacetime is not perfect at doing this, but it's the best that we have got at present.

The quantities that we are interested in studying are \textit{correlation functions}, which encode the correlation between separate points in space and time. It is precisely these patterns that observational cosmologists are looking for. A remarkable feature of inflation is that long-distance correlations are ``frozen'' in the early Universe, meaning that, once inflation ends, they remain unaffected by the subsequent epochs. Thus, these inflationary correlators can and have been observed in the sky, offering blueprints for the early Universe. From my perspective, I am interested in mathematically computing the long-distance behaviour of inflationary correlation functions using scalar quantum field theory in de Sitter spacetime. However, performing calculations using this theory in general is exceedingly difficult and so one must turn to other methods. The popular way to go about this is to introduce an \textit{effective theory}, which describes some sub-region of the full theory. For us, we are interested in the long-distance behaviour, so we will use an effective theory that is only valid at long distances. While we lose some generality by doing this, we often gain much more computational power as a result. 

The example that is at the heart of this thesis is the \textit{stochastic effective theory of the long-distance behaviour of scalar fields in de Sitter spacetime}. The idea is that, because we are in an expanding Universe, microscopic distances will be stretched to become macroscopic very quickly. Thus, long-distance correlation functions can actually be treated as classical objects, since we are now on large scales. Of course, not all distances will be stretched sufficiently far to make this approximation, but one can show that they can be treated as a statistical contribution to an otherwise classical, long-distance theory. Thus, the stochastic effective theory is a semi-classical theory, as opposed to a fully quantum one, which makes calculations significantly easier. Additionally, stochastic processes are prevalent in other areas of physics and mathematics, meaning that there is plenty of literature on the subject, which helps a huge amount!

And so we come to it at last: the achievements of this thesis. While the stochastic theory has been around since the early 90s, it has been focussed on very light (nearly-massless) scalar fields. My goal was to extend the theory beyond this limit. As it turns out, this was quite difficult to do! Instead of deriving the stochastic equations from the underlying quantum theory, as was originally done, we had to employ a \textit{matching procedure} to obtain meaningful results. The idea is as follows. We make an educated guess as to the form of the stochastic equations and use them to calculate stochastic correlation functions. We then compare the results with the equivalent quantities obtained using quantum theory, within the regime that such quantities are possible to calculate, and tweak our stochastic equations such that they match. Then, we use the robustness of the stochastic equations to compute these quantities more generally, thus going beyond the regime where quantum results are tractable. Further, our stochastic equations are not limited to nearly-massless fields and thus our model advances upon the original stochastic theory. In fact, as my work has unfolded, it has become increasingly clear that our new stochastic theory should be treated independently from the original. Thus, the crux of my PhD is that I have developed a new method with which one can compute correlation functions in the early Universe, catchily called the \textit{second-order stochastic effective theory of the long-distance behaviour of scalar fields in de Sitter}.

\section{Thesis Overview}

This thesis is broken down into five chapters, sandwiched between an introduction and conclusion. The idea is to weave a narrative, beginning with the motivation of cosmic inflation, before building a landscape of some of the current, well-established approximations for scalar fields in de Sitter, and finally introducing the latest of these: the second-order stochastic effective theory.

Chapter \ref{ch:inflation} introduces cosmic inflation, highlighting results that are important for this thesis. I will outline why we are interested in studying scalar fields in de Sitter for cosmology, and why the long-distance behaviour of quantum fluctuations in the early Universe is relevant for observations. This chapter can be seen as the motivation section of this thesis and will be the only place cosmology is discussed in technical detail. Following this, I will consider the oversimplified (but certainly not simple!) scenario of scalar quantum field theory (QFT) in de Sitter spacetime. While my motivation stems from inflationary cosmology, this work could be viewed as a study of QFT in de Sitter from a purely mathematical sense.

The next three chapters introduce three different approximations that can be used to compute the long-distance behaviour of scalar correlation functions in de Sitter spacetime: perturbative QFT, overdamped stochastic and second-order stochastic theory. I will develop them based on the model of a scalar field with mass $m$ and a quartic self-coupling $\lambda$, which exists on a de Sitter background with Hubble parameter $H$\footnote{For this thesis, we will ignore the effect the scalar field has on the dynamics of the spacetime. Thus, from a cosmological perspective, we are predominantly considering spectators. Further, we will assume $H$ is constant.}. For each approximation, I will compute the 2-pt and 4-pt functions, and subsequently use these objects to compare all three. This will allow us to consider the regions of the $(m,\lambda)$ parameter space in which each approximation is valid, thus putting constraints on our model due to theoretical (as opposed to observational) limitations.

Chapter \ref{ch:qft_dS} is dedicated to the perturbative expansion of scalar QFT in de Sitter. I will begin with a general overview of the de Sitter geometry and the process by which we quantise scalar fields in this spacetime via second quantisation. For free fields ($\lambda=0$), solutions to QFT in de Sitter are well-known, and no approximation is needed. I will detail the various types of 2-pt functions that emerge from such a theory, highlighting the Feynman propagator as the most important. Moving to the far more interesting case of interacting fields is challenging, which is why we require approximations. The standard first method to consider is to expand about the free field solution for small $\lambda$. General procedures are available for this, which stem from similar problems in Minkowski spacetime, and they can be used to good effect in de Sitter. I will outline how one can use this to compute the 2-pt and 4-pt functions to leading order in $\lambda$. However, when one examines the expansion carefully, one observes that the perturbative sum doesn't converge unless $\frac{\lambda H^4}{m^4}\ll1$. Thus, perturbation theory is limited to the regime in the parameter space where $\lambda\ll1$ and $\lambda\ll m^4/H^4$.

The limitations of perturbative QFT has led physicists to consider alternative methods. In this way, the stochastic approach was introduced as an effective theory of the long-distance behaviour of scalar fields in de Sitter. In Chapter \ref{ch:od_stochastic}, I introduce the stochastic approach, as it was originally presented in the 90s, namely in its overdamped form. Using this approximation allows one to derive a stochastic theory from the underlying QFT by separating field modes into long and short wavelength components. The theory describes the dynamics of the long wavelength modes, with the short wavelength modes offering a purely statistical contribution. Utilising this, I compute the 2-pt and 4-pt functions perturbatively, so that we can directly compare its results with that of perturbative QFT. Crucially, I will also outline a numerical method by which one can solve the stochastic equations non-perturbatively, overcoming the shortfalls of perturbative QFT. However, this approach is far from perfect; the overdamped approximation limits the model to the regime $m\ll H$ and $\lambda\ll m^2/H^2$.

The overdamped approximation is not an essential element of the stochastic approach, but rather a useful simplification. One can straightforwardly extend the overdamped stochastic equations to their full, second-order guise, but it is no longer possible to obtain the statistical contribution from the short wavelength modes by the same method. Thus, we seek a method by which we can compute this contribution. Chapter \ref{ch:second-order_stochastic} presents the solution - a second-order stochastic effective theory - which is the key achievement of this thesis and has been developed in the papers [K1], [K2] and [K3], listed below. I introduce generic second-order stochastic equations that resemble equations of motion for scalar fields in de Sitter and use them to obtain expressions for a stochastic 2-pt and 4-pt function. I present results both perturbatively, for comparison with perturbative QFT, and non-perturbatively, for comparison with the overdamped stochastic approach. However, at this stage the stochastic theory doesn't represent anything physical. To promote it to an effective theory of scalars fields in de Sitter, we compare the perturbative stochastic results with their equivalents from perturbative QFT at the level of the 2-pt and 4-pt functions, and select the form of the stochastic parameters such that the quantities match. In doing so, we obtain the effective theory we desire and, since we have non-perturbative results, the second-order stochastic theory goes beyond perturbative QFT. Note that, because we have to match the stochastic parameters via perturbative QFT, we still have some issues of convergence but these are less severe than its predecessor, limiting the stochastic theory to $\lambda^2\ll m^4/H^4$. This is a milder limitation on $\lambda$ compared to the overdamped stochastic approach because we include $\mathcal{O}(\lambda)$ corrections to our stochastic parameters, and we can now relax the condition on the mass such that $m\lesssim H$ \footnote{This limitation arises due to the underlying principles of the stochastic approach.}. Thus, we have extended the regime of the parameter space in which we can compute correlation functions beyond the two established models.

The final chapter of the bulk thesis, Chapter \ref{ch:comparison_models}, is a sense check. We perform a detailed analysis of the non-perturbative solutions to both stochastic approximations, comparing them to the perturbative QFT results. The goal here is to show that the second-order stochastic effective theory agrees with the other two approximations where it should - in the regime of the parameter space where they are valid - and disagrees where it should - in the regime where they break down. We find that this is indeed the case, which is an important first step in proving the validity of the second-order theory. Of course, as is the nature of such an effective theory, there will always be some doubt attached and there is always more that one could do to be more rigorous in the proof of its validity. For example, one could compare it with other approximations to QFT. However, we are content for the time being that our second-order stochastic theory has passed all the tests thrown at it thus far. Future work will decide whether it continues to do so.

In this section, I have purposefully been vague about pointing the reader to specific sections/equations that highlight the key achievements of this work. I wished here to be more holistic, giving the reader a picture of the structure and direction of this thesis. I have a more comprehensive summary in the final section of Chapter \ref{ch:conclusions}, which I refer the reader to if they are interested at this stage.

\section{Conventions}

In this section, I list some conventions and terminology that I use throughout the thesis.\\

{\large \textbf{Metric convention}}\\
I use the `mostly minus' metric convention of particle physics: $[+---]$.\\

{\large \textbf{Vectors}}\\
Scalars and vectors in 4-dimensions are denoted by unadorned letters e.g. $f$. Where index notation is necessary, I will use the Greek letters $\mu$, $\nu$,... 
Vectors in 3-dimensions are denoted by bold-type letters e.g. $\mathbf{f}$, and I will use the Latin letters $i$, $j$,... for their indices. For the magnitude of 3-vectors, I return to use unadorned letters e.g. $\abs{\mathbf{f}}=f$.\\

{\large \textbf{Derivatives}}\\
Derivatives with respect to cosmological time are denoted by a dot e.g. $\pdv{f}{t}=\Dot{f}$, while derivatives with respect to 3-dimensional space are denoted using nabla e.g. $\pdv{f}{\mathbf{x}}=\nabla f$. Apostrophes are used to denote derivatives with respect to the argument of a function e.g. $\dv{f(\phi)}{\phi}=f'(\phi)$.\\

{\large \textbf{Momentum space}}\\
Throughout this thesis, I refer to momentum space as $\mathbf{k}$-space. I use a `bar' to denote factors of $(2\pi)^3$ in the following instances:
$$\dbar^3\mathbf{k}=\frac{d^3\mathbf{k}}{(2\pi)^3},$$
$$\deltabar^{(3)}(\mathbf{k}-\mathbf{k}')=(2\pi)^3\delta^{(3)}(\mathbf{k}-\mathbf{k}').$$
Additionally, we denote functions in $\mathbf{k}$-space with a `tilde' where the equivalent function in coordinate space would not have a `tilde' e.g. $\Tilde{f}(\mathbf{k})$ corresponds to the $\mathbf{k}$-space equivalent of $f(x)$.

{\large \textbf{Mass terminology}}\\
Throughout this thesis, there are instances where the size of the mass $m$ of a field is important. I use the term \textit{light} to indicate fields $m\lesssim H$, while the term \textit{nearly-massless} refers to fields $m\ll H$, where $H$ is the Hubble parameter.

\section{Publications}

This thesis is accompanied by the following publications:

[K1] A. Cable and A. Rajantie. Free scalar correlators in de Sitter space via the stochastic approach beyond the slow-roll approximation. \textit{Phys. Rev. D}, 104:103511, 2021

[K2] A. Cable and A. Rajantie. Second-order stochastic theory for self-interacting scalar fields in de Sitter spacetime. \textit{Phys. Rev. D}, 106:123522, 2022

[K3] A. Cable and A. Rajantie. Perturbative corrections to stochastic parameters for self-interacting scalar fields in de Sitter spacetime. \textit{arXiv:}2310.07356, 2023

\chapter{Cosmic Inflation}
\label{ch:inflation}

\section{Introduction to Chapter \ref{ch:inflation}}

Cosmic inflation is the very early epoch when the Universe underwent a period of accelerated expansion. Built from theoretical foundations, it has gained traction over the years due to its simplistic ability to solve outstanding problems in modern cosmology, and has more recently become a mainstay of the cosmological model of the Universe, thanks to observations of anisotropies in the cosmic microwave background radiation. While much can be said about inflation, I will largely use it in this thesis to motivate the study of quantum fields in de Sitter. As we will see, the spacetime of the inflationary epoch is very nearly de Sitter and thus one often uses such a geometry as the underlying framework for inflationary models. 

In this chapter, I will offer a brief overview of inflation: its history in physics, how it works and why it is important in the context of this thesis. This is largely composed of textbook material; I used Ref. \cite{baumann_txtbk:2022} for the most part. I will start with a timeline of the early Universe in Sec. \ref{sec:history_universe}, to give the reader an overall picture of where inflation fits and why its important. I will then introduce some important features of an expanding Universe in Sec. \ref{sec:expanding_universe_geometry} before considering inflation proper in Sec. \ref{sec:physics_of_inflation}. Note that I will focus on single-field inflation for this chapter, though other models exist. I will end with a discussion of inflationary observables, which are the core quantities one wishes to study in inflation. They are also closely related to correlation functions in QFT, foreshadowing the bulk of the thesis.  

\section{A brief history of the early Universe}
\label{sec:history_universe}

Our Universe is 13.8 billion years old with a temperature of 2.7 K. However, as a cosmologist, most of the interesting epochs occurred within the first minute, when the Universe was extremely hot and dense. Throughout its lifetime, it has continuously expanded and cooled, allowing microscopic particles to coalesce into large scale structure before the arrival of life, including curious cosmologists!

\begin{table}[ht]
    \centering
    \begin{tabular}{c|c|c}
        Epoch & Time after the Big Bang & Temperature ($K$) \\ \hline &&\\
        Big Bang & 0 s & $\infty$ K?\\
        &&\\
        Inflation & $<10^{-34}$ s & $\lesssim10^{28}$ K\\
        &&\\
        Preheating/reheating &&\\
        \& EW phase transition & $10^{-11}$ s & $10^{15}$ K\\
        &&\\
        QCD phase transition & $10^{-5}$ s & $10^{12}$ K\\
        &&\\
        Neutrino decoupling & $1$ s & $10^{10}$ K\\
        &&\\
        Big Bang nucleosynthesis & $200$ s & $10^9$ K\\
        &&\\
        Recombination & $260000-380000$ yrs & $3400-2900$ K\\
        &&\\
        Large structure formation & $>10^8$ yrs & $<50$ K\\
        &&\\
    \end{tabular}
    \caption{A brief history of the Universe}
    \label{tab:history_universe}
\end{table}

In the beginning, there was the Big Bang. Little is known of the first moments, when the Universe as we know it came into existence. There are many hypotheses but it is largely uncharted territory as it requires the marriage of quantum mechanics and gravity in a Theory of Everything. Attempts to unify the two have thus far been unsuccessful and it remains one of the biggest unsolved mysteries in physics. The most famous example of an attempted Theory of Everything is string theory, but a lack of observational and experimental evidence has seen it lose some of its shine in recent years.

The earliest epoch we can make concrete statements about is also the focus of this chapter: cosmic inflation. During this period, the Universe underwent a period of accelerated expansion that far exceeds any other epoch. It was first introduced as a theory designed to solve some problems in cosmology, but has become a staple of the cosmological model thanks to evidence in cosmic microwave background (CMB) measurements. The end of inflation occurred when the energy driving the accelerated expansion was transferred to the Standard Model (SM) particles, causing the expansion to slow to a more reasonable rate. This allowed the SM particles to equilibrate, leaving a hot dense plasma of quarks, leptons, force-mediating particles and the Higgs boson. The process by which this energy transfer occurred and thermal equilibrium reached is called preheating/reheating\footnote{Historically, this was just reheating. More recent developments in the field have seen a distinction between the different processes, introducing the term preheating into the mix.}. This was immediately followed by the electroweak (EW) phase transition, where the EW and strong forces decoupled to appear as two distinct interaction types. During this time, spontaneous symmetry breaking also occurred such that the particles interacting with the Higgs boson gained mass. It is also during this time that we suspect other particle types decoupled from the SM, which would make up the mysterious dark matter candidates.  

Thus far, the Universe had existed for less than $10^{-5}$ s. At this stage, it was composed of the very basic building blocks of particle physics - the Standard Model - with particles happily in thermal equilibrium, interacting with one another but remaining independent and avoiding long term relationships. However, as the Universe continued to cool to about $10^{12}$ K, the quarks began to form bonds with each other to make hadrons. This is known as the quantum chromodynamics (QCD) phase transition, named for the theoretical structure of quark interactions. A second after the Big Bang, neutrinos fell out of thermal equilibrium, decoupling from the other particles such that they freely streamed through the otherwise-interacting plasma. These free-streaming neutrinos form the primordial relic known as the cosmic neutrino background (CNB), which in theory can be observed. In practice, neutrino observation is incredibly difficult and we are still some years off detecting the CNB, though attempts are being made - see e.g. \cite{betts2013development}.

The first second of the Universe is largely the subject of theoretical cosmology, with good observational and experimental evidence to back up the claims. The following epochs are the focus of observational physics and contain (rather ironically) some of the earliest discoveries of modern cosmology. A few minutes after the Big Bang was Big Bang nucleosynthesis, where light nuclei such as hydrogen and helium began to form. This process lasted for over 100,000 years before the arrival of the next big event: recombination. During this time, electrons began to orbit nuclei, forming the first atoms. Further, like the neutrinos before them, photons decoupled from matter and began freely streaming through the Universe to create the famous CMB. These photons cooled with the Universe, arriving at 2.7 K in the present day. Over this time period - the vast majority of the Universe's lifetime - large scale structure formed, beginning with the earliest stars that clustered to form galaxies, picking up pieces of debris to give us planets like Earth, where life came into being. This topic goes beyond the remit of cosmology, and certainly this thesis!

\begin{figure}
    \begin{subfigure}[ht]{0.455\textwidth}
        \includegraphics[width=\textwidth]{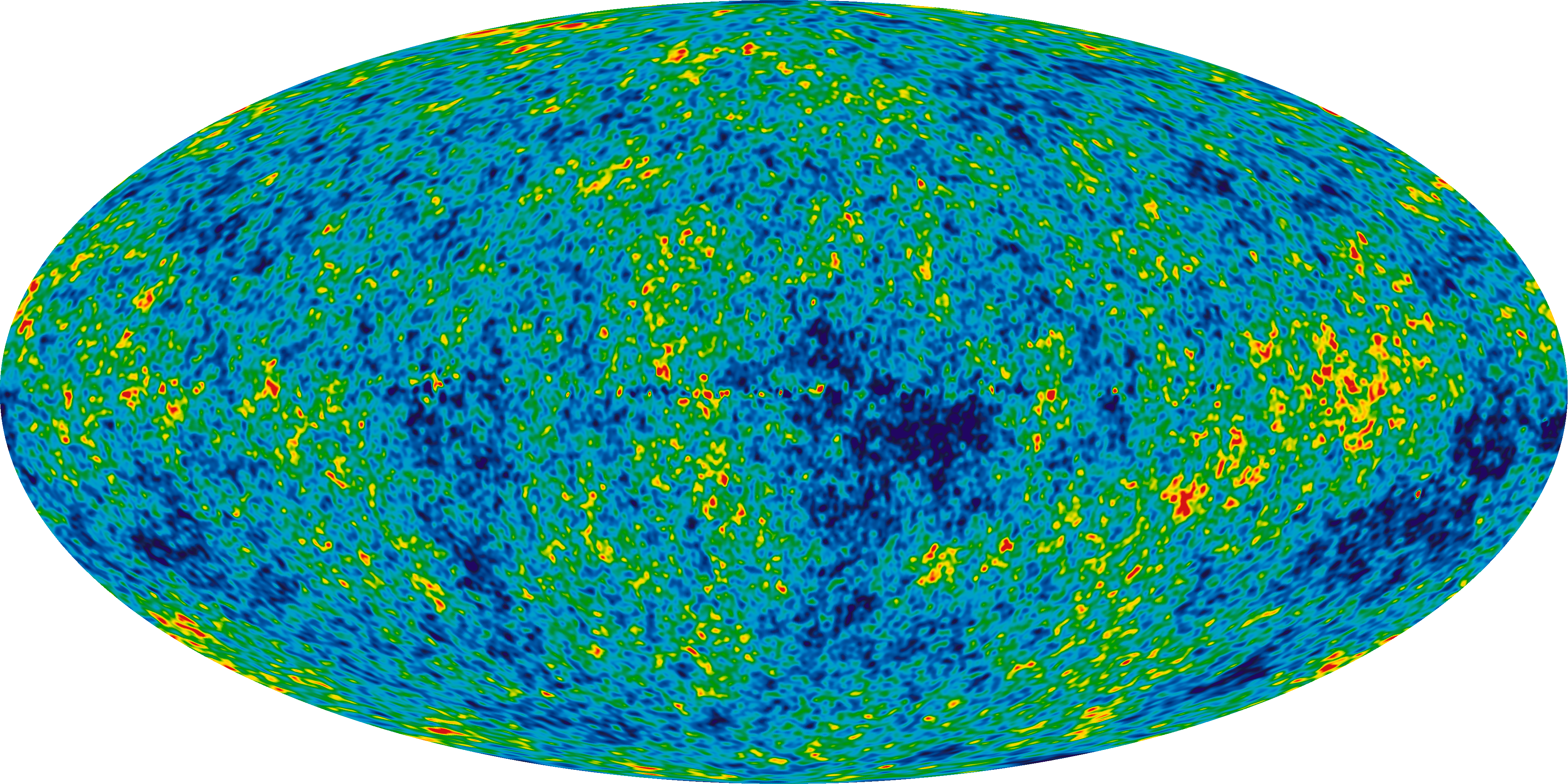}
        \caption{A map of the observable Universe from the perspective of the CMB. Red/blue indicates areas of slightly hotter/colder regions.}
        \label{fig:cmb_map}
    \end{subfigure}
    \begin{subfigure}[ht]{0.485\textwidth}
        \includegraphics[width=\textwidth]{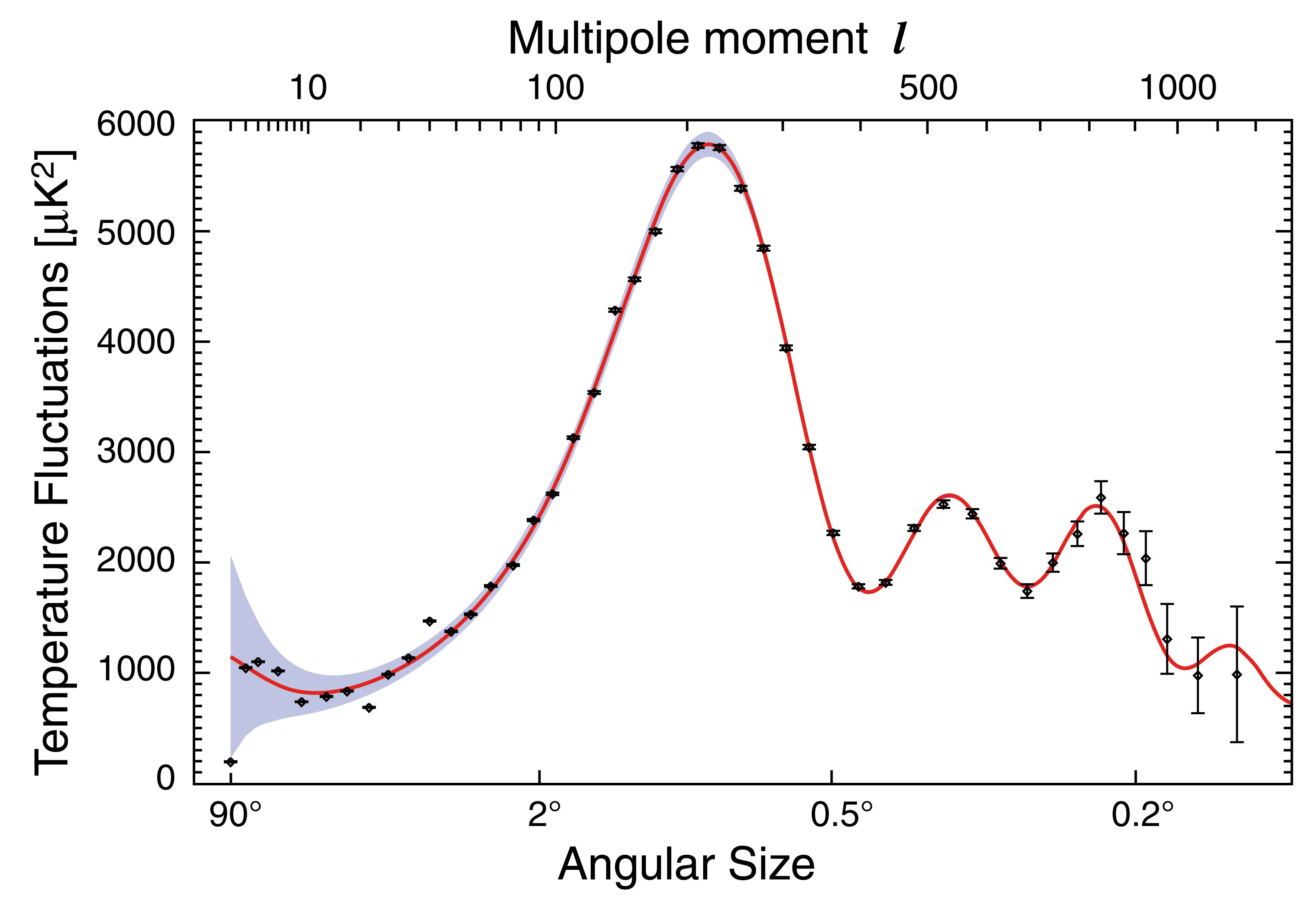}
        \caption{A plot of the temperature fluctuations of the CMB across the angular width of the observable Universe. The black points are observational data while the red line is the fit produced by theoretical models.}
        \label{fig:cmb_density}
    \end{subfigure}
    \caption{Observations of the CMB, as taken from WMAP \cite{Larson:2011,Bennett:2013}.}
    \label{fig:cmb}
\end{figure}

The crowning glory of modern cosmology is the map of the CMB (Fig. \ref{fig:cmb_map}) \cite{Bennett:2013}. This depicts the observable Universe from the perspective of the CMB as observed by the Wilkinson Microwave Anistropy Probe (WMAP), showing the temperature fluctuations to extreme precision. This is made more concrete in Fig. \ref{fig:cmb_density}, where we see a plot of the temperature fluctuations across the observable Universe \cite{Larson:2011}. These fluctuations are one part in 100,000, suggesting that the Universe is extremely close to thermal equilibrium. They give us an insight into the very early Universe and provide evidence for inflation. As will be discussed in more detail, quantum fluctuations during inflation become seeds for macroscopic growth in the late Universe. These primordial density perturbations become a feature of future epochs. In particular, they give rise to regions of higher and lower density in the baryon-photon plasma, which causes the propagation of sound waves. When recombination occurs, this behaviour is baked into the free-streaming photons. When one observes the CMB and its anistropies today, as in Fig. \ref{fig:cmb_density}, one observes an oscillatory behaviour in the temperature fluctuations that are directly related to the primordial sound waves. The triumph of inflationary models over other early Universe theories, such as cosmic strings, is that they produce coherent density fluctuations, which give rise to the observed oscillatory structure in the CMB temperature fluctuations.

\section{Geometry of the expanding Universe}
\label{sec:expanding_universe_geometry}

Modern cosmology is founded on the principles of universal homogeneity and isotropy. In other words, on large scales, the Universe is statistically the same at each point in space (homogeneous) and in all directions (isotropic). From this, we can write the most general cosmological metric: the \textit{Friedmann-Lemaitre-Robertson-Walker (FLRW) metric}
\begin{equation}
    \label{flrw_metric}
    ds^2=dt^2-a(t)^2\lb\frac{1}{1-\kappa \frac{r^2}{R_0^2}}dr^2+r^2d\Omega_2^2\rb,
\end{equation}
where $d\Omega_2^2=d\theta^2+\sin^2\theta d\phi^2$ is the metric of a unit 2-sphere, $a(t)$ is a time-dependent function known as the \textit{scale factor} and $\kappa=0,-1,+1$ determines whether the spatial part is flat, open or closed respectively, with $R_0$ representing the curvature scale. Current observations indicate that the Universe is very nearly flat so we will take $\kappa=0$ henceforth. We define the \textit{Hubble parameter} as
\begin{equation}
    \label{hubble_parameter}
    H=\frac{\Dot{a}(t)}{a(t)}.
\end{equation}
It is also convenient to introduce conformal time $d\eta=\frac{dt}{a(t)}$ such that the metric (\ref{flrw_metric}) can be written as
\begin{equation}
    \label{conformal_flrw_metric}
    ds^2=a(\eta)^2\lb d\eta^2-dr^2-r^2d\Omega_2^2\rb.
\end{equation}

\subsection{Cosmological horizons}
\label{subsec:cosmological_horizons}

An important concept in cosmology are horizons. The Universe began at a Big Bang singularity, which is a surface of constant time but not a single point in space. Take this time to be $t=0$ or in conformal time $\eta=\eta_i$. Consider an observer at a fixed spatial point at a time $\eta=\eta_0$, after the Big Bang. Imagine that they make a perfect observation whereby they receive all possible information about the history of the Universe. Unfortunately, they still won't know everything about the Big Bang. 

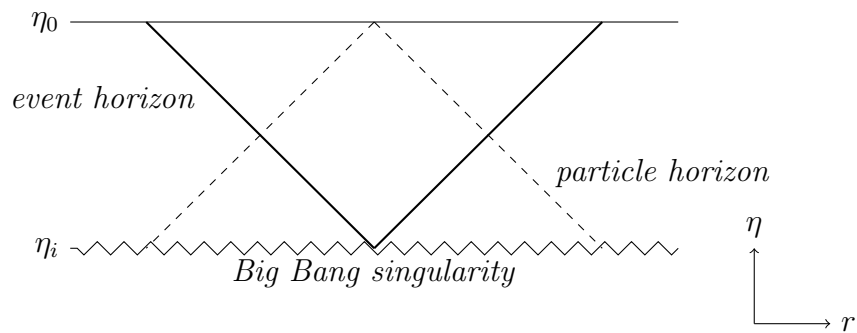
\begin{figure}[ht]
    \centering
    \begin{tikzpicture}
        \draw[decorate,decoration=zigzag] (8,0) -- (0,0) node[left]{$\eta_i$} node[midway,below]{\textit{Big Bang singularity}};
        \draw[black] (8,3) -- (0,3) node[left]{$\eta_0$};
        \draw[black,thick] (4,0) -- (7,3);
        \draw[black,thick] (4,0) -- (1,3);
        \draw[black,dashed] (4,3) -- (1,0);
        \draw[black,dashed] (4,3) --(7,0);
        \draw[black,->] (9,-1) -- (9,0) node[above]{$\eta$};
        \draw[black,->] (9,-1) -- (10,-1) node[right]{$r$};
        \draw[white] (1,2) -- (1.8,2) node[black,left]{\textit{event horizon}};
        \draw[white] (7.3,1) -- (8.3,1) node[black,midway]{\textit{particle horizon}};
    \end{tikzpicture}
    \caption{A spacetime diagram of cosmological horizons. The region enclosed by the particle horizon (solid, angled lines) indicates the spacetime region that can receive information from a fixed spatial point at the time of the Big Bang. The region enclosed by the event horizon (dotted, angled lines) indicates the spacetime region from which one can send information to a fixed spatial point at $\eta=\eta_0$.}
    \label{fig:cosmological_horizons}
\end{figure}

The reason is simple; information cannot be transferred faster than the speed of light. As we are considering a finite time, there will be some regions of the Big Bang singularity from which no information has been received; not enough time has passed. Such regions are \textit{causally disconnected}. The maximum distance from which an observer can receive information is called the \textit{particle horizon}. Similarly, one can define the \textit{event horizon} as the maximum distance an observer can send information. These are shown in Fig. \ref{fig:cosmological_horizons}.

The (comoving) particle horizon is given by
\begin{equation}
    \label{particle_horizon}
    R_H=\int_{t_i}^{t_0}dt\frac{1}{a(t)}
\end{equation}
For standard particle composition (see next section for more details), the scale factor is a monotonically increasing function of time. As such, $R_H\sim(aH)^{-1}$ for non-inflationary spacetimes and thus the particle horizon is often referred to as the Hubble radius.

\subsection{The Einstein equations}
\label{subsec:einstein_eq}

The Einstein equations are 
\begin{equation}
    \label{einstein_eq_cosmology}
    R_{\mu\nu}-Rg_{\mu\nu}=8\pi G T_{\mu\nu},
\end{equation}
where $R_{\mu\nu}$ and $R$ are the Ricci tensor and scalar, $g_{\mu\nu}$ is the metric, $G$ is Newton's constant and $T_{\mu\nu}$ is the stress-energy tensor. The left hand side of Eq. (\ref{einstein_eq_cosmology}) is purely a statement of the spacetime geometry - in our case, the FLRW spacetime - while the right hand side tells us about the matter existing in the spacetime and how it affects the geometry. For a homogeneous and isotropic spacetime, the stress-energy tensor is given by ${T^{\mu}}_{\nu}=\text{diag}\lsb\rho,-P,-P,-P\rsb$, where $\rho$ and $P$ are the energy density and pressure of our matter respectively.

Using the Einstein equations (\ref{einstein_eq_cosmology}) with the FLRW metric (\ref{flrw_metric}), we can compute the famous \textit{Friedmann} and \textit{continuity equations} as
\begin{subequations}
    \label{friedmann&continuity_eq}
    \begin{align}
        \label{friedmann1}
        H^2&=\lb\frac{\Dot{a}}{a}\rb^2=\frac{8\pi G}{3}\rho,\\
        \label{friedmann2}
        \frac{\Ddot{a}}{a}&=-\frac{4\pi G}{3}(\rho+3P),\\
        \label{continuity}
        0&=\Dot{\rho}+3H(\rho+3P).
    \end{align}
\end{subequations}
From these equations, we can understand how different types of particles behave in an expanding Universe. A single perfect fluid has the relation $P=w\rho$ such that
$\rho\propto a^{-3(1+w)}$, where $w$ is a constant that is dependent on the type of particles that the fluid is composed of. Some important examples are:
\begin{itemize}
    \item Matter: $w=0\quad\implies\quad P=0\quad\implies\rho\propto a^{-3}$
    \item Radiation: $w=1/3\quad\implies\quad P=\rho/3\quad\implies\rho\propto a^{-4}$
    \item Dark energy (cosmological constant): $w=-1\quad\implies\quad P=-\rho\quad\implies\rho=\text{constant}$
\end{itemize}
In reality, our Universe is made up of a combination of these; however, different epochs were dominated by different particles. After the EW phase transition, the Universe was dominated by radiation. As the Universe expanded and the scale factor increased, the energy density of radiation fell more sharply than that of matter. Eventually, when the photons decoupled from matter, we entered a matter-dominated Universe. However, the energy density of matter continued to fall whereas the dark energy density remains constant. Thus, we are now entering a phase of the Universe that is dark energy dominated. In this epoch, the Hubble parameter is roughly constant.

The present time is not the only epoch where the cosmological constant has dominated the energy budget; the period of inflation in the first moments following the Big Bang were also dominated by dark energy. It is the physics of this epoch that will be our focus for the remainder of this chapter.

\section{The physics of inflation}
\label{sec:physics_of_inflation}

\subsection{Motivating inflation}
\label{subsec:motivating_inflation}

Unlike many of the later epochs of the early Universe, cosmological inflation was a theoretical model long before there was observational evidence. It was initially introduced in the 1980s \cite{Guth:1981,Guth:1982,Starobinsky:1980,Starobinsky:1982,Hawking:1982} as a way to solve several outstanding problems with the cosmological model at the time. These are:
\begin{itemize}
    \item \underline{The Flatness Problem}\\
            \textit{Problem.} Observations suggest that the Universe is very close to being flat, which indicates that the early conditions must be very fine-tuned.\\
            \textit{Solution.} The accelerated expansion stretches the Universe so that a Hubble patch appears nearly flat.
    \item \underline{The Horizon Problem}\\
            \textit{Problem.} Observations suggest that causally disconnected patches of the Universe are in thermal equilibrium. How can this be true if they have never transferred information?\\
            \textit{Solution.} The early Universe was actually causally connected and therefore in thermal equilibrium, but the accelerated expansion separated these regions so they now appear to have forever been causally disconnected.
    \item \underline{The Monopole Problem}\\
            \textit{Problem.} Cosmological models suggest that large amounts of stable magnetic monopoles were produced in the early Universe, yet they've never been observed by particle accelerators. Where did they go?\\
            \textit{Solution.} The accelerated expansion caused the density of monopoles to drop exponentially, meaning they are no longer abundant.
\end{itemize} 
Since its inception, inflation has become a staple of the cosmological model, especially since the discovery that the anistropies in the CMB (Fig. \ref{fig:cmb_density}) can be explained by inflationary perturbations. Indeed, one can show that large scale structure (galaxies, stars etc.) that move the Universe away from homogeneity and isotropy can be explain by quantum fluctuations that occurred during inflation (see Sec. \ref{subsec:inflationary_observables} for elaboration).

\subsection{Inflationary models}

\subsubsection{Conditions for inflation}

The basic condition for inflation is that the \textit{expansion is accelerating}
\begin{equation}
    \label{accelerated_expansion}
    \Ddot{a}>0.
\end{equation}
Equivalently, our Hubble radius is decreasing
\begin{equation}
    \label{decreasing_Hubble_radius}
    \dv{t}(aH)^{-1}<0.
\end{equation}
By expanding the above, we find that $\dv{t}(aH)^{-1}=-\frac{1}{a}(1-\epsilon_1)$, where the \textit{first slow-roll parameter}
\begin{equation}
    \label{first_slow-roll_parameter}
    \epsilon_1=-\frac{\Dot{H}}{H^2}<1,
\end{equation}
so that Eq. (\ref{decreasing_Hubble_radius}) is satisfied\footnote{Note that, for inflationary perturbations, we actually require $\epsilon_1\ll1$.}. Thus, we have a further condition for inflation: \textit{the Hubble parameter is slowly-varying}. In the limit $\epsilon_1\rightarrow0$, the Hubble parameter becomes constant and the FLRW spacetime resembles that of the expanding patch of de Sitter. Thus, \textit{the geometry of inflation is quasi-de Sitter}. In Chapters \ref{ch:qft_dS}-\ref{ch:comparison_models}, we will consider an exact de Sitter spacetime as a first approximation of the inflationary epoch.

Additionally, we  require that inflation lasts a sufficiently long time such that the flatness, horizon and monopole problems are satisfactorily solved. Current estimations for the length of inflation are $40-60$ e-folds, where an e-fold is defined as
\begin{equation}
    \label{e-fold}
    \mathcal{N}=\ln a.
\end{equation}
To achieve this, we require that the first slow-roll parameter stays small for a long time i.e. that it too is slowly varying. This is parameterised by the \textit{second slow-roll parameter}
\begin{equation}
    \label{second_slow-roll_parameter}
    \epsilon_2=\frac{\Dot{\epsilon}_1}{H\epsilon_1}.
\end{equation}
For inflation to last a sufficient length of time, $\abs{\epsilon_2}<1$.

\subsubsection{Single-field inflation}

We will now consider how to construct a mechanism by which the accelerated expansion is driven. There are several things we need to bear in mind while we do this. The first, of course, is that we require the above conditions to be met. The second is that we require inflation to end i.e. we can't have a mechanism that eternally drives an accelerated expansion. Further, the conditions under which inflation ends must complement the future, well-established epochs and thus must include some mechanism under which the Standard Model particles arise after inflation ends. With these factors in mind, we will consider the simplest and most widely-used model:\textit{ single-field inflation}. We introduce a single scalar field $\Phi$ - the inflaton - in a scalar potential $V(\Phi)$ as a means of driving the accelerated expansion. The inflaton action is given by
\begin{equation}
    \label{inflaton_action}
    S[\Phi,a]=\int d^4x a(t)^3\lsb\frac{1}{2}\Dot{\Phi}^2+\frac{1}{2}\frac{(\nabla\Phi)^2}{a(t)^2}-V(\Phi)\rsb.
\end{equation}
By extremising the action, we find that the classical equations of motion are given by
\begin{equation}
    \label{inflaton_eom}
    \Ddot{\Phi}+3H\Dot{\Phi}+V'(\Phi)=0.
\end{equation}
Using the definition of the stress-energy tensor for a scalar field
\begin{equation}
    \label{stress-energy_tensor_scalar}
    T_{\mu\nu}=\pdv{\mathcal{L}}{\lb\partial^\mu\Phi\rb}\partial_\nu\Phi-g_{\mu\nu}\mathcal{L},
\end{equation}
where the Lagrangian is defined as $S=\int d^4x\mathcal{L}$, the energy density and pressure for the inflaton are given by
\begin{subequations}
    \label{inflaton_energy&pressure}
    \begin{align}
        \label{inflaton_energy_density}
        \rho_\Phi&=\frac{1}{2}\Dot{\Phi}^2+V(\Phi),\\
        P_\Phi&=\frac{1}{2}\Dot{\Phi}^2-V(\Phi).
    \end{align}
\end{subequations}
Note that for this section, where we are considering the inflaton to be classical, we drop the spatial gradient terms due to homogeneity and isotropy. Then, the first Friedmann equation (\ref{friedmann1}) for the inflaton is given by
\begin{equation}
    \label{inflaton_friedmann_1}
    H^2=\frac{8\pi G}{3}\lb\frac{1}{2}\Dot{\Phi}^2+V(\Phi)\rb.
\end{equation}
By combining the equation of motion (\ref{inflaton_eom}) and Friedmann equation (\ref{inflaton_friedmann_1}) with the definitions for the slow-roll parameters (\ref{first_slow-roll_parameter}) and (\ref{second_slow-roll_parameter}), we find that the slow-roll parameters are given by
\begin{subequations}
    \label{inflaton_slow-roll_parameters}
    \begin{align}
    \label{inflaton_first_SR}
    \epsilon_1&=3\frac{\frac{1}{2}\Dot{\Phi}^2}{\frac{1}{2}\Dot{\Phi}^2+V(\Phi)},\\
    \label{inflaton_second_SR}
    \epsilon_2&=2\epsilon_1+\frac{2\Ddot{\phi}}{H\Dot{\phi}}.
    \end{align}
\end{subequations}
The condition $\epsilon_1<1$ tells us that the kinetic term $\frac{1}{2}\Dot{\Phi}^2$ is small. In other words, the field is \textit{slowly rolling}. In this regime, the inflaton behaves akin to a dark energy dominated Universe, $\rho_\Phi\sim -P_\Phi$, and the Hubble parameter is nearly constant. Additionally, the condition $\abs{\epsilon_2}<1$ indicates that we also require that the acceleration of the field is sufficiently small i.e. $\abs{\Ddot{\Phi}\ll3H\Dot{\Phi}}$ \footnote{Note that this condition is the same as an overdamped approximation, where the friction term $3H\Dot{\Phi}$ dominates over the acceleration $\Ddot{\Phi}$. We will return to this in Chapter \ref{ch:od_stochastic}, when we introduce the overdamped stochastic approach.}. Thus, the inflaton's equation of motion (\ref{inflaton_eom}) under the slow-roll conditions becomes
\begin{equation}
    \label{slow-roll_inflaton_eom}
    3H\Dot{\Phi}+V'(\Phi)\simeq 0.
\end{equation}
Thus, we see that the inflaton drives the accelerated expansion, as required. There exist other fields during inflation, whose energy density is far smaller than that of the inflaton. \textit{Spectator fields} are those that don't contribute to the dynamics of the expansion as they don't couple directly to the inflaton. Primary examples of these are the SM fields, which will remain dormant during inflation, by which I mean particle condensates will not form. The Higgs scalar is the prominent example of a light scalar field that existed during inflation and motivates the study of spectator scalars, which is the focus of this thesis. Additional (scalar) fields could exist and these give rise to interesting examples of primordial dark matter candidates. 

The final important feature of single-field inflation is how it ends. Crucially, the Hubble parameter is only \textit{nearly} constant and thus the accelerated expansion will not be eternal. The idea is that the scalar potential $V(\Phi)$ consists of a regime where the inflaton is slowly-rolling towards a minimum. As it rolls, it gains speed; after all, the acceleration is small but non-zero. At some point, it will exit the regime of slow-roll, triggering the end of inflation. The field will oscillate about the minimum before settling at a final value. This oscillatory process is where the inflaton will transfer its energy to the other fields, such as SM, giving rise to particle condensates and setting the scene for the remaining epochs of the Universe. In doing so, the inflaton loses the energy it needs to drive the accelerated expansion, and the expansion of the Universe slows to a more reasonable rate for the purposes of structure formation. This end of inflation era is preheating/reheating.

While single-field inflation is the most popular inflationary model, it is certainly not the only one. I list some honorary mentions, though I won't go into any detail about them here:
\begin{itemize}
    \item \textit{Guth's inflation.} The original inflationary model in \cite{Guth:1981} proposed that the Universe underwent a vacuum phase transition which triggered inflation. It has largely been discounted due to the `graceful exit problem'.
    \item \textit{Higgs inflation.} The Higgs particle takes the role of the inflaton to drive inflation.
    \item \textit{Multi-field inflation.} Multiple scalar fields coexist to drive inflation.
    \item \textit{Hybrid inflation.} An extra ``waterfall'' scalar field triggers inflation to end.
\end{itemize}

\subsection{Inflationary observables}
\label{subsec:inflationary_observables}

\begin{figure}[ht]
    \centering
    \includegraphics[width=120mm]{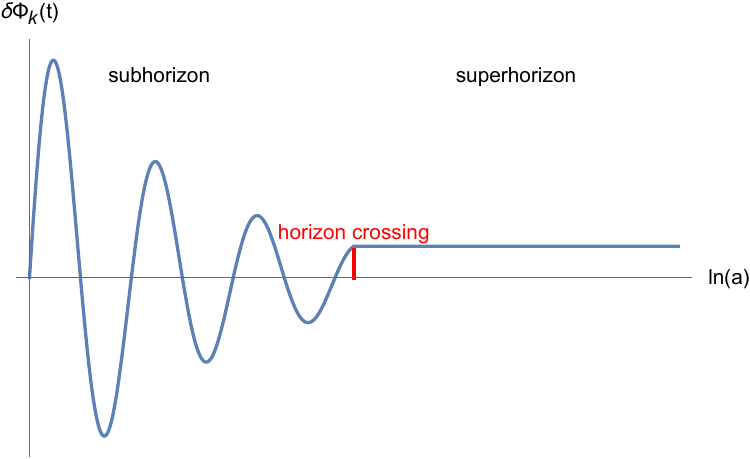}
    \caption{A schematic of the scalar perturbations $\delta\Phi_k(t)$ as a function of $\ln a$. One can see that the modes freeze at horizon crossing.}
    \label{fig:perturbation_modes}
\end{figure}

One of the great successes of inflation has been its description of large scale structure formation from a homogeneous and isotropic Universe. The idea is that quantum fluctuations of fields during inflation become macroscopic anisotropies that seed the growth of large scale structure. Consider the single-field inflation model above, where the inflaton potential $V(\Phi)=\frac{1}{2}m_\Phi^2\Phi^2$. For simplicity, we will consider a near-massless field such that $m_\Phi\ll H$. We perform a perturbative expansion of the field about the classical field as
\begin{equation}
    \label{field_perturbation}
    \Phi(t,\mathbf{x})=\overline{\Phi}(t)+\delta\Phi(t,\mathbf{x}),
\end{equation}
where $\overline{\Phi}$ represents the classical (homogeneous and isotropic) field from the previous section and $\delta\Phi$ is the quantum fluctuation. In $\mathbf{k}$-space, the equations of motion for the perturbations are given by
\begin{equation}
    \label{fluctuation_eom}
    \Ddot{\delta\Phi}_k+3H\Dot{\delta\Phi}_k+\frac{k^2}{a(t)^2}\delta\Phi_k=0.
\end{equation}
On subhorizon scales, $k\gg aH$ so
\begin{equation}
    \label{subhorizon_fluctuation_eom}
    \Ddot{\delta\Phi}_k+\frac{k^2}{a(t)^2}\delta\Phi_k=0
\end{equation}
and the solution is that of a dampening harmonic oscillator. On superhorizon scales, $k\ll aH$ and so
\begin{equation}
    \label{superhorizon_fluctuation_eom}
    0=\Ddot{\delta\Phi}_k+3H\Dot{\delta\Phi}_k,
\end{equation}
which gives a time-independent solution. The resulting behaviour can be seen in Fig. \ref{fig:perturbation_modes}. The key feature is that, upon horizon crossing, the modes are time-independent and become ``frozen out''. Additionally, one can compute the number density on superhorizon scales, $n_k\sim\lb\frac{k}{aH}\rb^{-3
}\gg1$ to discover that the modes behave classically. Thus, the initially-subhorizon quantum, microscopic perturbations grow with the expanding Universe to become macroscopic, classical fluctuations at superhorizon scales.

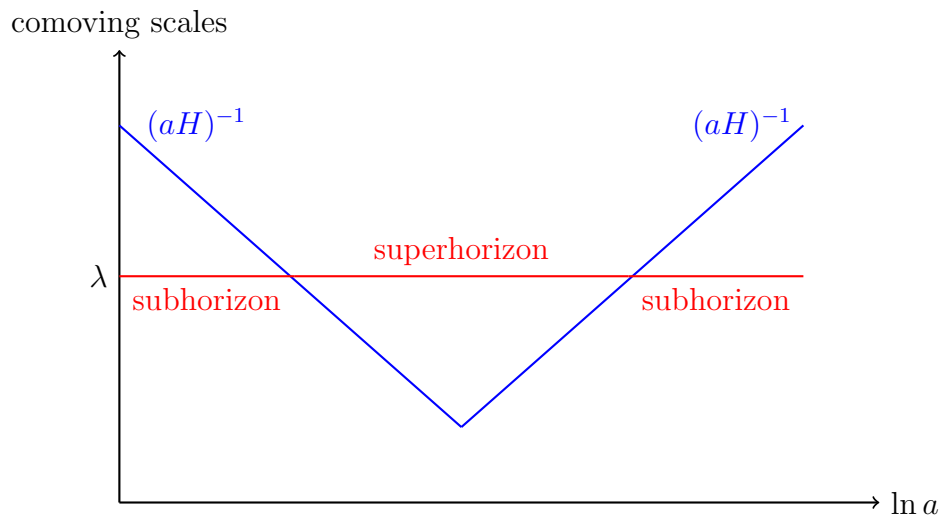
\begin{figure}[ht]
    \centering
    \begin{tikzpicture}
        \draw[black,thick,->] (0,0) -- (0,6) node[above]{comoving scales};
        \draw[black,thick,->] (0,0) -- (10,0) node[right] {$\ln a$};
        \draw[blue,thick] (4.5,1) -- (0,5) node[right] {\textcolor{white}{--}$(aH)^{-1}$};
        \draw[blue,thick] (4.5,1) -- (9,5) node[left] {$(aH)^{-1}$};
        \draw[red,thick] (2.3,3) -- (0,3) node[below,midway] {subhorizon} node[left,black] {$\lambda$};
        \draw[red,thick] (2.3,3) -- (6.7,3) node[above,midway] {superhorizon};
        \draw[red,thick] (6.7,3) -- (9,3) node[below,midway] {subhorizon};
    \end{tikzpicture}
    \caption{The journey of quantum fluctuations throughout the Universe, as it goes through the inflationary (decreasing $(aH)^{-1}$) and post-inflationary (increasing $(aH)^{-1}$) epochs. The red line indicates a specific wavelength $\lambda$.}
    \label{fig:quantum_fluctuations_reentry}
\end{figure}

I will not go into details about how these scalar perturbations are related to physical quantities here, but rather give a flavour of how they seed the large scale structure. The shrinking Hubble radius during inflation means that modes will naturally transition from sub- to superhorizon. As we have just seen, when quantum fluctuations cross the horizon during inflation, they become macroscopic, classical and (most importantly) frozen. In the post-inflationary epochs, the Hubble radius grows once again, meaning these fluctuations re-enter the horizon as macroscopic, classical quantities (see Fig. \ref{fig:quantum_fluctuations_reentry}). These quantities are then related to physical observables, such as the power spectrum. As I have stated, the great success of inflation is that it can describe the anisotropies in the CMB (see Fig. \ref{fig:cmb}) extremely well. The take-home point is this:

\begin{equation*}
\keybox{
    \begin{split}
    \text{Large scale ani}&\text{sotropies, observed in the present epoch, are a direct consequence}\\&\text{ of quantum fluctuations in the early Universe.}
    \end{split}
}
\end{equation*}

Quantum fluctuations during inflation are at the heart of this thesis. If one considers the inflationary spacetime to be pure de Sitter (which is largely a good approximation), one can compute the long distance behaviour of scalar correlation functions to obtain inflationary observables. With this in mind, we will leave the nuances of slow-roll inflation behind us and, for the rest of this thesis, focus on quantum fields in a pure de Sitter spacetime.

\chapter{Perturbative Quantum Field Theory in de Sitter Spacetime}
\label{ch:qft_dS}

\section{Introduction to Chapter \ref{ch:qft_dS}}
\label{sec:qft_dS_intro}

Quantum field theory (QFT) for scalar fields in de Sitter spacetime is the underlying theory that one uses to model inflation. For cosmologists, one is interested in using QFT to compute inflationary observables that, in theory and in practice, can one day be observed. The key quantities that one requires are correlation functions: expectation values of fields computed with respect to a vacuum state. For free fields in de Sitter, the computation of correlation functions is well-known and straightforward. However, the introduction of interactions means that calculations become extremely difficult to do in general; indeed, in most cases, solutions are unknown. One can use approximations to make some headway, three examples of which will be discussed in this thesis, beginning with the most well-known - perturbation theory - where one performs an expansion about the free solution to perturbatively add corrections from interactions.

In this chapter, I will begin with a review of the de Sitter geometry in Sec. \ref{sec:dS_geometry} before adding a free scalar field theory with mass $m$ and quantising in Sec. \ref{sec:free_scalars_dS}. This will introduce the standard method of second quantisation to establish the underlying QFT in de Sitter spacetime. It is predominantly textbook work; the key material that I used are \cite{Birrell_txtbk:1984,carroll:2003,Mukhanov:2007}, while the historical references are quoted in the relevant sections. 

I will then introduce interactions in Sec. \ref{sec:phi4_theory}, namely a quartic self-interaction term parameterised by the coupling $\lambda$. This is where the fun begins. I will pivot from second-quantisation to consider a path integral approach, known as the Schwinger-Kelydsh formalism in Sec. \ref{subsec:schwinger-keldysh}. This is not widely studied compared to second quantisation; it was only applied to curved spacetimes in the late 80s \cite{Calzetta:1987,Calzetta:1989,Calzetta_txtbk:2008}. These pioneering papers, alongside the textbook by the same authors, form the basis of my information on the subject. Additionally, Ref. \cite{Garbrecht:2011,Garbrecht:2014,Garbrecht:2015} use the Schwinger-Kelydsh approach in QFT to perform comparisons with the overdamped stochastic approach. This is of great relevance to this thesis, though I won't enter into the details here. From the Schwinger-Keldysh formalism, I will obtain a general expression for the correlation functions. Finally, I will perform our perturbative expansion to do some explicit calculations. In Sec. \ref{subsec:2-pt_function_one-loop}, I will compute the 2-point function to first-order in $\lambda$, including an outline of the UV renormalisation required. I will note the infrared problem that arises with perturbative QFT, which limits its regime of validity to $\lambda\ll m^4/H^4$. There is far more to be said about this than I will say here \cite{Tsamis:1993,Tsamis:2005,Urakawa:2009,Urakawa:2009_2,Suzuki:1994,Tokuda:2018,Tokuda:2018_2}. Finally in Sec. \ref{subsec:4-pt_functions}. with some additional comments in Appendix \ref{app:IR_lim_4-pt_func}, I will introduce the 4-point functions, focussing on the connected piece.

The key results of this chapter are Eq. (\ref{O(lambda)_2-pt_function_spacelike_long-distance}) and (\ref{spacelike_quantum_4-pt_func}): analytic expressions for the 2-point and connected 4-point functions at leading order in $\lambda$, for large spacetime separations\footnote{While we could write the 2-point function without taking the large spacetime separation limit, as we schematically do in Eq. (\ref{O(lambda)_2-pt_function_renormalised}), we will focus on this limit. This is because the stochastic approaches will only compute the 2-point function for asymptotically large spacetime separations. From a cosmology perspective, observables will also be in this limit as they are superhorizon during inflation.}. These results will fly the perturbative QFT flag when doing comparisons with the stochastic approximations. However, I will also draw attention to another novel result obtained in this chapter: analytic expressions for the renormalised mass and the UV-finite field variance at leading order in coupling using dimensional regularisation. These are given in Eq. (\ref{renormalised_mass_dim_reg_deSitter}) and (\ref{finite_field_variance_dim_reg_deSitter_ana}). While it is well-known how to remove the UV-divergent pieces via dimensional regularisation for scalar fields in de Sitter, to my knowledge, explicit expressions of the remaining finite pieces were not given in the literature until our work [K3].   

\section{de Sitter geometry}
\label{sec:dS_geometry}

\subsection{Global de Sitter}
\label{subsec:global_dS}

I will begin by discussing some features of the underlying geometry of de Sitter spacetime. This will not only be useful for our discussion of quantum fields in de Sitter in this chapter but also when we consider the stochastic approach in later chapters. I will keep to 4 spacetime dimensions for this discussion, although it is easy to generalise to $D$-dimensions if one wishes.

4-dimensional de Sitter spacetime can be viewed as that of a hyperboloid embedded in 5-dimensional Minkowski spacetime
\begin{equation}
    \label{5d_minkowksi}
    ds_{5d}^2=dX_0^2-dX_1^2-dX_2^2-dX_3^2-dX_4^2,
\end{equation}
defined by the equation
\begin{equation}
    \label{dS_hyperboloid}
    X_0^2-X_1^2-X_2^2-X_3^2-X_4^2=-\frac{1}{H^2},
\end{equation}
where $1/H$ is the hyperboloid radius. I have used $H$ here on purpose, to foreshadow its definition as the Hubble parameter. For now, it is sufficient simply to say that it is a constant scale defining the radius of the hyperboloid. It is evident from Eq. (\ref{dS_hyperboloid}) that the symmetry group for de Sitter - aptly named the ``de Sitter group'' - is SO(1,4).

\begin{figure}[ht]
    \centering
    \begin{tikzpicture}
        \draw[black,thick] (0,0) -- (8,0) node[below, midway]{$\Tilde{T}=-\pi/2$}; 
        \draw[black,thick] (0,8) -- (8,8) node[above, midway]{$\Tilde{T}=\pi/2$}; 
        \draw[black,thick] (0,0) -- (0,8) node[midway,above left,rotate = 90]{$\theta=\pi$}; 
        \draw[black,thick] (8,0) -- (8,8) node[midway,above right, rotate=-90]{$\theta=0$}; 
        \draw[red,thick] (0,0) -- (4,4); 
        \draw[red,thick] (4,4) -- (8,8) node[rotate=45,midway,above right]{horizon}; 
        \draw[red,thick] (0,8) -- (4,4) node[rotate=-45,midway,above left]{horizon}; 
        \draw[red,thick] (4,4) -- (8,0) ; 
        \draw[black,thick, ->] (9,-1) -- (9,1) node[anchor=south]{$\Tilde{T}$}; 
        \draw[black,thick, ->] (9,-1) -- (7,-1)  node[anchor=east]{$\theta$}; 
        \draw[blue] (0,1) -- (8,1); 
        \draw[blue] (0,1.5) -- (8,1.5); 
        \draw[blue] (0,2) -- (8,2); 
    \end{tikzpicture}
    \caption{The Penrose diagram for de Sitter. The top and bottom edges represent past and future infinity while the left and right edges are the south and north poles of the spatial 3-spheres respectively. The blue lines are constant $T$ slices. There exists a de Sitter horizon along the red lines.}
    \label{fig:dS_penrose_diag}
\end{figure}
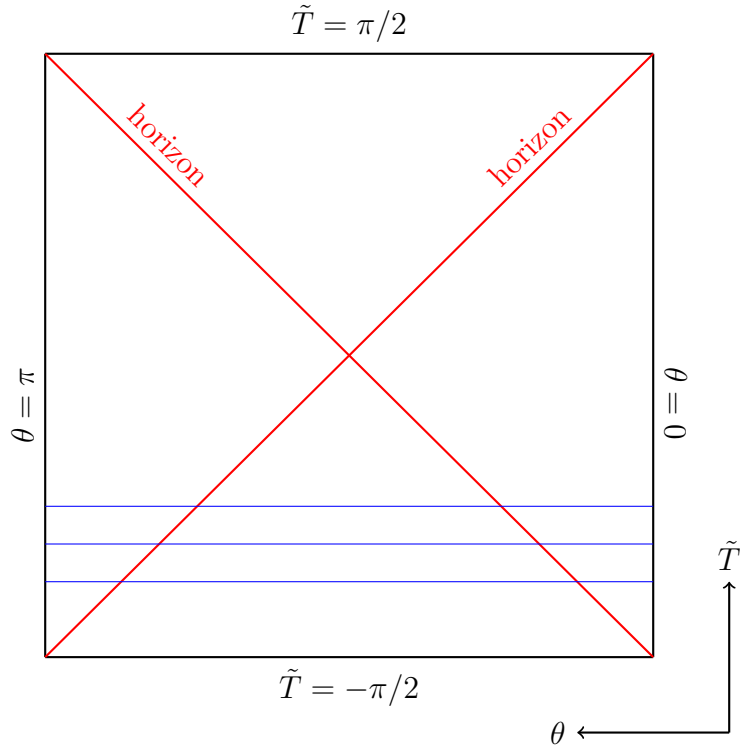

To write a 4-dimensional metric for de Sitter, we define
\begin{subequations}
    \label{global_dS_coords}
    \begin{align}
    X_0&=\frac{1}{H}\sinh\lb HT\rb,\\
    X_A&=\frac{1}{H}\cosh\lb HT\rb Z_A,
    \end{align}
\end{subequations}
where $\sum_A\lb Z_A\rb^2=1$ with $A\in\{1,2,3,4\}$. Thus, the coordinates $Z_A$ represent a unit 3-sphere and the \textit{global metric of de Sitter} is given by
\begin{equation}
    \label{global_dS_metric}
    ds^2=dT^2-\frac{1}{H^2}\cosh^2\lb HT\rb d\Omega_3^2,
\end{equation}
where the metric for a unit 3-sphere is
\begin{equation}
    \label{3-sphere_metric}
    \begin{split}
        d\Omega_3^2&=d\theta^2+\sin^2\theta d\Omega_2^2\\
        &=d\theta^2+\sin^2\theta \lb d\Phi^2+\sin^2\Phi d\varphi^2\rb.
    \end{split}
\end{equation}
The metric (\ref{global_dS_metric}) covers the entirety of the de Sitter spacetime in 4-dimensions, with $T\in\{-\infty,\infty\}$, $\theta,\Phi\in\{0,\pi\}$ and $\varphi\in\{0,2\pi\}$. One can draw the Penrose diagram for de Sitter by defining the coordinate $\tan\Tilde{T}=\sinh\lb HT\rb$ such that we can write the conformal global de Sitter metric as
\begin{equation}
    \label{conformal_global_dS_metric}
    ds^2=\frac{1}{H^2\cos^2\Tilde{T}}\qty(d\Tilde{T}^2-d\Omega_3^2).
\end{equation}
Now, $\Tilde{T}\in\{-\pi/2,\pi/2\}$ so we can immediately draw the Penrose diagram from this metric (Fig. \ref{fig:dS_penrose_diag}). It is a square, where each point on the diagram represents a 2-sphere. The top and bottom sides are at future and past temporal infinity while the right and left sides represent the poles of the spatial 3-spheres. Note that each point on the pole represents a single point in spacetime as opposed to a 2-sphere. Lines of constant $T$, depicted in blue in Fig. \ref{fig:dS_penrose_diag}, represent the spatial 3-spheres. Null geodesics travel at $45^o$ on the Penrose diagram and thus there exists a horizon (the red lines). 

\subsection{The de Sitter horizon and its temperature}

\begin{figure}
    \centering
    \begin{subfigure}[b]{0.45\textwidth}
    \centering
    \begin{tikzpicture}
        \draw[black,thick] (0,0)--(2,2)--(0,4)--cycle;
        \draw[black,thick] (4,0)--(2,2)--(0,0)--cycle;
        \draw[black,thick] (0,4)--(2,2)--(4,4)--cycle;
        \filldraw[color=black,fill=blue,thick] (4,4)--(2,2)--(4,0)--cycle;
    \end{tikzpicture}
    \caption{}
    \label{fig:static_Penrose}
     \end{subfigure}
    \begin{subfigure}[b]{0.45\textwidth}
    \centering
    \begin{tikzpicture}
        \draw[black,thick] (0,0)--(0,4)--(4,0)--cycle;
        \filldraw[color=black,fill=red,thick] (4,4)--(4,0)--(0,4)--cycle;
        \draw[black,thick] (0,0)--(4,4);
    \end{tikzpicture}
    \caption{}
    \label{fig:expanding_Penrose}
    \end{subfigure}
    \caption{The de Sitter Penrose diagram with (a) the static patch shaded in blue and (b) the expanding patch shaded in red.}
    \label{fig:patches_Penrose}
\end{figure}
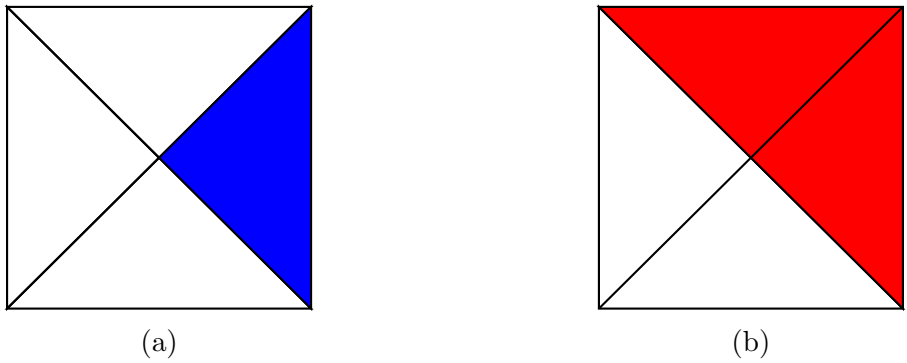

A crucial feature of de Sitter for this thesis, and more generally, is the existence of a horizon. To consider this more carefully, we consider the \textit{static patch} of de Sitter by defining the coordinates
\begin{subequations}
    \label{static_coords}
    \begin{align}
        X_0&=\frac{1}{H}\sqrt{1-H^2\rho^2}\sinh\lb H\tau\rb\\
        X_1&=\frac{1}{H}\sqrt{1-H^2\rho^2}\cosh\lb H\tau\rb\\
        X_i&=\rho Z_i
    \end{align}
\end{subequations}
where $i\in\{1,2,3\}$. The metric of the static patch is
\begin{equation}
    \label{static_dS_metric}
    ds^2=\lb1-H^2\rho^2\rb d\tau^2-\lb1-H^2\rho^2\rb^{-1}d\rho^2-\rho^2d\Omega_2^2.
\end{equation}
There exists a singularity in Eq. (\ref{static_dS_metric}) at $\rho=H^{-1}$, indicating the existence of the de Sitter horizon. This patch runs in the region $\tau\in\{-\infty,\infty\}$ and $\rho\in\{0,1/H\}$ and thus covers the right triangle of the Penrose diagram, as depicted in Fig. \ref{fig:static_Penrose}

The key point is this:
    \begin{center}
        \keybox{\text{There exists a de Sitter horizon at length scales $1/H$.}}
    \end{center}

As with black holes, the de Sitter horizon has an associated temperature \cite{Gibbons:1977}. The easiest way to see this is by rotating the metric (\ref{static_dS_metric}) to Euclidean space via $\tau_E=i\tau$ and $ds_E=ids$ and considering coordinates near the horizon via $\rho=H^{-1}(1-\epsilon^2)$ for some small $\epsilon$. Taking a constant $\Phi,\varphi$ slice, the Euclidean metric reads
\begin{equation}
    \label{euclidean_dS_metric}
    ds_E^2\simeq \frac{2}{H^2}\lb\epsilon^2d\lb H\tau_E\rb^2-d\epsilon^2\rb.
\end{equation}
Since these are just polar coordinates, this is non-singular provided we have a periodicity in $\tau_E$ of $\tau_E\rightarrow\tau_E+i\frac{2\pi}{H}$. Taking results from QFT at a finite temperature, we can relate periodicity in imaginary time to temperature, and thus the temperature of the de Sitter horizon is given by
\begin{equation}
    \label{dS_temperature}
    \keybox{T_{dS}=\frac{H}{2\pi}.}
\end{equation}

\subsection{The expanding de Sitter universe}

A generic feature of de Sitter is that it has both a contracting and expanding phase. One can confirm this by considering the 3-spheres on constant-$T$ surfaces. For $T<0$, we are in the contracting phase - the 3-spheres shrink as $T$ becomes more negative - while for $T>0$, we are in the expanding phase - the 3-spheres grow as $T$ increases. It is the latter phase that is most important in inflationary cosmology. To hone in on this region, we define the coordinates of the expanding patch as
\begin{subequations}
\label{expanding_dS_coords}
    \begin{align}
        X_0&=\frac{1}{H}\sinh\lb Ht\rb+\frac{1}{2}He^{Ht}x_i^2\\
        X_1&=\frac{1}{H}\cosh\lb Ht\rb-\frac{1}{2}He^{Ht}x_i^2\\
        X_i&=e^{Ht}x_i,
    \end{align}
\end{subequations}
where I have used the shorthand notation $x_i^2=\sum_{i=1}^3x_i^2$. In these coordinates, we have the \textit{expanding} or \textit{cosmological patch of de Sitter}, with metric
\begin{equation}
    \label{cosmological_dS_metric}
    \keybox{ds^2=dt^2-e^{2Ht}dx_i^2,}
\end{equation}
where $t,x_i\in\{-\infty,\infty\}$. This patch covers the upper right half of the Penrose diagram, as shown in Fig. \ref{fig:expanding_Penrose}\footnote{One can similarly define coordinates that cover the lower left half of the Penrose diagram, which would give us the contracting patch of de Sitter spacetime.}. One can see that, if we denote 
\begin{equation}
    \label{de_Sitter_scale_factor}
    a(t)=e^{Ht}, 
\end{equation}
this is just the FLRW metric (\ref{flrw_metric}) with an exponentially expanding scale factor. It is this region of de Sitter that is of most interest in inflationary cosmology as it naturally gives us a Universe undergoing an accelerated expansion. The rest of this thesis will focus on this region and use the metric (\ref{cosmological_dS_metric}) as the underlying geometry. Note that, in this set of coordinates, we can \textit{subhorizon} and \textit{superhorizon} length scales to be those less than and greater than $1/H$ respectively.

It is useful to note that there exists an additional coordinate system to describe the cosmological patch, which is obtained by the relation
\begin{equation}
    \label{conformal_time}
    \eta=-\frac{1}{H}e^{-Ht},
\end{equation}
where $\eta\in\{-\infty,0\}$. The metric associated with this coordinate change is given by
\begin{equation}
    \label{conformal_cosmological_dS_metric}
    ds^2=\frac{1}{\lb H\eta\rb^2}\lb d\eta^2-dx_i^2\rb.
\end{equation}
This is the conformal metric for the cosmological patch and thus $\eta$ is referred to as \textit{conformal time}. The scale factor written in terms of $\eta$ is just $a(\eta)=-\frac{1}{H\eta}$.

\section{Free scalar field theory in de Sitter}
\label{sec:free_scalars_dS}

\subsection{Classical field theory}
\label{subsec:classical_field_theory}

Now that the underlying geometry has been established, we will introduce scalar fields into the theory. Precisely, we will consider a single scalar field $\phi$ with a potential $V(\phi)$ living in the spacetime (\ref{cosmological_dS_metric}), with the metric $g_{\mu\nu}=\text{diag}\lsb1,-a(t),-a(t),-a(t)\rsb$. The action for such a theory is
\begin{equation}
    \label{scalar_field_action}
    S[\phi,g]=\int d^4x\sqrt{-g}\lb\frac{1}{16\pi G}R+\frac{1}{2}g^{\mu\nu}\partial_\mu\phi\partial_\nu\phi-\frac{1}{2}\xi R\phi^2-V(\phi)\rb,
\end{equation}
where $\sqrt{-g}=a(t)^3$, with $g$ the metric determinant, and $G$ is the gravitational constant. The first term is the Einstein-Hilbert action, dictating the dynamics of the spacetime independent of the scalar field. The term $\frac{1}{2}\xi R\phi^2$ represents the simplest non-minimal coupling of the scalar to gravity, with the strength of this interaction dictated by $\xi$. In de Sitter, $R=12H^2$ and so this is just an additional contribution to the mass term.

The equations of motion for $\phi$ and the Einstein equations can be found by extremising the action. They are given respectively by
\begin{subequations}
\begin{align}
    \label{scalar_eom}
    0&=\Ddot{\phi}+3H\Dot{\phi}-\frac{1}{a(t)^2}\nabla^2\phi+V'(\phi)+12\xi H^2\phi,\\
    \label{einstein_eq}
    H^2&=\frac{8\pi G}{3}\rho_\phi,
\end{align}
\end{subequations}
where the energy density of the scalar field 
\begin{equation}
    \label{energy_density_de_Sitter}
    \rho_\phi=\frac{1}{2}\Dot{\phi}^2+\frac{1}{2a(t)^2}\lb\nabla\phi\rb^2+V(\phi).
\end{equation}
The action (\ref{scalar_field_action}) is completely general for a single scalar field in de Sitter spacetime. From the perspective of inflationary cosmology, $\phi$ could represent the inflaton that drives the spacetime expansion, or some auxiliary field that exists during inflation. For the vast majority of this thesis, I will consider the scalar field to be a spectator, such that its energy density is subdominant to that of the inflaton. Thus, it has a negligible back reaction to gravity and we can ignore Eq. (\ref{einstein_eq}) in the analysis. Further, we can, for all intents and purposes, ignore the Einstein-Hilbert term in the action (\ref{scalar_field_action}). The action of this spectator field is then
\begin{equation}
    \label{spectator_action}
    S[\phi]=\int d^4x a(t)^3\lb\frac{1}{2}\Dot{\phi}^2-\frac{1}{2}\frac{(\nabla\phi)^2}{a(t)^2}-V(\phi)\rb,
\end{equation}
where we have now absorbed the non-minimal term into the scalar potential.

It will also prove convenient in the upcoming analysis to define the canonical momentum $\pi=\Dot{\phi}$. Thus, Eq. (\ref{scalar_eom}) can be written as a set of first order partial differential equations as
\begin{equation}
    \label{scalar_eom_2d}
    \keybox{
    \begin{pmatrix}\Dot{\phi}\\\Dot{\pi}\end{pmatrix}
    =
    \begin{pmatrix}
        \pi\\-3H\pi+\frac{\nabla^2\phi}{a(t)^2}-V'(\phi)
    \end{pmatrix}.
    }
\end{equation}
Let us now focus on free fields such that $V(\phi)=\frac{1}{2}m^2\phi^2$. I have absorbed the non-minimal coupling into the mass term via $m^2=m_0^2+12\xi H^2$, where $m_0^2$ is the scalar mass. We can write the solutions to the equations of motion as a mode expansion
\begin{subequations}
    \begin{align}
    \label{phi_mode_expansion}
    \phi(x)&=\int\dbar^3\mathbf{k}\lsb a_k\phi_k(t)e^{-i\mathbf{k}\cdot\mathbf{x}}+a_k^{\dagger}\phi_k^*(t)e^{i\mathbf{k}\cdot\mathbf{x}}\rsb,\\
    \label{pi_mode_expansion}
    \pi(x)&=\int\dbar^3\mathbf{k}\lsb a_k\pi_k(t)e^{-i\mathbf{k}\cdot\mathbf{x}}+a_k^{\dagger}\pi_k^*(t)e^{i\mathbf{k}\cdot\mathbf{x}}\rsb,
    \end{align}
\end{subequations}
where the mode functions obey the equations of motion
\begin{equation}
    \label{k-space_scalar_eom}
    \begin{pmatrix}\Dot{\phi}_k\\\Dot{\pi}_k\end{pmatrix}
    =
    \begin{pmatrix}
        \pi_k\\-3H\pi_k-\frac{k^2}{a(t)^2}\phi_k+m^2\phi_k
    \end{pmatrix}
\end{equation}
and are normalised by the condition
\begin{equation}
    \label{wronskian_normalisation}
    \frac{i}{a(t)^3}=\pi^*_k\phi_{k}-\phi^*_k\pi_{k}.
\end{equation}
$\{a_k,a_k^{\dagger}\}$ are a set of coefficients in the basis expansion; classically, they are just numbers. One set of solutions\footnote{Note that these solutions are not unique: more on this shortly.} to the mode function equation is
\begin{subequations}
\label{mode_functions}
    \begin{align}
    \label{phi_k_mode}
    \phi_k(t)&=\sqrt{\frac{\pi}{4Ha(t)^3}}\mathcal{H}^{(1)}_\nu\lb\frac{k}{a(t)H}\rb,\\
    \label{pi_k_mode}
    \pi_k(t)&=-\sqrt{\frac{\pi}{16Ha(t)^3}}\lsb3H\mathcal{H}^{(1)}_\nu\lb\frac{k}{a(t)H}\rb+\frac{k}{a(t)}\lb\mathcal{H}^{(1)}_{\nu-1}\lb\frac{k}{a(t)H}\rb-\mathcal{H}_{\nu+1}^{(1)}\lb\frac{k}{a(t)H}\rb\rb\rsb,
    \end{align}
\end{subequations}
where $\nu=\sqrt{\frac{9}{4}-\frac{m^2}{H^2}}$ and $\mathcal{H}_n^{(1)}(z)$ is the Hankel function of the first kind. There is similarly another set of solutions where one replaces $\mathcal{H}_n^{(1)}(z)$ with the Hankel function of the second kind, $\mathcal{H}_n^{(2)}(z)$. These are related by $\mathcal{H}_n^{(1)*}(z)=\mathcal{H}_n^{(2)}(z)$ for $n,z\in\RR$. It will prove useful to consider the form of these mode functions in the near-massless limit, when $\nu=3/2$. They are given by
\begin{subequations}
\label{mass_mode_functions}
    \begin{align}
        \label{phi_k_mode_massless}
        \phi_k(t)\eval_{\nu=3/2}&=-\frac{H}{\sqrt{2k^3}}\lb\frac{k}{a(t)H}+i\rb e^{i\frac{k}{a(t)H}},\\
        \pi_k(t)\eval_{\nu=3/2}&=\frac{i}{a(t)^2}\sqrt{\frac{k}{2}}e^{i\frac{k}{a(t)H}}.
        \label{pi_k_mode_massless}
    \end{align}
\end{subequations}
Note that, for much of this thesis, we will be considering the IR behaviour of these modes. Focussing on the field mode, the IR behaviour is found by expanding Eq. (\ref{phi_k_mode}) for $k\ll a(t)H$ to give
\begin{equation}
    \label{growing_decaying_modes}
    \phi_k(t)\eval_{k\ll a(t)H}=\frac{2^{-2-\nu}\pi}{a(t)^3H\Gamma(1+\nu)}k^\nu-i\frac{2^{-2+\nu}\Gamma(\nu)}{a(t)^3H}k^{-\nu}.
\end{equation}
One can see that the real part has a \textit{decaying} behaviour while the imaginary part is \textit{growing}. In cosmology, these are referred to as decaying and growing modes respectively.

\subsection{Canonical quantisation}
\label{subsec:quantisation}

We now quantise our fields by promoting them to operators $\{\phi,\pi\}\rightarrow\{\hat{\phi},\hat{\pi}\}$. We impose the equal time commutation relations
\begin{subequations}
    \label{equal_time_commuation_relations}
    \begin{align}
    \comm{\hat{\phi}(t,\mathbf{x})}{\hat{\pi}(t,\mathbf{x}')}&=\frac{i}{a(t)^3}\delta^{(3)}(\mathbf{x}-\mathbf{x}'),\\
    \comm{\hat{\phi}(t,\mathbf{x})}{\hat{\phi}(t,\mathbf{x}')}&=\comm{\hat{\pi}(t,\mathbf{x})}{\hat{\pi}(t,\mathbf{x}')}=0.
    \end{align}
\end{subequations}
This is achieved by promoting the mode coefficients to operators $\{a_k,a_k^{\dagger}\}\rightarrow\{\hat{a}_k,\hat{a}_k^{\dagger}\}$ such that they obey the usual commutation relations
\begin{subequations}
    \label{ladder_operators_comm_rel}
    \begin{align}
    \comm{\hat{a}_k}{\hat{a}_{k'}^{\dagger}}&=\deltabar^{(3)}(\mathbf{k}-\mathbf{k}'),\\
    \comm{\hat{a}_k}{\hat{a}_{k'}}&=\comm{\hat{a}_k^{\dagger}}{\hat{a}_{k'}^{\dagger}}=0.
    \end{align}
\end{subequations}
These are then the creation and annihilation, or ladder, operators, from which we can define our vacuum state $\ket{0_a}$ as $\hat{a}_k\ket{0_a}=0$ $\forall k$, and build our Hilbert space from there.

However, since we are in curved spacetime this vacuum is not uniquely defined. To see this, we can define a new basis $\{\Tilde{\phi}_k,\Tilde{\pi}_k\}$ that is related to our old one by a Bogolyubov transformation
\begin{equation}
    \label{bogolyubov_transformation}
    \Tilde{\phi}_k=A\phi_k+B\phi_k^*,
\end{equation}
where $A$ and $B$ are Bogolyubov coefficients. This new basis has a set of creation and annihilation operators associated with it, $\{\hat{b}_k,\hat{b}_k^{\dagger}\}$, from which we can similarly build a Hilbert space about the vacuum defined via $\hat{b}_k\ket{0_b}=0$ $\forall k$. There is no clear reason to prefer the vacuum state $\ket{0_a}$ over $\ket{0_b}$. We can place some restrictions on the vacuum state since the Bogolyubov coefficients must obey
\begin{equation}
    \label{Bogolyubov_coeff_constraint}
    1=\abs{A}^2-\abs{B}^2.
\end{equation}
The most general solution to this is 
\begin{subequations}
    \label{Bogolyubuv_coefficients}
    \begin{align}
    A&=\cosh\alpha,\\
    B&=e^{i\beta}\sinh\alpha,
    \end{align}
\end{subequations}
where $\alpha\in\{0,\infty\}$ and $\beta\in\{-\pi,\pi\}$. Further, the vacuum state must be invariant under the de Sitter group SO(1,4). One can show by examining the Green's functions - which must also be de Sitter invariants - that this is only true if $\beta=0$ \cite{Allen:1985,Brunetti:2005}. Thus, we have a one-parameter family of vacua, parameterised by $\alpha$, which are all viable candidates for the vacuum state of de Sitter. These are the famous \textit{$\alpha$-vacua}. The most popular choice is to take $\alpha=0$: the Euclidean or \textit{Bunch-Davies} vacuum \cite{chernikov:1968,bunch-davies:1978}. The mode functions (\ref{mode_functions}) are solutions in the Bunch-Davies vacuum. This is a useful choice because, in the far past $t\rightarrow-\infty$, the Bunch-Davies vacuum coincides with the Minkowski vacuum and the mode function is just the positive frequency mode\footnote{This is more clear if we use conformal time $\eta$ and write $\chi=a(\eta)\phi\simeq-\frac{1}{\sqrt{2k}}e^{-ik\eta}$.}
\begin{equation}
    \label{early_time_mode_function}
    \phi_k(t)\sim-\frac{1}{a}\frac{1}{\sqrt{2k}}e^{i\frac{k}{a(t)H}}.
\end{equation}
For $\alpha\ne0$, it will be some mix of positive and negative frequency modes. For the remainder of this thesis, I will consider the theory to be built from the Bunch-Davies vacuum.

\subsection{Two-point correlation functions for free fields}
\label{subsec:two-pt_funcs_free_fields}

The physical quantities that are of interest in cosmology are correlation functions: expectation values of the vacuum state. The most basic of these is the scalar two-point function, which is computed in $\mathbf{k}$-space using the mode expansion (\ref{phi_mode_expansion}) as
\begin{equation}
    \label{k-space_2-pt_func}
    \begin{split}
        \bra{0}\hat{\phi}(t,\mathbf{k})\hat{\phi}(t',\mathbf{k}')\ket{0}=&\phi_k(t)\phi_{k'}^*(t')
        \\=&
        \frac{\pi}{4Ha(t)^{3/2}a(t')^{3/2}}\mathcal{H}^{(1)}_\nu\lb\frac{k}{a(t)H}\rb\mathcal{H}^{(2)}_\nu\lb\frac{k'}{a(t')H}\rb,
    \end{split}
\end{equation}
where we have used the mode function (\ref{pi_k_mode}) and the commutation relations (\ref{ladder_operators_comm_rel}). We can perform the Fourier transform to obtain the 2-pt function in coordinate space as
\begin{equation}
    \label{2-pt_func_k-space}
    \bra{0}\hat{\phi}(t,\mathbf{x})\hat{\phi}(t',\mathbf{x}')\ket{0}=\int\dbar^3\mathbf{k}e^{-i\mathbf{k}\cdot(\mathbf{x}-\mathbf{x}')}\bra{0}\hat{\phi}(t,\mathbf{k})\hat{\phi}(t',\mathbf{k})\ket{0}.
\end{equation}
Computing this integral \cite{chernikov:1968,tagirov:1973,Schomblond:1976,Dowker:1976,bunch-davies:1978} results in the positive (+) and negative (-) frequency Wightman functions in the Bunch-Davies vacuum
\begin{subequations}
    \label{wightman_functions_pre_contour}
    \begin{align}
        \begin{split}
        \Delta^+(x,x')&:=\bra{0}\hat{\phi}(t,\mathbf{x})\hat{\phi}(t',\mathbf{x}')\ket{0}
        \\&=\frac{H^2}{16\pi^2}\Gamma(\alpha)\Gamma(\beta)_2F_1\lb\beta,\alpha,2;1+\frac{\lb\eta-\eta'-i\epsilon\rb^2-\abs{\mathbf{x}-\mathbf{x}'}^2}{4\eta\eta'}\rb,
        \end{split}\\
        \begin{split}
        \Delta^-(x,x')&:=\bra{0}\hat{\phi}(t',\mathbf{x}')\hat{\phi}(t,\mathbf{x})\ket{0}
        \\&=\frac{H^2}{16\pi^2}\Gamma(\alpha)\Gamma(\beta)_2F_1\lb\beta,\alpha,2;1+\frac{\lb\eta-\eta'+i\epsilon\rb^2-\abs{\mathbf{x}-\mathbf{x}'}^2}{4\eta\eta'}\rb,    
        \end{split}
    \end{align}
\end{subequations}
where $\Gamma(z)$ is the Euler-Gamma function, $_2F_1(a,b,c;z)$ is the hypergeometric function and $\alpha=3/2-\nu$, $\beta=3/2+\nu$ \footnote{The $\alpha$ and $\beta$ introduced here are unrelated to the ones introduced in the discussion of the $\alpha$-vacua.}. The $i\epsilon$ prescription indicates the pole about which we perform our contour integration in the complex plane. Note that I have used conformal time $\eta$ here, defined in Eq. (\ref{conformal_time}), for convenience. From the equation of motion (\ref{k-space_scalar_eom}) and (\ref{2-pt_func_k-space}), the Wightman functions obey the equation
\begin{equation}
    \label{Wightman_funcs_eom}
    \lb\Box_{dS}+m^2\mp i\epsilon\rb \Delta^{\pm}(x,x')=0,
\end{equation}
where $\Box_{dS}=\partial_t^2+3H\partial_t-\frac{\nabla_\mathbf{x}^2}{a(t)^2}$. 

Physical correlators must be invariant under the de Sitter group. This means that the behaviour of such correlators can be written purely in terms of a de Sitter invariant combination of the spacetime coordinates. The quantity in question is
\begin{equation}
    \label{de_Sitter_invariant}
    \begin{split}
        y(x,x')&=\frac{\lb\eta-\eta'\rb^2-\abs{\mathbf{x}-\mathbf{x}'}^2}{2\eta\eta'}\\
        &=\cosh\lb H(t-t')\rb-\frac{H^2}{2}e^{H(t+t')}\abs{\mathbf{x}-\mathbf{x}'}^2-1.
    \end{split}
\end{equation}
We can write the Wightman functions in terms of the de Sitter invariant by expanding about small $\epsilon$ to give
\begin{subequations}
    \label{wightman_functions}
    \begin{align}
        \Delta^+(x,x')=&\frac{H^2}{16\pi^2}\Gamma(\alpha)\Gamma(\beta)_2F_1\lb\beta,\alpha,2;1+\frac{y}{2}\rb+\frac{iH^2}{32\pi}(4\nu^2-1)_2F_1\lb\beta,\alpha,2;-\frac{y}{2}\rb\theta(y)\theta(t'-t),\\    
        \Delta^-(x,x')=&\frac{H^2}{16\pi^2}\Gamma(\alpha)\Gamma(\beta)_2F_1\lb\beta,\alpha,2;1+\frac{y}{2}\rb+\frac{iH^2}{32\pi}(4\nu^2-1)_2F_1\lb\beta,\alpha,2;-\frac{y}{2}\rb\theta(y)\theta(t-t'),
    \end{align}
\end{subequations}
where we have introduced the Heaviside function 
\begin{equation}
    \label{heaviside_function}
    \theta(z)=\begin{cases}1,&\text{if } z>0\\0,&\text{if }z<0\end{cases}.
\end{equation}
Hence, I will drop the $i\epsilon$ prescription. From the Wightman functions, we can build our other scalar 2-point correlators. For convenience, we define
\begin{subequations}
    \label{A&B_2-pt_amplitudes}
    \begin{align}
        A(y)&=\frac{H^2}{16\pi^2}\Gamma(\alpha)\Gamma(\beta)_2F_1\lb\beta,\alpha,2;1+\frac{y}{2}\rb,\\
        B(y)&=\frac{H^2}{32\pi}(4\nu^2-1)_2F_1\lb\beta,\alpha,2;-\frac{y}{2}\rb\theta(y).
    \end{align}
\end{subequations}
Then, we define various other 2-point functions in Table \ref{tab:2-pt_functions}.

\begin{table}[ht]
\begin{small}
    \centering
    \begin{tabular}{c|c|c|c|c}
        Name & Symbol & Correlator & Form & $\Box_{dS}+m^2=$\\ \hline &&&&\\
        Hadamard & $\Delta^H(x,x')$ & $\bra{0}\acomm{\hat{\phi}(t,\mathbf{x})}{\hat{\phi}(t',\mathbf{x}')}\ket{0}$ & $2A(y)+iB(y)$ & 0\\&&&&\\
        Causal & $\Delta^C(x,x')$ & $-i\bra{0}\comm{\hat{\phi(t,\mathbf{x})}}{\hat{\phi}(t',\mathbf{x}')}\ket{0}$ & $B(y)\lb\theta(t'-t)-\theta(t-t')\rb$ & 0 \\&&&&\\
        Advanced & $\Delta^A(x,x')$ & $-i\theta(t'-t)\bra{0}\comm{\hat{\phi}(t,\mathbf{x})}{\hat{\phi}(t',\mathbf{x}')}\ket{0}$ & $B(y)\theta(t'-t)$ & $\frac{1}{a(t)^3}\delta^{(4)}(x-x')$ \\&&&&\\
        Retarded & $\Delta^R(x,x')$ & $i\theta(t-t')\bra{0}\comm{\hat{\phi}(t,\mathbf{x})}{\hat{\phi}(t',\mathbf{x}')}\ket{0}$ & $B(y)\theta(t-t')$ & $\frac{1}{a(t)^3}\delta^{(4)}(x-x')$ \\&&&&\\
        Feynman & $i\Delta^F(x,x')$ & $\bra{0}T\hat{\phi}(t,\mathbf{x})\hat{\phi}(t',\mathbf{x}')\ket{0}$ & $A(y)$ & -$\frac{i}{a(t)^3}\delta^{(4)}(x-x')$ \\&&&&\\
        Dyson & $i\Delta^D(x,x')$ & $\bra{0}\Tilde{T}\hat{\phi}(t,\mathbf{x})\hat{\phi}(t',\mathbf{x}')\ket{0}$ & $A(y)+iB(y)$ & $\frac{i}{a(t)^3}\delta^{(4)}(x-x')$ 
    \end{tabular}
    \caption{The 2-point functions for free fields, built out of the Wightman functions (\ref{wightman_functions}).}
    \label{tab:2-pt_functions}
\end{small}
\end{table}

The most relevant 2-point function for this thesis is the Feynman propagator. Explicitly, the free field Feynman propagator is given by
\begin{equation}
\keybox{
    \label{free_feynman_propagator}
    i\Delta^F(x,x'):=\bra{0}T\hat{\phi}(t,\mathbf{x})\hat{\phi}(t',\mathbf{x}')\ket{0}=\frac{H^2}{16\pi^2}\Gamma(\alpha)\Gamma(\beta)_2F_1\lb\beta,\alpha,2;1+\frac{y}{2}\rb,
    }
\end{equation}
where $T(\Tilde{T})$ is the (anti-)time ordered operator, defined via the Heaviside function (\ref{heaviside_function}) as
\begin{subequations}
    \begin{align}
        \label{time-ordered_operator}
        Tf(t)g(t')&=f(t)g(t')\theta(t-t')+g(t')f(t)\theta(t'-t),\\
        \label{antitime-ordered_operator}
        \tilde{T}f(t)g(t')&=f(t)g(t')\theta(t'-t)+g(t')f(t)\theta(t-t').
    \end{align}
\end{subequations}
Note that for $m^2/H^2=2$ ($\nu=1/2$), the Feynman propagator is given by
\begin{equation}
    \label{conformal_mass_Feynman_propagator}
    i\Delta^F(x,x')=-\frac{1}{4\pi^2a(\eta)a(\eta')\lb(\eta-\eta')^2-\abs{\mathbf{x}-\mathbf{x}'}^2\rb}.
\end{equation}
This is the Feynman propagator for massless fields in Minkowski spacetime up to the factors of $a(\eta)$. Thus, scalar fields in de Sitter with this mass are conformal to massless scalars in Minkowski. $m^2=2H^2$ is known as the \textit{conformal mass}.

The Feynman propagator obeys the equation
\begin{equation}
    \label{Feynman_prop_eom}
    \lb\Box_{dS}+m^2\rb i\Delta^F(x,x')=-\frac{i}{a(t)^3}\delta^{(4)}(x-x'),
\end{equation}
which can be obtained by using the equations of motion (\ref{k-space_scalar_eom}) with (\ref{2-pt_func_k-space}), alongside the equal-time commutation relations (\ref{equal_time_commuation_relations}) and $\partial_t\theta(t-t')=\delta(t-t')$. We can obtain the time-ordered $\phi-\pi$, $\pi-\phi$ and $\phi-\phi$ correlators by taking time derivatives of Eq. (\ref{free_feynman_propagator}). The results are
\begin{subequations}
    \begin{align}
        \begin{split}
            \bra{0}T\hat{\phi}(t,\mathbf{x})\hat{\pi}(t',\mathbf{x}')\ket{0}=&\partial_{t'}\qty(i\Delta^F(x,x'))\\
            =&\qty(-H\sinh(H(t-t'))-\frac{H^3}{2}e^{H(t+t')}\abs{\mathbf{x}-\mathbf{x}'}^2)\\&\times\frac{H^2}{64\pi^2}\Gamma\qty(\frac{5}{2}+\nu)\Gamma\qty(\frac{5}{2}-\nu){_2}F_1\qty(\frac{5}{2}-\nu,\frac{5}{2}+\nu,3;1+\frac{y}{2})\\&
        \end{split}\\
        \begin{split}
            \bra{0}T\hat{\pi}(t,\mathbf{x})\hat{\phi}(t',\mathbf{x}')\ket{0}=&\partial_{t}\qty(i\Delta^F(x,x'))\\
            =&\qty(H\sinh(H(t-t'))-\frac{H^3}{2}e^{H(t+t')}\abs{\mathbf{x}-\mathbf{x}'}^2)\\&\times\frac{H^2}{64\pi^2}\Gamma\qty(\frac{5}{2}+\nu)\Gamma\qty(\frac{5}{2}-\nu){_2}F_1\qty(\frac{5}{2}-\nu,\frac{5}{2}+\nu,3;1+\frac{y}{2}),\\&
        \end{split}\\
        \begin{split}
            \bra{0}{T\hat{\pi}(t,\mathbf{x})\hat{\pi}(t',\mathbf{x}')\ket{0}}=&\partial_t\partial_{t'}\qty(i\Delta^F(x,x'))\\
            =&\qty(-H^2\cosh(H(t-t'))-\frac{H^4}{2}e^{H(t+t')}\abs{\mathbf{x}-\mathbf{x}'}^2)\\&\times\frac{H^2}{64\pi^2}\Gamma\qty(\frac{5}{2}+\nu)\Gamma\qty(\frac{5}{2}-\nu){_2}F_1\qty(\frac{5}{2}-\nu,\frac{5}{2}+\nu,3;1+\frac{y}{2}),
            \\&+\qty(H\sinh(H(t-t'))-\frac{H^3}{2}e^{H(t+t')}\abs{\mathbf{x}-\mathbf{x}'}^2)\\&\times\frac{H^2}{384\pi^2}\Gamma\qty(\frac{7}{2}+\nu)\Gamma\qty(\frac{7}{2}-\nu){_2}F_1\qty(\frac{7}{2}-\nu,\frac{7}{2}+\nu,4;1+\frac{y}{2}).
        \end{split}
    \end{align}
\end{subequations}
One can see that these quantities are not de Sitter invariant and therefore don't represent physical observables in cosmology. The purpose of writing them here is so that we can compare them with equivalent results from the stochastic approach. In lieu of this, we wish to consider long distance behaviour of the correlators. These are the important quantities in inflationary cosmology because they represent the modes that have exited the horizon and are thus important for observation. One can make this approximation by expanding the hypergeometric functions about large $y$. The Feynman propagator then goes as
\begin{equation}
    \label{free_feynman_propagator_long-distance_sum}
    \begin{split}
    i\Delta^F(x,x')=&\frac{H^2}{16\pi^2}\Bigg[\frac{\Gamma(-2\nu)\Gamma(1+2\nu)}{\Gamma\qty(\frac{1}{2}+\nu)\Gamma\qty(\frac{1}{2}-\nu)}\sum_{s=0}^{\infty}\frac{\Gamma\qty(\frac{3}{2}+\nu+s)\Gamma\qty(\frac{1}{2}+\nu+s)}{\Gamma\qty(1+2\nu+s)s!}\qty(-\frac{y}{2})^{-\frac{3}{2}-\nu-s}\\
    &+\frac{\Gamma(2\nu)\Gamma(1-2\nu)}{\Gamma\qty(\frac{1}{2}+\nu)\Gamma\qty(\frac{1}{2}-\nu)}\sum_{s=0}^{\infty}\frac{\Gamma\qty(\frac{3}{2}-\nu+s)\Gamma\qty(\frac{1}{2}-\nu+s)}{\Gamma\qty(1-2\nu+s)s!}\qty(-\frac{y}{2})^{-\frac{3}{2}+\nu-s}\Bigg].
    \end{split}
\end{equation}
\noindent Taking the leading order terms in the two sums gives
\begin{equation}
\keybox{
    \label{free_feynman_long-distance_leading}
    i\Delta^F(x,x')=\frac{H^2}{16\pi^2}\lsb\frac{\Gamma(2\nu)\Gamma(\frac{3}{2}-\nu)}{\Gamma(\frac{1}{2}+\nu)}\qty(-\frac{y}{2})^{-\frac{3}{2}+\nu}+\frac{\Gamma(-2\nu)\Gamma(\frac{3}{2}+\nu)}{\Gamma(\frac{1}{2}-\nu)}\qty(-\frac{y}{2})^{-\frac{3}{2}-\nu}\rsb.
    }
\end{equation}
\noindent For the range $3/2>\nu>0$, or $0<m^2/H^2< 9/4$, the long distance behaviour of the Feynman propagator is a decaying power law. The first term in the above expression is the leading term, while the second term is subleading but not necessarily next-to-leading order: this is true only when $3/2\ge\nu>1/2$. For $1/2>\nu\ge0$, subleading terms in the first sum of Eq. (\ref{free_feynman_propagator_long-distance_sum}) will be lower-order in the asymptotic expansion than this second term. Note that at $\nu=1/2$, we have the conformal result (\ref{conformal_mass_Feynman_propagator}) so we just have a $1/y$ behaviour. For heavier fields $m^2/H^2>9/4$, $\nu$ becomes imaginary and the propagator becomes oscillatory. We will focus on the light field case so that this behaviour is always considered to be decaying; for a discussion of the evolution of heavy fields, see Ref. \cite{Markkanen:2017}. In the cosmological language introduced earlier, the first and second terms represent the growing and decaying modes respectively.

The two limits of the spacetime regimes are equal-space (timelike) and equal-time (spacelike). The latter are particularly relevant in the context of observables. The timelike correlators to leading order in the two sums are
\begin{subequations}
\label{equal-space_correlators}
    \begin{align}
    \label{equal-space_phiphi_corr}
    \begin{split}
        \bra{0}T\hat{\phi}(t,\mathbf{x})\hat{\phi}(t',\mathbf{x})\ket{0}=&\frac{H^2}{16\pi^2}\frac{\Gamma(2\nu)\Gamma\qty(\alpha)(-4)^\alpha}{\Gamma\qty(\frac{1}{2}+\nu)}e^{-\alpha H(t-t')}\\&+\frac{H^2}{16\pi^2}\frac{\Gamma(-2\nu)\Gamma\qty(\beta)(-4)^\beta}{\Gamma\qty(\frac{1}{2}-\nu)}e^{-\beta H(t-t')},\\&\\
    \end{split}\\
    \label{equal-space_phipi_corr}
    \begin{split}
        \bra{0}T\hat{\phi}(t,\mathbf{x})\hat{\pi}(t',\mathbf{x})\ket{0}=&\frac{H^3}{16\pi^2}\frac{\alpha\Gamma(2\nu)\Gamma\qty(\alpha)(-4)^\alpha}{\Gamma\qty(\frac{1}{2}+\nu)}e^{-\alpha H(t-t')}\\&+\frac{H^3}{16\pi^2}\frac{\beta\Gamma(-2\nu)\Gamma\qty(\beta)(-4)^\beta}{\Gamma\qty(\frac{1}{2}-\nu)}e^{-\beta H(t-t')},\\&\\
    \end{split}\\
    \label{equal-space_piphi_corr}
    \begin{split}
        \bra{0}T\hat{\pi}(t,\mathbf{x})\hat{\phi}(t',\mathbf{x})\ket{0}=&-\frac{H^3}{16\pi^2}\frac{\alpha\Gamma(2\nu)\Gamma\qty(\alpha)(-4)^\alpha}{\Gamma\qty(\frac{1}{2}+\nu)}e^{-\alpha H(t-t')}\\&-\frac{H^3}{16\pi^2}\frac{\beta\Gamma(-2\nu)\Gamma\qty(\beta)(-4)^\beta}{\Gamma\qty(\frac{1}{2}-\nu)}e^{-\beta H(t-t')},\\&\\
    \end{split}\\
    \label{equal-space_pipi_corr}
    \begin{split}
        \bra{0}T\hat{\pi}(t,\mathbf{x})\hat{\pi}(t',\mathbf{x})\ket{0}=&\frac{H^4}{16\pi^2}\frac{\alpha^2\Gamma(2\nu)\Gamma\qty(\alpha)(-4)^\alpha}{\Gamma\qty(\frac{1}{2}+\nu)}e^{-\alpha H(t-t')}\\&+\frac{H^4}{16\pi^2}\frac{\beta^2\Gamma(-2\nu)\Gamma\qty(\beta)(-4)^\beta}{\Gamma\qty(\frac{1}{2}-\nu)}e^{-\beta H(t-t')},\\&\\
    \end{split}
    \end{align}
\end{subequations}
\noindent while the spacelike correlators are given by
\begin{subequations}
\label{equal-time_correlators}
    \begin{align}
    \label{equal-time_phiphi_corr}
    \begin{split}
        \bra{0}\hat{\phi}(t,\mathbf{0})\hat{\phi}(t,\mathbf{x})\ket{0}=&\frac{H^2}{16\pi^2}\frac{\Gamma(2\nu)\Gamma\qty(\alpha)4^\alpha}{\Gamma\qty(\frac{1}{2}+\nu)}\abs{Ha(t)\mathbf{x}}^{-2\alpha}\\&+\frac{H^2}{16\pi^2}\frac{\Gamma(-2\nu)\Gamma\qty(\beta)4^\beta}{\Gamma\qty(\frac{1}{2}-\nu)}\abs{Ha(t)\mathbf{x}}^{-2\beta},\\&\\
    \end{split}\\
    \label{equal-time_phipi_corr}
    \begin{split}
        \bra{0}\hat{\phi}(t,\mathbf{0})\hat{\pi}(t,\mathbf{x})\ket{0}=&-\frac{H^3}{16\pi^2}\frac{\alpha\Gamma(2\nu)\Gamma\qty(\alpha)4^\alpha}{\Gamma\qty(\frac{1}{2}+\nu)}\abs{Ha(t)\mathbf{x}}^{-2\alpha}\\&-\frac{H^3}{16\pi^2}\frac{\beta\Gamma(-2\nu)\Gamma\qty(\beta)4^\beta}{\Gamma\qty(\frac{1}{2}-\nu)}\abs{Ha(t)\mathbf{x}}^{-2\beta},\\&\\
    \end{split}\\
    \label{equal-time_piphi_corr}
    \begin{split}
        \bra{0}\hat{\pi}(t,\mathbf{0})\hat{\phi}(t,\mathbf{x})\ket{0}=&-\frac{H^3}{16\pi^2}\frac{\alpha\Gamma(2\nu)\Gamma\qty(\alpha)4^\alpha}{\Gamma\qty(\frac{1}{2}+\nu)}\abs{Ha(t)\mathbf{x}}^{-2\alpha}\\&-\frac{H^3}{16\pi^2}\frac{\beta\Gamma(-2\nu)\Gamma\qty(\beta)4^\beta}{\Gamma\qty(\frac{1}{2}-\nu)}\abs{Ha(t)\mathbf{x})}^{-2\beta},\\&\\
    \end{split}\\
    \label{equal-time_pipi_corr}
    \begin{split}
        \bra{0}\hat{\pi}(t,\mathbf{0})\hat{\pi}(t,\mathbf{x})\ket{0}=&\frac{H^4}{16\pi^2}\frac{\alpha^2\Gamma(2\nu)\Gamma\qty(\alpha)4^\alpha}{\Gamma\qty(\frac{1}{2}+\nu)}\abs{Ha(t)\mathbf{x}}^{-2\alpha}\\&+\frac{H^4}{16\pi^2}\frac{\beta^2\Gamma(-2\nu)\Gamma\qty(\beta)4^\beta}{\Gamma\qty(\frac{1}{2}-\nu)}\abs{Ha(t)\mathbf{x}}^{-2\beta}.\\&\\
    \end{split}
    \end{align}
\end{subequations}
The timelike and spacelike correlators are related by the analytic continuation $e^{H(t-t')}\leftrightarrow\abs{Ha(t)(\mathbf{x}-\mathbf{x}')}^2$ due to the symmetries of the de Sitter spacetime. The exceptions to this are the $\phi-\pi$ and $\pi-\phi$ correlators, which are antisymmetric under $\phi$ and $\pi$ for timelike correlators but symmetric for spacelike correlators. The symmetry of the spacelike correlators is explained by the equal time commutation relations (\ref{equal_time_commuation_relations}). These relations don't hold in the timelike case, but there is an antisymmetry here due to the time-ordering.

For the case where we have nearly massless fields, $m^2\ll H^2$, these correlators can be simplified further. Since the subleading term in the asymptotic expansion are also subdominant in the small mass expansion, we will focus on the leading term only. The spacelike correlators in the nearly massless limit are
\begin{subequations}
\label{free_light_field_spacelike_QFT_correlators}
    \begin{align}
        \bra{0}\hat{\phi}(t,\mathbf{0})\hat{\phi}(t,\mathbf{x})\ket{0}\eval_{m^2\ll H^2}&=\frac{3H^4}{8\pi^2m^2}\abs{Ha(t)\mathbf{x}}^{-\frac{2m^2}{3H^2}},\\
        \bra{0}\hat{\phi}(t,\mathbf{0})\hat{\pi}(t,\mathbf{x})\ket{0}\eval_{m^2\ll H^2}&=-\frac{H^3}{8\pi^2}\abs{Ha(t)\mathbf{x}}^{-\frac{2m^2}{3H^2}},\\
        \bra{0}\hat{\pi}(t,\mathbf{0})\hat{\pi}(t,\mathbf{x})\ket{0}\eval_{m^2\ll H^2}&=\frac{H^2m^2}{24\pi^2}\abs{Ha(t)\mathbf{x}}^{-\frac{2m^2}{3H^2}}.
    \end{align}
\end{subequations}
We can now build any 2-point functions from the $\phi-\phi$, $\phi-\pi$, $\pi-\phi$ and $\pi-\pi$ correlators using Wick's theorem. For example, the $\phi^2$ correlator is given by
\begin{equation}
    \label{phi^2_quantum_correlator}
    \bra{0}\hat{\phi}(t,\mathbf{x})^2\hat{\phi}(t',\mathbf{x}')^2\ket{0}=\expval{\hat{\phi}^2}^2+2\lb\bra{0}\hat{\phi}(t,\mathbf{x})\hat{\phi}(t',\mathbf{x}')\ket{0}\rb^2,
\end{equation}
where $\expval{\hat{\phi}^2}=\bra{0}\hat{\phi}(t,\mathbf{x})\hat{\phi}(t,\mathbf{x})\ket{0}$ is the equal-spacetime correlator, or the quantum variance. The quantum variances are given by
\begin{subequations}
    \label{quantum_variance}
    \begin{align}
    \label{quantum_phiphi_variance}
    \expval{\hat{\phi}^2}&=\frac{H^2}{16\pi^2}\Gamma\qty(\frac{3}{2}-\nu)\Gamma\qty(\frac{3}{2}+\nu){_2}F_1\qty(\frac{3}{2}+\nu,\frac{3}{2}-\nu,2;1),\\
    \label{quantum_phipi_variance}
    \expval{\hat{\phi}\hat{\pi}}&=0,\\
    \label{quantum_pipi_variance}
    \expval{\hat{\pi}^2}&=-\frac{\alpha\beta H^4}{64\pi^2}\Gamma\qty(\frac{3}{2}-\nu)\Gamma\qty(\frac{3}{2}+\nu){_2}F_1\qty(\frac{7}{2}+\nu,\frac{7}{2}-\nu,4;1).
    \end{align}\\
\end{subequations}
However, $_2F_1\lb\frac{3}{2}+\nu,\frac{3}{2}-\nu,2,1\rb$ and $_2F_1\lb\frac{7}{2}+\nu,\frac{7}{2}-\nu,4,1\rb$ are both divergent and thus these quantities are currently ill-defined. We can examine the structure of this behaviour by taking the small-$y$ limit\footnote{We take the limit from $y<0$, where points are spacelike separated. This avoids any issues with ordering the time coordinates. One can take from $y>0$ but one must be careful how they deal with the discontinuity generated by the $\theta$-functions. The simplest way to do this is by taking the limit using the Hadamard (symmetric) function.} of the two-point function
\begin{equation}
    \label{quantum_field_variance_UV_divergence}
    \begin{split}
    \expval{\hat{\phi}^2}=&\lim_{y\rightarrow0^-}i\Delta^F(x,x')\\=&-\frac{H^2}{8\pi^2y}+\frac{m^2-2H^2}{16\pi^2}\qty(\ln y-1+2\gamma_E-\ln2+\psi^{(0)}\qty(\frac{3}{2}-\nu)+\psi^{(0)}\qty(\frac{3}{2}+\nu)),
    \end{split}
\end{equation}
where $\psi^{(0)}(z)$ is the polygamma function and $\gamma_E$ is the Euler-Mascheroni constant. We then see that the divergent behaviour is of $\mathcal{O}\lb1/y\rb$ and $\mathcal{O}\lb\ln y\rb$. As we will see, these short-distance, or UV, divergences become increasingly prevalent when computing correlation functions for an interacting theory. Fortunately, they can be dealt with via standard renormalisation techniques.  

\section{Massive $\lambda\phi^4$ theory in de Sitter}
\label{sec:phi4_theory}

\subsection{The Schwinger-Keldysh formalism}
\label{subsec:schwinger-keldysh}

Having laid the foundations with free fields, we now turn our attention to the more interesting situation where we include interactions. The focus of this thesis will be a quartic self-interaction, parameterised by the coupling $\lambda$. The scalar potential becomes $V(\phi)=\frac{1}{2}m^2\phi^2+\frac{1}{4}\lambda\phi^4$. Clearly, this just reverts to the potential for free fields if one sets $\lambda=0$.

\begin{figure}[ht]
    \centering
    \begin{tikzpicture}
        \draw[black,thick,->] (0,4) -- (10.5,4) node[anchor=west]{$\Re t$}; 
        \draw[black,thick,->] (10,0) -- (10,8) node[anchor=south]{$\Im t$}; 
        \draw[black,thick,->] (0,6) -- (4,6) node[anchor=south]{$J^+$}; 
        \draw[black,thick] (4,6) -- (8,6); 
        \draw[black,thick,->] (8,2) -- (4,2) node[anchor=north]{$J^-$}; 
        \draw[black,thick] (0,2) -- (4,2); 
        \draw[black,thick,->] (8,6) -- (8,4);
        \draw[black,thick] (8,4) -- (8,2);
        \draw[black,fill=black] (0,6) circle (2pt) node[anchor=east]{$t\rightarrow-\infty$};
        \draw[black,fill=black] (0,2) circle (2pt) node[anchor=east]{$t\rightarrow-\infty$};
        \draw[black,dashed,->] (9,4) -- (9,6) node[midway, right]{$\epsilon$};
        \draw[black,dashed,->] (9,4) -- (9,2) node[midway,right]{$\epsilon$};
        \draw[black,dashed] (8,2) -- (8,1) node[anchor=north]{$t_*$};
    \end{tikzpicture}
    \caption{The closed time path of the in-in formalism. From the far past, the system evolves from the in state to $\ket{\psi(t_*)}$, sourced by $J^+$, before returning to the in state, sourced by $J^-$. The $i\epsilon$ prescription introduced in the Wightman functions (\ref{wightman_functions_pre_contour}) is made manifest here.}
    \label{fig:CTP_sketch}
\end{figure}
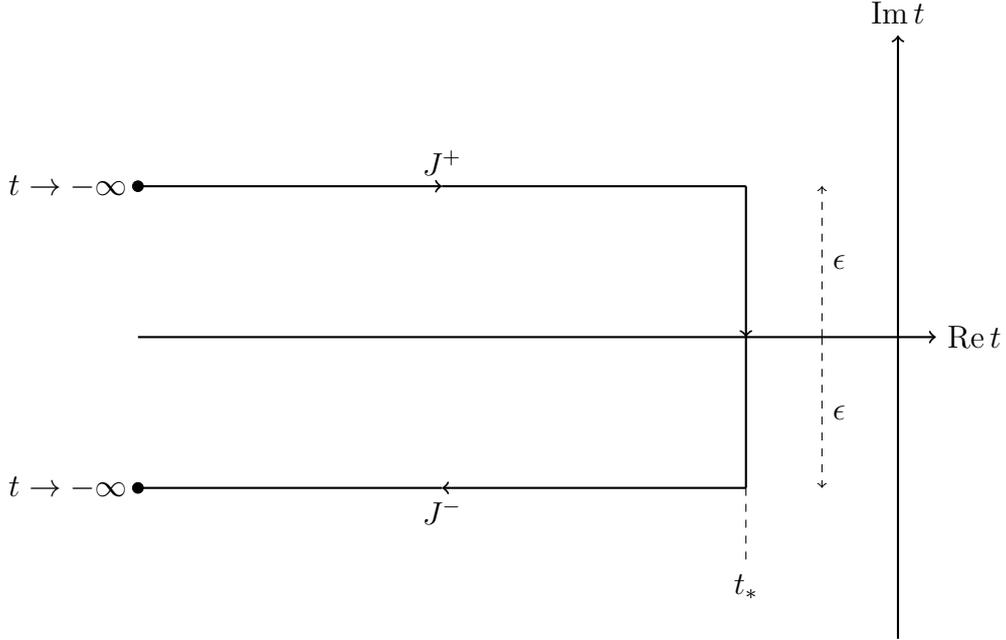

The addition of the interaction means we cannot straightforwardly compute the scalar correlators; the equations of motion (\ref{scalar_eom_2d}) cannot be solved analytically. Instead, we will consider a perturbative approach where one performs a small-$\lambda$ expansion about the free solution that we computed in the last section. To do this, it will be convenient to switch from the canonical quantisation approach of the last section to a path integral method. In Minkowski spacetime, one defines a \textit{generating functional} $Z[J]$ to be the probability of the vacuum state surviving its evolution from past infinity $\ket{0_-}$ (in vacuum state) to future infinity $\ket{0_+}$ (out vacuum state) when hit by some external source $J(x)$. It is given by
\begin{equation}
    \label{in-out_generating_functional}
    Z[J]=\braket{0_+}{0_-}_J=\bra{0_+}Te^{i\int d^4x J(x)\hat{\phi}(x)}\ket{0_-},
\end{equation}
where $Z[0]=1$. One can write the generating functional as a path integral 
\begin{equation}
    \label{in-out_gen_func_path_integral}
    Z[J]=\int D\phi e^{i\lb S[\phi]+\int d^4x J(x)\phi(x)\rb},
\end{equation}
from which all the scalar correlation functions can be computed via
\begin{equation}
    \label{in-out_scalar_correlators}
    \bra{0_+}T\hat{\phi}(x_1)...\hat{\phi}(x_n)\ket{0_-}=(-i)^n\frac{\delta^n Z[J]}{\delta J(x_1)...\delta J(x_n)}\eval_{J=0}.
\end{equation}
This is the \textit{in-out formalism}. This formalism is convenient in Minkowski spacetime because the in and out vacua are the same. However, in general curved spacetimes, this is not necessarily true and so the correlators (\ref{in-out_scalar_correlators}) will actually be matrix elements as opposed to the expectation values required for physical observables. This is true for the Bunch-Davies vacuum in de Sitter. To see this, consider the far past and far future of the mode function (\ref{phi_k_mode}) in the Bunch-Davies vacuum:
\begin{subequations}
    \label{phi_k_mode_limits}
    \begin{align}
        \label{far_past_phi_k}
        \phi_k(t)\eval_{t\rightarrow-\infty}&=-\frac{1}{a(t)}\frac{1}{\sqrt{2k}}e^{i\frac{k}{a(t)H}},\\
        \label{far_future_phi_k}
        \phi_k(t)\eval_{t\rightarrow\infty}&=-\frac{i}{2H\sqrt{\pi a(t)^3}}\Gamma(\nu)\lb\frac{k}{2a(t)H}\rb^{-\nu}.
    \end{align}
\end{subequations}
Evidently, these have different behaviours and therefore, in de Sitter, $\ket{0_-}\ne\ket{0_+}$. Instead, we will consider a path integral formalism where we begin in the in state, evolve our system to some intermediate state at time $t_*$, $\ket{\psi(t_*)}$ before evolving back to the in state (see Fig. \ref{fig:CTP_sketch}). This is the \textit{in-in formalism}, also known as the \textit{closed time path} or \textit{Schwinger-Keldysh formalism} (after the authors of the pioneering papers \cite{Schwinger:1960,Keldysh:1964})\footnote{Note that there are subtle differences between these methods but, for the purpose of this thesis, I will consider them to be synonymous.}.

Now, the system contains two external sources, $J^+$ and $J^-$, which source the evolution from $\ket{0_-}$ to $\ket{\psi(t_*)}$ and from $\ket{\psi(t_*)}$ to $\ket{0_-}$ respectively. The in-in generating functional is then defined by
\begin{equation}
    \label{in-in_generating_functional_def}
    \begin{split}
    Z[J^+,J^-]&=_{J^-}\braket{0_-}{0_-}_{J^+}
    \\&=\int D\psi\bra{0_-}\Tilde{T}e^{i\int^{-\infty}_{t_*}dt\int d^3\mathbf{x}\sqrt{-g}J^-(x)\hat{\phi}(x)}\ket{\psi}\bra{\psi}T e^{i\int^{t_*}_{-\infty}dt\int d^3\mathbf{x}\sqrt{-g}J^+(x)\hat{\phi}(x)}\ket{0_-},
    \end{split}
\end{equation}
where $Z[J,J]=1$ and we recall that $T$ and $\Tilde{T}$ are the time- and anti-time-ordered operators respectively. In the path integral representation, we introduce two auxiliary fields, $\phi^+$ and $\phi^-$, to differentiate the contributions from the two paths. The in-in generating functional is then given by
\begin{equation}
    \label{in-in_generating_functional_path_integral}
    Z[J^+,J^-]=\int D\phi^+\int D\phi^-e^{i\lb S^+[\phi^+]+J_x^+\phi_x^+-S^-[\phi^-]-J_x^-\phi_x^-\rb},
\end{equation}
where $S^\pm[\phi^\pm]=S[\phi^\pm]$ with the $\pm i\epsilon$ prescription. Further, I have introduce the de Witt condensed notation for convenience, where repeated indices represent integrals over the spacetime coordinate: for example, $J_x\phi_x=\int d^4x \sqrt{-g(x)}J(x)\phi(x)$. All scalar correlators can be built from the in-in generating functional via
\begin{equation}
    \label{in-in_scalar_correlators}
    \begin{split}
    \bra{0_-}\Tilde{T}\lsb\hat{\phi}(x_1)...\hat{\phi}(x_n)\rsb& T\lsb\hat{\phi}(x'_1)...\hat{\phi}(x'_m)\rsb\ket{0_-}\\&=(-i)^{n-m}\frac{\delta^{n+m}Z[J^+,J^-]}{\delta J^-(x_1)...\delta J^-(x_n)\delta J^+(x'_1)...\delta J^+(x'_m)}\eval_{J^\pm=0}.
    \end{split}
\end{equation}
Note that we can now see that the vacuum states considered in Sec. \ref{sec:free_scalars_dS}, $\ket{0}$, are really the in state $\ket{0_-}$. Henceforth, I will drop the subscript `-' on the vacuum state. 

To link the in-in formalism to what we have considered in Sec. \ref{sec:free_scalars_dS}, we return to free fields by setting $\lambda=0$ in our scalar potential. The free action can be written as
\begin{equation}
    \label{free_scalar_action_pm}
    S^\pm_0[\phi]=-\frac{1}{2}\int d^4x \sqrt{-g} \phi(x) \lb \Box_{dS}+m^2\mp i\epsilon\rb \phi(x),
\end{equation}
where we recall that $\Box_{dS}=\partial_t^2+3H\partial_t-\frac{\nabla_\mathbf{x}^2}{a(t)^2}$. The generating functional (\ref{in-in_generating_functional_path_integral}) now takes the form of a Gaussian and thus we can compute the functional integrals, resulting in the free generating functional of the form
\begin{equation}
    \label{free_in-in_gen_func_kernels}
    Z^{(0)}[J^+,J^-]=e^{\frac{i}{2}\lb J_x^+ K_{xx'}^{++} J_{x'}^+-J_x^+ K_{xx'}^{+-}J_{x'}^--J_x^- K_{xx'}^{-+}J_{x'}^++J_x^-K_{xx'}^{--}J_{x'}^-\rb},
\end{equation}
where the kernels obey the relations $K_{xx'}^{++}=K_{x'x}^{++}$, $K_{xx'}^{--}=K_{x'x}^{--}$ and $K_{xx'}^{+-}=K^{-+}_{x'x}$. Using Eq. (\ref{in-in_scalar_correlators}), we can relate the kernels to the free 2-point functions in Table \ref{tab:2-pt_functions}. Thus, the free in-in generating functional is given by
\begin{equation}
    \label{free_in-in_generating_functional}
    Z^{(0)}[J^+,J^-]=e^{-\frac{1}{2}\lb J_x^+ i\Delta^F_{xx'} J_{x'}^+-J_x^+ \Delta^-_{xx'}J_{x'}^--J_x^- \Delta^+_{xx'}J_{x'}^++J_x^-i\Delta^D_{xx'}J_{x'}^-\rb}.
\end{equation}
Note that we can simplify this further since $J_x^+ \Delta^-_{xx'}J_{x'}^-=J_x^- \Delta^+_{xx'}J_{x'}^+$. This is useful when we consider more complicated computations.

We can now re-introduce interactions such that our generating functional reads
\begin{equation}
    \label{interacting_in-in_gen_func_path_integral}
    Z[J^+,J^-]=\int D\phi^+\int D\phi^-e^{i\lb S_0^+[\phi^+]+J_x^+\phi_x^+-S_0^-[\phi^-]-J_x^-\phi_x^-\rb}e^{-i\frac{\lambda}{4}\int d^4x\sqrt{-g}\lb\lb\phi^+(x)\rb^4-\lb\phi^-(x)\rb^4\rb}.
\end{equation}
Using $F[\phi^\pm] e^{iJ^\pm_x\phi^\pm_x}=F\lsb -i\frac{\delta}{\delta J^\pm}\rsb e^{iJ^\pm_x\phi^\pm_x}$ for some arbitrary function $F[\phi]$, we can compute the functional integrals such that Eq. (\ref{interacting_in-in_gen_func_path_integral}) becomes
\begin{equation}
    \label{interacting_generating_functional}
    Z[J^+,J^-]=e^{-i\frac{\lambda}{4} d^4x \sqrt{-g}\lb\frac{\delta^4}{\delta J^+(x)^4}-\frac{\delta^4}{\delta J^-(x)^4}\rb}Z^{(0)}[J^+,J^-].
\end{equation}
Applying Eq. (\ref{in-in_scalar_correlators}) to (\ref{interacting_generating_functional}), we can compute scalar correlators for this self-interacting theory.

\subsection{Two-point QFT correlation functions to $\mathcal{O}(\lambda)$}
\label{subsec:2-pt_function_one-loop}

The quantity of most interest for this thesis is the time-ordered 2-point function. Eq. (\ref{free_feynman_propagator}) gives us this quantity for free fields; now, we will add corrections from interactions in a perturbative manner. By expanding Eq. (\ref{interacting_generating_functional}) to leading order in small coupling $\lambda$, the generating functional to $\mathcal{O}(\lambda)$ is given by
\begin{equation}
    \label{O(lambda)_generating_functional}
    Z[J^+,J^-]=\lb1-i\frac{\lambda}{4}\int d^4x \sqrt{-g}\lb\frac{\delta^4}{\delta J^+(x)^4}-\frac{\delta^4}{\delta^4J^-(x)}\rb\rb Z^{(0)}[J^+,J^-].
\end{equation}
Using Eq. (\ref{in-in_scalar_correlators}), the time-ordered 2-point function to $\mathcal{O}(\lambda)$ is
\begin{equation}
    \label{one-loop_UVdiv_2-pt_function}
    \begin{split}
    \bra{0}T\hat{\phi}(x_1)&\hat{\phi}(x_2)\ket{0}=i\Delta^F(x_1,x_2)\\&+3i\lambda \expval{\hat{\phi}^2}\int d^4z a(t_z)^3\lb i\Delta^F(z,x_1)i\Delta^F(z,x_2)-\Delta^+(z,x_1)\Delta^+(z,x_2)\rb\\&+\mathcal{O}(\lambda^2).
    \end{split}
\end{equation}
The first line is just the free Feynman propagator (\ref{free_feynman_propagator}) while the second line gives the contribution to $\mathcal{O}(\lambda)$, which is yet to be computed, and we have used the fact that $i\Delta^F(z,z)=i\Delta^D(z,z)=\expval{\hat{\phi}^2}$. Note that for the time-ordered correlation functions, the Feynman propagators are sourced by $J^+$ while the Wightman functions are sourced by $J^-$; in terms of the diagram in Fig. \ref{fig:CTP_sketch}, the Feynman and Wightman functions are associated with the upper and lower branches respectively. This is a key difference to the in-out formalism, where we would only have the Feynman propagators present\footnote{Of course, the in-out and in-in formalisms are equivalent so computations of physical quantities via either method will give the same result.}.

The $\mathcal{O}(\lambda)$ contribution to the 2-point function can be computed in a similar way to the standard procedure in Minkowski, by making a correction to the mass\footnote{In theory, the integral in Eq. (\ref{one-loop_UVdiv_2-pt_function}) can be computed numerically; however, it contains poles on the light cone that are not easy to deal with. Our (unsuccessful) attempts to do so are given in Appendix \ref{app:numerical_calculation_correlators}.}. Applying the operator $\Box_{dS}+m^2$ to the 2-point function (\ref{one-loop_UVdiv_2-pt_function}), we find that it obeys the equation
\begin{equation}
    \label{2-pt_func_eom_O(lambda)}
    \lb\Box_{dS}+m_B^2+3\lambda\expval{\hat{\phi}^2}\rb\bra{0}T\phi(x_1)\phi(x_2)\ket{0}=-\frac{i}{a(t)^3}\delta^{(4)}(x_1-x_2),
\end{equation}

where the bare mass $m_B^2=m_{0,B}^2+12\xi_B H^2$. This is an important distinction to make for interacting theories, where we need to consider renormalisation. This is the de Sitter equivalent to the self-energy correction that is prevalent in QFT in flat spacetime, which simply amounts to a mass correction when one performs renormalisation\footnote{There will be additional terms at higher orders in $\lambda$ that contribute further corrections to the mass, but there will also be other contributions that mean the renormalisation is not so simple.}. We infer that the 2-point function to $\mathcal{O}(\lambda)$ can be computed by making a mass correction to the Feynman propagator:
\begin{equation}
    \label{O(lambda)_2-pt_function}
    \bra{0}T\hat{\phi}(x_1)\hat{\phi}(x_2)\ket{0}=i\Delta^F(x_1,x_2)\eval_{m^2\rightarrow m_B^2+3\lambda\expval{\hat{\phi}^2}}.
\end{equation}
However, we are not yet in a position to write an explicit expression for the $\mathcal{O}(\lambda)$ 2-point function because, as observed in Eq. (\ref{quantum_field_variance_UV_divergence}), the quantum variance $\expval{\hat{\phi}^2}$ contains $\mathcal{O}(1/y)$ and $\mathcal{O}(\ln y)$ divergences. For the 2-point function, we need to ensure the mass correction
\begin{equation}
    \label{mass_correction_UV_div}
    m^2\rightarrow m^2_B+3\lambda\expval{\hat{\phi}^2}
\end{equation}
is UV-finite, which can be done by renormalisation. There are many ways one can do this but the method that is used for particle physics is the \textit{$\overline{MS}$ scheme of dimensional regularisation}, where one introduces a renormalisation scale that fixes the scale of the experiment. To ensure that our QFT aligns with particle physics, we will focus on dimensional regularisation \cite{Candelas:1975,Dowker:1976,Beneke:2013}. However, it is challenging to perform in de Sitter, and so we will also require some results from \textit{point-splitting regularisation} \cite{bunch-davies:1978} to get the job done. We will se that, by comparing the two methods with those from Minkowski spacetime (see Appendix \ref{app:renormalisation_minkowski}), we can obtain an analytic expression for the UV-finite mass correction using dimensional regularisation in the $\overline{\text{MS}}$ scheme.

\subsubsection{Point-splitting regularisation in de Sitter spacetime}

The only result we need from point-splitting is the expression for the field variance. Taking timelike separations to be exactly 0 such that $x=(t,\mathbf{x})$ and $x'=(t,\mathbf{0})$ in Eq. (\ref{quantum_field_variance_UV_divergence}), the UV divergent field variance becomes
\begin{equation}
    \label{field_variance_small-x_deSitter}
    \begin{split}
    \expval{\hat{\phi}^2}_{PS}:=&i\Delta^F(\mathbf{x},\mathbf{0})\eval_{\abs{Ha(t)\mathbf{x}}\rightarrow0}\\=&-\frac{H^2}{4\pi^2\abs{Ha(t)\mathbf{x}}^2}+\frac{2H^2-m_B^2}{16\pi^2}\Bigg(-2\ln\abs{Ha(t)\mathbf{x}}+1-2\gamma_E\\&+\ln4+\psi^{(0)}\qty(\frac{3}{2}-\nu_B)+\psi^{(0)}\qty(\frac{3}{2}+\nu_B)\Bigg),
    \end{split}
\end{equation}
where $\nu_B=\sqrt{\frac{9}{4}-\frac{m^2_B}{H^2}}$ and the subscript `PS' indicates it's computed using point-splitting. Note that these divergences are identical to that of Minkowski, which is discussed in Appendix \ref{app:renormalisation_minkowski}. We can then write the finite piece of the field variance defined by point-splitting as
\begin{equation}
    \label{field_variance_point-split_finite}
    \expval{\hat{\phi}^2}_{PS}^{(fin)}=\frac{2H^2-m_B^2}{16\pi^2}\Bigg(1-2\gamma_E+\ln4+\psi^{(0)}\qty(\frac{3}{2}-\nu_B)+\psi^{(0)}\qty(\frac{3}{2}+\nu_B)\Bigg).
\end{equation}

\subsubsection{Dimensional regularisation in de Sitter spacetime}

To perform dimensional regularisation, we consider the $\mathbf{k}$-space integral
\begin{equation}
    \label{k-space_field_variance}
    \begin{split}
    \expval{\hat{\phi}^2}=&\frac{\pi}{4Ha(t)^3}\int \dbar^3\mathbf{k}\abs{\mathcal{H}_{\nu}^{(1)}\lb\frac{k}{a(t)H}\rb}^2\\&\\
    =&\int\dbar^3\mathbf{k} \lb\frac{1}{2a(t)^2k}+\frac{2H^2-m_B^2}{4k^3}\rb+(\text{UV-finite terms}),
    \end{split}
\end{equation}
where the second line shows the UV-divergent behaviour in $\mathbf{k}$-space. One can easily verify that these divergences are equivalent to those in Eq. (\ref{field_variance_small-x_deSitter}). The idea behind dimensional regularisation is to shift the dimensions $D=3\rightarrow D=3-\epsilon$, where $\epsilon$ is some small number where the divergences will be isolated, such that the UV-divergent integral can be computed:
\begin{equation}
    \label{UV-infinite_dim_reg_integral}
    \int\dbar^3\mathbf{k} \lb\frac{1}{2a(t)^2k}+\frac{2H^2-m_B^2}{4k^3}\rb\rightarrow\mu^\epsilon\int\dbar^D\mathbf{k} \lb\frac{1}{2a(t)^2k}+\frac{2H^2-m_B^2}{4k^3}\rb,
\end{equation}
where $\mu$ is introduced as the regularisation scale. The problem is that the integral (\ref{UV-infinite_dim_reg_integral}) can't be computed analytically, even using this dimensional regularisation scheme. Instead, we shift the entire integral to $D$-dimensions such that the field variance is given by
\begin{equation}
    \label{field_variance_dim_reg}
    \expval{\hat{\phi}^2}_{DR}=\frac{\pi}{4Ha(t)^3}\int \dbar^D\mathbf{k}\abs{\mathcal{H}_{\nu}^{(1)}\lb\frac{k}{a(t)H}\rb}^2,
\end{equation}
where the subscript `DR' indicates that we have defined this integral via dimensional regularisation. Then, we can write an alternative integral that still contains the correct divergent behaviour but that can be solved by dimensional regularisation; for example,
\begin{equation}
    \label{scalar_field_variance_k_deSitter}
    \begin{split}
    \expval{\hat{\phi}^2}_{DR}=&\mu^{\epsilon}\int\dbar^Dk\qty(\frac{1}{2a(t)^2\sqrt{k^2+\delta^2}}+\frac{2H^2-m_B^2+\frac{\delta^2}{a(t)^2}}{4\qty(k^2+\delta^2)^{3/2}})
    \\&+\int\dbar^3k\qty(\frac{\pi}{4Ha(t)^3}\abs{\mathcal{H}_{\nu}^{(1)}\qty(\frac{k}{a(t)H})}^2-\frac{1}{2a(t)^2\sqrt{k^2+\delta^2}}-\frac{2H^2-m_B^2+\frac{\delta^2}{a(t)^2}}{4\qty(k^2+\delta^2)^{3/2}}),
    \end{split}
\end{equation}
where $\delta$ is an arbitrary energy scale. The first line gives the UV-divergent integral that can now be solved by dimensional regularisation. This computation will give a $\delta$-independent result. Note that the integrand will coincide with that of Eq. (\ref{UV-infinite_dim_reg_integral}) when one expands to leading order in small $\delta$. The second line in Eq. (\ref{scalar_field_variance_k_deSitter}) is the remaining UV-finite part, which can be computed numerically. Computing the divergent integral in the first line of Eq. (\ref{scalar_field_variance_k_deSitter}) and taking the limit $\epsilon\rightarrow0$, we obtain
\begin{equation}
    \label{dim_reg_integral_deSitter}
    \begin{split}
    \mu^{\epsilon}\int\dbar^Dk&\qty(\frac{1}{2a(t)^2\sqrt{k^2+\delta^2}}+\frac{2H^2-m_B^2+\frac{\delta^2}{a(t)^2}}{4\qty(k^2+\delta^2)^{3/2}})\\&=\frac{2H^2-m_B^2}{16\pi^2}\qty(\frac{2}{\epsilon}-\gamma_E+\ln(4\pi)+\ln\qty(\frac{\mu^2}{\delta^2}))-\frac{\delta^2}{16\pi^2a(t)^2}.
    \end{split}
\end{equation}
We can now absorb the divergences in the $\mathcal{O}(\lambda)$ 2-point function into the mass parameter. In the $\overline{\text{MS}}$ scheme, the renormalised mass is given by
\begin{equation}
\label{renormalised_mass_dim_reg_deSitter}
    \keybox{
    \begin{split}
        m_R^2&=m_{0,R}^2+12\xi_RH^2\\&=m_B^2+\frac{3\lambda\qty(2H^2-m_B^2)}{16\pi^2}\qty(\frac{2}{\epsilon}-\gamma_E+\ln(\frac{4\pi\mu^2}{M^2}))+\mathcal{O}\qty(\lambda^2),
    \end{split}
    }
\end{equation}
where $M$ is the renormalisation scale. Explicitly, we must renormalise both the scalar mass and the non-minimal coupling respectively as
\begin{subequations}
    \begin{align}
    \label{renormalised_mass_0}
        m_{0,R}^2&=m_{0,B}^2-\frac{3\lambda m_{0,B}^2}{16\pi^2}\qty(\frac{2}{\epsilon}-\gamma_E+\ln(\frac{4\pi\mu^2}{M^2}))+\mathcal{O}\qty(\lambda^2),\\
    \label{renormalised_non-minimal_coupling}
        \xi_R&=\xi_B+\frac{3\lambda\qty(\frac{1}{6}-\xi_B)}{16\pi^2}\qty(\frac{2}{\epsilon}-\gamma_E+\ln(\frac{4\pi\mu^2}{M^2}))+\mathcal{O}\qty(\lambda^2).
    \end{align}
\end{subequations}
We see that it is crucial to include the non-minimal coupling term for this to be a renormalisable theory. We can write the field variance in the $\overline{\text{MS}}$ scheme as
\begin{equation}
    \label{field_variance_dim_reg_deSitter}
    \begin{split}
    \expval{\hat{\phi}^2}_{\overline{MS}}=\frac{2H^2-m_R^2}{16\pi^2}\lb\frac{2}{\epsilon}-\gamma_E+\ln\lb\frac{4\pi\mu^2}{M^2}\rb\rb+\expval{\hat{\phi}^2}_{\overline{MS}}^{(fin)},  
    \end{split}
\end{equation}
where the finite piece is given by
\begin{equation}
    \label{finite_field_variance_dim_reg_deSitter}
    \begin{split}
    \expval{\hat{\phi}^2}_{\overline{MS}}^{(fin)}=&\frac{2H^2-m_R^2}{16\pi^2}\ln\qty(\frac{M^2}{\delta^2})-\frac{\delta^2}{16\pi^2a(t)^2}+\int\dbar^3k\Bigg(\frac{\pi}{4Ha(t)^3}\abs{\mathcal{H}_{\nu}^{(1)}\qty(\frac{k}{a(t)H})}^2\\&-\frac{1}{2a(t)^2\sqrt{k^2+\delta^2}}-\frac{2H^2-m_R^2+\frac{\delta^2}{a(t)^2}}{4\qty(k^2+\delta^2)^{3/2}}\Bigg)+\mathcal{O}(\lambda).   
    \end{split}
\end{equation}
The subscript `$\overline{MS}$' indicates that it is computed in the $\overline{\text{MS}}$ scheme of dimensional regularisation. Note that the explicit $M$-dependence in Eq. (\ref{finite_field_variance_dim_reg_deSitter}) will cancel with the implicit dependence in $m_R^2$. Since $m_R$ is now the mass parameter for our theory, we will keep $M$-dependence in the final result, which sets the scale of any physical observable we compute. The integral in Eq. (\ref{finite_field_variance_dim_reg_deSitter}) can be computed numerically. Since this integral goes over an infinite number of $k$-modes, which is impractical when performing the numerics, we choose a large value for $k$ at which we truncate. The only limitation of this is that we must keep $\delta$ sufficiently small so that it will not come to dominate in the large $k$ regime, where we have truncated. Since we know that the finite field variance will not depend on $\delta$, we can readily notice when this problem arises simply by checking the $\delta$-dependence explicitly. We find that, for a truncated $\frac{k}{a(t)H}=10^5$, there becomes a degree of $\delta$-dependence when $\Delta^2=\frac{\delta^2}{a(t)^2H^2}$ is of the order $10^2$. Thus, when it comes to choosing a value of $\delta$, we just need to stay below this value.

\begin{figure}
    \begin{subfigure}[ht]{0.455\textwidth}
        \includegraphics[width=\textwidth]{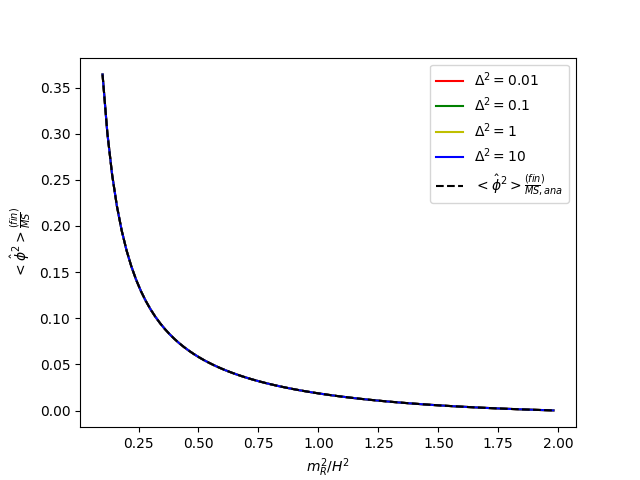}
        \caption{Numerical (solid) and analytic (black, dashed) expressions the field variance as a function of $m_R^2/H^2$.}
        \label{fig:variance_v_m2}
    \end{subfigure}
    \begin{subfigure}[ht]{0.485\textwidth}
        \includegraphics[width=\textwidth]{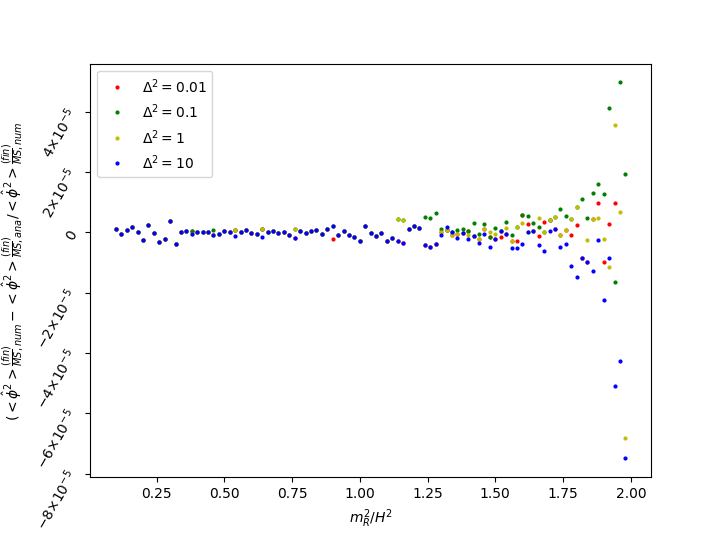}
        \caption{Relative difference between the numerical and analytic field variances as a function of $m_R^2/H^2$}
        \label{fig:ana_v_num_var_diff}
    \end{subfigure}
    \caption{Plots indicating the degree with which the analytic field variance agrees with the numerical solutions. Numerical solutions plotted for 4 different values of $\Delta^2=\qty(\frac{\delta}{a(t)H})^2$: 0.01 (red), 0.1 (green), 1 (yellow) and 10 (blue). We take $M=a(t)H$.}
    \label{fig:analytic_v_numeric_variance}
\end{figure}

We have plotted the numerical expression for the finite field variance as a function of $m_R^2/H^2$ in Fig. \ref{fig:variance_v_m2}. We see that the curves completely coincide with each other for all values of mass plotted and that there is next to no $\delta$-dependence, as expected. Note that we have plotted up to the conformal mass value $m_R^2/H^2=2$. On the QFT side, we can go beyond this to where $\nu_R$ becomes imaginary. For this work, we are content to keep the fields below this conformal mass such that we can compare the results with the stochastic theory, which requires the fields to be light.

Now that we have a numerical solution, we can make an ansatz for the analytic expression for the finite field variance by making the observation that de Sitter asymptotes to Minkowski in the limit $H\rightarrow0$. Thus, we expect the finite field variance of de Sitter to be the same as that of Minkowski in this limit. Extrapolating this, we expect the difference between the finite terms obtained via dimensional regularisation and point splitting to be of a similar form in de Sitter as to that of Eq. (\ref{finite_field_variance_difference_minkowski}), the difference in flat spacetime. Namely, we expect the difference in de Sitter to be
\begin{equation}
    \label{finite_field_variance_difference_deSitter}
    \expval{\hat{\phi}^2}^{(fin)}_{\overline{MS}}-\expval{\hat{\phi}^2}^{(fin)}_{PS}=\frac{2H^2-m_R^2}{16\pi^2}\qty(2\gamma_E-\ln4+\ln\qty(\frac{M^2}{a(t)^2H^2})),
\end{equation}
where we have used the subscript $\overline{MS}$ and $PS$ to indicate that these are the finite terms using the $\overline{\text{MS}}$ scheme of dimensional regularisation and point splitting respectively. Since we know what the finite terms are in point-splitting regularisation, see Eq. (\ref{field_variance_point-split_finite}), we can solve this to find an analytic expression for the finite part of the field variance in dimensional regularisation. Our ansatz for this expression is thus
\begin{equation}
    \label{finite_field_variance_dim_reg_deSitter_ana}
    \expval{\hat{\phi}^2}_{\overline{MS}}^{(fin)}=\frac{2H^2-m_R^2}{16\pi^2}\qty(1-\psi^{(0)}\qty(3/2-\nu_R)-\psi^{(0)}\qty(3/2+\nu_R)+\ln\qty(\frac{M^2}{a(t)^2H^2})).
\end{equation}
We can see that this reduces to the Minkowski expression, Eq. (\ref{finite_field_variance_mink_dim_reg}) in the limit $H\rightarrow0$, as expected. We can also check the accuracy of this expression by comparing it with the numerical solutions. The black, dashed curve in Fig. \ref{fig:variance_v_m2} is Eq. (\ref{finite_field_variance_dim_reg_deSitter_ana}) as a function of $m_R^2/H^2$. We can see that the agreement with the other curves is near exact. We can be more concrete about this by plotting the relative difference between the numerical and analytic results as a function of $m^2_R/H^2$, as in Fig. \ref{fig:ana_v_num_var_diff}. One can see that the relative error is very small, on the order of $10^{-5}$. Indeed, for the majority of masses the error is even smaller than this, with the largest errors occurring as $m_R^2/H^2$ approaches 2, where the field variance tends towards zero and thus the relative error is amplified. We suggest that the source of these errors is from the numerical integration. Thus, we have found the finite terms in the field variance following renormalisation by dimensional regularisation. To obtain our $\mathcal{O}(\lambda)$ 2-point functions, we simply make the replacement
\begin{equation}
\keybox{
    \label{renormalised_mass_correction}
    \begin{split}
        m^2\rightarrow& m_R^2+3\lambda\expval{\hat{\phi}^2}_{\overline{MS}}^{(fin)}
        \\\rightarrow& m_R^2+\frac{2H^2-m_R^2}{16\pi^2}\qty(1-\psi^{(0)}\qty(3/2-\nu_R)-\psi^{(0)}\qty(3/2+\nu_R)+\ln\qty(\frac{M^2}{a(t)^2H^2})).
    \end{split}
    }
\end{equation}

\subsubsection{The infrared problem in perturbative QFT}

We can now explicitly write the UV-renormalised scalar 2-point function to $\mathcal{O}(\lambda)$. Using Eq. (\ref{renormalised_mass_correction}), the 2-point function is given by
\begin{equation}
    \keybox{
    \label{O(lambda)_2-pt_function_renormalised}
    \bra{0}T\hat{\phi}(x_1)\hat{\phi}(x_2)\ket{0}=i\Delta^F(x_1,x_2)\eval_{m^2\rightarrow m_R^2+3\lambda \expval{\hat{\phi}^2}_{\overline{MS}}^{(fin)}}.
    }
\end{equation}
For comparison with the stochastic approach, we will be interested in the long-distance behaviour of the 2-point function. Focussing on spacelike separations, the leading term in the asymptotic expansion about long distances is
\begin{equation}
    \label{O(lambda)_2-pt_function_spacelike_long-distance}
    \begin{split}
    \bra{0}\hat{\phi}(t,\mathbf{0})\hat{\phi}(t,\mathbf{x})\ket{0}=&\frac{H^2}{16\pi^2}\frac{\Gamma\qty(3/2-\nu_R)\Gamma\qty(2\nu_R)4^{3/2-\nu_R}}{\Gamma\qty(1/2+\nu_R)}\Bigg[1+\frac{3 \lambda \qty(2H^2-m_R^2)}{32\pi^2\nu_RH^2}\\&\times\qty(\ln4+\psi^{(0)}\qty(3/2-\nu_R)-2\psi^{(0)}\qty(2\nu_R)+\psi^{(0)}\qty(1/2+\nu_R))\\&\times\qty(1-\psi^{(0)}\qty(3/2-\nu_R)-\psi^{(0)}\qty(3/2+\nu_R)+\ln\qty(\frac{M^2}{a(t)^2H^2}))\Bigg]\\&\times\abs{Ha(t)\mathbf{x}}^{-\frac{2\Lambda_1^{(QFT)}}{H}}+\mathcal{O}(\lambda^2),
    \end{split}
\end{equation}
where the exponent\footnote{We introduce the notation $\Lambda_1^{(QFT)}$ as a precursor to that used for the spectral expansion method in the stochastic theories.} is
\begin{equation}
    \label{QFT_exponent}
    \begin{split}
    \Lambda_1^{(QFT)}=&\qty(\frac{3}{2}-\nu_R)H+\frac{3\lambda (2H^2-m_R^2)}{32\pi^2\nu_RH}\Bigg(1-\psi^{(0)}\qty(3/2-\nu_R)-\psi^{(0)}\qty(3/2+\nu_R)\\&+\ln\qty(\frac{M^2}{a(t)^2H^2})\Bigg)+\mathcal{O}\qty(\lambda^2).
    \end{split}
\end{equation}
This is a first-order result in perturbation theory. In theory, we can now repeat the process for higher orders in $\lambda$, albeit with increasing levels of complexity. However, there is a problem that stems from the IR limit hidden amongst our results. To see this, we expand the 2-point function (\ref{O(lambda)_2-pt_function_spacelike_long-distance}) to leading order in light fields $m^2\ll H^2$ to give
\begin{equation}
    \label{O(lambda)_2-point_function_light_fields}
    \bra{0}\hat{\phi}(t,\mathbf{0})\hat{\phi}(t,\mathbf{x})\ket{0}=\lb\frac{3H^4}{8\pi^2m^2_R}-\frac{27\lambda H^8}{64\pi^4m^6_R}\rb\abs{Ha(t)\mathbf{x}}^{-\frac{2m^2_R}{3H^2}-\frac{3\lambda H^2}{4\pi^2m_R^2}}.
\end{equation}
One can see that both corrections to the amplitude and exponent are of relative order $\frac{\lambda H^4}{m_R^4}$ in this light field expansion. In order for the sum to converge at higher orders in $\lambda$, we require $\frac{\lambda H^4}{m_R^4}\ll1$ \footnote{Note that this is an IR effect because it stems from the lower limit of the $k$-integral for the field variance (\ref{k-space_field_variance}).}. This is not a priori true since our perturbation theory takes $\lambda$ to be the small parameter about which we expand. Thus, perturbative QFT is limited to the following region in the parameter space:
\begin{equation}
    \label{perturbative_QFT_limits}
    \lambda\ll1 \quad , \quad \lambda\ll m^4/H^4.
\end{equation}
The blue left-hashed region in Fig. \ref{fig:model_limitations_comparison} shows the region in the parameter space where perturbative QFT is valid. Therefore, to leading order in $\lambda H^4/m^4$, the long-distance behaviours of the spacelike field correlators (including the $\phi-\pi$ and $\pi-\pi$ correlators for completeness) are
\begin{subequations}
    \label{O(lambdaH4m4)_spacelike_QFT_correlators}
    \begin{align}
    \label{phi-phi_O(lambdaH4m4}
        \begin{split}
        \bra{0}\hat{\phi}(t,\mathbf{0})\hat{\phi}(t,\mathbf{x})\ket{0}=&\lb\frac{H^2}{16\pi^2}\frac{\Gamma\qty(3/2-\nu_R)\Gamma\qty(2\nu_R)4^{3/2-\nu_R}}{\Gamma\qty(1/2+\nu_R)}-\frac{27\lambda H^8}{64\pi^4m^6_R}+\mathcal{O}\lb\frac{\lambda H^6}{m^4_R}\rb\rb\\&\times\abs{Ha(t)\mathbf{x}}^{-(3-2\nu_R)-\frac{3\lambda H^2}{4\pi^2m_R^2}+\mathcal{O}(\lambda)},
        \end{split}\\
        \begin{split}
            \bra{0}\hat{\phi}(t,\mathbf{0})\hat{\pi}(t,\mathbf{x})\ket{0}=&\lb-\frac{H^3}{16\pi^2}\frac{\alpha\Gamma(2\nu)\Gamma\qty(\alpha)4^\alpha}{\Gamma\qty(\frac{1}{2}+\nu)}+\mathcal{O}\lb\frac{\lambda H^5}{m^2_R}\rb\rb\\&\times\abs{Ha(t)\mathbf{x}}^{-(3-2\nu_R)-\frac{3\lambda H^2}{4\pi^2m_R^2}+\mathcal{O}(\lambda)},
        \end{split}\\
        \begin{split}
            \bra{0}\hat{\pi}(t,\mathbf{0})\hat{\pi}(t,\mathbf{x})\ket{0}=&\lb\frac{H^4}{16\pi^2}\frac{\alpha^2\Gamma(2\nu)\Gamma\qty(\alpha)4^\alpha}{\Gamma\qty(\frac{1}{2}+\nu)}+\frac{3\lambda H^6}{64\pi^4m^2_R}+\mathcal{O}(\lambda H^4)\rb\\&\times\abs{Ha(t)\mathbf{x}}^{-(3-2\nu_R)-\frac{3\lambda H^2}{4\pi^2m_R^2}+\mathcal{O}(\lambda)}.
        \end{split}        
    \end{align}
\end{subequations}

This is as far as perturbative QFT will take us for 2-point correlation functions. In order to go beyond this, we must employ alternative methods, such as the stochastic effective theory of scalar fields in de Sitter. 

\subsection{Four-point QFT correlation functions to $\mathcal{O}(\lambda)$}
\label{subsec:4-pt_functions}

To round out our discussion of perturbative QFT, we will briefly consider the 4-pt functions. In this thesis, I will largely consider them as a tool for computing stochastic parameters and so won't go into a huge amount of detail. However, it is important to recognise that these objects are computationally challenging and our work has raised some questions surrounding this, which I will touch on at the end of this section.

Using the Schwinger-Keldysh formalism outlined in Sec. \ref{subsec:schwinger-keldysh}, we can combine Eq. (\ref{in-in_scalar_correlators}) with (\ref{free_in-in_generating_functional}) and (\ref{interacting_generating_functional}) to obtain the time-ordered 4-pt scalar correlation function to $\mathcal{O}(\lambda)$ as
\begin{equation}
    \label{4-pt_function_Scwhinger-Keldysh}
    \begin{split}
        \bra{0}T\hat{\phi}(x_1)&\hat{\phi}(x_2)\hat{\phi}(x_3)\hat{\phi}(x_4)\ket{0}\\=&i\Delta^F(x_1,x_2)i\Delta^F(x_3,x_4)+[\text{2 more permutations}]
        \\&
        \begin{split}
        +3i\lambda\expval{\hat{\phi}^2}i\Delta^F(x_3,x_4)\int d^4z a(t_z)^3( &i\Delta^F(z,x_1)i\Delta^F(z,x_2)\\&-\Delta^+(z,x_1)\Delta^+(z,x_2))
        \end{split}
        \\&+[\text{5 more permutations}]
        \\&
        \begin{split}
        -6i\lambda\int d^4z a(t_z)^3(& i\Delta^F(z,x_1)i\Delta^F(z,x_2)i\Delta^F(z,x_3)i\Delta^F(z,x_4)\\&-\Delta^+(z,x_1)\Delta^+(z,x_2)\Delta^+(z,x_3)\Delta^+(z,x_4)).
        \end{split}
    \end{split}
\end{equation}
The first line after the equality sign is the free part, composed of a combination of Feynman propagators. The next three lines are a similar combination but this time occur at $\mathcal{O}(\lambda)$, thus containing the $\mathcal{O}(\lambda)$ piece of the 2-pt function (\ref{one-loop_UVdiv_2-pt_function}), multiplied by the free Feynman propagator. The final lines indicate the new contribution to the 4-pt function that first appears at $\mathcal{O}(\lambda)$. It is often referred to as the connected piece in Minkowski, stemming from its diagrammatic representation\footnote{The use of Feynman diagrams in de Sitter is trickier than in Minkowski because the in-in formalism - which is now preferable over in-out - introduces many more 2-pt functions as ``legs''. It can be done - see Ref. \cite{Garbrecht:2011,Garbrecht:2014,Garbrecht:2015,Bounakis:2020} - but it is slightly less intuitive.}. For the purposes of this work, I will focus on these final lines and will refer to the object as the \textit{connected 4-point function}. Explicitly, this is related to the 2-point functions by
\begin{equation}
    \label{connected_4-pt_func_QFT}
    \begin{split}
    \bra{0}T\hat{\phi}(x_1)\hat{\phi}(x_2)\hat{\phi}(x_3)\hat{\phi}(x_4)\ket{0}_C=&\bra{0}T\hat{\phi}(x_1)\hat{\phi}(x_2)\hat{\phi}(x_3)\hat{\phi}(x_4)\ket{0}\\&-\bra{0}T\hat{\phi}(x_1)\hat{\phi}(x_2)\ket{0}\bra{0}T\hat{\phi}(x_3)\hat{\phi}(x_4)\ket{0}\\&-\bra{0}T\hat{\phi}(x_1)\hat{\phi}(x_3)\ket{0}\bra{0}T\hat{\phi}(x_2)\hat{\phi}(x_4)\ket{0}\\&-\bra{0}T\hat{\phi}(x_1)\hat{\phi}(x_4)\ket{0}\bra{0}T\hat{\phi}(x_2)\hat{\phi}(x_3)\ket{0},
    \end{split}
\end{equation}

where the subscript `\textit{C}' stands for `connected'. To compute this quantity, we must perform the $z$-integral. Since the integrand is composed of a series of hypergeometric functions, doing an analytic calculation is extremely difficult. Moreover, attempts at a numerical computation have proved fruitless due to the existence of poles in the integrand. Indeed, these poles stem from the same source as those that arise in the 2-pt calculation; see Appendix \ref{app:numerical_calculation_correlators} for details. Unfortunately, unlike its 2-pt counterpart, the 4-pt integral cannot be solved by a mass redefinition. We can make some progress by moving from position to momentum $\mathbf{k}$-space, as outlined in Ref. \cite{Serreau:2013,Serreau:2014,Nacir:2019}, where computations simplify and the pole structure is no longer a problem. For this purpose, we will focus on equal-time 4-pt functions. The Fourier transform goes as
\begin{equation}
    \label{fourier_transform}
    G^{(4)}_C(\eta,\{\mathbf{x}_i\})=\prod_{i=1}^4\lsb\int\dbar^3\mathbf{k}_ie^{-i\mathbf{k_i}\cdot\mathbf{x}_i}\rsb \Tilde{G}^{(4)}_C(\eta,\{\mathbf{k}_i\})\deltabar^{(3)}\lb\sum_{i=1}^4\mathbf{k}_i\rb,
\end{equation}
where we use the shorthand notation for the connected 4-pt function $G^{(4)}_C(\eta,\{\mathbf{x}_i\})$, with $\{\mathbf{x}_i\}=(\mathbf{x}_1,\mathbf{x}_2,\mathbf{x}_3,\mathbf{x}_4)$, and the `tilde' indicates equivalent quantities in $\mathbf{k}$-space. Note that we will use conformal time in the following calculations, as defined in Eq. (\ref{conformal_time}). The equal-time connected 4-pt function in $\mathbf{k}$-space is given by
\begin{equation}
    \label{4-pt_func_k-space_general}
    \begin{split}
    \Tilde{G}^{(4)}_C(\eta,\{\mathbf{k}_i\})=&-6i\lambda\int^0_{-\infty} d\eta_z \frac{1}{(H\eta_z)^4}\\&
    \begin{split}
    \times\Bigg(&i\Tilde{\Delta}^F(\eta_z,\eta,\mathbf{k}_1)i\Tilde{\Delta}^F(\eta_z,\eta,\mathbf{k}_2)i\Tilde{\Delta}^F(\eta_z,\eta,\mathbf{k}_3)i\Tilde{\Delta}^F(\eta_z,\eta,\mathbf{k}_4)\\&-\Tilde{\Delta}^+(\eta_z,\eta,\mathbf{k}_1)\Tilde{\Delta}^+(\eta_z,\eta,\mathbf{k}_2) \Tilde{\Delta}^+(\eta_z,\eta,\mathbf{k}_3) \Tilde{\Delta}^+(\eta_z,\eta,\mathbf{k}_4)\Bigg), 
    \end{split}
    \end{split}
\end{equation}
where the Wightman function in $\mathbf{k}$-space is given by Eq. (\ref{k-space_2-pt_func}) and the Feynman propagator is found by using its definition in Table \ref{tab:2-pt_functions}. Using the time-ordering definition (\ref{time-ordered_operator}), one can simplify the integral such that
\begin{equation}
    \label{4-pt_func_k-space_Wightman}
    \begin{split}
    \Tilde{G}^{(4)}_C(\eta,\{\mathbf{k}_i\})=&-6i\lambda\int^\eta_{-\infty} d\eta_z \frac{1}{(H\eta_z)^4}\\&
    \begin{split}
    \times\Bigg(&\Tilde{\Delta}^-(\eta_z,\eta,\mathbf{k}_1)\Tilde{\Delta}^-(\eta_z,\eta,\mathbf{k}_2)\Tilde{\Delta}^-(\eta_z,\eta,\mathbf{k}_3)\Tilde{\Delta}^-(\eta_z,\eta,\mathbf{k}_4)\\&-\Tilde{\Delta}^+(\eta_z,\eta,\mathbf{k}_1)\Tilde{\Delta}^+(\eta_z,\eta,\mathbf{k}_2) \Tilde{\Delta}^+(\eta_z,\eta,\mathbf{k}_3) \Tilde{\Delta}^+(\eta_z,\eta,\mathbf{k}_4)\Bigg), 
    \end{split}
    \end{split}
\end{equation}
where $\Tilde{\Delta}^-(\eta_z,\eta,\mathbf{k})$ is the complex conjugate of $\Tilde{\Delta}^+(\eta_z,\eta,\mathbf{k})$.

This integral is hard to solve in general. Analytic solutions are difficult because the integrand is a product of Hankel functions whilst the oscillatory behaviour of the integrand in the limit $\eta_z\rightarrow-\infty$ make numerical computations challenging. However, for the purposes of this thesis, we are interested in the leading IR behaviour of this quantity so we can make use of the asymptotic expansion of the Hankel functions to write the Wightman functions in the limit $-k\eta\ll 1$ as
\begin{equation}
    \label{Wightman_k-space_Hankels_kll1}
    \begin{split}
    \Tilde{\Delta}^{\pm}(\eta_z,\eta,\mathbf{k})\simeq\frac{\pi}{4Ha(\eta)^{3/2}a(\eta_z)^{3/2}}\mathcal{H}_{\nu_R}^{(1)/(2)}(-k\eta_z)\Bigg(&\frac{2^{-\nu_R}}{\nu_R\Gamma(\nu_R)}(-k\eta)^{\nu_R}\\&\pm i\frac{2^{\nu_R}\Gamma(\nu_R)}{\pi}(-k\eta_z)^{-\nu_R}\Bigg),
    \end{split}
\end{equation}
which can be plugged into Eq. (\ref{4-pt_func_k-space_Wightman}) to obtain a simpler integrand. Note that we introduce the momentum scale $k$ to represent the order of $k_i$ $\forall i\in\{1,2,3,4\}$. 

This integral is still not easy to solve. To go further, we will focus on the upper bound of the integrand, such that we only perform the integral up to some intermediate time $\eta_0=-\frac{\Lambda}{k}$, where $\Lambda\ll1$ such that $-k\eta_0\ll1$ i.e. $\int_{-\infty}^\eta=\int_{\eta_0}^\eta+\int^{\eta_0}_{-\infty}\sim\int_{\eta_0}^\eta$ in the IR limit. Then, we can make the approximation $-k\eta_z\ll1$ and thus write the Wightman functions as 
\begin{equation}
    \label{k_Wightman_IR_lim}
    \begin{split}
    \Tilde{\Delta}^\pm(\eta_z,\eta,\mathbf{k})\simeq \frac{\pi}{4Ha(\eta)^{3/2}a(\eta_z)^{3/2}}\Bigg[&\frac{4^{\nu_R}\Gamma(\nu_R)^2}{\pi^2}\lb\eta_z\eta\rb^{-\nu_R}k^{-2\nu_R}\\&\pm i\frac{1}{\pi\nu_R}\lb\lb\frac{\eta}{\eta_z}\rb^{\nu_R}-\lb\frac{\eta_z}{\eta}\rb^{\nu_R}\rb\Bigg].
    \end{split}
\end{equation}
Plugging Eq. (\ref{k_Wightman_IR_lim}) into (\ref{4-pt_func_k-space_Wightman}), one can compute the integral to give
\begin{equation}
    \label{k-space_4-pt_func}
    \begin{split}
    \Tilde{G}_C^{(4)}(\eta,\{\mathbf{k}_i\})&\simeq \mathcal{O}(k^{-3-4\nu_R})-\frac{3\lambda}{4H^5}\lb\frac{4^{\nu_R}\Gamma(\nu_R)^2}{2\pi}\rb^3\frac{1}{(3-4\nu_R)(3-2\nu_R)}(-H\eta)^{9-6\nu_R}
    \\&\times\lsb\lb\frac{k_1k_2k_3}{H^3}\rb^{-2\nu_R}+\lb\frac{k_1k_3k_4}{H^3}\rb^{-2\nu_R}+\lb\frac{k_2k_3k_4}{H^3}\rb^{-2\nu_R}+\lb\frac{k_1k_2k_4}{H^3}\rb^{-2\nu_R}\rsb,
    \end{split}
\end{equation}
where the extra contribution $\mathcal{O}(k^{-3-4\nu_R})$ comes from the $\eta_0$ limit. This is actually the leading contribution in the IR limit for light fields, over the term $\mathcal{O}(k^{-6\nu_R})$. Note that this is the only other dominant IR contribution that enters the integral; all other contributions are subdominant to $k^{-6\nu_R}$. For a deeper discussion of this, see Appendix \ref{app:IR_lim_4-pt_func}, where we discuss the limit $\eta\rightarrow-\infty$ more carefully.

It is challenging to get analytic results for the leading term $\mathcal{O}(k^{-3-4\nu_R})$, especially in coordinate space, because it will depend on all 4 $k_i$s simultaneously and thus the $\delta$-function arising in the Fourier transform (\ref{fourier_transform}) will result in a mixing of momenta. On the other hand, the $\mathcal{O}(k^{-6\nu_R})$ contribution will deal with the $\delta$-function trivially because each term only ever depends on 3 of the 4 momenta. For this thesis, it is sufficient to have an analytic expression for one of the leading IR terms so that we can do a comparison with the stochastic approach. However, this does leave the door open for more careful analysis of the 4-pt functions.

To convert Eq. (\ref{k-space_4-pt_func}) to coordinate space, we can use the Fourier transform (\ref{fourier_transform}), using the result\cite{Nacir:2019}
\begin{equation}
    \label{Nacir_identity}
    \int \dbar^3\mathbf{k}e^{-i\mathbf{k}\cdot\mathbf{x}}k^{w-3}=\frac{1}{(2\pi)^3}\frac{2^{3-2\nu_R}\pi^{3/2}\Gamma\lb\frac{5}{2}-\nu_R\rb}{\lb\frac{3}{2}-\nu_R\rb\Gamma(\nu_R)}x^{-w},
\end{equation}
to obtain the equal-time connected 4-pt function in coordinate space as
\begin{equation}
    \label{spacelike_quantum_4-pt_func_filler}
    \begin{split}
    \bra{0}\hat{\phi}(t,\mathbf{x}_1)&\hat{\phi}(t,\mathbf{x}_2)\hat{\phi}(t,\mathbf{x}_3)\hat{\phi}(t,\mathbf{x}_4)\ket{0}\\&=...+\frac{3\lambda H^4}{4\pi^{15/2}}\frac{\Gamma(\nu_R)^3\Gamma\lb\frac{5}{2}-\nu_R\rb^3}{\lb4\nu_R-3\rb\lb3-2\nu_R\rb^4}\\&\times\Bigg(\abs{Ha(t)(\mathbf{x}_1-\mathbf{x}_4)}^{-3+2\nu_R}\abs{Ha(t)(\mathbf{x}_2-\mathbf{x}_4)}^{-3+2\nu_R}\abs{Ha(t)(\mathbf{x}_3-\mathbf{x}_4)}^{-3+2\nu_R}\\&+\abs{Ha(t)(\mathbf{x}_1-\mathbf{x}_2)}^{-3+2\nu_R}\abs{Ha(t)(\mathbf{x}_3-\mathbf{x}_2)}^{-3+2\nu_R}\abs{Ha(t)(\mathbf{x}_4-\mathbf{x}_2)}^{-3+2\nu_R}\\&+\abs{Ha(t)(\mathbf{x}_2-\mathbf{x}_1)}^{-3+2\nu_R}\abs{Ha(t)(\mathbf{x}_3-\mathbf{x}_1)}^{-3+2\nu_R}\abs{Ha(t)(\mathbf{x}_4-\mathbf{x}_1)}^{-3+2\nu_R}\\&+\abs{Ha(t)(\mathbf{x}_1-\mathbf{x}_3)}^{-3+2\nu_R}\abs{Ha(t)(\mathbf{x}_2-\mathbf{x}_3)}^{-3+2\nu_R}\abs{Ha(t)(\mathbf{x}_4-\mathbf{x}_3)}^{-3+2\nu_R}\Bigg),
    \end{split}
\end{equation}
where the `$...+$' indicates the other leading IR contribution $\mathcal{O}(\abs{Ha(t)\mathbf{x}}^{-6+4\nu_R})$. For $\abs{\mathbf{x}}=\abs{\mathbf{x}_i-\mathbf{x}_j}$ $\forall i\ne j$, $i,j\in\{1,2,3,4\}$, the equal-time connected 4-pt function is given by
\begin{equation}
    \label{spacelike_quantum_4-pt_func}
    \keybox{
    \bra{0}\hat{\phi}(t,\mathbf{x}_1)\hat{\phi}(t,\mathbf{x}_2)\hat{\phi}(t,\mathbf{x}_3)\hat{\phi}(t,\mathbf{x}_4)\ket{0}=...+\frac{3\lambda H^4}{\pi^{15/2}}\frac{\Gamma(\nu_R)^3\Gamma\lb\frac{5}{2}-\nu_R\rb^3}{\lb4\nu_R-3\rb\lb3-2\nu_R\rb^4}\abs{Ha(t)\mathbf{x}}^{-9+6\nu_R},
    }
\end{equation}
 Note that, for light fields, this contribution is given by 
\begin{equation}
    \label{spacelike_quantum_4-pt_func_light_fields}
    \bra{0}\hat{\phi}(t,\mathbf{x}_1)\hat{\phi}(t,\mathbf{x}_2)\hat{\phi}(t,\mathbf{x}_3)\hat{\phi}(t,\mathbf{x}_4)\ket{0}\eval_{m^2\ll H^2}=...+\frac{81\lambda H^{12}}{128\pi^6m_R^8}\abs{Ha(t)\mathbf{x}}^{-\frac{2m^2_R}{H^2}}.
\end{equation}
\chapter{Overdamped Stochastic Theory of Near-Massless Scalar Fields in de Sitter}
\label{ch:od_stochastic}

\section{Introduction to the stochastic approach}
\label{stochastic_approach_intro}

In the previous chapter, I discussed in some detail how to compute scalar correlation functions via perturbation theory of QFT in de Sitter spacetime. I showed that, for a scalar field with mass $m$ and quartic self-interaction $\lambda$, the perturbative sum will only converge when $\lambda\ll m^4/H^4$. This limitation of perturbation theory has led physicists to explore other methods to perform computations in QFT in de Sitter \cite{Hu:1987,boyanovsky:2006,Boyanovsky:2012,Boyanovsky:2016,Arai:2012,Serreau:2011,Gautier:2013,Gautier:2015,Herranen:2014_2,Guilleux:2015,Guilleux:2016,Nacir:2016,Nacir:2019,Akhmedov:2012,Youssef:2014,Burgess:2010,Kaya:2013,van_der_Meulen:2007}. The focus of the next two chapters will be one such method: the stochastic approach. Pioneered by Starobinsky and Yokoyama in the late 80s \cite{starobinsky:1986,Starobinsky-Yokoyama:1994}, it has become a common method through which one can study inflation \cite{morikawa:1990,Finelli:2009,Finelli:2010,Vennin:2015,Grain:2017,Pattison:2019,Hardwick:2019,Tokuda:2018,Tokuda:2018_2,Glavan:2018,Cruces:2019,Firouzjahi:2019,Pinol:2019,Pinol:2020,Moreau:2020,Moreau:2020_2,Garbrecht:2011,Garbrecht:2014,Garbrecht:2015,Rigopoulos:2013,Rigopoulos:2016,Moss:2017,Prokopec:2018,Bounakis:2020,Andersen:2022,Markkanen:2019,Markkanen:2020,Tomberg:2022}. The idea is this:

\begin{equation*}
\keybox{
    \begin{split}
    \text{ Quantum beh}& \text{aviour is summarised as a statistical contribution }\\
    &\text{to the classical equations of motion.}
    \end{split}
}
\end{equation*}

The existence of the de Sitter horizon naturally gives rise to a separation between long and short wavelength modes. Due to the expanding spacetime, the short wavelength modes are continuously stretched to become superhorizon, joining their long wavelength compatriots. The long wavelength modes are also amplified and it can be shown that, after a sufficient period of time, a sufficient amount of squeezing occurs such that these modes can be considered classical \cite{Albrecht:1994,Martin:2022}. Thus, if one wishes to consider the behaviour of long wavelength modes - as one does for inflationary observables - one can focus on the classical equations of motion. To make this statement more concrete, we consider the $\phi-\pi$ correlator in $k$-space, which is found to be
\begin{equation}
    \label{phi-pi_k-space}
    \begin{split}
    \phi_k^*(t)\pi_k(t)=&-\frac{\pi}{8Ha(t)^3}\mathcal{H}^{(2)}_\nu\lb\frac{k}{a(t)H}\rb\\&\times\lsb3H\mathcal{H}^{(1)}_\nu\lb\frac{k}{a(t)H}\rb+\frac{k}{a(t)}\lb\mathcal{H}^{(1)}_{\nu-1}\lb\frac{k}{a(t)H}\rb-\mathcal{H}_{\nu+1}^{(1)}\lb\frac{k}{a(t)H}\rb\rb\rsb
    \end{split}
\end{equation}
where we have used Eq. (\ref{mode_functions}). Expanding to leading order in $k\ll a(t)H$,
\begin{equation}
    \label{phi-pi_k-space_IR_lim}
    \phi_k^*(t)\pi_k(t)=\frac{2^{-3+2\nu}\Gamma(\nu)\lb-3\Gamma(\nu)+2\Gamma(1+\nu)\rb}{a(t)^3\pi}k^{-2\nu}-\frac{i}{2a(t)^3},
\end{equation}
one can see that the real part will dominate over the imaginary part in the IR regime because we have a growing mode $\mathcal{O}(k^{-2\nu})$; the modes are squeezed such that they are approximately 1-dimensional. Thus, the classical behaviour of the field modes, which is governed by the symmetric part, will come to dominate over the quantum behaviour, which is governed by the antisymmetric part (\ref{wronskian_normalisation});. Note that this happens quickest for massless fields; the growing behaviour is then $\sim k^{-3}$. As the mass increases, the rate of growth decreases until one goes beyond $m^2/H^2=3/2$, when it is no longer a power law but instead oscillatory. Thus, the region where one can consider the long wavelength modes to be classical moves deeper into the IR for more massive fields. Once the behaviour becomes oscillatory, the fields are too heavy and they can no longer be approximated as classical, even in the far IR.

The Starobinsky-Yokoyama stochastic approach goes a step further by making the approximation that the long and short wavelengths are decoupled, with the only contribution from short-distance behaviour coming from modes crossing the horizon in the long-distance regime. As it turns out, this coarse-graining of the horizon-crossing modes contributes a statistical noise term to the (classical) long wavelength mode equation. Thus, the stochastic approach is considered a long-distance, or infrared (IR), effective theory of scalar QFT in de Sitter.

From this basic description, we can immediately observe some limitations to our stochastic approach. The first is that we require the fields to be sufficiently light, $m\lesssim H$, so that the expanding spacetime amplifies the modes to become classical. We can be a little more precise; modes are not amplified at the conformal mass so we require $m<\sqrt{2}H$ for there to be any degree of amplification. Since we need this amplification to be sufficiently strong, we can infer that $m\lesssim H$ is a reasonable approximation for the stochastic approach. The other limitation of the stochastic approach is that it will never include long-distance quantum effects, such as entanglement, because we assume decoupling between the short and long wavelength modes. Therefore, there is a limit to the types of behaviour one can expect to compute using this approach.

However, there are certainly many reasons to use the stochastic approach. The main attraction is that the stochastic equations can be solved non-perturbatively, probing a parameter space that cannot be reached using perturbative QFT. Additionally, stochastic equations are significantly more straightforward to solve than those in QFT; indeed, it is for this very reason that non-perturbative methods are available. As we will discuss in more detail in this chapter, the stochastic equations for a scalar field in de Sitter are of the same form as that of Brownian motion, which has been extensively studied: for example, see this textbook on the subject \cite{pavliotis_txtbk:2014s}.

\section{Introduction to Chapter \ref{ch:od_stochastic}}
\label{sec:od_stochastic_intro}

The original stochastic approach, proposed by Starobinsky and Yokoyama \cite{starobinsky:1986,Starobinsky-Yokoyama:1994}, was considered in the context of single-field inflation, where the inflaton exists in slow-roll. As we saw in Chapter \ref{ch:inflation}, using such an approximation simplifies the inflaton's equations of motion to first-order differential equations. One can then introduce a strict cut-off between sub- and superhorizon modes, from which stochastic equations can be derived from the underlying QFT. This approach has since been extensively used in many inflationary scenarios, both for the inflaton and spectator fields during inflation (see the references in the previous section) to great effect. I dub this stochastic approach the \textit{overdamped stochastic approach} to distinguish it from the second-order stochastic approach I will introduce in the next chapter.

In this chapter, I will outline the overdamped stochastic approach, starting from the derivation of the overdamped stochastic equations and continuing to compute correlation functions, to be compared with the results from perturbative QFT. This chapter can be considered a literature review of the overdamped stochastic approach. In Sec. \ref{sec:od_stochastic_eq}, I will derive the overdamped stochastic equations using the same method that Starobinsky and Yokoyama introduced in their original work \cite{starobinsky:1986,Starobinsky-Yokoyama:1994}. I will then introduce the Fokker-Planck equation - the equation describing the time-evolution of the probability distribution function - in Sec. \ref{sec:od_fokker-planck} and solve it via a spectral expansion, with both perturbative, analytical and non-perturbative, numerical results. I have largely followed the work of Ref. \cite{Markkanen:2019} for this, though I have tweaked the spectral expansion to align with the methods in the next chapter. Finally in Sec. \ref{sec:od_correlators}, I compute the stochastic correlation functions and compare their perturbative form to that of perturbative QFT, showing that they agree. 

\section{The overdamped stochastic equation}
\label{sec:od_stochastic_eq}

We will begin with the same theory as the one considered in Chapter \ref{ch:qft_dS}; namely a spectator scalar field $\phi$ with a scalar potential $V(\phi)=\frac{1}{2}m^2\phi^2+\frac{1}{4}\lambda\phi^4$ in a de Sitter spacetime, which obeys the equation of motion (\ref{scalar_eom})
\begin{equation}
    \label{scalar_eom_repeated}
    \Ddot{\phi}+3H\Dot{\phi}-\frac{1}{a(t)^2}\nabla^2\phi+V'(\phi)=0,
\end{equation}
where the term involving the non-minimal coupling between the scalar field and gravity, $\frac{1}{2}\xi R\phi^2$, is absorbed into the scalar potential as an additional mass term. The parameters $m$, $\xi$ and $\lambda$ are the same ones as in QFT. Notably, when we have our interactions turned on ($\lambda\ne0$), $m$ and $\xi$ will be the renormalised quantities (\ref{renormalised_mass_dim_reg_deSitter}), though I will drop the subscript `$R$' here. For this chapter, we will consider near-massless fields $m\ll H$.

One can then quantise this theory in the same way as outlined in Sec. \ref{subsec:quantisation}. The field operator is given by
\begin{equation}
    \label{field_operator}
    \begin{split}
    \hat{\phi}(t,\mathbf{x})&=\int\dbar^3\mathbf{k}\hat{\phi}_k(t,\mathbf{x})
    \\&=\int \dbar^3\mathbf{k}\lb\hat{a}_k\phi_k(t)e^{-i\mathbf{k}\cdot\mathbf{x}}+\hat{a}_k^{\dagger}\phi_k^*(t)e^{i\mathbf{k}\cdot\mathbf{x}}\rb,
    \end{split}
\end{equation}
where the nearly-massless field modes are given in Eq. (\ref{phi_k_mode_massless}), repeated here for ease, as
\begin{equation}
    \label{phi_k_mode_od}
    \phi_k(t)=-\frac{H}{\sqrt{2k^3}}\lb\frac{k}{a(t)H}+i\rb e^{i\frac{k}{a(t)H}}.
\end{equation}
In the seminal work of Starobinsky and Yokoyama \cite{starobinsky:1986,Starobinsky-Yokoyama:1994}, the stochastic equations are derived by splitting the field modes explicitly via the introduction of a window function $W_k(t)$ into long, superhorizon modes and short, subhorizon modes:
\begin{equation}
    \label{mode_split}
    \hat{\phi}=\phi_L+\hat{\phi}_S.
\end{equation}
The long wavelength modes $\phi_L$ are considered to be classical, while the short wavelength modes, which remain quantum, are given by
\begin{equation}
    \label{short_modes_window_function}
    \hat{\phi}_S(t,\mathbf{x})=\int\dbar^3\mathbf{k}W_k(t)\hat{\phi}_k(t,\mathbf{x}),
\end{equation}
such that the integral only contains the $k>a(t)H$ (subhorizon) modes, courtesy of the window function. These modes are also to be approximated as non-interacting, such that $V(\phi_S)=\frac{1}{2}m^2\phi_S^2$. 

For this chapter, we will consider the \textit{slow-roll} (SR) or \textit{overdamped} (OD) approximations\footnote{The differing terminology depends on which angle you are coming from. The term `slow-roll' comes directly from inflationary cosmology, as discussed. The term `overdamped' stems from stochastic processes in Brownian motion, where a particle exists in a friction-dominated environment. Equivalently, this is the overdamped case of a simple harmonic oscillator.}
\begin{equation}
    \label{slow-roll_approx}
    \Ddot{\phi}\ll 3H\Dot{\phi} \quad , \quad V''(\phi)\ll H^2.
\end{equation}
The latter condition gives us two constraints on our theory's parameters, which will limit the OD stochastic approach to the following region of the parameter space:
\begin{equation}
    \label{od_stochastic_limits}
    m\ll H \quad , \quad \lambda \ll m^2/H^2.
\end{equation}
This region is represented in Fig. \ref{fig:model_limitations_comparison} by the green, right-hashed region. Additionally, we must drop the gradient term in the equations of motion\footnote{In $\mathbf{k}$-space, this term will be $\frac{k^2}{a(t)^2}\phi$, which is small for the long wavelength modes.}. Using this approximation, Eq. (\ref{scalar_eom_repeated}) becomes
\begin{equation}
    \label{od_scalar_eom}
    3H\Dot{\phi}+V'(\phi)=0.
\end{equation}

Substituting Eq. (\ref{mode_split}) into the OD equation (\ref{od_scalar_eom}), we obtain
\begin{equation}
    \label{od_stochastic_eq_deriv}
    3H\Dot{\phi}_L+V'(\phi_L)+\hat{\xi}_{OD}(t,\mathbf{x})=0,
\end{equation}
where
\begin{equation}
    \label{od_xi_quantum}
    \hat{\xi}_{OD}(t,\mathbf{x})=3H\int\dbar^3\mathbf{k}\Dot{W}_k(t)\hat{\phi}_k(t,\mathbf{x}).
\end{equation}
The spatial gradient term in Eq. (\ref{od_stochastic_eq_deriv}) is negligible for the long wavelength modes. The crux of the stochastic approach stems from the nature of the subhorizon modes. For this derivation, we will consider a strict cut-off such that the window function is given by
\begin{equation}
    \label{window_function}
    W_k(t)=\theta\lb k-\epsilon a(t)H\rb,
\end{equation}
where the cut-off parameter $\epsilon$ is small such that $\mathcal{O}(\epsilon^2)$ is negligible. This window function is the simplest and most commonly used; one can use smoother functions, with restrictions \cite{Andersen:2022}. The statistics of the operator $\hat{\xi}_{OD}$ match that of a white noise, stochastic contribution:
\begin{subequations}
    \label{od_quantum_noise_statistics}
    \begin{align}
        \label{od_1-pt_quantum_noise_function}
        \bra{0}\hat{\xi}_{OD}(t,\mathbf{x})\ket{0}&=0,\\
        \label{od_quantum_noise_variance}
        \bra{0}\hat{\xi}_{OD}(t,\mathbf{x})\hat{\xi}_{OD}(t',\mathbf{x}')\ket{0}&=\frac{9H^5}{4\pi^2}(1+\epsilon^2)\delta(t-t')\frac{\sin\lb\epsilon a(t)H\abs{\mathbf{x}-\mathbf{x}'}\rb}{\epsilon a(t)H\abs{\mathbf{x}-\mathbf{x}'}}.
    \end{align}
\end{subequations}
Then, we make the assumption that $\phi_L$ can be considered as a stochastic quantity, obeying the OD \textit{stochastic} or \textit{Langevin equation}
\begin{equation}
\keybox{
    \label{od_stochastic_eq}
    3H\Dot{\phi}_L+V'(\phi_L)=\xi_{OD}
    }
\end{equation}
with a white noise contribution that has the 1-pt function
\begin{equation}
\keybox{
    \label{od_1-pt_noise_func}
    \expval{\xi_{OD}(t)}=0
    }
    \end{equation}
and 2-pt function
\begin{equation}
\keybox{
    \label{od_noise_variance}
    \expval{\xi_{OD}(t)\xi_{OD}(t')}=\frac{9H^5}{4\pi^2}\delta(t-t'),
    }
\end{equation}
where we have taken the $\mathcal{O}(\epsilon^2)$ term to be negligible. Thus, the noise amplitude (\ref{od_noise_variance}) is $\epsilon$-independent. In other words, provided one takes $\epsilon$ to be sufficiently small so that the long wavelength modes are indeed superhorizon, results obtained using the stochastic equations will not depend on ``how superhorizon'' one considers the long wavelength modes to be. This can be explained by considering the 2-pt function at late times. Using Eq. (\ref{far_future_phi_k}), the integral
\begin{equation}
    \label{2-pt_func_k-integral_late_times}
    \int\dbar^3\mathbf{k}\abs{\phi_k(t)}^2\eval_{t\rightarrow\infty}\sim \lb\frac{k}{a(t)H}\rb^{3-2\nu}\sim1,
\end{equation}
for near-massless fields $m\ll H$, as is the case here. So we see that, at late times, there is no $k$-dependence. Note that, for heavier fields, this becomes a mild power law and so the same logic will not hold, as we shall see in the next chapter.  

We can now frame all of our calculations as a stochastic system revolving around the stochastic equation (\ref{od_stochastic_eq}) with a white noise (\ref{od_noise_variance}). It is useful to note that these equations take the form of a friction-dominated system in a thermal bath with de Sitter temperature (\ref{dS_temperature}). These equations are akin to (1+1)-dimensional Brownian motion of a particle in a potential $V(x)$
\begin{equation}
    \label{brownian_motion_eq}
    \xi_{BM}=\gamma\Dot{x}+V'(x)
\end{equation}
with
\begin{equation}
    \label{brownian_noise_variance}
    \expval{\xi_{BM}(t)\xi_{BM}(t')}=2\gamma T \delta(t-t'),
\end{equation}
where $\gamma$ is the friction coefficient and $T$ is the temperature of the system \cite{pavliotis_txtbk:2014s}. We can make the comparison with the OD stochastic theory concrete: $3H$ is the friction coefficient and $\frac{H}{2\pi}\times\frac{3H^3}{4\pi}$ is the temperature density of our (3+1)-dimensional system \cite{Rigopoulos:2016}. Thus, we can directly use techniques garnered from the study of Brownian motion for our upcoming calculations of OD stochastic correlation functions.

\section{The overdamped Fokker-Planck equation}
\label{sec:od_fokker-planck}

Since the field $\phi$ is now considered to be a stochastic quantity, we can define an associated one-point probability distribution function (1PDF), $P^{(OD)}(\phi;t)$, from which we can compute stochastic correlation functions. Its time-evolution is described by the Fokker-Planck equation
\begin{equation}
    \label{od_fokker-planck_eq}
    \begin{split}
    \partial_t P^{(OD)}(\phi;t)=&\lb\frac{V''(\phi)}{3H}+\frac{V'(\phi)}{3H}\partial_\phi+\frac{H^3}{8\pi^2}\partial_\phi^2\rb P^{(OD)}(\phi;t)
    \\=&\mathcal{L}_{OD} P^{(OD)}(\phi;t),
    \end{split}
\end{equation}
which is associated with the stochastic equation (\ref{od_stochastic_eq}), where $\mathcal{L}_{OD}$ is the OD Fokker-Planck operator. For two arbitrary functions $f(\phi)$ and $g(\phi)$ that exist in the space $\{f|(f,f)_{OD}<\infty\}$, we define the inner product
\begin{equation}
    \label{od_inner_product}
    (f,g)_{OD}=\int_{-\infty}^{\infty} d\phi f(\phi)g(\phi).
\end{equation}
Then, the adjoint OD Fokker-Planck operator $\mathcal{L}_{OD}^*$ is defined via
\begin{equation}
    \label{od_fp_operator)adjoint_defn}
    \lb f,\mathcal{L}_{OD}g\rb_{OD}=\lb\mathcal{L}_{OD}^*f,g\rb_{OD}.
\end{equation}
Explicitly,
\begin{equation}
    \label{adjoint_od_fokker-planck_operator}
    \mathcal{L}_{OD}^*=-\frac{V'(\phi)}{3H}\partial_\phi+\frac{H^3}{8\pi^2}\partial_\phi^2
\end{equation}
We also observe that there exists an equilibrium solution to the Fokker-Planck equation (\ref{od_fokker-planck_eq}), defined as $\partial_t P^{(OD)}_{eq}(\phi)=0$, which is given by
\begin{equation}
    \label{od_equilibrium_1PDF}
    P^{(OD)}_{eq}(\phi)\propto e^{-\frac{8\pi^2V(\phi)}{3H^4}}.
\end{equation}
It is normalised as $\int d\phi P^{(OD)}_{eq}(\phi)=1$. Further, the OD Fokker-Planck operator is related to its adjoint by
\begin{equation}
    \label{od_fp_operator->adjoint}
    \mathcal{L}_{OD}\lb e^{-\frac{8\pi^2V(\phi)}{3H^4}f(\phi)}\rb=e^{-\frac{8\pi^2V(\phi)}{3H^4}}\mathcal{L}_{OD}^*f(\phi).
\end{equation}
We can solve the OD Fokker-Planck equation (\ref{od_fokker-planck_eq}) via a spectral expansion. We start by writing the 1PDF as
\begin{equation}
    \label{1PDF_spectral_expansion}
    P^{(OD)}(\phi;t)=\varphi^*_0(\phi)\sum_{n=0}^{\infty}c_n^{(OD)}\varphi_n(\phi)e^{-\Lambda_n^{(OD)}t},
\end{equation}
where the eigenvalues $\Lambda_n^{(OD)}$ and (adjoint) eigenstates $\varphi_n^{(*)}(\phi)$ obey the eigenequations
\begin{subequations}
    \label{od_eigenequation}
    \begin{align}
    \mathcal{L}_{OD}\varphi_n(\phi)&=-\Lambda_n^{(OD)}\varphi_n(\phi),\\
    \mathcal{L}_{OD}^*\varphi^*_n(\phi)&=-\Lambda_n^{(OD)}\varphi^*_n(\phi),
    \end{align}
\end{subequations}
with the biorthonormality and completeness relations
\begin{subequations}
    \label{od_biorthnormality&completeness}
    \begin{align}
        \label{od_biorthonormality}
        (\varphi_n,\varphi^*_{n'})_{OD}=&\delta_{n'n},\\
        \label{od_completeness}
        \sum_{n=0}^{\infty}\varphi_n(\phi)\varphi^*_n(\phi')=&\delta(\phi-\phi').
    \end{align}
\end{subequations}
Note that all sums of this form run from 0 to $\infty$, so I will also drop these limits henceforth. The coefficients $c_n^{(OD)}$ give us an infinite number of solutions, though they are constrained by the normalisation $c_0=1$. The specific choice of these coefficients is irrelevant for our purposes. The eigenvalues are the same for both eigenequations, which can be seen by substituting the eigenequations (\ref{od_eigenequation}) into (\ref{od_fp_operator)adjoint_defn}), with the aid of the biorthonormality relation (\ref{od_biorthonormality}).

From this, we can construct our stochastic correlators from the eigenspectrum, as we will do in the next section. For the remainder of this section, we will discuss how one solves the eigenequation (\ref{od_eigenequation}) in order to compute the eigenspectrum. In general, this cannot be solved analytically and we will turn to numerical methods to perform a complete calculation. First, we will consider a perturbative solution, which can be used to directly compare OD stochastic correlators with those computed from perturbative QFT in the previous chapter.

\subsection{Perturbative solution of the OD eigenspectrum}
\label{subsec:perturbative_OD_eigenspectrum}

Before considering an interacting theory, we compute the eigenspectrum for free fields. Setting $\lambda=0$ in our scalar potential, the OD Fokker-Planck operators are given by
\begin{subequations}
\label{free_OD_fokker-planck_operators}
    \begin{align}
    \label{free_OD_fokker-planck_operator}
    \mathcal{L}_{OD}^{(0)}&=\frac{m^2}{3H}+\frac{m^2}{3H}\phi\partial_\phi+\frac{H^3}{8\pi^2}\partial_\phi^2\\
    \label{free_adjoint_OD_fokker-planck_operator}
    \mathcal{L}_{OD}^{(0)*}&=-\frac{m^2}{3H}\phi\partial_\phi+\frac{H^3}{8\pi^2}\partial_\phi^2.
\end{align}
\end{subequations}
The free eigenspectrum is simply solved in terms of the Hermite polynomials $H_n(z)$:
\begin{subequations}
    \label{free_OD_eigenspectrum}
    \begin{align}
    \label{free_OD_eigenvalue}
        \Lambda_n^{(OD)(0)}&=n\frac{m^2}{3H},\\
    \label{free_OD_eigenstate}
        \varphi_n^{(0)}(\phi)&=\frac{1}{\sqrt{2^nn!}}\lb\frac{4\pi m^2}{3H^4}\rb^{1/4}H_n\lb\sqrt{\frac{4\pi^2m^2}{3H^4}}\phi\rb e^{-\frac{4\pi^2}{3H^4}m^2\phi^2},\\
    \label{free_OD_adjoint_eigenstate}
        \varphi_n^{(0)*}(\phi)&=\frac{1}{\sqrt{2^nn!}}\lb\frac{4\pi m^2}{3H^4}\rb^{1/4}H_n\lb\sqrt{\frac{4\pi^2m^2}{3H^4}}\phi\rb,
    \end{align}
\end{subequations}
where $n\in\ZZ^+$. Switching our interaction back on such that $\lambda\ne0$, we perform a small-$\lambda$ perturbation about the free solution as
\begin{subequations}
    \label{OD_perturbative_expansion}
    \begin{align}
        \label{OD_FP_operator_expansion}
        \mathcal{L}_{OD}^{(*)}&=\mathcal{L}_{OD}^{(0)(*)}+\lambda\mathcal{L}_{OD}^{(1)(*)},\\
        \label{OD_eigenvalue_expansion}
        \Lambda_n^{(OD)}&=\Lambda_n^{(OD)(0)}+\lambda\Lambda_n^{(OD)(1)}+\mathcal{O}(\lambda^2),\\
        \label{OD_eigenstate_expansion}
        \varphi_n^{(*)}(\phi)&=\varphi_n^{(0)(*)}(\phi)+\lambda\varphi_n^{(1)(*)}(\phi)+\mathcal{O}(\lambda^2),
    \end{align}
\end{subequations}
where
\begin{subequations}
    \label{O(lambda)_OD_FP_operators}
    \begin{align}
    \label{O(lambda)_OD_FP_operator}
        \mathcal{L}_{OD}^{(1)}&=\frac{1}{H}\phi^2+\frac{1}{3H}\phi^3\partial_\phi+\frac{H^3}{8\pi^2}\partial_\phi^2,\\
    \label{O(lambda)_OD_adjoint_FP_operator}
        \mathcal{L}_{OD}^{(1)*}&=-\frac{1}{3H}\phi^3\partial_\phi.
\end{align}
\end{subequations}

Using the eigenequations (\ref{od_eigenequation}) alongside the biorthonormality and completeness relations (\ref{od_biorthnormality&completeness}), the $\mathcal{O}(\lambda)$ corrections to the eigenspectrum are in general given by (see e.g. Ref. \cite{griffiths_book:2018})
\begin{subequations}
    \label{O(lambda)_eigenspectrum_gen}
    \begin{align}
        \label{O(lambda)_eigenvalue_gen}
        \Lambda_n^{(OD)(1)}&=-\lb\varphi_n^{(0)},\mathcal{L}_{OD}^{*(1)}\varphi_n^{(0)}\rb_{OD},\\
        \label{o(lambda)_eigenstate_gen}
        \varphi_n^{(1)}(\phi)&=\sum_{n'\ne n}\varphi_{n'}^{(0)}(\phi)\frac{\lb\varphi_{n'}^{(0)},\mathcal{L}_{OD}^{*(1)}\varphi_n^{(0)}\rb_{OD}}{\Lambda_{n'}^{(OD)(0)}-\Lambda_n^{(OD)(0)}}.
    \end{align}
\end{subequations}
Using the free eigenspectrum (\ref{free_OD_eigenspectrum}), the $\mathcal{O}(\lambda)$ eigenvalues are given explicitly by
\begin{subequations}
    \label{O(lambda)_eigenspectrum}
    \begin{align}
        \label{O(lambda)_eigenvalue}
        \begin{split}
        \Lambda_n^{(OD)(1)}=&\frac{3H^3n^2}{8\pi^2m^2},
        \end{split}\\
        \label{O(lambda)_eigenstates}
        \begin{split}
        \varphi_n^{(1)}(\phi)=&-\frac{3H^4}{8\pi^2m^4\sqrt{2^nn!}}\lb\frac{4\pi m^2}{3H^4}\rb^{1/4}e^{-\frac{4\pi^2}{3H^4}m^2\phi^2}\\&\Bigg[\frac{1}{16}H_{n+4}\lb\sqrt{\frac{4\pi^2m^2}{3H^4}}\phi\rb+\frac{3}{4}(n+1)H_{n+2}\lb\sqrt{\frac{4\pi^2m^2}{3H^4}}\phi\rb\\&-n(n-1)(n-2)H_{n-2}\lb\sqrt{\frac{4\pi^2m^2}{3H^4}}\phi\rb\Bigg],
        \end{split}\\
        \label{O(lambda)_adjoint_eigenstates}
        \begin{split}
        \varphi_n^{(1)*}(\phi)=&-\frac{3nH^4}{8\pi^2m^4\sqrt{2^nn!}}\lb\frac{4\pi m^2}{3H^4}\rb^{1/4}\\&\Bigg[\frac{1}{4}H_{n+2}\lb\sqrt{\frac{4\pi^2m^2}{3H^4}}\phi\rb-3(n-1)^2H_{n-2}\lb\sqrt{\frac{4\pi^2m^2}{3H^4}}\phi\rb\\&-(n-3)(n-2)(n-1)H_{n-4}\lb\sqrt{\frac{4\pi^2m^2}{3H^4}}\phi\rb\Bigg].
        \end{split}
    \end{align}
\end{subequations}

\subsection{Numerical solution of the OD eigenspectrum}
\label{subsec:numerical_sol_OD}

The perturbative solution to the eigenspectrum will be useful for a direct comparison with perturbative QFT, but such solutions will still contain the same IR problems that we have already seen. The real strength of the stochastic approach is that one can compute the eigenspectrum, and thus correlation functions, non-perturbatively via simple numerical techniques. For the OD stochastic approach, we will use the \textit{overshoot/undershoot method}. To do this in a straightforward manner, we will slightly redefine our eigenspectrum by defining the operator
\begin{equation}
    \label{od_numerical_fokker-planck_operator}
    \hat{\mathcal{H}}_{OD}=\frac{1}{2}\partial_\phi^2-\frac{1}{2}W(\phi)
\end{equation}
where
\begin{equation}
    \label{potential_W(phi)}
    W(\phi)=-\frac{4\pi^2}{3H^4}V''(\phi)+\lb\frac{4\pi^2}{3H^4}V'(\phi)\rb^2.
\end{equation}
This is related to the OD Fokker-Planck operator (\ref{od_fokker-planck_eq}) by
\begin{equation}
    \label{OD_FP_operator-->numerical_operator}
    \mathcal{L}_{OD}\lb e^{-\frac{4\pi^2}{3H^4}V(\phi)}f(\phi)\rb=e^{-\frac{4\pi^2}{3H^4}V(\phi)}\hat{\mathcal{H}}_{OD}f(\phi).
\end{equation}
The eigenspectrum of interest is now
\begin{equation}
    \label{OD_numerical_eigenequation}
    \hat{\mathcal{H}}_{OD}\Tilde{\varphi}_n(\phi)=-\Tilde{\Lambda}_n^{(OD)}\Tilde{\varphi}_n(\phi),
\end{equation}
where this new eigenspectrum is related to the original (\ref{od_eigenequation}) by
\begin{subequations}
    \label{old-->new_od_eigenspectrum}
    \begin{align}
        \label{old-->new_od_eigenvalues}
        \Tilde{\Lambda}_n^{(OD)}&=\frac{4\pi^2}{3H^4}\Lambda_n^{(OD)},\\
        \label{old-->new_od_eigenstates}
        \Tilde{\varphi}_n(\phi)&=e^{-\frac{4\pi^2}{3H^4}V(\phi)}\varphi_n^*(\phi).
    \end{align}
\end{subequations}
The operator (\ref{od_numerical_fokker-planck_operator}) is of the same form as the Hamiltonian operator in the time-independent Schrodinger equation, and so we can use our knowledge of this well-studied equation here. The boundary conditions are
\begin{subequations}
    \label{od_eigenstate_boundary_conditions}
    \begin{align}
        \Tilde{\varphi}_n(\pm\infty)=&0\\
        \Tilde{\varphi}_n'(\pm\infty)=&0.
    \end{align}
\end{subequations}
Further, for our massive $\phi^4$ theory, the quasi-potential\footnote{I mean this in the sense that $W(\phi)$ plays the role of a potential term in an equivalent time-independent Schrodinger equation.} 
\begin{equation}
    \label{W(phi)_phi4}
    W(\phi)=-\frac{4\pi^2}{3H^4}\lb m^2+3\lambda \phi^2\rb+\frac{16\pi^4}{9H^8}\lb m^4\phi^2+2\lambda m^2\phi^4+\lambda^2\phi^6\rb
\end{equation}
is symmetric. Thus, we know that our eigenstates are (anti-)symmetric for (odd)even $n$.

\begin{figure}[ht]
    \centering
    \includegraphics[width=150mm]{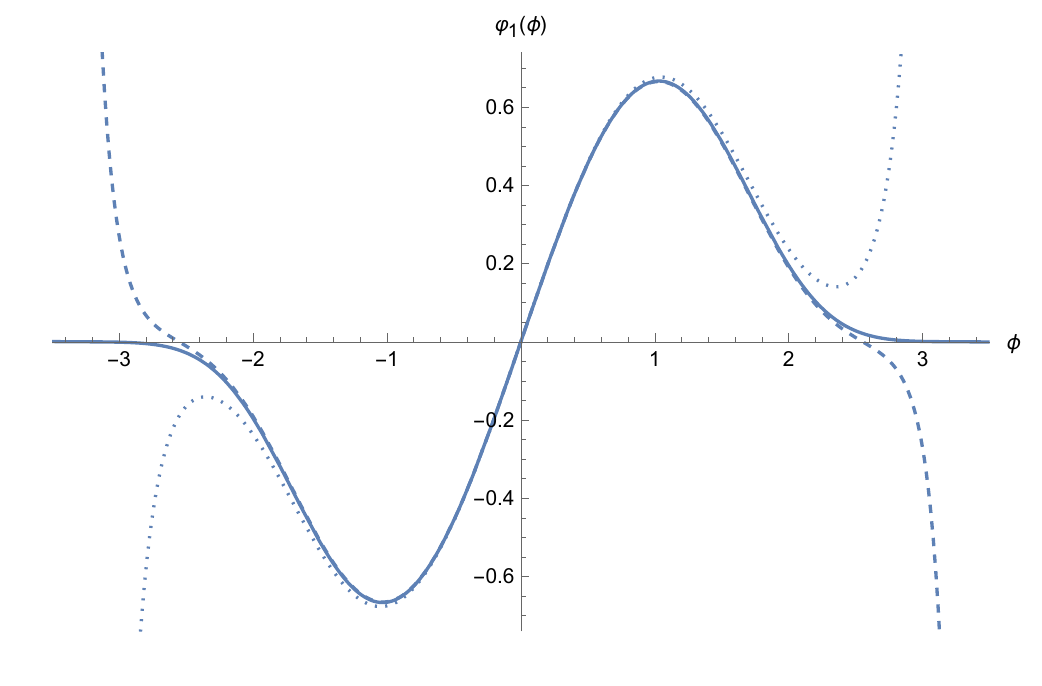}
    \caption{An example of the overshoot/undershoot method to compute the first-excited eigenvalue for Eq. (\ref{OD_numerical_eigenequation}), for the parameters $H=1$, $m^2=0.035$ and $\lambda=0.02$. The solid line shows the correct boundary conditions for the eigenstate, with the correct eigenvalue $\Lambda_1^{(OD)}=0.02088496$. The dashed and dotted lines show the cases where we have overshot - the guessed value for $\Lambda_1^{(OD)}$ was too large - and undershot - the guessed value for $\Lambda_1^{(OD)}$ was too small - respectively.}
    \label{fig:overshoot-undershoot_example}
\end{figure}

Now we can implement the overshoot/undershoot method. Let's focus on the first excited state ($n=1$) for this analysis. The idea is that we guess some numerical value of our eigenvalue and solve the differential equation (\ref{OD_numerical_eigenequation}) numerically for the eigenstate, fixing our solution at some field value. For this analysis, we choose to fix our eigenstate at $\phi=0$. We plot this solution as a function of $\phi$ and observe whether the remaining unfixed boundary conditions are satisfied. Unless one is extremely lucky, one will observe that the solution has either overshot - $\Tilde{\varphi}_1(\phi)$ becomes large and negative - or undershot -  $\Tilde{\varphi}_1(\phi)$ becomes large and positive - as $\phi$ becomes large. These are not the boundary conditions expected, suggesting that the eigenvalue guess is incorrect. One adjusts the guess, increasing/decreasing the eigenvalue for undershooting/overshooting, and repeats the process until the chosen eigenvalue offers a solution that satisfies all the correct boundary conditions. An example of this method in action is given in Fig. \ref{fig:overshoot-undershoot_example}.  

\subsubsection{A comment on massless fields}

For perturbative QFT, one can never consider a completely massless field because of the IR problem. However, since this is merely a failure of perturbation theory, one would expect the massless case to work when considering the stochastic approach, and indeed it does. In fact, for the OD stochastic approach, we can make some concrete statements about the behaviour of the eigenspectrum and hence correlation functions in such a limit. In the massless limit, the eigenequation (\ref{OD_numerical_eigenequation}) becomes
\begin{equation}
    \label{od_eigenequation_massless}
    \frac{\sqrt{\lambda}}{H^2}\lb\frac{1}{2}\partial_x^2+2\pi^2 x^2-\frac{8\pi^4}{9}x^6\rb\Tilde{\varphi}_n(x)=-\frac{4\pi^2}{3H^4}\Lambda_n^{(OD)}\Tilde{\varphi}_n(x),
\end{equation}
where $x=\frac{\lambda^{1/4}}{H}\phi$. Thus, we see that the eigenvalue must behave as\cite{Starobinsky-Yokoyama:1994,Markkanen:2019}
\begin{equation}
    \label{massless_od_eigenvalue}
    \Lambda_n^{(OD)}=b_n\sqrt{\lambda}H,
\end{equation}
where $b_n$ are numerical coefficients, which can be computed using the overshoot/undershoot method. For example, the lowest three eigenvalues are \cite{Markkanen:2019}
\begin{subequations}
    \begin{align}
        \Lambda_0^{(OD)}&=0,\\
        \Lambda_1^{(OD)}&\simeq 0.08892\sqrt{\lambda}H,\\
        \Lambda_2^{(OD)}&\simeq 0.28938\sqrt{\lambda}H.
    \end{align}
\end{subequations}
This result for the eigenvalue is evidently a non-perturbative effect; an example of a situation where the stochastic approach trumps perturbative QFT.

\section{Overdamped stochastic correlation functions}
\label{sec:od_correlators}

Having computed the eigenspectrum, all that remains is to translate these results to correlation functions. We introduce a transfer matrix for the OD equations, $U^{(OD)}(\phi_0,\phi;t-t_0)$, between $\phi_0=\phi(t_0)$ and $\phi=\phi(t)$, which is defined as the Green's function of the Fokker-Planck equation
\begin{equation}
    \label{od_transfer_matrix_Greens_func}
    \partial_tU^{(OD)}(\phi_0,\phi;t-t_0)=\mathcal{L}_{OD}U^{(OD)}(\phi_0,\phi;t-t_0),
\end{equation}
for all values of $\phi_0$, with the boundary condition $U^{(OD)}(\phi_0,\phi;0)=\delta(\phi-\phi_0)$. Then, the time-dependence of the OD 1PDF is given by
\begin{equation}
    \label{od_tranfer_matrix-->1PDF}
    P^{(OD)}(\phi;t)=\int d\phi_0 P^{(OD)}(\phi_0;t_0)U^{(OD)}(\phi_0,\phi;t-t_0).
\end{equation}
From Eq. (\ref{1PDF_spectral_expansion}), the OD transfer matrix is given in terms of our eigenspectrum as
\begin{equation}
    \label{od_transfer_matrix_spectral_expansion}
    U^{(OD)}(\phi_0,\phi;t-t_0)=\frac{\varphi_0^*(\phi)}{\varphi_0^*(\phi_0)}\sum_n\varphi_n^*(\phi_0)\varphi_n(\phi)e^{-\Lambda_n^{(OD)}(t-t_0)}.
\end{equation}

\subsection{Two-point overdamped correlation functions}
\label{subsec:2-pt_OD_correlators}

To compute the 2-point OD stochastic correlators, we define an OD 2-point probability distribution function (2PDF) as
\begin{equation}
    \label{OD_2PDF}
    \begin{split}
    P^{(OD)}_2(\phi_0,\phi;t-t_0):=P^{(OD)}(\phi_0;t_0)U^{(OD)}(\phi_0,\phi;t-t_0)
    \end{split}
\end{equation}
Assuming that the initial state is in equilibrium\footnote{Note that because we make this assumption, the coefficients $c_n^{(OD)}$ don't feature, other than $c_0^{(OD)}=1$.} (which we will do for the remainder of the thesis), $P^{(OD)}(\phi_0;t_0)=P_{eq}^{(OD)}(\phi_0)$, we can use the spectral expansion to write
\begin{equation}
    \label{OD_2PDF_spectral}
    P^{(OD)}_2(\phi_0,\phi;t-t_0)=\varphi_0^*(\phi)\varphi_0(\phi_0)\sum_n\varphi_n^*(\phi_0)\varphi_n(\phi)e^{-\Lambda_n^{(OD)}(t-t_0)}.
\end{equation}
Thus, we can write a 2-point stochastic correlator with temporal separations between some functions $f(\phi_0)$ and $g(\phi)$ as
\begin{equation}
    \label{OD_timelike_stochastic_correlator_fg}
    \begin{split}
        \expval{f(\phi_0)g(\phi)}:=&\int d\phi_0\int d\phi P^{(OD)}_2(\phi_0,\phi;t-t_0)f(\phi_0)g(\phi)\\
        =&\sum_n f^{(OD)}_{0n}g^{(OD)}_{n0}e^{-\Lambda_n^{(OD)}(t-t_0)},
    \end{split}
\end{equation}
where
\begin{equation}
    \label{OD_f_nl}
    f_{nn'}^{(OD)}=\lb\varphi_n,f\varphi_{n'}^*\rb_{OD}.
\end{equation}
We can now use the computed eigenspectrum from the previous section to compute the stochastic correlators. To see this in action and write some analytic results, we compute the $\phi-\phi$, $\phi-\pi$, $\pi-\phi$ and $\pi-\pi$ correlators to $\mathcal{O}(\lambda)$ using the perturbed eigenspectrum, (\ref{free_OD_eigenspectrum}) and (\ref{O(lambda)_eigenspectrum}):
\begin{subequations}
    \label{OD_timelike_stochastic_field_correlators}
    \begin{align}
        \expval{\phi(0)\phi(t)}&=\lb\frac{3H^4}{8\pi^2m^2}-\frac{27\lambda H^8}{64\pi^4m^6}+\mathcal{O}(\lambda^2)\rb e^{-\lb\frac{m^2}{3H}+\frac{3\lambda H^3}{8\pi^2m^2}+\mathcal{O}(\lambda^2)\rb t},\\
        \expval{\phi(0)\pi(t)}&=\lb\frac{H^3}{8\pi^2}+\mathcal{O}(\lambda^2)\rb e^{-\lb\frac{m^2}{3H}+\frac{3\lambda H^3}{8\pi^2m^2}+\mathcal{O}(\lambda^2)\rb t},\\
        \expval{\pi(0)\phi(t)}&=-\lb\frac{H^3}{8\pi^2}+\mathcal{O}(\lambda^2)\rb e^{-\lb\frac{m^2}{3H}+\frac{3\lambda H^3}{8\pi^2m^2}+\mathcal{O}(\lambda^2)\rb t},\\
        \expval{\pi(0)\pi(t)}&=\lb\frac{m^2H^2}{24\pi^2}+\frac{3\lambda H^6}{64\pi^4m^2}+\mathcal{O}(\lambda^2)\rb e^{-\lb\frac{m^2}{3H}+\frac{3\lambda H^3}{8\pi^2m^2}+\mathcal{O}(\lambda^2)\rb t}.
    \end{align}
\end{subequations}
The stochastic equations are only for time-separations. However, we can compute 2-point correlation functions with general spacetime coordinates by computing a specific type of 3-point function. Following the original Starobinsky-Yokoyama paper \cite{Starobinsky-Yokoyama:1994}, we consider a field value $\phi_r$ that exists on a surface of constant time
\begin{equation}
    \label{space_time_coordinate}
    t_r=-\frac{1}{H}\ln\lb H\abs{\mathbf{x}_1-\mathbf{x}_2}\rb,
\end{equation}
where $\mathbf{x}_1$ and $\mathbf{x}_2$ are spatial coordinates within the same Hubble volume. We then evolve this field value independently to time-separated points $t_1$ and $t_2$ - denoted $\phi_1$ and $\phi_2$ respectively - using the time-evolution operator. To do this, we define the OD 3-point probability distribution function
\begin{equation}
    \label{OD_3PDF}
    \begin{split}
    P_3^{(OD)(S)}(\phi_0,\phi_1,\phi_2;t_0,t_1,t_2)=&P^{(OD)}(\phi_0;t_0)U(\phi_0,\phi_1;t_1-t_0)U^{(OD)}(\phi_0,\phi_2;t_2-t_0)\\
    =&\frac{\varphi_0(\phi_0)\varphi_0^*(\phi_1)\varphi_0^*(\phi_2)}{\varphi_0^*(\phi_0)}\sum_n\varphi_n^*(\phi_0)\varphi_n(\phi_1)\\&\times\sum_{n'}\varphi_{n'}^*(\phi_0)\varphi_{n'}(\phi_2)e^{-\Lambda_n^{(OD)}(t_1-t_0)-\Lambda_{n'}^{(OD)}(t_2-t_0)}
    \end{split}
\end{equation}
where the superscript $(S)$ indicates that it is the 3PDF used to define spacelike correlators\footnote{We could similarly define a 3PDF for computing timelike 3-pt functions, where the evolution would be chronological from $\phi_0$ to $\phi_1$ to $\phi_2$.}. Focussing on equal-times such that $t_1=t_2$, we compute the OD spacelike 2-point correlator between two functions $f(\phi(t,\mathbf{x}_1)$ and $g(\phi(t,\mathbf{x}_2)$ as
\begin{equation}
    \label{OD_spacelike_2-pt_function}
    \begin{split}
        \expval{f(\phi;t,\mathbf{x}_1)g(\phi;t,\mathbf{x}_2)}=&\int d\phi_r\int d\phi_1\int d\phi_2 P_3^{(OD)(S)}(\phi_r,\phi_1,\phi_2;t_r,t,t)\\
        =&\int d\phi_r\frac{\varphi_0(\phi_r)}{\varphi_0^*(\phi_r)}\sum_{nn'}\varphi_n^*(\phi_r)\varphi_{n'}^*(\phi_r)f_n^{(OD)} g_{n'}^{(OD)}\\&\times\abs{Ha(t)(\mathbf{x}_1-\mathbf{x}_2)}^{-\frac{\Lambda_n^{(OD)}+\Lambda_{n'}^{(OD)}}{H}}.
    \end{split}
\end{equation}
Using the perturbative results of spectral expansion outlined in Sec. \ref{subsec:perturbative_OD_eigenspectrum}, the OD stochastic spacelike correlators to $\mathcal{O}(\lambda)$ are given by
\begin{subequations}
    \label{OD_spacelike_stochastic_field_correlators}
    \begin{align}
        \expval{\phi(t,\mathbf{0})\phi(t,\mathbf{x})}&=\lb\frac{3H^4}{8\pi^2m^2}-\frac{27\lambda H^8}{64\pi^4m^6}+\mathcal{O}(\lambda^2)\rb \abs{Ha(t)\mathbf{x}}^{-\lb\frac{2m^2}{3H^2}+\frac{3\lambda H^2}{4\pi^2m^2}+\mathcal{O}(\lambda^2)\rb},\\
        \expval{\phi(t,\mathbf{0})\pi(t,\mathbf{x})}&=\lb-\frac{H^3}{8\pi^2}+\mathcal{O}(\lambda^2)\rb  \abs{Ha(t)\mathbf{x}}^{-\lb\frac{2m^2}{3H^2}+\frac{3\lambda H^2}{4\pi^2m^2}+\mathcal{O}(\lambda^2)\rb},\\
        \expval{\pi(t,\mathbf{0})\pi(t,\mathbf{x})}&=\lb\frac{m^2H^2}{24\pi^2}+\frac{3\lambda H^6}{64\pi^4m^2}+\mathcal{O}(\lambda^2)\rb \abs{Ha(t)\mathbf{x}}^{-\lb\frac{2m^2}{3H^2}+\frac{3\lambda H^2}{4\pi^2m^2}+\mathcal{O}(\lambda^2)\rb}.
    \end{align}
\end{subequations}

By comparing the timelike and spacelike correlators, (\ref{OD_timelike_stochastic_field_correlators}) and  (\ref{OD_spacelike_stochastic_field_correlators}), one can see that they are related by the continuation $e^{Ht}\rightarrow\abs{Ha(t)\mathbf{x}}^2$ \footnote{In QFT, there is a complex factor here as well, but this will not be picked up by the stochastic approach.}. This is the same analytic continuation that one uses in QFT, resulting from the symmetries of de Sitter. It is useful to note that this property is preserved in the OD stochastic approach.

Comparing Eq. (\ref{OD_spacelike_2-pt_function}) with (\ref{O(lambdaH4m4)_spacelike_QFT_correlators}), we see that the OD stochastic approach reproduces the equivalent results from QFT, noting that the free part for near-massless fields is given in the QFT by Eq. (\ref{free_light_field_spacelike_QFT_correlators}). This relationship holds in the limits $m\ll H$ (OD stochastic limitation) and $\lambda\ll m^4/H^4$ (perturbative QFT limitation). We can go beyond the latter limitation by computing the eigenspectrum non-perturbatively, as outlined in Sec. \ref{subsec:numerical_sol_OD}, extending the regime of validity in which scalar correlators can be computed.

\subsection{Four-point overdamped stochastic correlation functions}
\label{4-pt_OD_correlators}

To round out our analysis of the OD stochastic approach, we will compute the $\mathcal{O}(\lambda)$ OD 4-point function. Unlike in QFT, the full expression can be computed analytically through the use of the spectral expansion method. We can then isolate and compare the OD connected 4-point function, which can be compared with the quantum connected 4-point function (\ref{spacelike_quantum_4-pt_func_light_fields}) computed using perturbative QFT. Defining the OD equilibrium 4-point probability distribution function (4PDF) between the points $\phi_i=\phi(t_i)$, $i\in\{1,2,3,4\}$, as
\begin{equation}
    \label{OD_4PDF}
    \begin{split}
        P_4^{(OD)}&(\phi_1,\phi_2,\phi_3,\phi_4;t_1,t_2,t_3,t_4)\\:&=P^{(OD)}(\phi_1;t_1)U^{(OD)}(\phi_1,\phi_2;t_2-t_1)U^{(OD)}(\phi_2,\phi_3;t_3-t_2)U^{(OD)}(\phi_3,\phi_4;t_4-t_3)\\
        &=\varphi_0(\phi_1)\varphi^*_0(\phi_4)\sum_{n}\varphi_n^*(\phi_1)\varphi_n(\phi_2)\sum_{n'}\varphi_{n'}^*(\phi_2)\varphi_{n'}(\phi_3)\sum_{n''}\varphi_{n''}^*(\phi_3)\varphi_{n''}(\phi_4),
    \end{split}
\end{equation}
where we assumed $t_1<t_2<t_3<t_4$ and $P^{(OD)}(\phi_1;t_1)=P^{(OD)}_{eq}(\phi_1)$, the 4-point correlation function is given by
\begin{equation}
    \label{od_4-pt_function_general}
    \begin{split}
        \langle f_1(\phi_1)f_2&(\phi_2)f_3(\phi_3)f_4(\phi_4)\rangle\\&=\prod_{i=1}^4\int d\phi_i P_4^{(OD)}(\phi_1,\phi_2,\phi_3,\phi_4;t_1,t_2,t_3,t_4)f(\phi_1)f_2(\phi_2)f_3(\phi_3)f_4(\phi_4)\\
        &=\sum_{n''n'n}(f_1)_{0n}^{(OD)}(f_2)_{nn'}^{(OD)}(f_3)_{n'n''}^{(OD)}(f_1)_{n''0}^{(OD)}e^{-\Lambda_n^{(OD)}(t_2-t_1)-\Lambda_{n'}^{(OD)}(t_3-t_2)-\Lambda_{n''}^{(OD)}(t_4-t_3)},
    \end{split}
\end{equation}
where $(f_i)_{nn'}^{(OD)}$ is given in Eq. (\ref{OD_f_nl}). Focussing on the 4-point field correlator, such that $f_i(\phi)=\phi$ $\forall i$, the leading non-zero contributions are given by
\begin{equation}
    \label{OD_4-pt_func_leading_pieces}
    \begin{split}
    \expval{\phi(t_1)\phi(t_2)\phi(t_3)\phi(t_4)}=&\phi_{01}^{(OD)}\phi_{10}^{(OD)}\phi_{01}^{(OD)}\phi_{10}^{(OD)}e^{-\Lambda_1^{(OD)}(t_2-t_1)-\Lambda_0^{(OD)}(t_3-t_2)-\Lambda_1^{(OD)}(t_4-t_3)}
    \\&+\phi_{01}^{(OD)}\phi_{12}^{(OD)}\phi_{21}^{(OD)}\phi_{10}^{(OD)}e^{-\Lambda_1^{(OD)}(t_2-t_1)-\Lambda_2^{(OD)}(t_3-t_2)-\Lambda_1^{(OD)}(t_4-t_3)}
    \\&+\phi_{01}^{(OD)}\phi_{12}^{(OD)}\phi_{23}^{(OD)}\phi_{30}^{(OD)}e^{-\Lambda_1^{(OD)}(t_2-t_1)-\Lambda_2^{(OD)}(t_3-t_2)-\Lambda_3^{(OD)}(t_4-t_3)}
    \\&+\phi_{03}^{(OD)}\phi_{32}^{(OD)}\phi_{21}^{(OD)}\phi_{10}^{(OD)}e^{-\Lambda_3^{(OD)}(t_2-t_1)-\Lambda_2^{(OD)}(t_3-t_2)-\Lambda_1^{(OD)}(t_4-t_3)}.
    \end{split}
\end{equation}
Using the explicit expressions for the eigenspectrum (\ref{free_OD_eigenspectrum}) and (\ref{O(lambda)_eigenspectrum}), the OD timelike 4-point correlation function to $\mathcal{O}(\lambda)$ is given by
\begin{equation}
    \label{OD_4-pt_function}
    \begin{split}
    \expval{\phi(t_1)\phi(t_2)\phi(t_3)\phi(t_4)}=&\lb\frac{9H^8}{64\pi^4m^4}-\frac{81\lambda H^{12}}{256\pi^6m^8}\rb e^{-\frac{m^2}{3H}(t_2-t_1)}e^{-\frac{m^2}{3H}(t_4-t_3)}
    \\&+\lb\frac{9H^8}{64\pi^4m^4}-\frac{81\lambda H^{12}}{256\pi^6m^8}\rb e^{-\frac{m^2}{3H}(t_3-t_1)}e^{-\frac{m^2}{3H}(t_4-t_2)}
    \\&+\lb\frac{9H^8}{64\pi^4m^4}-\frac{81\lambda H^{12}}{256\pi^6m^8}\rb e^{-\frac{m^2}{3H}(t_4-t_1)}e^{-\frac{m^2}{3H}(t_3-t_2)}
    \\&-\frac{81\lambda H^{12}}{128\pi^6m^8}e^{-\frac{m^2}{3H}(t_4+t_3-t_2-t_1)}
    \\&+\frac{81\lambda H^{12}}{512\pi^6m^8}\lb e^{-\frac{m^2}{3H}(t_4+t_3+t_2-3t_1)}+e^{-\frac{m^2}{3H}(3t_4-t_3-t_2-t_1)}\rb.
    \end{split}
\end{equation}
We notice that the first three lines are just the different permutations of the square of the 2-point functions (\ref{OD_timelike_stochastic_field_correlators}). Thus, we can define the OD connected 4-point function in the same way as for the QFT, namely as
\begin{equation}
    \label{OD_connected_4-pt_func_def}
    \begin{split}
    \expval{\phi(t_1)\phi(t_2)\phi(t_3)\phi(t_4)}_C=&\expval{\phi(t_1)\phi(t_2)\phi(t_3)\phi(t_4)}
    \\&-\expval{\phi(t_1)\phi(t_2)}\expval{\phi(t_3)\phi(t_4)}
    \\&-\expval{\phi(t_1)\phi(t_3)}\expval{\phi(t_2)\phi(t_4)}
    \\&-\expval{\phi(t_1)\phi(t_4)}\expval{\phi(t_2)\phi(t_3)},
    \end{split}
\end{equation}
where the subscript `$C$' stands for connected. Explicitly to $\mathcal{O}(\lambda)$, it is
\begin{equation}
    \label{connected_OD_4pt_function}
    \keybox{
    \begin{split}
    \expval{\phi(t_1)\phi(t_2)\phi(t_3)\phi(t_4)}_C=&\frac{81\lambda H^{12}}{512\pi^6m^8}\lb e^{-\frac{m^2}{3H}(t_4+t_3+t_2-3t_1)}+e^{-\frac{m^2}{3H}(3t_4-t_3-t_2-t_1)}\rb
    \\&-\frac{81\lambda H^{12}}{128\pi^6m^8}e^{-\frac{m^2}{3H}(t_4+t_3-t_2-t_1)}.
    \end{split}
    }
\end{equation}
In a similar way to the 2-point function, we can compute the spacelike 4-point functions by defining a OD ``spacelike'' equilibrium 5-point probability distribution function (5PDF) as
\begin{equation}
    \label{5PDF_OD}
    \begin{split}
    P_5^{(OD)(S)}&(\phi_0,\phi_1,\phi_2,\phi_3,\phi_4;t_0,t_1,t_2,t_3,t_4)
    \\:=&P^{(OD)}(\phi_0;t_0)U^{(OD)}(\phi_0,\phi_1;t_1-t_0)U^{(OD)}(\phi_0,\phi_2;t_2-t_0)U^{(OD)}(\phi_0,\phi_3;t_3-t_0)\\&\times U^{(OD)}(\phi_0,\phi_4;t_4-t_0)
    \\=&\frac{\varphi_0(\phi_0)}{\varphi_0^*(\phi_0)^3}\prod_{i=1}^4\lsb\varphi_0^*(\phi_i)\sum_n\varphi_n^*(\phi_0)\varphi_n(\phi_i)\rsb.
    \end{split}
\end{equation}
This method requires the spatial separations to be equal, $\abs{\mathbf{x}}=\abs{\mathbf{x}_i-\mathbf{x}_j}$ $\forall i\ne j$. We could be more general, but we would have to use a different PDF and account for all the various separations that one could use. This is not necessary for this thesis, so we will stick to equal spatial separations. Using the $t_r$ coordinate defined in Eq. (\ref{space_time_coordinate}), the OD spacelike 4-point function between some functions $f_i(\phi)$, $i\in\{1,2,3,4\}$, is given by
\begin{equation}
    \label{OD_spacelike_4-pt_func_fi}
    \begin{split}
    \langle f_1&(\phi_1)f_2(\phi_2)f_3(\phi_3)f_4(\phi_4)\rangle
    \\=&\int d\phi_r\prod_{i=1}^4\int d\phi_i P_5^{(OD)(S)}(\phi_r,\phi_1,\phi_2,\phi_3,\phi_4;t_r,t_1,t_2,t_3,t_4)f_1(\phi_1)f_2(\phi_2)f_3(\phi_3)f_4(\phi_4)
    \\=&\int d\phi_r\frac{\varphi_0(\phi_r)}{\varphi_0^*(\phi_r)^3}\prod_{i=1}^4\lsb\sum_n \varphi_n^*(\phi_r)(f_i)_{n0}^{(OD)}\abs{Ha(t)\mathbf{x}}^{-\frac{\Lambda_n}{H}}\rsb.
    \end{split}
\end{equation}
We can now compute the OD spacelike 4-point function for the field using the above expression with $f_i=\phi$ $\forall i$. The leading terms are
\begin{equation}
    \label{OD_spacelike_4-pt_func_leading}
    \begin{split}
    \expval{\phi(\mathbf{x}_1)\phi(\mathbf{x}_2)\phi(\mathbf{x}_3)\phi(\mathbf{x}_4)}=&\int d\phi_r\frac{\varphi_0(\phi_r)}{\varphi_0^*(\phi_r)^3}\lb\varphi_1^*(\phi_r)\phi_{10}^{(OD)}\rb^4\abs{Ha(t)\mathbf{x}}^{-\frac{4\Lambda_1^{(OD)}}{H}}
    \\&+\int d\phi_r\frac{\varphi_0(\phi_r)}{\varphi_0^*(\phi_r)^3}\varphi_3^*(\phi_r)\phi_{30}^{(OD)}\lb\varphi_1^*(\phi_r)\phi_{10}^{(OD)}\rb^3\\&\times\abs{Ha(t)\mathbf{x}}^{-\frac{\Lambda_3^{(OD)}+3\Lambda_1^{(OD)}}{H}},
    \end{split}
\end{equation}
which are computed explicitly using the eigenspectrum (\ref{free_OD_eigenspectrum}) and (\ref{O(lambda)_eigenspectrum}) to give
\begin{equation}
    \label{OD_spacelike_4-pt_func_explicit}
    \begin{split}
        \expval{\phi(\mathbf{x}_1)\phi(\mathbf{x}_2)\phi(\mathbf{x}_3)\phi(\mathbf{x}_4)}=&\lb\frac{27H^8}{64\pi^4m^4}-\frac{243\lambda H^{12}}{128\pi^6m^8}\rb\abs{Ha(t)\mathbf{x}}^{-\frac{4m^2}{3H^2}}
        \\&+\frac{81\lambda H^{12}}{128\pi^6m^8} \abs{Ha(t)\mathbf{x}}^{-\frac{2m^2}{H^2}}.
    \end{split}
\end{equation}
Using the definition for the connected 4-point function given in Eq. (\ref{OD_connected_4-pt_func_def}), where we replace $t_i$ with $\mathbf{x}_i$, we write the OD connected 4-point function to $\mathcal{O}(\lambda)$ as
\begin{equation}
    \label{spacelike_connected_OD_4pt_function}
    \keybox{
    \begin{split}
    \expval{\phi(\mathbf{x}_1)\phi(\mathbf{x}_2)\phi(\mathbf{x}_3)\phi(\mathbf{x}_4)}_C=&\frac{81\lambda H^{12}}{128\pi^6m^8} \abs{Ha(t)\mathbf{x}}^{-\frac{2m^2}{H^2}}
    \\&-\frac{243\lambda H^{12}}{256\pi^6m^8}\abs{Ha(t)\mathbf{x}}^{-\frac{4m^2}{3H^2}},
    \end{split}
    }
\end{equation}
The first line is the same as the analytic expression from perturbative QFT for near-massless fields, given in Eq. (\ref{spacelike_quantum_4-pt_func_light_fields}), as one would expect. The second is the other contribution to the connected piece, denoted by the `...+' in Eq. (\ref{spacelike_quantum_4-pt_func_light_fields}), which is prevalent in perturbative QFT but cannot be calculated explicitly. Thus, the OD stochastic approach gives us an indication as to what this contribution should be.

\section{Limitations of the overdamped stochastic approach}

The OD stochastic is a useful effective theory for scalar fields in de Sitter but it has its limitations. The first comes from the fact that we use OD stochastic equations, which limits the theory to the regime $m\ll H$ and $\lambda\ll m^2/H^2$. This is represented by the green, right hashed regions in Fig. \ref{fig:model_limitations_comparison}. Additionally, we are limited by the approximation that the short wavelength modes are free, while the long wavelength modes only interact amongst themselves. This is important for the cut-off procedure to work but means that the only subhorizon effect we have enters when the short wavelength modes cross the horizon. However, this means we are neglecting any UV effects that manifest themselves in the IR regime - for example, through UV renormalisation - which can be important. Both of these limitations will be mitigated in Chapter \ref{ch:second-order_stochastic}, where we consider second-order stochastic equations via a matching procedure.

\chapter{Second-order Stochastic Effective Theory of Light Scalar Fields in de Sitter}
\label{ch:second-order_stochastic}

\section{Introduction to Chapter \ref{ch:second-order_stochastic}}
\label{sec:so_stochastic_intro}

In the previous chapter, I introduced the original and well-studied stochastic approach, dubbed the overdamped stochastic approach. As discussed, there are reasons why the overdamped approach may be required, such as for near-massless fields. However, such an approximation is not a tenet of the stochastic approach provided sufficient squeezing occurs and so it begs the question: can one extend the stochastic approach beyond the overdamped limit? This question has been central to my PhD work, and it is the question I shall address in this chapter. This chapter is my own work, with its key results published in [K1], [K2] and [K3].

The starting point is the second-order equations of motion for a scalar field in de Sitter spacetime. We find that, if one follows the cut-off procedure used in the overdamped stochastic approach, the stochastic parameters are explicitly dependent on the cut-off parameter $\epsilon$. Thus, this cut-off method of deriving the stochastic approach cannot be used beyond the near-massless limit. Instead, we propose a method whereby one starts with a stochastic theory based off the second-order scalar equations of motion but without specifying the stochastic parameters. These parameters are then obtained on the level of correlation functions, by requiring stochastic and perturbative QFT 2-point and 4-point functions are equal within the regime that perturbative QFT is valid. This promotes our general stochastic theory to a stochastic effective theory of the long-distance behaviour of scalar fields in de Sitter.

I will begin the chapter by explicitly following the cut-off procedure to compute second-order stochastic equations with $\epsilon$-dependent noise in Sec. \ref{sec:so_stochastic_equations}. I will show that this is not a useful way to consider the second-order theory and instead use it to inspire a general stochastic theory. Working purely with stochastic methods, without trying to relate our stochastic theory to QFT, I will outline a method for computing stochastic correlation functions in Sec. \ref{sec:stochastic_correlators}. Indeed, this is the same method introduced in Sec. \ref{sec:od_correlators} for the overdamped stochastic approach. Sec. \ref{sec:so_stochastic_free_fields} is where I will first relate our stochastic theory with results from QFT. This will be done for free fields, where the exact QFT correlators are known analytically. By comparing our stochastic and QFT correlators, we will compute the stochastic parameters required to reproduce the physical results, promoting our stochastic theory to an effective theory of free scalar fields in de Sitter. In Sec. \ref{sec:interacting_stochastic_theory}, we will continue this analysis to interacting fields, computing the stochastic parameters for a quartic self-interacting theory to leading order in coupling, using the results from perturbative QFT. This promotes the stochastic theory to an effective theory of interacting scalar fields in de Sitter.   

\section{The second-order stochastic equations}
\label{sec:so_stochastic_equations}

As in both Chapters \ref{ch:qft_dS} and \ref{ch:od_stochastic}, we will consider a spectator scalar field $\phi$ in a scalar potential $V(\phi)=\frac{1}{2}m^2\phi^2+\frac{1}{4}\lambda\phi^4$ in a de Sitter spacetime with the equations of motion (\ref{scalar_eom_2d}) in their 2-dimensional guise
\begin{equation}
    \label{scalar_eom_2d_repeat}
    \begin{pmatrix}\Dot{\phi}\\\Dot{\pi}\end{pmatrix}=\begin{pmatrix}\pi\\-3H\pi+\frac{1}{a(t)^2}\nabla^2\phi-V'(\phi)\end{pmatrix},
\end{equation}
where, again, the non-minimal coupling between the scalar field and gravity is hidden in the mass term, and we have defined the canonical momentum $\pi=\Dot{\phi}$. Quantising the theory gives us the field (\ref{field_operator}) and canonical momentum operators as 
\begin{subequations}
    \label{field&momentum_operators}
    \begin{align}
        \label{field_operator_repeat}
        \begin{split}
            \hat{\phi}(t,\mathbf{x})&=\int\dbar^3\mathbf{k}\hat{\phi}_k(t,\mathbf{x})\\
            &=\int\dbar^3\mathbf{k}\lb\hat{a}_k\phi_k(t)e^{-i\mathbf{k}\cdot\mathbf{x}}+\hat{a}^{\dagger}_k\phi^*_k(t)e^{i\mathbf{k}\cdot\mathbf{x}}\rb,
        \end{split}\\
        \label{momentum_operator}
        \begin{split}
            \hat{\pi}(t,\mathbf{x})&=\int\dbar^3\mathbf{k}\hat{\pi}_k(t,\mathbf{x})\\
            &=\int\dbar^3\mathbf{k}\lb\hat{a}_k\pi_k(t)e^{-i\mathbf{k}\cdot\mathbf{x}}+\hat{a}^{\dagger}_k\pi^*_k(t)e^{i\mathbf{k}\cdot\mathbf{x}}\rb,
        \end{split}\\
    \end{align}
\end{subequations}
where the modes are given in Eq. (\ref{mode_functions}) as
\begin{subequations}
\label{mode_functions_repeated}
    \begin{align}
    \label{phi_k_mode_repeated}
    \phi_k(t)&=\sqrt{\frac{\pi}{4Ha(t)^3}}\mathcal{H}^{(1)}_\nu\lb\frac{k}{a(t)H}\rb,\\
    \label{pi_k_mode_repeated}
    \pi_k(t)&=-\sqrt{\frac{\pi}{16Ha(t)^3}}\lsb3H\mathcal{H}^{(1)}_\nu\lb\frac{k}{a(t)H}\rb+\frac{k}{a(t)}\lb\mathcal{H}^{(1)}_{\nu-1}\lb\frac{k}{a(t)H}\rb-\mathcal{H}_{\nu+1}^{(1)}\lb\frac{k}{a(t)H}\rb\rb\rsb.
    \end{align}
\end{subequations}
We will now attempt to derive the second-order stochastic equations in the same way as was done in the overdamped case: by introducing a window function to separate sub- and superhorizon modes. Indeed, this is the more general case, where we are not limiting ourselves to slow-roll. The only restriction we are putting on the parameters of our theory thus far is that $m\lesssim H$, which is a fundamental limitation of the stochastic approach. We begin by using the window function $W_k(t)$ to separate the sub- and superhorizon modes
\begin{subequations}
    \label{mode_split_repeat}
    \begin{align}
        \label{phi_mode_split_repeat}
        \hat{\phi}&=\phi_L+\hat{\phi}_S,\\
        \label{pi_mode_split}
        \hat{\pi}&=\pi_L+\hat{\pi}_S,
    \end{align}
\end{subequations}
where $L$ and $S$ stand for long (superhorizon) and short (subhorizon) modes respectively. Note that we drop the hat on the long modes because we consider them to be classical: the first assumption of the stochastic approach. The subhorizon modes are given by
\begin{subequations}
    \label{subhorizon_modes}
    \begin{align}
        \label{phi_subhorizon}
        \hat{\phi}_S(t,\mathbf{x})&=\int\dbar^3\mathbf{k}W_k(t)\hat{\phi}_k(t,\mathbf{x}),\\
        \label{pi_subhorizon}
        \hat{\pi}_S(t,\mathbf{x})&=\int\dbar^3\mathbf{k}W_k(t)\hat{\pi}_k(t,\mathbf{x}),
    \end{align}
\end{subequations}
where the window function ensures only modes with $k>a(t)H$ are included in the integral. Plugging this split into the equation of motion (\ref{scalar_eom_2d_repeat}) results in
\begin{equation}
    \label{2d_eom_split_repeat}
    \begin{pmatrix}\Dot{\phi}_L\\\Dot{\pi}_L\end{pmatrix}=\begin{pmatrix}\pi_L\\-3H\pi_L+\frac{1}{a(t)^2}\nabla^2\phi_L-V'(\phi_L)\end{pmatrix}+\begin{pmatrix}\hat{\xi}_{\phi}(t,\mathbf{x})\\\hat
    {\xi}_\pi(t,\mathbf{x})\end{pmatrix}
\end{equation}
where
\begin{subequations}
    \label{quantum_xi}
    \begin{align}
        \label{quantum_xi_phi_repeat}
        \hat{\xi}_\phi(t,\mathbf{x})&=-\int\dbar^3\mathbf{k}\Dot{W}_k(t)\hat{\phi}_k(t,\mathbf{x}),\\
        \label{quantum_xi_phi}
        \hat{\xi}_\pi(t,\mathbf{x})&=-\int\dbar^3\mathbf{k}\Dot{W}_k(t)\hat{\pi}_k(t,\mathbf{x}).
    \end{align}
\end{subequations}
For the long wavelength modes, the spatial gradient term in Eq. (\ref{2d_eom_split_repeat}) is negligible and will be dropped henceforth. As before, we choose the window function to be $W_k(t)=\theta(k-\epsilon a(t)H)$ for some small number $\epsilon$ such that the statistics of $\hat{\xi}_{\phi,\pi}$ are
\begin{subequations}
    \label{so_xi_statistics}
    \begin{align}
        \label{so_xi_1-pt}
        \bra{0}\hat{\xi}_i(t,\mathbf{x})\ket{0}&=0,\\
        \label{so_xi_2-pt}
        \frac{1}{2}\bra{0}\{\hat{\xi}_i(t,\mathbf{x}),\hat{\xi}_j(t',\mathbf{x}')\}\ket{0}&=\sigma_{cut,ij}^2\delta(t-t')\frac{\sin\lb\epsilon a(t)H\abs{\mathbf{x}-\mathbf{x}'}\rb}{\epsilon a(t)H\abs{\mathbf{x}-\mathbf{x}'}},
    \end{align}
\end{subequations}
where $i,j\in\{\phi,\pi\}$, mirror those of white noise. Note that we are interested in the symmetric part of the correlator so that our stochastic noise will be real. The amplitudes are given by
\begin{subequations}
    \label{noise_amplitudes_modes}
    \begin{align}
         \sigma_{cut,\phi\phi}^2=&\frac{\epsilon^3 a(t)^3 H^4}{2\pi^2}\abs{\phi_{\epsilon a(t) H}(t,\mathbf{x})}^2\\
         =&\frac{H^3 \epsilon^3}{8\pi}\mathcal{H}_{\nu}^{(1)}(\epsilon)\mathcal{H}_{\nu}^{(2)}(\epsilon),\nonumber\\&\nonumber\\
        \sigma_{cut,\phi\pi}^2=&\sigma_{cut,\pi\phi}^2=\frac{\epsilon^3 a(t)^3 H^4}{4\pi^2}\qty(\phi_{\epsilon a(t) H}(t,\mathbf{x})\pi^*_{\epsilon a(t)H}(t,\mathbf{x})+\pi_{\epsilon a(t) H}(t,\mathbf{x})\phi^*_{\epsilon a(t)H}(t,\mathbf{x}))\\
        =&-\frac{H^4\epsilon^3}{32\pi}\Bigg[\epsilon\qty(\mathcal{H}_{\nu-1}^{(1)}(\epsilon)-\mathcal{H}_{\nu+1}^{(1)}(\epsilon))\mathcal{H}_{\nu}^{(2)}(\epsilon)\nonumber\\
        &+\mathcal{H}_{\nu}^{(1)}(\epsilon)\qty(\epsilon\mathcal{H}_{\nu-1}^{(2)}(\epsilon)+6\mathcal{H}_{\nu}^{(2)}(\epsilon)-\epsilon\mathcal{H}_{\nu+1}^{(2)}(\epsilon))\Bigg],\nonumber\\&\nonumber\\
        \sigma_{cut,\pi\pi}^2&=\frac{\epsilon^3 a(t)^3 H^4}{2\pi^2}\abs{\pi_{\epsilon a(t) H}(t,\mathbf{x})}^2\\
        =&\frac{H^5\epsilon^3}{32\pi}\qty(\epsilon\mathcal{H}_{\nu-1}^{(1)}(\epsilon)+3\mathcal{H}_{\nu}^{(1)}(\epsilon)-\epsilon\mathcal{H}_{\nu+1}^{(1)}(\epsilon))\qty(\epsilon\mathcal{H}_{\nu-1}^{(2)}(\epsilon)+3\mathcal{H}_{\nu}^{(2)}(\epsilon)-\epsilon\mathcal{H}_{\nu+1}^{(2)}(\epsilon)).\nonumber
    \end{align}
\end{subequations}
where the subscript `$cut$' indicates we have derived these equations via the cut-off method. We postulate that we can approximate Eq. (\ref{2d_eom_split_repeat}) as a stochastic equation
\begin{equation}
    \label{2d_stochastic_eq_cut-off}
    \begin{pmatrix}\Dot{\phi_L}\\\Dot{\pi_L}\end{pmatrix}=\begin{pmatrix}\pi_L\\-3H\pi_L-V'(\phi_L)\end{pmatrix}+\begin{pmatrix}\xi_{cut,\phi}(t,\mathbf{x})\\\xi_{cut,\pi}(t,\mathbf{x})\end{pmatrix}
\end{equation}
with a white noise contribution
\begin{subequations}
    \label{white_noise_stats}
    \begin{align}
        \label{1-pt_xi_noise}
        \expval{\xi_{cut,i}(t)}&=0,\\
        \label{2-pt_xi_noise}
        \expval{\xi_{cut,i}(t)\xi_{cut,j}(t')}&=\sigma_{cut,ij}^2\delta(t-t'),
    \end{align}
\end{subequations}
This system will be called the \textit{second-order cut-off stochastic approach}. As this is a long-distance theory, I will henceforth drop the subcript `L'. This system contains an explicit dependence on the choice of cut-off, even for small $\epsilon$. To see this, consider the leading order contribution in $\epsilon$ of $\sigma_{cut,\phi\phi}^2$:  
\begin{equation}
    \label{sigma_phiphi_cut_epsilon}
    \sigma_{cut,\phi\phi}^2\sim\frac{H^3\Gamma(\nu)^2}{2^{3-2\nu}\pi^3}\epsilon^{3-2\nu}.
\end{equation}
This is not surprising because, unlike their near-massless counterparts, correlators of massive fields decay at late times (see Eq. (\ref{2-pt_func_k-integral_late_times})). Thus, results for massive fields will naturally depend on the choice of cut-off. Note that previous work has considered stochastic inflation in a 2-dimensional phase space in a similar way we have described above \cite{Grain:2017,Tolley:2008,Tolley:2009}. The noise amplitudes computed in these works are entirely equivalent to Eq. (\ref{noise_amplitudes_modes}), which shows the explicit dependence on the cut-off parameter. However, we will see that this is problematic when it comes to computing stochastic correlation functions because the results will depend on our choice of cut-off. Thus, it appears that this cut-off method, used to such good effect in the OD case, is not a viable way to construct a second-order stochastic theory that focusses on computing correlation functions that relate to those from QFT in de Sitter\footnote{That is, beyond the near-massless limit: see Appendix \ref{app:so_stochastic_light_field} for a discussion.}.

Instead, we propose a `matching' procedure, where we compute the stochastic parameters at the level of correlation functions. We make an ansatz about the nature of the stochastic equations, based on Eq. (\ref{2d_stochastic_eq_cut-off}); namely that the \textit{second-order stochastic equations} are
\begin{equation}
\keybox{
    \label{2d_stochastic_eq}
    \begin{pmatrix}\Dot{\phi}\\\Dot{\pi}\end{pmatrix}=\begin{pmatrix}\pi\\-3H\pi-V'(\phi)\end{pmatrix}+\begin{pmatrix}\xi_{\phi}(t,\mathbf{x})\\\xi_{\pi}(t,\mathbf{x})\end{pmatrix}
    }
\end{equation}
with a white noise contribution
\begin{equation}
    \label{white_noise_general}
    \keybox{
    \expval{\xi_i(t)\xi_j(t')}=\sigma_{ij}^2\delta(t-t').
    }
\end{equation}
The noise amplitudes $\sigma_{ij}^2$ are left unspecified for the time being, other than the fact that they do not depend on the spacetime coordinates and that they are symmetric, preserving the reality of the noise. We note that this is the simplest type of noise contribution one can write down: additive and white. Other work is available, which discusses alternative noise forms, such as coloured and/or multiplicative, which gives different stochastic (and therefore Fokker-Planck) equations \cite{Cohen:2020,Cohen:2021,Green:2022,Cohen:2023}. Further, the parameters hidden in the potential $V(\phi)$ will now be labelled as $m_S$ and $\lambda_S$, indicating that they do not \textit{a priori} equal the parameters of the underlying QFT. The form of the \textit{stochastic parameters} $m_S$, $\lambda_S$ and $\sigma_{ij}^2$ will be determined by comparing stochastic correlators with their perturbative QFT counterparts. We will choose these quantities to be the same, promoting our stochastic theory from something general to an effective theory of QFT.

\section{The second-order stochastic correlators}
\label{sec:stochastic_correlators}

We will now compute stochastic correlation functions using the general stochastic theory (\ref{2d_stochastic_eq}), though this method will also work for the cut-off stochastic approach of Eq. (\ref{2d_stochastic_eq_cut-off}). The spectral expansion method introduced in Sec. \ref{sec:od_fokker-planck} and its application to stochastic correlators in Sec. \ref{sec:od_correlators} will re-emerge; indeed, the methods are almost identical, though for clarity's sake I will repeat it here. 

The one-point probability distribution function (1PDF) $P(\phi,\pi;t)$ will be our starting point. Its time-evolution is described by the Fokker-Planck equation
\begin{equation}
    \label{phi-pi_fokker-planck_eq}
    \begin{split}
        \partial_t P(\phi,\pi;t)=&\Bigg[3H-\pi\partial_\phi+(3H\pi+V'(\phi))\partial_\pi+\frac{1}{2}\sigma_{\phi\phi}^2\partial_\phi^2+\sigma_{\phi\pi}^2\partial_\phi\partial_\pi+\frac{1}{2}\sigma_{\pi\pi}^2\partial_\pi^2\Bigg]P(\phi,\pi;t)\\
        =&\mathcal{L}_{FP}P(\phi,\pi;t),
    \end{split}
\end{equation}
where $\mathcal{L}_{FP}$ is the Fokker-Planck operator. For a space of functions $\{f|(f,f)<\infty\}$ with the inner product
\begin{equation}
    \label{scalar_product}
    (f,g)=\int_{-\infty}^{\infty} d\phi\int_{-\infty}^{\infty} d\pi f(\phi,\pi)g(\phi,\pi),
\end{equation}
we define the adjoint of the Fokker-Planck operator, $\mathcal{L}_{FP}^*$, as
\begin{equation}
    \label{adjoint_fp_op_definition}
    \qty(\mathcal{L}_{FP}f,g)=\qty(f,\mathcal{L}_{FP}^*g).
\end{equation}
Note that all integrals over $\phi$ and $\pi$ have the limits $(-\infty,\infty)$ unless otherwise stated. Explicitly,
\begin{equation}
    \label{adjoint_fokker-planck_operator}
    \begin{split}
        \mathcal{L}_{FP}^*=&\pi\partial_{\phi}-\qty(3H\pi+V'(\phi))\partial_{\pi}+\frac{1}{2}\sigma_{\phi\phi}^2\partial_{\phi}^2+\sigma_{\phi\pi}^2\partial_{\phi}\partial_{\pi}+\frac{1}{2}\sigma_{\pi\pi}^2\partial_{\pi}^2.
    \end{split}    
\end{equation}
The 1PDF can be written as a spectral expansion
\begin{equation}
    \label{spectral_expansion_1PDF}
    P(\phi,\pi;t)=\Psi_{0}^*(\phi,\pi)\sum_{N=0}^{\infty}c_N\Psi_{N}(\phi,\pi)e^{-\Lambda_{N}t},
\end{equation}
where $\Lambda_{N}$ and $\Psi_{N}^{(*)}(\phi,\pi)$ are the respective eigenvalues and (adjoint) eigenstates to the (adjoint) Fokker-Planck operator
\begin{subequations}
    \label{FP_eigenequations}
    \begin{align}
        \mathcal{L}_{FP}\Psi_{N}(\phi,\pi)&=-\Lambda_{N}\Psi_{N}(\phi,\pi),\\
        \mathcal{L}_{FP}^*\Psi_{N}^*(\phi,\pi)&=-\Lambda_{N}\Psi^*_{N}(\phi,\pi),
    \end{align}
\end{subequations}
and $c_N$ are coefficients. We consider eigenstates that obey the biorthogonality and completeness relations
\begin{subequations}
    \label{biorthogonal&completeness_relations}
    \begin{align}
        \qty(\Psi_{N}^*,\Psi_{N'})&=\delta_{N'N},\\
        \sum_{N}\Psi_{N}^*(\phi,\pi)\Psi_{N}(\phi',\pi')&=\delta(\phi-\phi')\delta(\pi-\pi'),
    \end{align}
\end{subequations}
\noindent and there exists an equilibrium state $P_{eq}(\phi,\pi)=\Psi_{0}^*(\phi,\pi)\Psi_{0}(\phi,\pi)$ obeying $\partial_tP_{eq}(\phi,\pi)=0$. All sums of this form run from $N=0$ to $N=\infty$. Unlike the overdamped case, an explicit expression for this equilibrium state can't be written analytically for a general potential.

To obtain stochastic correlators, we introduce the transfer matrix $U(\phi_0,\phi,\pi_0,\pi;t-t_0)$ between $(\phi_0,\pi_0)=(\phi(t_0,\mathbf{x}),\pi(t_0,\mathbf{x}))$ and $(\phi,\pi)=(\phi(t,\mathbf{x}),\pi(t,\mathbf{x}))$, which is defined as the Green's function of the Fokker-Planck equation 
\begin{equation}
    \label{time-evolution_op_FP_equation}
    \partial_tU(\phi_0,\phi,\pi_0,\pi;t-t_0)=\mathcal{L}_{FP}U(\phi_0,\phi,\pi_0,\pi;t-t_0)
\end{equation}
for all values of $\phi_0$ and $\pi_0$. Then, the time-dependence of the 1PDF is given by
\begin{equation}
    \label{time-dependence_1PDF}
    P(\phi,\pi;t)=\int d\phi_0\int d\pi_0 P(\phi_0,\pi_0;t_0)U(\phi_0,\phi,\pi_0,\pi;t-t_0).
\end{equation}
From Eq. (\ref{spectral_expansion_1PDF}), making use of the relations (\ref{biorthogonal&completeness_relations}), we find that the transfer matrix can be written with the spectral expansion as
\begin{equation}
    \label{spectral_expansion_time-ev_operator}
    U(\phi_0,\phi,\pi_0,\pi;t-t_0)=\frac{\Psi_{0}^*(\phi,\pi)}{\Psi_{0}^*(\phi_0,\pi_0)}\sum_{N}\Psi_{N}^*(\phi_0,\pi_0)\Psi_{N}(\phi,\pi)e^{-\Lambda_{N}(t-t_0)}.
\end{equation}

\subsection{Two-point stochastic correlation functions}
\label{subsec:2-pt_2O_correlators}

We can write an equilibrium 2-point probability distribution function (2PDF) as
\begin{equation}
    \label{2PDF}
    \begin{split}
        P_2(\phi_0,\phi,\pi_0,\pi;t-t_0)&=P(\phi_0,\pi_0;t_0)U(\phi_0,\phi,\pi_0,\pi;t-t_0)\\
        &=\Psi_{0}^*(\phi,\pi)\Psi_{0}(\phi_0,\pi_0)\sum_{N}\Psi_{N}^*(\phi_0,\pi_0)\Psi_{N}(\phi,\pi)e^{-\Lambda_{N}(t-t_0)},
    \end{split}
\end{equation}
where we take $P(\phi_0,\pi_0;t_0)=P_{eq}(\phi_0,\pi_0)$. Then, the 2-point timelike (equal-space) stochastic correlator between some functions $f(\phi_0,\pi_0)$ and $g(\phi,\pi)$ is given by
\begin{equation}
    \label{2pt_timelike_stochastic_correlator_fg}
    \begin{split}
        \expval{f(\phi_0,\pi_0)g(\phi,\pi)}=&\int d\phi_0\int d\phi\int d\pi_0\int d\pi P_2(\phi_0,\phi,\pi_0,\pi;t-t_0)f(\phi_0,\pi_0)g(\phi,\pi)\\
        =&\sum_{N}f_{0N}g_{N0}e^{-\Lambda_{N}(t-t_0)},
    \end{split}
\end{equation}
where
\begin{equation}
    \label{fnl}
        f_{NN'}=\qty(\Psi_{N},f\Psi_{N'}^*).
\end{equation}
We can compute spacelike correlators in the same way as we did for the OD stochastic approach, by defining a equilibrium 3-point probability distribution function (3PDF), where we evolve both $(\phi_1,\pi_1)$ and $(\phi_2,\pi_2)$ to $(\phi_0,\pi_0)$ independently, as 
\begin{equation}
    \label{3PDF_spectral_expansion}
    \begin{split}
        P_3^{(S)}(\phi_0,\phi_1,\phi_2,\pi_0,\pi_1,\pi_2;&t_0,t_1,t_2)\\=&P(\phi_0,\pi_0;t_0)U(\phi_0,\phi_1,\pi_0,\pi_1;t_1-t_0)U(\phi_0,\phi_2,\pi_0,\pi_2;t_2-t_0)\\
        =&\frac{\Psi_{0}(\phi_0,\pi_0)\Psi^*_{0}(\phi_1,\pi_1)\Psi^*_{0}(\phi_2,\pi_2)}{\Psi^*_{0}(\phi_0,\pi_0)}\sum_{N}\Psi_{N}^*(\phi_0,\pi_0)\Psi_{N}(\phi_1,\pi_1)\\&\times\sum_{N'}\Psi_{N'}^*(\phi_0,\pi_0)\Psi_{N'}(\phi_2,\pi_2)e^{-\qty(\Lambda_{N}(t_1-t_0)+\Lambda_{N'}(t_2-t_0))},
    \end{split}
\end{equation}
where the superscript $(S)$ indicates that it is the 3PDF used to define spacelike correlators\footnote{Again, we could similarly define a 3PDF for computing timelike correlators, where the evolution would be from $(\phi_0,\pi_0)$ to $(\phi_1,\pi_1)$ to $(\phi_2,\pi_2)$ i.e. chronologically along a line of constant spatial coordinate (assuming $t_0<t_1<t_2$).}. To evaluate the spacelike (equal-time) stochastic correlators, we compute the 3-point function between two timelike separated points $t_1$ and $t_2$ and some intermediate time coordinate $t_r$, defined in Eq. (\ref{space_time_coordinate}). The spacelike stochastic correlator between the functions $f(\phi(t,\mathbf{x}_1),\pi(t,\mathbf{x}_1))$ and $g(\phi(t,\mathbf{x}_2),\pi(t,\mathbf{x}_2))$ is given by integrating over $\phi_r$ and $\pi_r$ as
\begin{equation}
    \label{2pt_spacelike_stochastic_correlators_fg}
    \begin{split}
        \langle& f(\phi,\pi;t,\mathbf{x}_1)g(\phi,\pi;t,\mathbf{x}_2)\rangle\\=&\int d\phi_r\int d\phi_1\int d\phi_2\int d\pi_r\int d\pi_1\int d\pi_2 P_3(\phi_r,\phi_1,\phi_2,\pi_r,\pi_1,\pi_2;t_r,t_1,t_2)\\&\times f(\phi_1,\pi_1)g(\phi_2,\pi_2)
        \\=&\int d\phi_r\int d\pi_r \frac{\Psi_{0}(\phi_r,\pi_r)}{\Psi_{0}^*(\phi_r,\pi_r)}\sum_{NN'}\Psi_{N}^*(\phi_r,\pi_r)\Psi_{N'}^*(\phi_r,\pi_r)f_{N}g_{N'}\abs{Ha(t)(\mathbf{x}_1-\mathbf{x}_2)}^{-\frac{\Lambda_{N}+\Lambda_{N'}}{H}}.
    \end{split}
\end{equation}

\subsection{Four-point stochastic correlation functions}
\label{subsec:4-pt_2o_stochastic_correlator}

For the timelike 4-point functions, we write the equilibrium 4-point probability distribution function (4PDF) as
\begin{equation}
    \label{4PDF}
    \begin{split}
        P_4^{(T)}&(\phi_1,\phi_2,\phi_3,\phi_4,\pi_1,\pi_2,\pi_3,\pi_4;t_1,t_2,t_3,t_4)\\:=&P(\phi_1,\pi_1;t_1)U(\phi_1,\phi_2,\pi_1,\pi_2;t_2-t_1)U(\phi_2,\phi_3,\pi_2,\pi_3;t_3-t_2)U(\phi_3,\phi_4,\pi_3,\pi_4;t_4-t_3)\\
        =&\Psi_0(\phi_1,\pi_1)\Psi^*_0(\phi_4,\pi_4)\sum_{N}\Psi_N^*(\phi_1,\pi_1)\Psi_N(\phi_2,\pi_2)\sum_{N'}\Psi_{N'}^*(\phi_2,\pi_2)\Psi_{N'}(\phi_3,\pi_3)\\&\times\sum_{N''}\Psi_{N''}^*(\phi_3,\pi_3)\Psi_{N''}(\phi_4,\pi_4),
    \end{split}
\end{equation}
where the superscript $(T)$ indicates we are using this 4PDF to compute timelike correlators. Assuming that $t_1<t_2<t_3<t_4$, and that $P(\phi_1,\pi_1;t_1)=P_{eq}(\phi_1,\pi_1)$, the timelike 4-point correlation function is given by
\begin{equation}
    \label{4-pt_stochastic_function_general}
    \begin{split}
        \langle f_1(\phi_1,\pi_1)f_2&(\phi_2,\pi_2)f_3(\phi_3,\pi_3)f_4(\phi_4,\pi_4)\rangle\\=&\prod_{i=1}^4\int d\phi_i\int d\pi_i P_4^{(T)}(\phi_1,\phi_2,\phi_3,\phi_4,\pi_1,\pi_2,\pi_3,\pi_4;t_1,t_2,t_3,t_4)\\&\times f_1(\phi_1,\pi_1)f_2(\phi_2,\pi_2)f_3(\phi_3,\pi_3)f_4(\phi_4,\pi_4)\\
        =&\sum_{N''N'N}(f_1)_{0N}(f_2)_{NN'}(f_3)_{N'N''}(f_4)_{N''0}e^{-\Lambda_N(t_2-t_1)-\Lambda_{N'}(t_3-t_2)-\Lambda_{N''}(t_4-t_3)}.
    \end{split}
\end{equation}
In a similar computation to the 2-point function, we can compute the spacelike 4-point function. Now, we define the ``spacelike'' equilibrium 5-point probability distribution function (5PDF) as
\begin{equation}
    \label{5PDF}
    \begin{split}
        P_5^{(S)}&(\phi_0,\phi_1,\phi_2,\phi_3,\phi_4,\pi_1,\pi_2,\pi_3,\pi_4;t_0,t_1,t_2,t_3,t_4)\\:=&P(\phi_0,\pi_0;t_0)U(\phi_0,\phi_1,\pi_0,\pi_1;t_1-t_0)U(\phi_1,\phi_2,\pi_1,\pi_2;t_2-t_0)U(\phi_2,\phi_3,\pi_2,\pi_3;t_3-t_0)\\&\times U(\phi_3,\phi_4,\pi_3,\pi_4;t_4-t_0)\\
        =&\frac{\Psi_0(\phi_0,\pi_0)}{\Psi_0^*(\phi_0,\pi_0)^3}\prod_{i=1}^4\lsb\Psi_0^*(\phi_i,\pi_i)\sum_N\Psi_N^*(\phi_0,\pi_0)\Psi_N(\phi_i,\pi_i)\rsb.
    \end{split}
\end{equation}
Using the $t_r$ coordinate in Eq. (\ref{space_time_coordinate}) and assuming $\abs{\mathbf{x}}=\abs{\mathbf{x}_i-\mathbf{x}_j}\text{ }\forall i\ne j$, the spacelike stochastic 4-point function between some functions $f_i(\phi,\pi)$, $i\in\{1,2,3,4\}$, is given by
\begin{equation}
    \label{spacelike_4-pt_correlator_f}
    \begin{split}
        \langle& f_1(\phi_1,\pi_1)f_2(\phi_2,\pi_2)f_3(\phi_3,\pi_3)f_4(\phi_4,\pi_4)\rangle\\
        =&\int d\phi_r\int d\pi_r\prod_{i=1}^4\int d\phi_i\int d\pi_i P_5^{(S)}(\phi_r,\phi_1,\phi_2,\phi_3,\phi_4,\pi_r,\pi_1,\pi_2,\pi_3,\pi_4;t_r,t_1,t_2,t_3,t_4)\\&\times f_1(\phi_1,\pi_1)f_2(\phi_2,\pi_2)f_3(\phi_3,\pi_3)f_4(\phi_4,\pi_4)\\
        =&\int d\phi_r\int d\pi_r\frac{\Psi_0(\phi_r,\pi_r)}{\Psi_0^*(\phi_r,\pi_r)^3}\prod_{i=1}^4\lsb\sum_N\Psi_N^*(\phi_r,\pi_r)(f_i)_{N0}\abs{Ha(t)\mathbf{x}}^{-\frac{\Lambda_N}{H}}\rsb.
    \end{split}
\end{equation}

\section{Free fields in stochastic theory}
\label{sec:so_stochastic_free_fields}

\subsection{The free eigenspectrum}
\label{subsec:free_eigenspectrum}

Now that we have set up the formalism for the second-order stochastic correlators, we can consider the question: what stochastic parameters are needed (if any) to consider it an effective theory of scalar QFT in de Sitter? We will start by answering this question for free fields by computing the 2-point stochastic correlators explicitly and comparing them with exact results from QFT\footnote{Throughout this section on free fields, our stochastic parameters should really have a superscript $(0)$ - $m_S^{(0)}$, $\sigma_{ij}^{2(0)}$ - to indicate that they are for free fields. I omit them because it clogs up the equations.}. We write the stochastic equations for free fields as
\begin{equation}
    \label{free_stochastic_equations}
        \begin{pmatrix}\Dot{\phi}\\\Dot{\pi}\end{pmatrix}=\begin{pmatrix}\pi\\-3H\pi-m^2_S\phi\end{pmatrix}+\begin{pmatrix}\xi_{\phi}(t,\mathbf{x})\\\xi_{\pi}(t,\mathbf{x})\end{pmatrix},
\end{equation}

where the subscript `S' represents a stochastic quantity and the stochastic noise contributions $\xi_{\phi,\pi}$ are left undetermined. From this equation, we can use the formalism developed above to compute correlators. The first thing to note is that the equilibrium solution for free fields can be computed as
\begin{equation}
    \label{1PDF_equilibrium_solution}
    P^{(0)}_{eq}(\phi,\pi)\propto 
    e^{-\frac{3H\qty(((9H^2+m_S^2)\sigma_{\phi\phi}^2+6H\sigma_{\phi\pi}^2+\sigma_{\pi\pi}^2)\pi^2+6Hm_S^2\sigma_{\phi\phi}^2\phi\pi+(m_S^2\sigma_{\phi\phi}^2+\sigma_{\pi\pi}^2)m_S^2\phi^2)}{(m_S^2\sigma_{\phi\phi}^2+3H\sigma_{\phi\pi}^2+\sigma_{\pi\pi}^2)^2+9H^2(\sigma_{\phi\phi}^2\sigma_{\pi\pi}^2-\sigma_{\phi\pi}^4)}},
\end{equation}\\
with the normalisation condition $\int d\phi\int d\pi P_{eq}(\phi,\pi)=1$. To solve beyond the equilibrium state, we define a new set of variables $(q,p)$, which are related to $(\phi,\pi)$ by
\begin{equation}
    \label{phi,pi-->q,p}
    \begin{pmatrix}p\\q\end{pmatrix}=\frac{1}{\sqrt{1-\frac{\alpha_S}{\beta_S}}}\begin{pmatrix}1&\alpha_S H\\\frac{1}{\beta_S H}&1\end{pmatrix}\begin{pmatrix}\pi\\\phi\end{pmatrix},
\end{equation}
where $\alpha_S=\frac{3}{2}-\nu_S$ and $\beta_S=\frac{3}{2}+\nu_S$, with $\nu_S=\sqrt{\frac{9}{4}-\frac{m_S^2}{H^2}}$. The inverse transformation is given by
\begin{equation}
    \label{p,q-->pi,phi}
    \begin{pmatrix}\pi\\\phi\end{pmatrix}=\frac{1}{\sqrt{1-\frac{\alpha_S}{\beta_S}}}\begin{pmatrix}1&-\alpha_S H\\-\frac{1}{\beta_S H}&1\end{pmatrix}\begin{pmatrix}p\\q\end{pmatrix}.
\end{equation}
In these new variables, the stochastic equation (\ref{2d_stochastic_eq}) can be written as
\begin{equation}
    \label{langevin_p,q}
    \begin{pmatrix}\Dot{q}\\\Dot{p}\end{pmatrix}=\begin{pmatrix}-\alpha_S H q\\-\beta_S H p\end{pmatrix}+\begin{pmatrix}\xi_q\\\xi_p\end{pmatrix},
\end{equation}
where $\xi_q=\frac{1}{\sqrt{1-\frac{\alpha_S}{\beta_S}}}\qty(\frac{1}{\beta_S H}\xi_{\pi}+\xi_{\phi})$ and $\xi_p=\frac{1}{\sqrt{1-\frac{\alpha_S}{\beta_S}}}\qty(\xi_{\pi}+\alpha_S H \xi_{\phi})$, such that $\xi_{q,p}$ have the same statistics (\ref{white_noise_general}) as $\xi_{\phi,\pi}$. The noise amplitudes are related by
\begin{subequations}
    \label{noise_amplitudes_transformation}
    \begin{equation}
        \label{sigma_q,p-->sigma_phi,pi}
        \begin{split}
            \sigma_{qq}^2=&\frac{1}{1-\frac{\alpha_S}{\beta_S}}\qty(\frac{1}{\beta_S^2H^2}\sigma_{\pi\pi}^2+\frac{2}{\beta_S H}\sigma_{\phi\pi}^2+\sigma_{\phi\phi}^2),\\
            \sigma_{qp}^2=&\frac{1}{1-\frac{\alpha_S}{\beta_S}}\qty(\frac{1}{\beta_S H}\sigma_{\pi\pi}^2+\qty(1+\frac{\alpha_S}{\beta_S })\sigma_{\phi\pi}^2+\alpha_S H\sigma_{\phi\phi}^2),\\
            \sigma_{pp}^2=&\frac{1}{1-\frac{\alpha_S}{\beta_S}}\qty(\sigma_{\pi\pi}^2+2\alpha_S H\sigma_{\phi\pi}^2+\alpha_S^2H^2\sigma_{\phi\phi}^2);
        \end{split}
    \end{equation}\\
    \begin{equation}
        \label{sigma_phi,pi-->sigma_q,p}
        \begin{split}
            \sigma_{\phi\phi}^2=&\frac{1}{1-\frac{\alpha_S}{\beta_S}}\qty(\frac{1}{\beta_S^2H^2}\sigma_{pp}^2-\frac{2}{\beta_S H}\sigma_{qp}^2+\sigma_{qq}^2),\\
            \sigma_{\phi\pi}^2=&\frac{1}{1-\frac{\alpha_S}{\beta_S}}\qty(-\frac{1}{\beta_S H}\sigma_{pp}^2+\qty(1+\frac{\alpha_S}{\beta_S })\sigma_{qp}^2-\alpha_S H\sigma_{qq}^2),\\
            \sigma_{\pi\pi}^2=&\frac{1}{1-\frac{\alpha_S}{\beta_S}}\qty(\sigma_{pp}^2-2\alpha_S H\sigma_{qp}^2+\alpha_S^2H^2\sigma_{qq}^2).   
        \end{split}
    \end{equation}
\end{subequations}
Thus, we have two 1-dimensional Langevin equations that are only related by the correlated noise. These directions align with the growing and decaying modes, $q$ and $p$ respectively. The resulting Fokker-Planck equation for the $(q,p)$ 1PDF $P(q,p;t)$ is
\begin{equation}
    \label{p,q_fokker-planck}
    \begin{split}
    \partial_tP(q,p;t)=&\alpha_S H P(q,p;t)+\alpha_S H q\partial_q P(q,p;t)+\frac{1}{2}\sigma_{qq}^2\partial_q^2P(q,p;t)\\&+\beta_S H P(q,p;t)+\beta_S H p\partial_pP(q,p;t)+\frac{1}{2}\sigma_{pp}^2\partial_p^2P(q,p;t)\\&+\sigma_{qp}^2\partial_p\partial_qP(q,p;t) ,
    \end{split}
\end{equation}
The equilibrium solution to this Fokker-Planck equation is
\begin{equation}
    \label{p,q_equilibrium_1PDF}
    P_{eq}(q,p)=\frac{3H}{\pi}\sqrt{\frac{\alpha_S \beta_S}{9\sigma_{qq}^2\sigma_{pp}^2-4\alpha_S\beta_S \sigma_{qp}^4}}e^{-\frac{9H(\sigma_{qq}^2\beta_S p^2-\frac{4}{3}\sigma_{qp}^2\alpha_S\beta_S q p+\sigma_{pp}^2\alpha_S q^2)}{9\sigma_{qq}^2\sigma_{pp}^2-4\alpha_S\beta_S \sigma_{qp}^4}},
\end{equation}\\
where we have included the normalisation found by the condition $\int dp \int dq P_{eq}(q,p)=1$. 

This Fokker-Planck equation can now be solved using the spectral expansion outlined in Sec. \ref{sec:stochastic_correlators}, simply replacing $(\phi,\pi)$ with $(q,p)$ using their relation (\ref{p,q-->pi,phi}). The eigenequations in question are
\begin{subequations}
    \label{free_eigenequations}
    \begin{align}
        \mathcal{L}_{FP}^{(0)}\Psi_{rs}^{(0)}(q,p)&=-\Lambda_{rs}^{(0)}\Psi_{rs}^{(0)}(q,p)\\
        \mathcal{L}_{FP}^{(0)*}\Psi_{rs}^{(0)*}(q,p)&=-\Lambda_{rs}^{(0)}\Psi_{rs}^{(0)*}(q,p),
    \end{align}
\end{subequations}
where we separate $N$ into two integers $(r,s)\in\ZZ^+$, with $r$ and $s$ corresponding to $p$ and $q$ respectively. While the eigenstates don't strictly separate because $\sigma_{qp}^2\ne0$, it is still convenient to make this adjustment. They are solved to give
\begin{subequations}
    \label{free_eigenspectrum_solved_general_noise}
    \begin{align}
    \label{free_eigenvalue}
        \Lambda_{rs}^{(0)}&=\lb s\alpha_S+r\beta_S\rb H,\\
    \label{free_eigenstate_general_noise}
        \Psi_{rs}^{(0)}(q,p)&=e^{-\frac{9H(\sigma_{qq}^2\beta_S p^2-\frac{4}{3}\sigma_{qp}^2\alpha_S\beta_S q p+\sigma_{pp}^2\alpha_S q^2)}{9\sigma_{qq}^2\sigma_{pp}^2-4\alpha_S\beta_S \sigma_{qp}^4}}\sum_{r's'}a_{rsr's'}q^{r'}p^{s'},\\ 
    \label{free_adjoint_eigenstate_general_noise}
        \Psi_{rs}^{(0)*}(q,p)&=\sum_{r's'}a^*_{rsr's'}q^{r'}p^{s'},
    \end{align}
\end{subequations}
where $a_{rsr's'}^{(*)}$ are some coefficients that depend on the stochastic parameters. Unfortunately, there doesn't exist a general solution and these coefficients must be computed systematically for each values of $(r,s)$. There exists two important cases where the eigenstates can be computed explicitly: for $\sigma_{qp}^2=0$ and for $\sigma_{qp}^2=\sigma_{pp}^2=0$ \footnote{There is also a method by which one can analytically solve the Fokker-Planck equation when $\sigma_{\phi\phi}^2=\sigma_{\phi\pi}^2=0$. Instead of a spectral expansion, one can introduce ladder operators to solve for the 1PDF. I won't go into details of this here; for further details, see Ref. \cite{pavliotis_txtbk:2014s}}. For $\sigma_{qp}^2=0$, the $(q,p)$ stochastic equations (\ref{langevin_p,q}) completely decouple and can be solved independently. The result is that the eigenstates are given by
\begin{subequations}
    \label{free_eigenstates_sqp=0}
    \begin{align}
        \label{free_non-adjoint_estate_sqp=0}
        \begin{split}
        \Psi_{rs}^{(0)}(q,p)&=P_r(p)Q_s(q),
        \end{split}\\
        \label{free_adjoint_estatesqp=0}
        \begin{split}
        \Psi_{rs}^{(0)*}(q,p)&=P_r^*(p)Q_s^*(q),
        \end{split}
    \end{align}
\end{subequations}
where
\begin{subequations}
    \label{Ps(p)&Qs(q)}
    \begin{align}
        \begin{split}
            Q_s(q)&=\frac{1}{\sqrt{2^{s}s!}}\qty(\frac{\alpha_S H}{\pi\sigma_{qq}^{2}})^{1/4}H_s\qty(\sqrt{\frac{\alpha_S H}{\sigma_{qq}^{2}}}q)e^{-\frac{\alpha_S H}{\sigma_{qq}^{2}}q^2},\\
            Q_s^*(q)&=\frac{1}{\sqrt{2^{s}s!}}\qty(\frac{\alpha H}{\pi\sigma_{qq}^{2}})^{1/4}H_s\qty(\sqrt{\frac{\alpha_S H}{\sigma_{qq}^{2}}}q),
        \end{split}\\&\nonumber\\
        \begin{split}
            P_r(p)&=\frac{1}{\sqrt{2^{r}r!}}\qty(\frac{\beta_S H}{\pi\sigma_{pp}^{2}})^{1/4}H_r\qty(\sqrt{\frac{\beta_S H}{\sigma_{pp}^{2}}}q)e^{-\frac{\beta_S H}{\sigma_{pp}^{2}}p^2},\\
            P_r^*(p)&=\frac{1}{\sqrt{2^{r}r!}}\qty(\frac{\beta_S H}{\pi\sigma_{pp}^{2}})^{1/4}H_r\qty(\sqrt{\frac{\beta_S H}{\sigma_{pp}^{2}}}q).
        \end{split}\
    \end{align}
\end{subequations}
For the case where we also set $\sigma_{pp}^2=0$, the eigenstates can be written as\footnote{To take the limit, we have used the identity $\lim_{\epsilon\rightarrow0}\frac{(-1)^{-n}(\sqrt{2}\epsilon)^{n-1}}{\sqrt{\pi}}H_n\qty(\frac{x}{\sqrt{2}\epsilon})e^{-\frac{x^2}{2\epsilon^2}}=\delta^{(n)}(x)$.}
\begin{subequations}
    \label{free_eigenstates_spp=0}
    \begin{align}
        \label{free_non-adjoint_estate_spp=0}
        \lim_{\sigma_{pp}^2\rightarrow0}\Psi_{rs}^{(0)}(q,\Tilde{p})&=\frac{(-1)^{-r}}{\sqrt{2^{r+s}r!s!}}\qty(\frac{\alpha_S H}{\sigma_{qq}^2})^{1/4}\delta^{(r)}(\Tilde{p})H_s\qty(\sqrt{\frac{\alpha_S H}{\sigma_{qq}^2}q})e^{-\frac{\alpha_S H}{\sigma_{qq}^2}q^2},\\
        \label{free_adjoint_estate_spp=0}
        \lim_{\sigma_{pp}^2\rightarrow0}\Psi_{rs}^{(0)*}(q,\Tilde{p})&=\sqrt{\frac{2^r}{2^sr!s!}}\qty(\frac{\alpha_S H}{\pi^2\sigma_{qq}^2})^{1/4}\Tilde{p}^rH_s\qty(\sqrt{\frac{\alpha_S H}{\sigma_{qq}^2}q}),
    \end{align}
\end{subequations}
where $\Tilde{p}=\sqrt{\frac{\beta_S H}{\sigma_{pp}^2}}p$ and superscript $(r)$ indicates we are taking the $r$th derivative of the $\delta$-function. These are well behaved eigenstates if we use $(q,\Tilde{p})$ as our variables, with which we have the biorthogonality and completeness relations. These two cases will become important later. 

\subsection{Free two-point stochastic correlators}
\label{subsec:free_2-pt_stochastic_correlators}

For now, we seek a way of solving the eigenequations whilst keeping our noise general without having to resort to a case-by-case basis. We can do this if we initially only consider the $q-q$, $q-p$, $p-q$ and $p-p$ 2-point correlation functions. For these, we only need to evolve $p$ or $q$ forward in time for any given correlator, not both simultaneously. Therefore, one only needs to use the 1-dimensional transfer matrices $U_q(q_0,q;t-t_0)$ and $U_p(p_0,p;t-t_0)$
defined by
\begin{subequations}
    \label{time-ev_1PDF}
    \begin{equation}
        \label{q_time-ev_1pdf}
        P_q(q;t)=\int dq_0 P_q(q_0;t_0)U_q(q_0,q;t-t_0),
    \end{equation}
    \begin{equation}
        \label{p_time-ev_1pdf}
        P_p(p;t)=\int dp_0 P_p(p_0;t_0)U_p(p_0,p;t-t_0).
    \end{equation}
\end{subequations}
They obey the 1-dimensional Fokker-Planck equations
\begin{subequations}
    \label{time-ev_pde}
    \begin{align}
        \label{q_time-ev_pde}
        \begin{split}
        \partial_t U_q(q_0,q;t-t_0)=&\alpha_SHU_q(q_0,q;t-t_0)+\alpha_SHq\partial_qU_q(q_0,q;t-t_0)\\&+\frac{1}{2}\sigma_{qq}^2\partial_q^2U_q(q_0,q;t-t_0),
        \end{split}\\
        \label{p_time-ev_pde}
        \begin{split}
        \partial_t U_p(p_0,p;t-t_0)=&\beta_SHU_p(p_0,p;t-t_0)+\beta_SHp\partial_pU_p(p_0,p;t-t_0)\\&+\frac{1}{2}\sigma_{pp}^2\partial_p^2U_p(p_0,p;t-t_0),
        \end{split}
    \end{align}
\end{subequations}
which can be derived in the standard way from the two components of the Langevin equation (\ref{langevin_p,q}). Alternatively, they can also be obtained from the two-dimensional Fokker-Planck equation (\ref{time-evolution_op_FP_equation}), by first observing that they can be expressed in terms of the two-dimensional time-evolution operators as
\begin{subequations}
    \begin{align}
        U_q(q_0,q;t-t_0)&=\int dp\, U(q_0,q,p_0,p;t-t_0),\\
        U_p(p_0,p;t-t_0)&=\int dq\, U(q_0,q,p_0,p;t-t_0),
    \end{align}
\end{subequations}
which do not depend on $p_0$ and $q_0$, respectively.
Integrating Eq. (\ref{time-evolution_op_FP_equation}) over $p$ and $q$, respectively, and integrating the relevant terms by parts gives Eq. (\ref{q_time-ev_pde}) and (\ref{p_time-ev_pde}). Then, using the spectral expansion method with the eigenstates (\ref{Ps(p)&Qs(q)}), the 1-dimensional transfer matrices are given by
\begin{subequations}
    \label{time-ev_solution}
    \begin{align}
        \label{q_time-ev_solution}
        U_q(q_0,q;t-t_0)&=\frac{Q_0^*(q)}{Q_0^*(q_0)}\sum_s Q^*_s(q_0)Q_s(q)e^{-s H\alpha_S (t-t_0)},\\
        \label{p_time-ev_solution}
        U_p(p_0,p;t-t_0)&=\frac{P_0^*(p)}{P_0^*(p_0)}\sum_r P_r^*(p_0)P_r(p)e^{-rH\beta_S(t-t_0)}.
    \end{align}
\end{subequations}
It is a special property of the coordinates $(q,p)$ that one obtains time-evolution equations that only depend on one variable. This is because the stochastic equations (\ref{langevin_p,q}) are diagonal i.e. the time derivative of $q$ is not dependent on $p$ and, similarly, the time derivative of $p$ is not dependent on $q$. It is this property that allows us to analytically evaluate the 2-point correlators using the 1-dimensional time-evolution operators. Note, however, that in order to calculate the higher-order correlators, one needs to evaluate the full time-evolution operator.

\subsubsection{2-point timelike stochastic correlators}

The timelike 2-point $(q,p)$ correlators are given by
\begin{subequations}
\label{p,q-correlators_time-ev}
\begin{align}
        \expval{q(0)q(t)}&=\int dp_0\int dq\int dq_0 P_{eq}(q_0,p_0)U_q(q_0,q;t)q_0 q=\frac{\sigma_{qq}^2}{2\alpha_S H}e^{-\alpha_S Ht},\\
        \expval{p(0)q(t)}&=\int dp_0\int dq\int dq_0 P_{eq}(q_0,p_0)U_q(q_0,q;t)p_0 q=\frac{\sigma_{qp}^2}{3H}e^{-\alpha_S Ht},\\
        \expval{q(0)p(t)}&=\int dq_0 \int dp \int dp_0 P_{eq}(q_0,p_0)U_p(p_0,p;t)q_0 p=\frac{\sigma_{qp}^2}{3H}e^{-\beta_S Ht},\\
        \expval{p(0)p(t)}&=\int dp\int dp_0\int dq_0 P_{eq}(q_0,p_0)U_p(p_0,p;t)p_0p=\frac{\sigma_{pp}^2}{2\beta_S H}e^{-\beta_S Ht}.
\end{align}
\end{subequations}
One can see explicitly here that the $q-q$ and $p-p$ correlators represent the growing and decaying modes respectively. Using Eq. (\ref{phi,pi-->q,p}), we can compute the $(\phi,\pi)$ 2-point functions from their $(q,p)$ counterparts using
\begin{small}
\begin{subequations}
\label{phi,pi_correlators_pq}
\begin{align}
        \label{phi-phi_pq}
        \expval{\phi(0)\phi(t)}&=\frac{1}{1-\frac{\alpha_S}{\beta_S}}\qty(\frac{1}{\beta_S^2H^2}\expval{p(0)p(t)}-\frac{1}{\beta_S H}\qty(\expval{q(0)p(t)}+\expval{p(0)q(t)})+\expval{q(0)q(t)}),\\
        \label{phi-pi_pq}
        \expval{\phi(0)\pi(t)}&=\frac{1}{1-\frac{\alpha_S}{\beta_S}}\qty(-\frac{1}{\beta_S H}\expval{p(0)p(t)}+\expval{q(0)p(t)}+\frac{\alpha_S}{\beta_S}\expval{p(0)q(t)}-\alpha_S H\expval{q(0)q(t)}),\\
        \label{pi-phi_pq}
        \expval{\pi(0)\phi(t)}&=\frac{1}{1-\frac{\alpha_S}{\beta_S}}\qty(-\frac{1}{\beta_S H}\expval{p(0)p(t)}+\frac{\alpha_S}{\beta_S}\expval{q(0)p(t)}+\expval{p(0)q(t)}-\alpha_S H\expval{q(0)q(t)}),\\
        \label{pi-pi_pq}
        \expval{\pi(0)\pi(t)}&=\frac{1}{1-\frac{\alpha_S}{\beta_S}}\qty(\expval{p(0)p(t)}-\alpha_S H\qty(\expval{q(0)p(t)}+\expval{p(0)q(t)})+\alpha_S^2H^2\expval{q(0)q(t)}).
\end{align}
\end{subequations}
\end{small}
Thus, the timelike 2-point $(\phi,\pi)$ stochastic correlators are
\begin{subequations}
\label{equal-space_(phi,pi)_stochastic_correlators}
\begin{align}
    \label{timelike_phi-phi_stochastic_correlator}
    \expval{\phi(0)\phi(t)}&=\frac{1}{1-\frac{\alpha_S}{\beta_S}}\qty[\qty(\frac{\sigma_{qq}^2}{2H\alpha_S}-\frac{\sigma_{qp}^2}{3H^2\beta_S})e^{-\alpha_S H t}+\qty(\frac{\sigma_{pp}^2}{2H^3\beta_S^3}-\frac{\sigma_{qp}^2}{3H^2\beta_S})e^{-\beta_S H t}],\\
    \label{timelike_phi-pi_stochastic_correlator}
    \expval{\phi(0)\pi(t)}&=\frac{1}{1-\frac{\alpha_S}{\beta_S}}\qty[\qty(-\frac{\sigma_{qq}^2}{2}+\frac{\alpha_S\sigma_{qp}^2}{3H\beta_S})e^{-\alpha_S H t}+\qty(-\frac{\sigma_{pp}^2}{2H^2\beta_S^2}+\frac{\sigma_{qp}^2}{3H})e^{-\beta_S H t}],\\
    \label{timelike_pi-phi_stochastic_correlator}
    \expval{\pi(0)\phi(t)}&=\frac{1}{1-\frac{\alpha_S}{\beta_S}}\qty[\qty(-\frac{\sigma_{qq}^2}{2}+\frac{\sigma_{qp}^2}{3H})e^{-\alpha_S H t}+\qty(-\frac{\sigma_{pp}^2}{2H^2\beta_S^2}+\frac{\alpha_S\sigma_{qp}^2}{3H\beta_S})e^{-\beta_S H t}],\\
    \label{timelike_pi-pi_stochastic_correlator}
    \expval{\pi(0)\pi(t)}&=\frac{1}{1-\frac{\alpha_S}{\beta_S}}\qty[\qty(\frac{\alpha_S H\sigma_{qq}^2}{2}-\frac{\alpha_S\sigma_{qp}^2}{3})e^{-\alpha_S H t}+\qty(\frac{\sigma_{pp}^2}{2H\beta_S}-\frac{\alpha_S\sigma_{qp}^2}{3})e^{-\beta_S H t}].
\end{align}
\end{subequations}
For completeness, we can also evaluate the stochastic variances. They are given by
\begin{subequations}
\label{stochastic_variance}
    \begin{align}
        \label{stoch_phiphi_variance}
        \expval{\phi^2}&=\frac{1}{1-\frac{\alpha_S}{\beta_S}}\qty(\expval{q^2}-\frac{2}{\beta_S H}\expval{qp}+\frac{1}{\beta_S^2H^2}\expval{p^2}),\\
        \label{stoch_phipi_variance}
        \expval{\phi\pi}&=\frac{1}{1-\frac{\alpha_S}{\beta_S}}\qty(-\frac{1}{\beta_S H}\expval{p^2}+\qty(1+\frac{\alpha_S}{\beta_S})\expval{qp}-\alpha_S H \expval{q^2}),\\
        \label{stoch_pipi_variance}
        \expval{\pi^2}&=\frac{1}{1-\frac{\alpha_S}{\beta_S}}\qty(\expval{p^2}-2\alpha_S H \expval{qp}+\alpha_S^2H^2\expval{q^2}),
    \end{align}
\end{subequations}
where
\begin{subequations}
    \label{p,q_variance}
    \begin{align}
        \label{<q^2>}
        \expval{q^2}&=\int dq\int dp P_{eq}(p,q)q^2=\frac{\sigma_{qq}^2}{2H\alpha_S},\\
        \label{<pq>}
        \expval{qp}&=\int dq \int dp P_{eq}(q,p)pq=\frac{\sigma_{qp}^2}{3H},\\
        \label{<p^2}
        \expval{p^2}&=\int dq\int dp P_{eq}(q,p)p^2=\frac{\sigma_{pp}^2}{2H\beta_S}.
    \end{align}
\end{subequations}
Thus, the stochastic variances in terms of a general noise term are given by
\begin{subequations}
    \label{stoch_variance_noise}
    \begin{align}
        \label{stoch_phiphi_variance_noise}
        \expval{\phi^2}&=\frac{1}{1-\frac{\alpha}{\beta}}\qty(\frac{\sigma_{qq}^2}{2H \alpha_S}-\frac{2\sigma_{qp}^2}{3\beta_S H^2}+\frac{\sigma_{pp}^2}{2\beta_S^3H^3}),\\
        \label{stoch_phipi_variance_noise}
        \expval{\phi\pi}&=\frac{1}{1-\frac{\alpha_S}{\beta_S}}\qty(-\frac{\sigma_{qq}^2}{2}+\qty(1+\frac{\alpha_S}{\beta_S})\frac{\sigma_{qp}^2}{3H}-\frac{\sigma_{pp}^2}{2\beta^2H^2}),\\
        \label{stoch_pipi_variance_noise}
        \expval{\pi^2}&=\frac{1}{1-\frac{\alpha}{\beta}}\qty(\frac{\alpha_S H \sigma_{qq}^2}{2}-\frac{2\alpha_S\sigma_{qp}^2}{3}+\frac{\sigma_{pp}^2}{2\beta_S H}).
    \end{align}
\end{subequations}
Note that these are the same as taking $t\rightarrow0$ in the timelike 2-point functions (\ref{equal-space_(phi,pi)_stochastic_correlators}). We can compare the stochastic variance with the QFT variance (\ref{quantum_variance}); since these are short-distance quantities, we don't expect the stochastic approach to reproduce them. Indeed, we have seen that these quantities are UV divergent, a property that won't be reproduced in an effective theory of the IR regime, as is the case with the stochastic approach. However, unequal-time correlators should agree.

Now that we have calculated all the timelike 2-point $(\phi,\pi)$ stochastic correlators and their associated variance, we can write a general solution for the 2PDF, bypassing our inability to generally compute the eigenstates:
\begin{equation}
    \label{joint_PDF_time}
    P_2(\phi_0,\phi,\pi_0,\pi;t-t_0)= N(t) e^{-\frac{1}{2}\Phi^T M(t-t_0)^{-1}\Phi},
\end{equation}
where
\begin{subequations}
    \label{matrix&vector}
    \begin{align}
        \label{vector}
        \Phi&=(\pi_0,\phi_0,\pi,\phi)^T,\\
        \label{matrix}
        M(t-t_0)&=\begin{pmatrix}\expval{\pi_0\pi_0}&\expval{\pi_0\phi_0}&\expval{\pi_0\pi}&\expval{\pi_0\phi}\\\expval{\phi_0\pi_0}&\expval{\phi_0\phi_0}&\expval{\phi_0\pi}&\expval{\phi_0\phi}\\\expval{\pi\pi_0}&\expval{\pi\phi_0}&\expval{\pi\pi}&\expval{\pi\phi}\\\expval{\phi\pi_0}&\expval{\phi\phi_0}&\expval{\phi\pi}&\expval{\phi\phi}\end{pmatrix}.
    \end{align}\\
\end{subequations}
The constant $N(t)$ is found by the condition $\int d\phi \int d\phi_0 \int d\pi \int d\pi_0 P_2(\phi_0,\phi,\pi_0,\pi;t)=1$. Note that Eq. (\ref{joint_PDF_time}) is a solution to the Fokker-Planck equation (\ref{phi-pi_fokker-planck_eq}) for any value of $\phi_0$ and $\pi_0$. This allows us to calculate higher-order timelike 2-point correlators directly. Let's consider the example of the $\phi^2$ correlator. Using the first line of Eq. (\ref{2pt_timelike_stochastic_correlator_fg}) with the 2PDF (\ref{joint_PDF_time}),
\begin{equation}
    \label{phi2_free_timelike_stochastic_correlator}
    \begin{split}
        \expval{\phi(0)^2\phi(t)^2}=&\frac{1}{1-\frac{\alpha_S}{\beta_S}}\Bigg[\qty(\frac{\sigma_{qq}^2}{2H \alpha_S}-\frac{2\sigma_{qp}^2}{3\beta_S H^2}+\frac{\sigma_{pp}^2}{2\beta_S^3H^3})^2+2\Bigg(\qty(\frac{\sigma_{qq}^2}{2H\alpha_S}-\frac{\sigma_{qp}^2}{3H^2\beta_S})e^{-\alpha_S H t}\\&+\qty(\frac{\sigma_{pp}^2}{2H^3\beta_S^3}-\frac{\sigma_{qp}^2}{3H^2\beta_S})e^{-\beta_S H t}\Bigg)^2\Bigg]\\
        =&\expval{\phi^2}^2+2\expval{\phi(0)\phi(t)}^2.
    \end{split}
\end{equation}
The convenient form of the second line is nothing other than a manifestation of Wick's theorem. Since the 2PDF (\ref{joint_PDF_time}) is Gaussian, Wick's theorem can be used to compute 2-point correlators of arbitrary functions of $\phi$ and $\pi$. We note that this mirrors the equivalent property apparent in QFT. 

\subsubsection{2-point spacelike stochastic correlators}

We will now compute the 2-point spacelike correlators by using Eq. (\ref{2pt_spacelike_stochastic_correlators_fg}). We will follow a similar line of reasoning as the timelike correlators, whereby we calculate the spacelike $(q,p)$ correlators using the 1-dimensional transfer matrices (\ref{time-ev_solution}) and then move to the spacelike $(\phi,\pi)$ correlators using Eq. (\ref{phi,pi_correlators_pq}). The spacelike $(q,p)$ correlators are given by
\begin{subequations}
    \begin{align}
        \begin{split}
            \expval{q(t,\mathbf{x})q(t,\mathbf{x}')}&=\int dq_r \int dp_r P_{eq}(q_r,p_r)\int dq U_q(q_r,q;t-t_r)q\int dq' U_q(q_r,q';t-t_r)q'\\
            &=\frac{\sigma_{qq}^2}{2\alpha_S H}\qty(Ha(t)\abs{\mathbf{x}-\mathbf{x}'})^{-2\alpha_S},
        \end{split}\\
        \begin{split}
            \expval{p(t,\mathbf{x})q(t,\mathbf{x}')}&=\int dq_r \int dp_r P_{eq}(q_r,p_r)\int dp U_p(p_r,p;t-t_r)p\int dq' U_q(q_r,q';t-t_r)q'\\
            &=\frac{\sigma_{qp}^2}{3H}\qty(Ha(t)\abs{\mathbf{x}-\mathbf{x}'})^{-3},
        \end{split}\\
        \begin{split}
            \expval{q(t,\mathbf{x})p(t,\mathbf{x}')}&=\int dq_r \int dp_r P_{eq}(q_r,p_r)\int dq U_q(q_r,q;t-t_r)q\int dp' U_p(p_r,p';t-t_r)p'\\
            &=\frac{\sigma_{qp}^2}{3 H}\qty(Ha(t)\abs{\mathbf{x}-\mathbf{x}'})^{-3},
        \end{split}\\
        \begin{split}
            \expval{p(t,\mathbf{x})p(t,\mathbf{x}')}&=\int dq_r \int dp_r P_{eq}(q_r,p_r)\int dp U_p(p_r,p;t-t_r)p\int dp' U_p(p_r,p';t-t_r)\\
            &=\frac{\sigma_{pp}^2}{2\beta_S H}\qty(Ha(t)\abs{\mathbf{x}-\mathbf{x}'})^{-2\beta_S}.
        \end{split}
    \end{align}
\end{subequations}Thus, using the spacelike version of Eq. (\ref{phi,pi_correlators_pq}), the spacelike $(\phi,\pi)$ stochastic 2-point correlators are given by
\begin{subequations}
\label{equal-time_(phi,pi)_stochastic_correlators}
\begin{align}
    \label{spacelike_phi-phi_stochastic_correlator}
    \begin{split}
    \expval{\phi(t,\mathbf{0})\phi(t,\mathbf{x})}=&\frac{1}{1-\frac{\alpha_S}{\beta_S}}\Bigg[\frac{\sigma_{qq}^2}{2H\alpha_S}\abs{Ha(t)\mathbf{x}}^{-2\alpha_S}+\frac{\sigma_{pp}^2}{2H^3\beta_S^3}\abs{Ha(t)\mathbf{x}}^{-2\beta_S}\\&-\frac{2\sigma_{qp}^2}{3H^2\beta_S}\abs{Ha(t)\mathbf{x}}^{-3}\Bigg],
    \end{split}\\
    \label{spacelike_phi-pi_stochastic_correlator}
    \begin{split}
    \expval{\phi(t,\mathbf{0})\pi(t,\mathbf{x})}=&\frac{1}{1-\frac{\alpha_S}{\beta_S}}\Bigg[-\frac{\sigma_{qq}^2}{2}\abs{Ha(t)\mathbf{x}}^{-2\alpha_S}-\frac{\sigma_{pp}^2}{2H^2\beta_S^2}\abs{Ha(t)\mathbf{x}}^{-2\beta_S}\\&+\frac{\sigma_{qp}^2}{H\beta_S}\abs{Ha(t)\mathbf{x}}^{-3}\Bigg],
    \end{split}\\
    \label{spacelike_pi-phi_stochastic_correlator}
    \begin{split}
    \expval{\pi(t,\mathbf{0})\phi(t,\mathbf{x})}=&\frac{1}{1-\frac{\alpha_S}{\beta_S}}\Bigg[-\frac{\sigma_{qq}^2}{2}\abs{Ha(t)\mathbf{x}}^{-2\alpha_S}-\frac{\sigma_{pp}^2}{2H^2\beta_S^2}\abs{Ha(t)\mathbf{x}}^{-2\beta_S}\\&+\frac{\sigma_{qp}^2}{H\beta_S}\abs{Ha(t)\mathbf{x}}^{-3}\Bigg],
    \end{split}\\
    \label{spacelike_pi-pi_stochastic_correlator}
    \begin{split}
    \expval{\pi(t,\mathbf{0})\pi(t,\mathbf{x})}=&\frac{1}{1-\frac{\alpha_S}{\beta_S}}\Bigg[\frac{\alpha_S H\sigma_{qq}^2}{2}\abs{Ha(t)\mathbf{x}}^{-2\alpha_S}+\frac{\sigma_{pp}^2}{2H\beta_S}\abs{Ha(t)\mathbf{x}}^{-2\beta_S}\\&-\frac{2\alpha_S\sigma_{qp}^2}{3}\abs{Ha(t)\mathbf{x}}^{-3}\Bigg].
    \end{split}
\end{align}
\end{subequations}
Currently, it is not clear that one can compute the higher-order spacelike 2-point correlators in the same way as before because the 3PDF is not so easy to write down. Certainly, it is not clear that Wick's theorem will be obeyed. Further, it is clear that the symmetry of de Sitter is not reproduced for arbitrary noise parameters in this stochastic theory; the timelike and spacelike correlators are separate, unrelated quantities. In fact, the spacelike correlators are really a specific type of 3-point function. However, when we choose our stochastic parameters in the next section, we will see that this property is reinstated.

\subsection{Free stochastic parameters}
\label{subsec:free_stochastic_parameters}

From Eq. (\ref{2d_stochastic_eq}) until now, this chapter has been dedicated to developing a stochastic theory that is only related to the physical system by the postulated form of the stochastic equations. We will now promote this second-order stochastic theory to a \textit{second-order stochastic effective theory of QFT in de Sitter} by giving explicit forms for the stochastic parameters. This is done on the level of correlation functions. We compare the expressions for the stochastic timelike and spacelike correlators, Eq. (\ref{equal-space_(phi,pi)_stochastic_correlators}) and (\ref{equal-time_(phi,pi)_stochastic_correlators}) respectively, with their free QFT equivalents, Eq. (\ref{equal-space_correlators}) and (\ref{equal-time_correlators}). For them to agree, we choose our stochastic parameters in such a way that these quantities are equal\footnote{That is, up to a complex factor.}. It is sufficient to only consider these 8 2-point functions because higher-order correlators are found via Wick's theorem in both cases\footnote{The caveat to this is that the quantum and stochastic variances don't agree. Higher-order correlators that contain the variances will also be UV divergent in the QFT, behaviour that will not be reproduced by the IR effective stochastic theory. This problem will be dealt with for interacting fields, when we can incorporate UV renormalisation.}. Since there are 4 stochastic parameters $m_S$, $\sigma_{\phi\phi}^2$, $\sigma_{\phi\pi}^2$ and $\sigma_{\pi\pi}^2$, we require 4 conditions. They are as follows:
\begin{enumerate}[(i)]
    \item The exponents of the leading terms in the stochastic and QFT 2-point correlators are equal.
    \item The prefactors of the leading terms in the stochastic and QFT 2-point correlators are equal.
    \item The timelike and spacelike 2-point correlators are related via the continuation $e^{H(t-t')}\leftrightarrow \abs{Ha(t)(\mathbf{x}-\mathbf{x}')}^2$ as per the de Sitter symmetry.
    \item The subleading terms are equal? The subleading terms vanish? Some other condition? 
\end{enumerate}
The first three conditions are concrete and essential. As one can see, the final entry is not so much a condition but a choice. Since the stochastic theory is an effective theory of the IR behaviour, we only require it to reproduce the leading term in the asymptotic expansion of the QFT correlators. For free fields, we can push this further and capture the subleading contribution as well\footnote{Recall that this is the leading term of the second sum in the asymptotic expansion (\ref{free_feynman_propagator_long-distance_sum}). This is only subleading when $m<\sqrt{2}H$, but we expect this to be the regime of validity where the stochastic approach holds. For heavier fields, the expansion of the spacetime won't stretch the modes sufficiently to consider this semiclassical approximation.}. However, it is not clear that this will hold when one introduces interactions, and it is certainly not essential, so it may be more convenient to just choose it to vanish instead. For now, we will just focus on these two possibilities.

There is a one-to-one correspondence between the 4 conditions and our 4 stochastic parameters if we consider the $(q,p)$ noise amplitudes; (i)-(iv) corresponds to $m_S$, $\sigma_{qq}^2$, $\sigma_{qp}^2$ and $\sigma_{pp}^2$ respectively. Imposing the conditions gives the \textit{matched stochastic parameters}
\begin{subequations}
    \label{free_matched_stochastic_parameters}
\begin{align}
        \label{matched_mass}
        m_S^{(0)}&=m\\
        \label{matched_sigma_qq}
        \sigma_{qq}^{(0)2}&=\sigma_{Q,qq}^{2(0)}=\frac{H^3\alpha\nu}{4\pi^2\beta}\frac{\Gamma(2\nu)\Gamma(\frac{3}{2}-\nu)4^{\frac{3}{2}-\nu}}{\Gamma(\frac{1}{2}+\nu)},\\
        \label{matched_sigma_qp}
        \sigma_{qp}^{(0)2}&=\sigma_{Q,qp}^{2(0)}=0,\\
        \label{matched_sigma_pp}
        \sigma_{pp}^{(0)2}&=\sigma_{Q,pp}^{2(0)}=\begin{cases}\sigma_{pp}^{2(NLO)(0)}=\frac{H^5\beta^2\nu}{4\pi^2}\frac{\Gamma(-2\nu)\Gamma(\frac{3}{2}+\nu)4^{\frac{3}{2}+\nu}}{\Gamma(\frac{1}{2}-\nu)}\\0\end{cases}.
\end{align}
\end{subequations}
I denote the subscript `$Q$' to indicate that these have been chosen such that the stochastic theory is promoted to an effective theory of the QFT and I have reinstated the superscript `$(0)$' to make it clear that these are the stochastic parameters for free fields. Introducing interactions will change their form. The two cases for $\sigma_{Q,pp}^{2(0)}$ are the two choices for the final `condition' (iv): to reproduce the subleading term (top) or to have the subleading term vanish (bottom). Note also that, for free fields, the mass parameters are equal; this will not necessarily hold for interacting theories. The key result is this:

\begin{equation*}
\keybox{
\begin{split}
\text{Using the } &\text{stochastic parameters (\ref{free_matched_stochastic_parameters}) elevates the stochastic theory (\ref{2d_stochastic_eq}) to }\\& \text{an IR effective theory of free scalar QFT in de Sitter.}
\end{split}
}
\end{equation*}

\subsubsection{Comparison between the cut-off stochastic and quantum correlators}

\begin{figure}[ht]
    \centering
    \includegraphics[width=150mm]{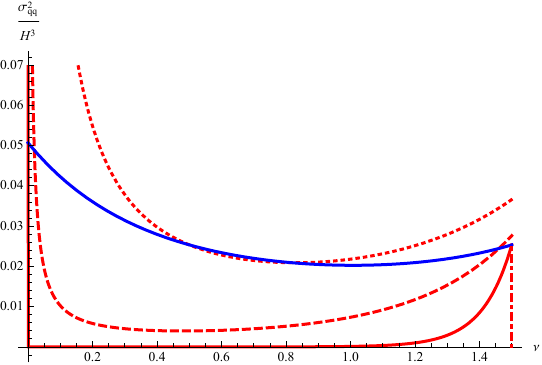}
    \caption{The quantum (blue) $\sigma_{Q,qq}^2$ and the cut-off (red) noises $\sigma_{cut,qq}^2$ with $\epsilon=0$ (dot-dashed), $\epsilon=0.01$ (solid), $\epsilon=0.5$ (dashed) and $\epsilon=0.99$ (dotted) is plotted as a function of $\nu$. We see that the two approaches don't agree for any value of $\epsilon$ and therefore the cut-off procedure is unable to reproduce the quantum field correlator for all masses.}
    \label{fig:noise_plot}
\end{figure}

Before we proceed to interacting theories, we will compare the original cut-off procedure, introduced at the start of the chapter, with the second-order stochastic effective theory and QFT. Consider first the spacelike field 2-point correlator. By substituting the cut-off noise amplitudes of Eq. (\ref{noise_amplitudes_modes}) into Eq. (\ref{sigma_q,p-->sigma_phi,pi}), we can express the $(q,p)$ noise amplitudes in terms of Hankel functions and hence the cut-off stochastic field 2-point correlator is given by
\begin{equation}
    \label{stochastic_field_mode_correlator}
    \begin{split}
        \expval{\phi(t,\mathbf{0})\phi(t,\mathbf{x})}_{cut}=&\frac{H^2\epsilon^3}{256\nu^2\pi\alpha}\abs{\epsilon\mathcal{H}_{\nu-1}^{(1)}(\epsilon)-2\nu\mathcal{H}_{\nu}^{(1)}(\epsilon)-\epsilon\mathcal{H}_{\nu+1}^{(1)}(\epsilon)}^2\abs{Ha(t)\mathbf{x}}^{-3+2\nu}
        \\&
        +\frac{H^2\epsilon^3}{256\nu^2\pi\beta}\abs{\epsilon\mathcal{H}_{\nu-1}^{(1)}(\epsilon)+2\nu\mathcal{H}_{\nu}^{(1)}(\epsilon)-\epsilon\mathcal{H}_{\nu+1}^{(1)}(\epsilon)}^2\abs{Ha(t)\mathbf{x}}^{-3-2\nu}.
    \end{split}
\end{equation}
Note that we exclude the $\abs{Ha(t)\mathbf{x}}^{-3}$ term in what follows as that is an additional stochastic term that does not appear in the quantum correlator. Comparing this expression with Eq. (\ref{equal-time_phiphi_corr}), we find that the cut-off approach will reproduce the leading order term of the quantum correlator if
\begin{equation}
    \label{SY=quantum_alpha}
    \begin{split}
    \frac{H^2\epsilon^3}{256\nu^2\pi\alpha}\abs{\epsilon\mathcal{H}_{\nu-1}^{(1)}(\epsilon)-2\nu\mathcal{H}_{\nu}^{(1)}(\epsilon)-\epsilon\mathcal{H}_{\nu+1}^{(1)}(\epsilon)}^2=\frac{H^2}{16\pi^2}\Bigg[\frac{\Gamma(\frac{3}{2}-\nu)\Gamma(2\nu)4^{\frac{3}{2}-\nu}}{\Gamma(\frac{1}{2}+\nu)}\Bigg].
    \end{split}
\end{equation}
To draw further comparisons, we plot the cut-off (\ref{noise_amplitudes_modes}) and matched (\ref{free_matched_stochastic_parameters}) noise as a function of the mass in Fig. \ref{fig:noise_plot}. We can see that, unsurprisingly, there is a strong dependence on $\epsilon$ from the cut-off and there is no one choice of $\epsilon$ that fully reproduces the leading term in the QFT correlator. Of course, we could make $\epsilon$ mass-dependent so that the two match but the whole idea of the cut-off procedure is that one hopes to derive a stochastic effective theory from the underlying QFT. This purpose is undermined if one ultimately needs to choose $\epsilon$ in a given way based on knowledge of the QFT correlation functions. Further, the cut-off approach will only give one degree of freedom and hence only one of the 4 conditions outlined above can be satisfied consistently. This is more problematic when one considers interacting fields.

Note that the results converge for light fields $m\ll H$ and for small $\epsilon$. This confirms that the cut-off procedure is useful in this regime. This aligns with the analysis of the previous chapter, where we used overdamped stochastic equations. However, if one is interested in more massive fields, the cut-off method for computing the noise is no longer reliable and one must use the stochastic parameters obtained via the matching procedure.

\section{Interacting fields in stochastic theory}
\label{sec:interacting_stochastic_theory}

Having established our second-order stochastic effective theory for free fields, we will now introduce quartic self-interactions by considering the stochastic equations
\begin{equation}
    \label{stochastic_eq_quartic}
        \begin{pmatrix}\Dot{\phi}\\\Dot{\pi}\end{pmatrix}=\begin{pmatrix}\pi\\-3H\pi-m^2_S\phi-\lambda_S\phi^3\end{pmatrix}+\begin{pmatrix}\xi_{\phi}(t,\mathbf{x})\\\xi_{\pi}(t,\mathbf{x})\end{pmatrix}.
\end{equation}
However, the stochastic parameters (\ref{free_matched_stochastic_parameters}) are only applicable for free fields; we need to update these to include interactions. We will do this by computing stochastic 2-pt and 4-pt functions perturbatively about the free solution, once again using general forms for the stochastic parameters. We can then match the results to their counterparts of perturbative QFT, obtained in Sec. \ref{sec:phi4_theory}, thus promoting our stochastic approach to a stochastic effective theory of quartic self-interacting scalar fields.

\subsection{Two-point stochastic correlators to $\mathcal{O}(\lambda)$}
\label{subsec:perturbative_stochastic_2-pt_func}

We begin by solving the eigenequations (\ref{FP_eigenequations}) to $\mathcal{O}(\lambda_S)$ in much the same way as was done in the overdamped theory in Sec. \ref{subsec:perturbative_OD_eigenspectrum}. Since we are perturbing about the free theory, we will continue to use the $(q,p)$ variables in this analysis. We perturb the eigensystem as
\begin{subequations}
    \label{perturbed_eigensolutions}
    \begin{align}
        \mathcal{L}_{FP}^{(*)}&=\mathcal{L}_{FP}^{(0)(*)}+\lambda_S\mathcal{L}_{FP}^{(1)(*)}\\
        \Lambda_{rs}=&\Lambda_{rs}^{(0)}+\lambda_S \Lambda_{rs}^{(1)}+\mathcal{O}(\lambda_S^2),\\
        \Psi_{rs}^{(*)}(q,p)=&\Psi_{rs}^{(0)(*)}(q,p)+\lambda_S\Psi_{rs}^{(1)(*)}(q,p)+\mathcal{O}(\lambda_S)^2,
    \end{align}
\end{subequations}
where the $\mathcal{O}(\lambda_S)$  Fokker-Planck operator and its adjoint are
\begin{subequations}
    \label{O(lambda)_FP_operator_q,p}
    \begin{align}
    \mathcal{L}_{FP}^{(1)}&=\frac{1}{\lb1-\frac{\alpha_S}{\beta_S}\rb^2}\lb-\frac{1}{\beta_S H}p+q\rb^3\lb\partial_p+\frac{1}{\beta_SH}\partial_q\rb\\
    \mathcal{L}_{FP}^{(1)*}&=-\frac{1}{\lb1-\frac{\alpha_S}{\beta_S}\rb^2}\lb-\frac{1}{\beta_S H}p+q\rb^3\lb\partial_p+\frac{1}{\beta_SH}\partial_q\rb
    \end{align}
\end{subequations}
and the $\mathcal{O}(\lambda_S)$ eigenspectrum is
\begin{subequations}
    \label{O(lambda)_eigenvalues&eigenstates}
    \begin{align}
    \label{O(lambda)_eigenvalues_repeated}
        \Lambda_{rs}^{(1)}&=-\qty(\Psi_{rs}^{(0)*},\mathcal{L}_{FP}^{(1)}\Psi_{rs}^{(0)}),\\
    \label{O(lambda)_eigenstates_repeated}  
        \Psi_{rs}^{(1)}(q,p)&=\sum_{r's'}\Psi_{r's'}^{(0)}(q,p)\frac{\qty(\Psi_{r's'}^{(0)*},\mathcal{L}_{FP}^{(1)}\Psi_{rs}^{(0)})}{\Lambda_{r's'}^{(0)}-\Lambda_{rs}^{(0)}},\\ 
    \label{O(lambda)_adjoint_eigenstates_repeated}   
        \Psi_{rs}^{(1)*}(q,p)&=\sum_{r's'}\Psi_{r's'}^{(0)*}(q,p)\frac{\qty(\Psi_{r's'}^{(0)},\mathcal{L}_{FP}^{(1)*}\Psi_{rs}^{(0)*})}{\Lambda_{r's'}^{(0)}-\Lambda_{rs}^{(0)}}, 
    \end{align}
\end{subequations}
where for Eq. (\ref{O(lambda)_eigenstates_repeated}) and (\ref{O(lambda)_adjoint_eigenstates_repeated}), $r'\ne r$ and $s'\ne s$. 
By applying the expansion (\ref{perturbed_eigensolutions}) to Eq. (\ref{2pt_timelike_stochastic_correlator_fg}), we can write the timelike correlator between two functions $f(q_0,p_0)$ and $g(q,p)$ to $\mathcal{O}(\lambda_S)$ as
\begin{equation}
    \label{O(lambda)_timelike_stochastic_correlator}
    \begin{split}
    \expval{f(q_0,p_0)g(q,p)}=&\sum_{rs}\qty[f^{(0)}_{00rs}g_{rs00}^{(0)}+\lambda_S\qty(f_{00rs}^{(0)}g_{rs00}^{(1)}+f_{00rs}^{(1)}g_{rs00}^{(0)})+\mathcal{O}(\lambda_S^2)]\\&\times e^{-\qty(\Lambda_{rs}^{(0)}+\lambda_S\Lambda_{rs}^{(1)}+\mathcal{O}(\lambda_S^2))(t-t_0)},
    \end{split}
\end{equation}
where
\begin{subequations}
    \begin{align}
        \label{gnl_perturbative_expansion}
        f_{rsr's'}^{(0)}&=\qty(\Psi_{rs}^{(0)},f\Psi_{r's'}^{(0)*}),\\
        f_{rsr's'}^{(1)}&=\qty(\Psi_{rs}^{(1)},f\Psi_{r's'}^{(0)*})+\qty(\Psi_{rs}^{(0)},f\Psi_{r's'}^{(1)}).
    \end{align}
\end{subequations}
Note that these are taken from the definition (\ref{fnl}) such that $f_{NN'}=f_{NN'}^{(0)}+\lambda_S f_{NN'}^{(1)}$ for $N=(r,s)$. A similar expression for the spacelike correlator can be written using Eq. (\ref{2pt_spacelike_stochastic_correlators_fg}) as
\begin{equation}
\label{O(lambda)_spacelike_stochastic_correlator}
    \begin{split}
        \expval{f(q_1,p_1)g(q_2,p_2)}=&\int dq_r\int dp_r\sum_{r'rs's}\Bigg[\frac{\Psi_{00}^{(0)}(q_r,p_r)}{\Psi_{00}^{(0)*}(q_r,p_r)}\Psi_{rs}^{(0)*}(q_r,p_r)\Psi_{r's'}^{(0)*}(q_r,p_r)f_{rs}^{(0)}g_{r's'}^{(0)}\\&
        \begin{split}
        +\lambda_S\Bigg(&\frac{\Psi_{00}^{(1)}(q_r,p_r)}{\Psi_{00}^{(0)*}(q_r,p_r)}\Psi_{rs}^{(0)*}(q_r,p_r)\Psi_{r's'}^{(0)*}(q_r,p_r)f_{rs00}^{(0)}g_{r's'00}^{(0)}
        \\&-\frac{\Psi_{00}^{(1)*}(q_r,p_r)\Psi_{00}^{(0)}(q_r,p_r)}{\Psi_{00}^{(0)*}(q_r,p_r)^2}\Psi_{rs}^{(0)*}(q_r,p_r)\Psi_{r's'}^{(0)*}(q_r,p_r)f_{rs00}^{(0)}g_{r's'00}^{(0)}
        \\&+\frac{\Psi_{00}^{(0)}(q_r,p_r)}{\Psi_{00}^{(0)*}(q_r,p_r)}\Psi_{rs}^{(1)*}(q_r,p_r)\Psi_{r's'}^{(0)*}(q_r,p_r)f_{rs00}^{(0)}g_{r's'00}^{(0)}
        \\&+\frac{\Psi_{00}^{(0)}(q_r,p_r)}{\Psi_{00}^{(0)*}(q_r,p_r)}\Psi_{rs}^{(0)*}(q_r,p_r)\Psi_{r's'}^{(1)*}(q_r,p_r)f_{rs00}^{(0)}g_{r's'00}^{(0)}
        \\&+\frac{\Psi_{00}^{(0)}(q_r,p_r)}{\Psi_{00}^{(0)*}(q_r,p_r)}\Psi_{rs}^{(0)*}(q_r,p_r)\Psi_{r's'}^{(0)*}(q_r,p_r)f_{rs00}^{(1)}g_{r's'00}^{(0)}
        \\&+\frac{\Psi_{00}^{(0)}(q_r,p_r)}{\Psi_{00}^{(0)*}(q_r,p_r)}\Psi_{rs}^{(0)*}(q_r,p_r)\Psi_{r's'}^{(0)*}(q_r,p_r)f_{rs00}^{(0)}g_{r's'00}^{(1)}\Bigg)
        \end{split}
        \\&+\mathcal{O}(\lambda_S^2)\Bigg]\abs{Ha(t)(\mathbf{x}_1-\mathbf{x}_2)}^{-\qty(\frac{\Lambda_{rs}^{(0)}+\Lambda_{r's'}^{(0)}}{H}+\lambda\frac{\Lambda_{rs}^{(1)}+\Lambda_{r's'}^{(1)}}{H}+\mathcal{O}(\lambda_S^2))}.
    \end{split}
\end{equation}
We can now use the free eigenspectrum computed in the previous section in the above 2-point functions. However, recall that it was not possible to write an expression for the eigenstates in a general form for general noise (see. Eq. (\ref{free_eigenspectrum_solved_general_noise})). Fortunately, we only need to compute the eigenspectrum for the indices $(r,s)=(0,0)$, $(r,s)=(0,1)$ and $(r,s)=(1,0)$ in order to compute the 2-point functions because the higher-order terms vanish. Since the free eigenstates (\ref{free_eigenspectrum_solved_general_noise}) can be computed on a case-by-case basis, we can compute these results and hence obtain the $\mathcal{O}(\lambda_S)$ correlation functions. The $q-q$, $q-p$, $p-q$ and $p-p$ timelike 2-point correlators are
    \begin{subequations}
    \label{(q,p)_timelike_stochastic_correlators_gen_noise}
    \begin{align}
        \label{qq_corr_gen_noise}
        \begin{split}
        \expval{q(0,\mathbf{x})q(t,\mathbf{x})}=&\Bigg[\frac{\sigma_{qq}^{2}}{2H\alpha_S}\\&+\lambda_S\frac{\qty(\alpha_S\sigma_{qp}^{2}-3\nu_S H\sigma_{qq}^{2})\qty(3\alpha_S\sigma_{pp}^{2}-4H\alpha_S\beta_S^2\sigma_{qp}^{2}+3H^2\beta_S^3\sigma_{qq}^{2})}{48\nu_S^3H^7\alpha_S^3\beta_S^2}
        \\&+\mathcal{O}(\lambda_S^2)\Bigg] e^{-\qty(\alpha_S H+\lambda_S\frac{3\alpha_S\sigma_{pp}^{2}-4H\alpha_S\beta_S^2\sigma_{qp}^{2}+3H^2\beta_S^3\sigma_{qq}^{2}}{8\nu_S^2H^4\alpha_S\beta_S^2}+\mathcal{O}(\lambda_S^2)) t}\\
            &+\qty[\lambda_S\frac{\sigma_{qp}^{2}\qty(-3\alpha_S \sigma_{pp}^{2}+4H\alpha_S \beta_S^2\sigma_{qp}^{2}-3H^2\beta_S^3\sigma_{qq}^{2})}{48\nu_S^3H^7\alpha_S\beta_S^3}+\mathcal{O}(\lambda_S^2)]\\&\times e^{-\qty(\beta_S H-\lambda_S\frac{3\alpha_S\sigma_{pp}^{2}-4H\alpha_S\beta_S^2\sigma_{qp}^{2}+3H^2\beta_S^3\sigma_{qq}^{2}}{8\nu_S^2H^4\alpha_S\beta_S^2}+\mathcal{O}(\lambda_S^2))t},\\&
        \end{split}\\
        \label{pq_corr_gen_noise}
        \begin{split}
            \expval{p(0,\mathbf{x})q(t,\mathbf{x})}=&\Bigg[\frac{\sigma_{qp}^{2}}{3H}\\&+\lambda_S\frac{\qty(\nu_S H^2\beta_S^2\sigma_{qq}^{2}-\alpha_S \sigma_{pp}^{2})\qty(3\alpha_S\sigma_{pp}^{2}-4H\alpha_S\beta_S^2\sigma_{qp}^{2}+3H^2\beta_S^3\sigma_{qq}^{2})}{48\nu^3H^7\alpha_S^2\beta_S^3}\\&+\mathcal{O}(\lambda_S^2)\Bigg] e^{-\qty(\alpha_S H+\lambda_S\frac{3\alpha_S\sigma_{pp}^{2}-4H\alpha_S\beta_S^2\sigma_{qp}^{2}+3H^2\beta_S^3\sigma_{qq}^{2}}{8\nu_S^2H^4\alpha_S\beta_S^2}+\mathcal{O}(\lambda_S^2)) t}
            \\&
            +\qty[\lambda_S\frac{\sigma_{pp}^{2}(-3\alpha_S\sigma_{pp}^{2}+4H\alpha_S\beta_S^2\sigma_{qp}^{2}-3H^2\beta_S^3\sigma_{qq}^{2})}{32\nu_S^3H^7\alpha_S\beta_S^4}+\mathcal{O}(\lambda_S^2)]\\&\times e^{-\qty(\beta_S H-\lambda_S\frac{3\alpha_S\sigma_{pp}^{2}-4H\alpha_S\beta_S^2\sigma_{qp}^{2}+3H^2\beta_S^3\sigma_{qq}^{2}}{8\nu_S^2H^4\alpha_S\beta_S^2}+\mathcal{O}(\lambda_S^2))t},\\&
        \end{split}\\
        \label{qp_corr_gen_noise}
        \begin{split}
            \expval{q(0,\mathbf{x})p(t,\mathbf{x})}=&\Bigg[\lambda_S\frac{\sigma_{qq}^{2}\qty(3\alpha_S\sigma_{pp}^{2}-4H\alpha_S\beta_S^2\sigma_{qp}^{2}+3H^2\beta_S^3\sigma_{qq}^{2})}{32\nu_S^3H^5\alpha_S^2\beta_S}+\mathcal{O}(\lambda_S^2)\Bigg]\\&\times e^{-\qty(\alpha_S H+\lambda_S\frac{3\alpha_S\sigma_{pp}^{2}-4H\alpha_S\beta_S^2\sigma_{qp}^{2}+3H^2\beta_S^3\sigma_{qq}^{2}}{8\nu_S^2H^4\alpha_S\beta_S^2}+\mathcal{O}(\lambda_S^2)) t}
            \\&
            +\Bigg[\frac{\sigma_{qp}^{2}}{3H}\\&+\lambda_S\frac{\qty(\nu_S\sigma_{pp}^{2}+H^2\beta_S^3\sigma_{qq}^{2})\qty(3\alpha_S\sigma_{pp}^{2}-4H\alpha_S\beta_S^2\sigma_{qp}^{2}+3H^2\beta_S^3\sigma_{qq}^{2})}{48\nu_S^3H^7\alpha_S\beta_S^4}\\&+\mathcal{O}(\lambda_S^2)\Bigg] e^{-\qty(\beta_S H-\lambda_S\frac{3\alpha_S\sigma_{pp}^{2}-4H\alpha_S\beta_S^2\sigma_{qp}^{2}+3H^2\beta_S^3\sigma_{qq}^{2}}{8\nu_S^2H^4\alpha_S\beta_S^2}+\mathcal{O}(\lambda_S^2))t},
            \\&
        \end{split}\\
        \begin{split}
            \label{pp_corr_gen_noise}
            \expval{p(0,\mathbf{x})p(t,\mathbf{x})}=&\qty[\lambda_S\qty(\frac{\sigma_{qp}^{2}\qty(-3\alpha_S\sigma_{pp}^{2}+4H\alpha_S\beta_S^2\sigma_{qp}^{2}-3H^2\beta_S^3\sigma_{qq}^{2})}{48\nu_S^3H^5\alpha_S\beta_S})+\mathcal{O}(\lambda_S^2)]\\&\times e^{-\qty(\alpha_S H+\lambda_S\frac{3\alpha_S\sigma_{pp}^{2}-4H\alpha_S\beta_S^2\sigma_{qp}^{2}+3H^2\beta_S^3\sigma_{qq}^{2}}{8\nu_S^2H^4\alpha_S\beta_S^2}+\mathcal{O}(\lambda_S^2)) t}\\&
            +\Bigg[\frac{\sigma_{pp}^{2}}{2H\beta_S}\\&+\lambda_S\frac{\qty(3\nu_S\sigma_{pp}^{2}+H\beta_S^2\alpha_S\sigma_{qp}^{2})\qty(3\alpha_S\sigma_{pp}^{2}-4H\alpha_S\beta_S^2\sigma_{qp}^{2}+3H^2\beta_S^3\sigma_{qq}^{2})}{48\nu_S^3H^6\alpha_S\beta_S^4}\\&+\mathcal{O}(\lambda_S^2)\Bigg]e^{-\qty(\beta_S H-\lambda_S\frac{3\alpha_S\sigma_{pp}^{2}-4H\alpha_S\beta_S^2\sigma_{qp}^{2}+3H^2\beta_S^3\sigma_{qq}^{2}}{8\nu_S^2H^4\alpha_S\beta_S^2}+\mathcal{O}(\lambda_S^2))t}
        \end{split}
    \end{align}
    \end{subequations}
and their spacelike counterparts are
\begin{subequations}
    \label{(q,p)_spacelike_stochastic_correlators_gen_noise}
    \begin{align}
    \label{spacelike_qq_gen_noise}
        \begin{split}
            \expval{q(t,\mathbf{0})q(t,\mathbf{x})}=&\Bigg[\frac{\sigma_{qq}^{2}}{2H\alpha_S}\\&+\lambda_S\frac{\qty(3\alpha_S\sigma_{pp}^{2}-4H\alpha_S\beta_S^2\sigma_{qp}^{2}+3H^2\beta_S^3\sigma_{qq}^{2})\qty(3\alpha_S\sigma_{qp}^{2}-3\nu_S H\beta_S\sigma_{qq}^{2})}{48\nu_S^3H^7\alpha_S^3\beta_S^3}\\&+\mathcal{O}(\lambda_S^2)\Bigg]\abs{Ha(t)\mathbf{x}}^{-2\alpha_S-\lambda_S\frac{3\alpha_S\sigma_{pp}^{2}-4H\alpha_S\beta_S^2\sigma_{qp}^{2}+3H^2\beta_S^3\sigma_{qq}^{2}}{4\nu_S^2H^5\alpha_S\beta_S^2}+\mathcal{O}(\lambda_S^2)}\\&
            +\qty[\lambda_S\frac{\sigma_{qp}^{2}(-3\alpha_S\sigma_{pp}^{2}+4H\alpha_S\beta_S^2\sigma_{qp}^{2}-3H^2\beta_S^3\sigma_{qq}^{2})}{24\nu_S^3H^7\alpha_S\beta_S^3}+\mathcal{O}(\lambda_S^2)]\\&\times \abs{Ha(t)\mathbf{x}}^{-3+\mathcal{O}(\lambda_S^2)},
        \end{split}\\
    \label{spacelike_qp_gen_noise}
        \begin{split}
            \expval{q(t,\mathbf{0})p(t,\mathbf{x})}=&\expval{p(t,\mathbf{0})q(t,\mathbf{x})}\\=
            &\qty[\lambda_S\frac{\sigma_{qq}^{2}\qty(-3\alpha_S\sigma_{pp}^{2}+4H\alpha_S\beta_S^2\sigma_{qp}^{2}-3H^2\beta_S^3\sigma_{qq}^{2})}{32\nu_S^3H^5\alpha_S^2\beta_S}+\mathcal{O}(\lambda_S^2)]\\&\times\abs{Ha(t)\mathbf{x}}^{-2\alpha_S-\lambda_S\frac{3\alpha_S\sigma_{pp}^{2}-4H\alpha_S\beta_S^2\sigma_{qp}^{2}+3H^2\beta_S^3\sigma_{qq}^{2}}{4\nu_S^2H^5\alpha_S\beta_S^2}+\mathcal{O}(\lambda_S^2)}\\&
            +\qty[\lambda_S\frac{\sigma_{pp}^{2}\qty(-3\alpha_S\sigma_{pp}^{2}+4H\alpha_S\beta_S^2\sigma_{qp}^{2}-3H^2\beta_S^3\sigma_{qq}^{2})}{32\nu_S^3H^7\alpha_S\beta_S^4}+\mathcal{O}(\lambda_S^2)]\\&\times \abs{Ha(t)\mathbf{x}}^{-2\beta_S+\lambda_S\frac{3\alpha_S\sigma_{pp}^{2}-4H\alpha_S\beta_S^2\sigma_{qp}^{2}+3H^2\beta_S^3\sigma_{qq}^{2}}{4\nu_S^2H^5\alpha_S\beta_S^2}}\\&
            +\Bigg[\frac{\sigma_{qp}^{2}}{3H}+\lambda_S\frac{\qty(\sigma_{pp}^{2}+H^2\beta_S^2\sigma_{qq}^{2})\qty(3\alpha_S\sigma_{pp}^{2}-4H\alpha_S\beta_S^2\sigma_{qp}^{2}+3H^2\beta_S^3\sigma_{qq}^{2})}{48\nu_S^3H^7\alpha_S\beta_S^3}\\&+\mathcal{O}(\lambda_S^2)\Bigg]\abs{Ha(t)\mathbf{x}}^{-3+\mathcal{O}(\lambda_S^2)},
        \end{split}\\
    \label{spacelike_pp_gen_noise}
        \begin{split}
            \expval{p(t,\mathbf{0})p(t,\mathbf{x})}=&\Bigg[\frac{\sigma_{pp}^{2}}{2H\beta_S}\\&+\lambda_S\frac{\qty(3\nu_S\sigma_{pp}^{2}+3H\beta_S^2\sigma_{qp}^{2})\qty(3\alpha_S\sigma_{pp}^{2}-4H\alpha_S\beta_S^2\sigma_{qp}^{2}+3H^2\beta_S^3\sigma_{qq}^{2}))}{48\nu_S^3H^6\alpha_S\beta_S^4}\\&+\mathcal{O}(\lambda_S^2)\Bigg] \abs{Ha(t)\mathbf{x}}^{-2\beta_S+\lambda_S\frac{3\alpha_S\sigma_{pp}^{2}-4H\alpha_S\beta_S^2\sigma_{qp}^{2}+3H^2\beta_S^3\sigma_{qq}^{2}}{4\nu_S^2H^5\alpha_S\beta_S^2}+\mathcal{O}(\lambda_S^2)}
            \\&\qty[\lambda_S\frac{\sigma_{qp}^{2}(-3\alpha_S\sigma_{pp}^{2}+4H\alpha_S\beta_S^2\sigma_{qp}^{2}-3H^2\beta_S^3\sigma_{qq}^{2})}{24\nu_S^3H^5\alpha_S\beta_S}+\mathcal{O}(\lambda_S)^2]\\&\times \abs{Ha(t)\mathbf{x}}^{-3+\mathcal{O}(\lambda_S^2)}.
        \end{split}
    \end{align}
\end{subequations}
Substituting these expressions into Eq. (\ref{phi-phi_pq}), we obtain an expression for the $\phi-\phi$ stochastic 2-point correlator to $\mathcal{O}(\lambda_S)$. The timelike version is
\begin{equation}
    \label{phi-phi_timelike_stochastic_correlator_gen_noise}
    \begin{split}
        \expval{\phi(0,\mathbf{x})\phi(t,\mathbf{x})}=&\frac{1}{1-\frac{\alpha_S}{\beta_S}}\Bigg[\frac{\sigma_{qq}^{2}}{2H\alpha_S }-\frac{\sigma_{qp}^{2}}{3H^2\beta_S}+\lambda_S\Bigg(\Bigg(2H\alpha_S^4\beta_S\sigma_{qp}^{2}\qty(-3\sigma_{pp}^{2}+4H\beta_S^2\sigma_{qp}^{2})\\&+9H^4\beta_S^7(\sigma_{qq}^{2})^2+9H^2\alpha_S\beta_S^4\sigma_{qq}^{2}\qty(-\sigma_{pp}^{2}+2H\beta_S^2\qty(\sigma_{qp}^{2}+H\sigma_{qq}^{2}))\\&-3\alpha_S^3\qty(6(\sigma_{pp}^{2})^2+6H^3\beta_S^4\sigma_{qp}^{2}\sigma_{qq}^{2}-H\beta_S^2\sigma_{pp}^{2}\qty(8\sigma_{qp}^{2}+3H\sigma_{qq}^{2}))\\&+H\alpha_S^2\beta_S^3\qty(6\sigma_{pp}^{2}\sigma_{qp}^{2}+H\beta_S^2\qty(-8(\sigma_{qp}^{2})^2-24H\sigma_{qp}^{2}\sigma_{qq}^{2}+9H^2(\sigma_{qq}^{2})^2))\Bigg)\\&/\qty(288\nu_S^3H^8\alpha_S^3\beta_S^4)\Bigg)+\mathcal{O}(\lambda_S^2)\Bigg]\\&\times e^{-\qty(\alpha_S H+\lambda_S\frac{3\alpha_S\sigma_{pp}^{2}-4H\alpha_S\beta_S^2\sigma_{qp}^{2}+3H^2\beta_S^3\sigma_{qq}^{2}}{8\nu_S^2H^4\alpha_S\beta_S^2}+\mathcal{O}(\lambda_S^2)) t}
        \\&+\\&
        \frac{1}{1-\frac{\alpha_S}{\beta_S}}\Bigg[\frac{\sigma_{pp}^{2}}{2H^3\beta_S^3}-\frac{\sigma_{qp}^{2}}{3H^2\beta_S}+\lambda_S\Bigg( \Bigg(9\alpha_S\qty(-\alpha_S^2+2\alpha_S \beta_S+\beta_S^2)(\sigma_{pp}^{2})^2\\&+6H\alpha_S \beta_S^2\qty(3\alpha_S^2-4\alpha_S\beta_S-3\beta_S^2)\sigma_{qp}^{2}\sigma_{pp}^{2}\\&+6H^3\beta_S^5\qty(\alpha_S^2+4\alpha_S\beta_S-\beta_S^2)\sigma_{qp}^{2}\sigma_{qq}^{2}-18H^4\beta_S^7(\sigma_{qq}^{2})^2\\&+H^2\beta_S^3\qty(\beta_S^2-\alpha_S^2)\qty(8\alpha_S\beta_S(\sigma_{qp}^{2})^2+9\sigma_{pp}^{2}\sigma_{qq}^{2})\Bigg)/\qty(288\nu_S^3H^8\alpha_S\beta_S^6)\Bigg)\\&+\mathcal{O}(\lambda_S^2)\Bigg]
         e^{-\qty(\beta_S H-\lambda_S\frac{3\alpha_S\sigma_{pp}^{2}-4H\alpha_S\beta_S^2\sigma_{qp}^{2}+3H^2\beta_S^3\sigma_{qq}^{2}}{8\nu_S^2H^4\alpha_S\beta_S^2}+\mathcal{O}(\lambda_S^2))t}
    \end{split}
\end{equation}
while the spacelike version is
\begin{equation}
    \label{phi-phi_spacelike_stochastic_correlator_gen_noise}
    \begin{split}
        \expval{\phi(t,\mathbf{0})\phi(t,\mathbf{x})}=&\frac{1}{1-\frac{\alpha_S}{\beta_S}}\Bigg[\frac{\sigma_{qq}^{2}}{2H\alpha_S}+\lambda_S\Bigg(\qty(H\beta_S(3\alpha_S-\beta_S)\sigma_{qq}^{2}+2\alpha_S\sigma_{qp}^{2})\\&\times\qty(3\alpha_S\sigma_{pp}^{2}-4H\alpha_S\beta_S^2\sigma_{qp}^{2}+3H^2\beta_S^3\sigma_{qq}^{2})/(32\nu_S^3H^7\alpha_S^3\beta_S^3)\Bigg)+\mathcal{O}(\lambda_S^2)\Bigg]\\&\times \abs{Ha(t)\mathbf{x}}^{-2\alpha_S-\lambda_S\frac{3\alpha_S\sigma_{pp}^{2}-4H\alpha_S\beta_S^2\sigma_{qp}^{2}+3H^2\beta_S^3\sigma_{qq}^{2}}{4\nu_S^2H^5\alpha_S\beta_S^2}+\mathcal{O}(\lambda_S^2)}\\&
        +\\&
        \frac{1}{1-\frac{\alpha_S}{\beta_S}}\Bigg[\frac{\sigma_{pp}^{2}}{2H^3\beta_S^3}+\lambda_S\Bigg((\qty(3\beta_S-\alpha_S)\sigma_{pp}^{2}+2H\beta_S^2\sigma_{qp}^{2})\\&\times\qty(3\alpha_S\sigma_{pp}^{2}-4H\alpha_S\beta_S^2\sigma_{qp}^{2}+3H^2\beta_S^3\sigma_{qq}^{2})/(32\nu_S^3H^8\alpha_S\beta_S^6)\Bigg)+\mathcal{O}(\lambda_S^2)\Bigg]\\&\times \abs{Ha(t)\mathbf{x}}^{-2\beta_S+\lambda_S\frac{3\alpha_S\sigma_{pp}^{2}-4H\alpha_S\beta_S^2\sigma_{qp}^{2}+3H^2\beta_S^3\sigma_{qq}^{2}}{4\nu_S^2H^5\alpha_S\beta_S^2}+\mathcal{O}(\lambda_S^2)}\\&
        +\\&
        \frac{1}{1-\frac{\alpha_S}{\beta_S}}\Bigg[-\frac{2\sigma_{qp}^{2}}{3H^2\beta_S}+\lambda_S\Bigg(\Bigg(2H\alpha_S^2\beta_S\sigma_{qp}^{2}\qty(4H\beta_S^2\sigma_{qp}^{2}-3\sigma_{pp}^{2})\\&-3H^2\beta_S^3\sigma_{qq}^{2}\qty(3\sigma_{pp}^{2}+H\beta_S^2\qty(2\sigma_{qp}^{2}+3H\sigma_{qq}^{2}))-\alpha_S(9(\sigma_{pp}^{2})^2\\&-3H\beta_S^2\sigma_{pp}^{2}\qty(2\sigma_{qp}^{2}-3H\sigma_{qq}^{2})-2H^2\beta_S^4\sigma_{qp}^{2}\qty(4\sigma_{qp}^{2}+3H\sigma_{qq}^{2}))\Bigg)\\&/\qty(72\nu_S^3H^8\alpha_S\beta_S^4)\Bigg)+\mathcal{O}(\lambda_S^2)\Bigg]\abs{Ha(t)\mathbf{x}}^{-3+\mathcal{O}(\lambda_S^2)}.
    \end{split}
\end{equation}
Similar expressions can be found for the $\phi-\pi$, $\pi-\phi$ and $\pi-\pi$ 2-point functions. Here, we will focus on the $\phi-\phi$ correlator.

\subsection{Four-point stochastic correlators to $\mathcal{O}(\lambda)$}
\label{subsec:4-pt_stochastic_functions_O(lambda)}

Using the perturbative eigenspectrum computed above, we can also compute the 4-point function. For the purposes of this thesis, we are using the 4-point function to find the relationship between $\lambda_S$ and the quantum coupling $\lambda$, as will be discussed in more detail in the next section. Thus, we are only interested in the contribution that first arises at $\mathcal{O}(\lambda)$, which will be labelled with a subscript $C$ and referred to as ``connected'', defined in Eq. (\ref{OD_connected_4-pt_func_def}). Further, we can use the free stochastic parameters from Eq. (\ref{free_matched_stochastic_parameters}) since this is an $\mathcal{O}(\lambda_S)$ contribution and, for simplicity, we will make the choice $\sigma_{Q,pp}^2=0$.

In general, the equal-space stochastic 4-point functions are given by Eq. (\ref{4-pt_stochastic_function_general}). As has so often been the case in this chapter, we will switch from $(\phi,\pi)$ to $(q,p)$, using Eq. (\ref{phi,pi-->q,p}). The only non-zero 4-point functions that are relevant are 
\begin{subequations}
    \label{(q,p)_4-point_functions}
    \begin{align}
        \begin{split}
            \expval{q(t_1)q(t_2)q(t_3)q(t_4)}=&q_{00,01}q_{01,00}q_{00,01}q_{01,00}e^{-\Lambda_{01}(t_2-t_1)-\Lambda_{00}(t_3-t_2)-\Lambda_{01}(t_4-t_3)}
            \\&+q_{00,01}q_{01,02}q_{02,01}q_{01,00}e^{-\Lambda_{01}(t_2-t_1)-\Lambda_{02}(t_3-t_2)-\Lambda_{01}(t_4-t_3)}
            \\&+q_{00,03}q_{03,02}q_{02,01}q_{01,00}e^{-\Lambda_{03}(t_2-t_1)-\Lambda_{02}(t_3-t_2)-\Lambda_{01}(t_4-t_3)}
            \\&+q_{00,01}q_{01,02}q_{02,03}q_{03,00}e^{-\Lambda_{01}(t_2-t_1)-\Lambda_{02}(t_3-t_2)-\Lambda_{03}(t_4-t_3)},
        \end{split}\\
        \begin{split}
            \expval{p(t_1)q(t_2)q(t_3)q(t_4)}=&p_{00,01}q_{01,00}q_{00,01}q_{01,00}e^{-\Lambda_{01}(t_2-t_1)-\Lambda_{00}(t_3-t_2)-\Lambda_{01}(t_4-t_3)}
            \\&+p_{00,01}q_{01,02}q_{02,01}q_{01,00}e^{-\Lambda_{01}(t_2-t_1)-\Lambda_{02}(t_3-t_2)-\Lambda_{01}(t_4-t_3)}
            \\&+p_{00,03}q_{03,02}q_{02,01}q_{01,00}e^{-\Lambda_{03}(t_2-t_1)-\Lambda_{02}(t_3-t_2)-\Lambda_{01}(t_4-t_3)}
            \\&+p_{00,01}q_{01,02}q_{02,03}q_{03,00}e^{-\Lambda_{01}(t_2-t_1)-\Lambda_{02}(t_3-t_2)-\Lambda_{03}(t_4-t_3)},
        \end{split}\\
        \begin{split}
            \expval{q(t_1)p(t_2)q(t_3)q(t_4)}=&q_{00,01}p_{01,00}q_{00,01}q_{01,00}e^{-\Lambda_{01}(t_2-t_1)-\Lambda_{00}(t_3-t_2)-\Lambda_{01}(t_4-t_3)}
            \\&+q_{00,01}p_{01,02}q_{02,01}q_{01,00}e^{-\Lambda_{01}(t_2-t_1)-\Lambda_{02}(t_3-t_2)-\Lambda_{01}(t_4-t_3)}
            \\&+q_{00,03}p_{03,02}q_{02,01}q_{01,00}e^{-\Lambda_{03}(t_2-t_1)-\Lambda_{02}(t_3-t_2)-\Lambda_{01}(t_4-t_3)}
            \\&+q_{00,01}p_{01,02}q_{02,03}q_{03,00}e^{-\Lambda_{01}(t_2-t_1)-\Lambda_{02}(t_3-t_2)-\Lambda_{03}(t_4-t_3)},
        \end{split}\\
        \begin{split}
            \expval{q(t_1)q(t_2)p(t_3)q(t_4)}=&q_{00,01}q_{01,00}p_{00,01}q_{01,00}e^{-\Lambda_{01}(t_2-t_1)-\Lambda_{00}(t_3-t_2)-\Lambda_{01}(t_4-t_3)}
            \\&+q_{00,01}q_{01,02}p_{02,01}q_{01,00}e^{-\Lambda_{01}(t_2-t_1)-\Lambda_{02}(t_3-t_2)-\Lambda_{01}(t_4-t_3)}
            \\&+q_{00,03}q_{03,02}p_{02,01}q_{01,00}e^{-\Lambda_{03}(t_2-t_1)-\Lambda_{02}(t_3-t_2)-\Lambda_{01}(t_4-t_3)}
            \\&+q_{00,01}q_{01,02}p_{02,03}q_{03,00}e^{-\Lambda_{01}(t_2-t_1)-\Lambda_{02}(t_3-t_2)-\Lambda_{03}(t_4-t_3)},
        \end{split}\\
        \begin{split}
            \expval{q(t_1)q(t_2)q(t_3)p(t_4)}=&q_{00,01}q_{01,00}q_{00,01}p_{01,00}e^{-\Lambda_{01}(t_2-t_1)-\Lambda_{00}(t_3-t_2)-\Lambda_{01}(t_4-t_3)}
            \\&+q_{00,01}q_{01,02}q_{02,01}p_{01,00}e^{-\Lambda_{01}(t_2-t_1)-\Lambda_{02}(t_3-t_2)-\Lambda_{01}(t_4-t_3)}
            \\&+q_{00,03}q_{03,02}q_{02,01}p_{01,00}e^{-\Lambda_{03}(t_2-t_1)-\Lambda_{02}(t_3-t_2)-\Lambda_{01}(t_4-t_3)}
            \\&+q_{00,01}q_{01,02}q_{02,03}p_{03,00}e^{-\Lambda_{01}(t_2-t_1)-\Lambda_{02}(t_3-t_2)-\Lambda_{03}(t_4-t_3)}.
        \end{split}\\
    \end{align}
\end{subequations}
Using the perturbative eigenspectrum (\ref{free_eigenvalue}) and (\ref{free_eigenstates_spp=0}), one can compute these explicitly as
\begin{subequations}
    \label{(q,p)_4-point_functions_explicit}
    \begin{align}
        \begin{split}
            \expval{q(t_1)q(t_2)q(t_3)q(t_4)}=&\lb\lb\frac{\sigma_{Q,qq}^2}{2\alpha H}\rb^2-\frac{6\lambda_S\beta}{H^2\alpha(\alpha-\beta)^2}\lb\frac{\sigma_{Q,qq}^2}{2\alpha H}\rb^3\rb e^{-\alpha H(t_4+t_2-t_3-t_1)}
            \\&+\lb2\lb\frac{\sigma_{Q,qq}^2}{2\alpha H}\rb^2-\frac{24\lambda_S\beta}{H^2\alpha(\alpha-\beta)^2}\lb\frac{\sigma_{Q,qq}^2}{2\alpha H}\rb^3\rb e^{-\alpha H(t_4+t_3-t_2-t_1)}
            \\&+\frac{3\lambda_S\beta(\sigma_{Q,qq}^2)^3}{32H^5\alpha^4\nu^2}e^{-\alpha H(t_4+t_3+t_2-3t_1)}\\&+\frac{3\lambda_S\beta(\sigma_{Q,qq}^2)^3}{32H^5\alpha^4\nu^2}e^{-\alpha H(3t_4-t_4-t_2-t_1)},
        \end{split}\\
        \begin{split}
            \expval{p(t_1)q(t_2)q(t_3)q(t_4)}=&-\frac{\lambda_S\beta^2}{H(\alpha-\beta)^2}\lb\frac{\sigma_{Q,qq}^2}{2\alpha H}\rb^3e^{-\alpha H(t_4+t_2-t_3-t_1)}
            \\&-\frac{2\lambda_S\beta^2}{H(\alpha-\beta)^2}\lb\frac{\sigma_{Q,qq}^2}{2\alpha H}\rb^3e^{-\alpha H(t_4+t_3-t_2-t_1)}
            \\&-\frac{3\lambda_S\beta^2(\sigma_{Q,qq}^2)^3}{32H^5\alpha^3(3\alpha+\beta)\nu^2}e^{-\alpha H(t_4+t_3+t_2-3t_1)},
        \end{split}\\
        \begin{split}
            \expval{q(t_1)p(t_2)q(t_3)q(t_4)}=&\frac{3\lambda_S\beta^2}{H(\alpha-\beta)^3}\lb\frac{\sigma_{Q,qq}^2}{2\alpha H}\rb^3e^{-\alpha H(t_4+t_2-t_3-t_1)}
            \\&-\frac{4\lambda_S\beta^2}{H(\alpha-\beta)^2}\lb\frac{\sigma_{Q,qq}^2}{\alpha H}\rb^3e^{-\alpha H(t_4+t_3-t_2-t_1)},
        \end{split}\\
        \begin{split}
            \expval{q(t_1)q(t_2)p(t_3)q(t_4)}=&\frac{3\lambda_S\beta^2}{H(\alpha-\beta)^3}\lb\frac{\sigma_{Q,qq}^2}{\alpha H}\rb^3e^{-\alpha H(t_4+t_2-t_3-t_1)}
            \\&+\frac{12\lambda_S\beta^2}{H(\alpha-\beta)^3}\lb\frac{\sigma_{Q,qq}^2}{2\alpha H}\rb^3 e^{-\alpha H(t_4+t_3-t_2-t_1)},
        \end{split}\\
        \begin{split}
            \expval{q(t_1)q(t_2)q(t_3)p(t_4)}=&
            -\frac{\lambda_S\beta^2}{H(\alpha-\beta)^2}\lb\frac{\sigma_{Q,qq}^2}{\alpha H}\rb^3e^{-\alpha H(t_4+t_2-t_3-t_1)}
            \\&+\frac{6\lambda_S\beta^2}{H(\alpha-\beta)^3}\lb\frac{\sigma_{Q,qq}^2}{2\alpha H}\rb^3 e^{-\alpha H(t_4+t_3-t_2-t_1)}
            \\&+\frac{3\lambda_S\beta^3(\sigma_{Q,qq}^2)^3}{16H^4\alpha^3(3\alpha-\beta)\nu^2}e^{-\alpha H(3t_4-t_3-t_2+t_1)}.
        \end{split}
    \end{align}
\end{subequations}
We can then compute the timelike stochastic field 4-point function using Eq. (\ref{p,q-->pi,phi}) to get
\begin{equation}
    \label{field_stochastic_4pt_func}
    \begin{split}
    \expval{\phi(t_1)\phi(t_2)\phi(t_3)\phi(t_4)}=&
    \frac{\lambda_S\beta^3(2\beta^2+21\beta-72)(\sigma_{Q,qq}^2)^3}{4H^5\alpha^4(\alpha-\beta)^5}e^{-\alpha H(t_2-t_1+t_4-t_3)}
    \\&+\frac{3\lambda_S\beta^3(\beta^2+\beta-6)(\sigma_{Q,qq}^2)^3}{2H^5\alpha^4(\alpha-\beta)^5}e^{-\alpha H(t_3-t_1+t_4-t_2)}
    \\&+\frac{3\lambda_S\beta^3(4\alpha+\beta)(\sigma_{Q,qq}^{2(0)})^3}{128H^5\alpha^4(3\alpha+\beta)\nu^4}e^{-\alpha H(t_4+t_3+t_2-3t_1)}\\&-\frac{3\lambda_S\beta^3(\sigma_{Q,qq}^{2(0)})^3}{64H^5\alpha^4(3\alpha-\beta)\nu^3}e^{-\alpha H(3t_4-t_3-t_2+t_1)}.
    \end{split}
\end{equation}
Using the definition of the stochastic connected 4-point function (\ref{OD_connected_4-pt_func_def}), we can write the timelike stochastic connected field 4-point function as
\begin{equation}
    \label{stochastic_timelike_4pt_func}
    \keybox{
    \begin{split}
    \expval{\phi(t_1)\phi(t_2)\phi(t_3)\phi(t_4)}_C=&
    \frac{\lambda_S\beta^3(2\beta^2+21\beta-81)(\sigma_{Q,qq}^2)^3}{4H^5\alpha^4(\alpha-\beta)^5}e^{-\alpha H(t_2-t_1+t_4-t_3)}
    \\&+\frac{3\lambda_S\beta^3(\beta^2+\beta-9)(\sigma_{Q,qq}^2)^3}{2H^5\alpha^4(\alpha-\beta)^5}e^{-\alpha H(t_3-t_1+t_4-t_2)}
    \\&+\frac{3\lambda_S\beta^3(4\alpha+\beta)(\sigma_{Q,qq}^{2(0)})^3}{128H^5\alpha^4(3\alpha+\beta)\nu^4}e^{-\alpha H(t_4+t_3+t_2-3t_1)}\\&-\frac{3\lambda_S\beta^3(\sigma_{Q,qq}^{2(0)})^3}{64H^5\alpha^4(3\alpha-\beta)\nu^3}e^{-\alpha H(3t_4-t_3-t_2+t_1)}.
    \end{split}
    }
\end{equation}
Note that in the near-massless limit $m\ll H$, the 4-point function becomes
\begin{equation}
    \label{connected_timelike_SO_4pt_function_light}
    \begin{split}
    \expval{\phi(t_1)\phi(t_2)\phi(t_3)\phi(t_4)}_C\eval_{m\ll H}=&\frac{81\lambda_S H^{12}}{512\pi^6m^8}\lb e^{-\frac{m^2}{3H}(t_4+t_3+t_2-3t_1)}+e^{-\frac{m^2}{3H}(3t_4-t_3-t_2-t_1)}\rb
    \\&-\frac{81\lambda_S H^{12}}{128\pi^6m^8}e^{-\frac{m^2}{3H^2}(t_4+t_3-t_2-t_1)},
    \end{split}
\end{equation}
which coincides with the OD connected 4-point function (\ref{connected_OD_4pt_function}) for $\lambda_S=\lambda$.

We can also compute the equal-time stochastic connected 4-point function at $\mathcal{O}(\lambda)$. Using Eq. (\ref{spacelike_4-pt_correlator_f}), and recalling that $\abs{\mathbf{x}}=\abs{\mathbf{x}_j-\mathbf{x}_i}$ $\forall i\ne j$, we find that the only non-zero spacelike $(q,p)$ correlators that are relevant are
\begin{subequations}
    \label{spacelike_(q,p)_4-pt_functions}
    \begin{align}
        \begin{split}
        \expval{q(\mathbf{x}_1)q(\mathbf{x}_2)q(\mathbf{x}_3)q(\mathbf{x}_4)}=&\int dq_r\int dp_r \frac{\Psi_{0}(q_r,p_r)}{\Psi_0^*(q_r,p_r)^3}\lb\Psi_1^*(q_r,p_r)q_{10}\rb^4\abs{Ha(t)\mathbf{x}}^{-\frac{4\Lambda_1}{H}}
        \\&+\int dq_r\int dp_r \frac{\Psi_{0}(q_r,p_r)}{\Psi_0^*(q_r,p_r)^3}\lb\Psi_1^*(q_r,p_r)q_{10}\rb^3\lb\Psi_3^*(q_r,p_r)q_{30}\rb\\&\times\abs{Ha(t)\mathbf{x}}^{-\frac{3\Lambda_1+\Lambda_3}{H}},
        \end{split}\\
        \begin{split}
        \expval{q(\mathbf{x}_1)q(\mathbf{x}_2)q(\mathbf{x}_3)p(\mathbf{x}_4)}=&\expval{q(\mathbf{x}_1)q(\mathbf{x}_2)p(\mathbf{x}_3)q(\mathbf{x}_4)}
        \\=&\expval{q(\mathbf{x}_1)p(\mathbf{x}_2)q(\mathbf{x}_3)q(\mathbf{x}_4)}
        \\=&\expval{p(\mathbf{x}_1)q(\mathbf{x}_2)q(\mathbf{x}_3)q(\mathbf{x}_4)}
        \\=&\int dq_r\int dp_r \frac{\Psi_{0}(q_r,p_r)}{\Psi_0^*(q_r,p_r)^3}\lb\Psi_1^*(q_r,p_r)q_{10}\rb^3\lb\Psi_3^*(q_r,p_r)p_{10}\rb\\&\times\abs{Ha(t)\mathbf{x}}^{-\frac{4\Lambda_1}{H}}
        \\&+\int dq_r\int dp_r \frac{\Psi_{0}(q_r,p_r)}{\Psi_0^*(q_r,p_r)^3}\lb\Psi_1^*(q_r,p_r)q_{10}\rb^3\lb\Psi_3^*(q_r,p_r)p_{30}\rb\\&\times\abs{Ha(t)\mathbf{x}}^{-\frac{3\Lambda_1+\Lambda_3}{H}}.
        \end{split}
    \end{align}
\end{subequations}
Then we can use the eigenspectrum (\ref{free_eigenvalue}) and (\ref{free_eigenstates_spp=0}) to obtain
\begin{subequations}
    \begin{align}
        \label{qqqq_spacelike_correlator}
        \begin{split}
            \expval{q(\mathbf{x}_1)q(\mathbf{x}_2)q(\mathbf{x}_3)q(\mathbf{x}_4)}
            =&\lb\frac{3(\sigma_{Q,qq}^{2})^2}{4H^2\alpha^2}-\frac{3\lambda_S\beta(\sigma_{Q,qq}^2)^3}{H^5\alpha^4(\alpha-\beta)^2}\rb\abs{Ha(t)\mathbf{x}}^{-4\alpha}
            \\&+\frac{3\lambda_S\beta\lb\sigma_{Q,qq}^{2}\rb^3}{2H^5\alpha^4(\alpha-\beta)^2}\abs{Ha(t)\mathbf{x}}^{-6\alpha},
        \end{split}\\
        \begin{split}
            \expval{p(\mathbf{x}_1)q(\mathbf{x}_2)q(\mathbf{x}_3)q(\mathbf{x}_4)}
            =&\frac{9\lambda_S\beta^2\lb\sigma_{Q,qq}^2\rb^3}{8H^4\alpha^3(\alpha-\beta)^3}\abs{Ha(t)\mathbf{x}}^{-4\alpha}
            \\&+\frac{3\lambda_S\beta^2\lb\sigma_{Q,qq}^{2}\rb^3}{4H^4\alpha^3(\alpha-\beta)^2(3\alpha-\beta)}\abs{Ha(t)\mathbf{x}}^{-6\alpha}.
        \end{split}
    \end{align}
\end{subequations}
Thus, the spacelike stochastic field 4-point function is
\begin{equation}
    \label{stochastic_4pt_func}
    \begin{split}
        \expval{\phi(\mathbf{x}_1)\phi(\mathbf{x}_2)\phi(\mathbf{x}_3)\phi(\mathbf{x}_4)}=&\lb\frac{3\beta^2\lb\sigma_{Q,qq}^2\rb^2}{4H^2\alpha^2(\alpha-\beta)^2}-\frac{9\lambda_S\beta^3(2\alpha-\beta)\lb\sigma_{Q,qq}^2\rb^3}{2H^5\alpha^4(\alpha-\beta)^5}\rb\abs{Ha(t)\mathbf{x}}^{-4\alpha}
        \\&+\frac{3\lambda_S\beta^6(\sigma_{Q,qq}^{2})^3}{16H^5\alpha^4\nu^3(\beta-3\alpha)}\abs{Ha(t)\mathbf{x}}^{-6\alpha}.
    \end{split}
\end{equation}
Then we can use Eq. (\ref{OD_connected_4-pt_func_def}) (replacing $t_i$ with $\mathbf{x}_i$) to get the connected spacelike stochastic 4-pt function to $\mathcal{O}(\lambda_S)$ as
\begin{equation}
    \label{stochastic_spacelike_4pt_func}
    \keybox{
    \begin{split}
        \expval{\phi(\mathbf{x}_1)\phi(\mathbf{x}_2)\phi(\mathbf{x}_3)\phi(\mathbf{x}_4)}_C=&-\frac{9\lambda_S\beta^3\lb\sigma_{Q,qq}^2\rb^3}{4H^5\alpha^4(\alpha-\beta)^4}\abs{Ha(t)\mathbf{x}}^{-6+4\nu}
        \\&+\frac{3\lambda_S\beta^6(\sigma_{Q,qq}^{2})^3}{16H^5\alpha^4\nu^3(\beta-3\alpha)}\abs{Ha(t)\mathbf{x}}^{-9+6\nu}.
    \end{split}
    }
\end{equation}
In the light field limit $m\ll H$, the spacelike 4-point function is given by
\begin{equation}
    \label{spacelike_4pt_func_light_fields}
    \begin{split}
    \expval{\phi(\mathbf{x}_1)\phi(\mathbf{x}_2)\phi(\mathbf{x}_3)\phi(\mathbf{x}_4)}_C\eval_{m^2\ll H^2}=&\frac{81\lambda_S H^{12}}{128\pi^6m^6}\abs{Ha(t)\mathbf{x}}^{-\frac{2m^2}{H^2}}
    \\&-\frac{243\lambda_SH^{12}}{256\pi^6m^8}\abs{Ha(t)\mathbf{x}}^{-\frac{4m^2}{3H^2}},
    \end{split}
\end{equation}
which is the same as the OD spacelike 4-point function (\ref{spacelike_connected_OD_4pt_function}) for $\lambda_S=\lambda$.

It is worth noting that we have not introduced any non-Gaussianities here, for example via a 4-point noise correlator $\expval{\xi(t_1)\xi(t_2)\xi(t_3)\xi(t_4)}$. We will see in the next section that it is not necessary because the stochastic 4-point function correctly reproduces its quantum counterpart without having to non-trivially match the stochastic parameters beyond the level of 2-point functions. However, it is not clear from this analysis whether we will need to include non-Gaussian behaviour at higher-orders in $\lambda$ so that the stochastic theory can be extended.

\subsection{Stochastic parameters to $\mathcal{O}(\lambda)$}
\label{subsec:perturbative_stochastic_parameters}

We can now compute the $\mathcal{O}(\lambda)$ stochastic parameters by comparing the $\mathcal{O}(\lambda_S)$ stochastic 2-point and 4-point functions with their QFT counterparts. The procedure for matching our 2-point functions remains unchanged from the free case; we again have our three conditions (plus a choice) for the stochastic results to match perturbative QFT: (i)  the leading exponents match, (ii) the leading prefactors match, (iii) the analytic continuation between spacelike and timelike correlators is preserved and (iv) either the subleading terms match or it vanishes in the stochastic correlators. We will consider both cases presented by (iv), though as stated these are not unique choices. For the following, I will consider the latter choice; the former is dealt with in Appendix \ref{app:stochastic_parameters_NLO}. We will see that the choice doesn't affect the physical results, as indeed it shouldn't.

However, we now have a fifth stochastic parameter to contend with - $\lambda_S$ - which can be matched to its QFT counterpart by considering the connected 4-point function. By equating the spacelike stochastic 4-point function (\ref{stochastic_spacelike_4pt_func}), using the free stochastic parameters (\ref{free_matched_stochastic_parameters}), to the spacelike quantum 4-point function (\ref{spacelike_quantum_4-pt_func}), we find that the terms $\mathcal{O}(\abs{Ha(t)\mathbf{x}}^{-9+6\nu}$ are equal for any value of $m$ if
\begin{equation}
    \label{matched_stochastic_lambda}
    \lambda_S=\lambda+\mathcal{O}(\lambda^2).
\end{equation}
Thus, at this order, the $\lambda$ parameters in both quantum and stochastic theories are the same and we will drop the subscript $S$ henceforth. We note that, as in the OD stochastic approximation, the second-order stochastic theory also gives us an expression of $\mathcal{O}\lb\abs{Ha(t)\mathbf{x}}^{-6+4\nu}\rb$, which we know also appears in the QFT counterpart (\ref{spacelike_quantum_4-pt_func}), denoted by the `...+'. Thus, the stochastic theory gives us a way of computing this term explicitly, which is difficult to do in perturbative QFT.

We can now turn our attention to the 2-point functions, and the other 4 $\mathcal{O}(\lambda)$ stochastic parameters. Making a perturbative expansion of our stochastic parameters about $\mathcal{O}(\lambda)$, where the free parameters are given in Eq. (\ref{free_matched_stochastic_parameters}) and we take $\sigma_{Q,pp}^{2(0)}=0$,
\begin{subequations}
    \label{perturbative_expansion_stochastic_parameters}
    \begin{align}
    m_S^2&=m_R^2+\lambda m_S^{2(1)},\\
    \sigma_{qq}^2&=\frac{H^3\lb3/2-\nu_R\rb\nu_R}{4\pi^2\lb3/2+\nu_R\rb}\frac{\Gamma(2\nu_R)\Gamma\lb\frac{3}{2}-\nu_R\rb4^{\frac{3}{2}-\nu_R}}{\Gamma\lb\frac{1}{2}+\nu_R\rb}+\lambda\sigma_{qq}^{2(1)}+\mathcal{O}(\lambda^2),\\
    \sigma_{qp}^2&=\lambda\sigma_{qp}^{2(1)}+\mathcal{O}(\lambda^2),\\
    \sigma_{pp}^2&=\lambda\sigma_{pp}^{2(1)}+\mathcal{O}(\lambda^2),
    \end{align}
\end{subequations}
where we now have to use the renormalised mass $m_R$ as we are in the interacting theory. Note that, because we are considering a renormalised mass, we expect that $m_S^{2(1)}$ will be non-zero. From Eq. (\ref{equal-time_(phi,pi)_stochastic_correlators}) and (\ref{equal-space_(phi,pi)_stochastic_correlators}), the timelike and spacelike stochastic 2-point functions for this choice of noise are
\begin{equation}
    \label{O(lambda)_timelike_stochastic_2pt_func_sigma1}
    \begin{split}
        \expval{\phi(0,\mathbf{x})\phi(t,\mathbf{x})}=&\Bigg[\frac{H^2}{16\pi^2}\frac{\Gamma(2\nu_R)\Gamma\lb\frac{3}{2}-\nu_R\rb4^{\alpha_R}}{\Gamma\lb\frac{1}{2}+\nu_R\rb}\\&+\lambda\Bigg(\frac{\beta_R\sigma_{qq}^{2(1)}}{4H\nu_R\alpha_R}-\frac{\sigma_{qp}^{2(1)}}{6H^2\nu_R}+\frac{9H^2}{1024\pi^4\alpha_R\nu_R^2}\lb\frac{\Gamma\lb\frac{3}{2}-\nu_R\rb\Gamma(2\nu_R)4^{\frac{3}{2}-\nu_R}}{\Gamma\lb\frac{1}{2}+\nu_R\rb}\rb^2\\&+\frac{(4\nu_R^2+12\nu_R-9)\nu^{(1)}_S}{16H\alpha_R^2\nu_R^2}\frac{H^3\alpha_R\nu_R}{4\pi^2\beta_R}\frac{\Gamma(2\nu_R)\Gamma\lb\frac{3}{2}-\nu_R\rb4^{\frac{3}{2}-\nu_R}}{\Gamma\lb\frac{1}{2}+\nu_R\rb}\Bigg)\\&+\mathcal{O}(\lambda^2)\Bigg]\times e^{-\lb\alpha_RH-\lambda\lb\nu_S^{(1)}H+\frac{3H}{32\pi^2\nu_R}\frac{\Gamma(2\nu_R)\Gamma\lb\frac{3}{2}-\nu_R\rb4^{\frac{3}{2}-\nu_R}}{\Gamma\lb\frac{1}{2}+\nu_R\rb}\rb+\mathcal{O}(\lambda^2)\rb}
        \\&+\\&
        \Bigg[\lambda\Bigg(\frac{\sigma_{pp}^{2(1)}}{4H^3\nu_R\beta_R^2}-\frac{\sigma_{qp}^{2(1)}}{6H^2\nu_R}-\frac{H^2\alpha_R}{512\pi^4\nu_R^2}\lb\frac{\Gamma\lb\frac{3}{2}-\nu_R\rb\Gamma(2\nu_R)}{\Gamma\lb\frac{1}{2}+\nu_R\rb^2}\rb\Bigg)\\&+\mathcal{O}(\lambda^2)\Bigg]
        e^{-\lb\beta_RH+\lambda\lb\nu_S^{(1)}H+\frac{3H}{32\pi^2\nu_R}\frac{\Gamma(2\nu_R)\Gamma\lb\frac{3}{2}-\nu_R\rb4^{\frac{3}{2}-\nu_R}}{\Gamma\lb\frac{1}{2}+\nu_R\rb}\rb+\mathcal{O}(\lambda^2)\rb}
    \end{split}
\end{equation}
and
\begin{equation}
    \label{O(lambda)_spacelike_stochastic_2pt_func_sigma1}
    \begin{split}
        \expval{\phi(t,\mathbf{0})\phi(t,\mathbf{x})}=&\Bigg[\frac{H^2}{16\pi^2}\frac{\Gamma\qty(\frac{3}{2}-\nu_R)\Gamma\qty(2\nu_R)4^{\alpha_R}}{\Gamma\qty(\frac{1}{2}+\nu_R)}
        \\&+\lambda\Bigg(\frac{\beta_R\sigma_{qq}^{2(1)}}{4H\nu_R\alpha_R}+\frac{3(3-4\nu_R)H^4\Gamma(\nu_R)^2\Gamma\qty(\frac{3}{2}-\nu_R)^2}{32\pi^5\nu_R m_R^2}
        \\&+\frac{(4\nu_R^2+12\nu_R-9)\nu^{(1)}_S}{16H\alpha_R^2\nu_R^2}\frac{H^3\alpha_R\nu_R}{4\pi^2\beta_R}\frac{\Gamma(2\nu_R)\Gamma\lb\frac{3}{2}-\nu_R\rb4^{\frac{3}{2}-\nu_R}}{\Gamma\lb\frac{1}{2}+\nu_R\rb}\Bigg)+\mathcal{O}(\lambda^2)\Bigg]
        \\&\times\abs{Ha(t)\mathbf{x}}^{-\lb2\alpha_R-\lambda\lb2\nu_S^{(1)}+\frac{3}{16\pi^2\nu_R}\frac{\Gamma(2\nu_R)\Gamma\lb\frac{3}{2}-\nu_R\rb4^{\frac{3}{2}-\nu_R}}{\Gamma\lb\frac{1}{2}+\nu_R\rb}\rb+\mathcal{O}(\lambda^2)\rb}
        \\&+\lsb\frac{\lambda \sigma_{pp}^{2(1)}}{4H^3\nu_R\beta_R^2}+\mathcal{O}(\lambda^2)\rsb\\&\times\abs{Ha(t)\mathbf{x}}^{-\lb2\beta_R+\lambda\lb2\nu_S^{(1)}+\frac{3}{16\pi^2\nu_R}\frac{\Gamma(2\nu_R)\Gamma\lb\frac{3}{2}-\nu_R\rb4^{\frac{3}{2}-\nu_R}}{\Gamma\lb\frac{1}{2}+\nu_R\rb}\rb+\mathcal{O}(\lambda^2)\rb}\\&+\lsb-
        \lambda\qty(\frac{\sigma_{qp}^{2(1)}}{3H^2\nu_R}+\frac{H^4\Gamma(\nu_R)^2\Gamma\qty(\frac{5}{2}-\nu_R)^2}{8\pi^5\nu_R m_R^2})+\mathcal{O}(\lambda^2)\rsb\\&\times\abs{Ha(t)\mathbf{x}}^{-3+\mathcal{O}(\lambda^2)},
    \end{split}
\end{equation}
where $\nu_S^{(1)}=-\frac{m_S^{2(1)}}{H^2\nu_R}$. These reproduce the perturbative QFT 2-point function (\ref{O(lambda)_2-pt_function_spacelike_long-distance}) for the parameters
\begin{subequations}
    \label{one-loop_qp_stochastic_parameters}
    \begin{align}
        \label{one-loop_stochastic_mass}
        \begin{split}
         m_S^2=&m_R^2+\frac{3\lambda H^2}{16\pi^2}\Bigg[-\frac{4\Gamma\qty(3/2-\nu_R)\Gamma\qty(\nu_R)}{\sqrt{\pi}}\\&+\qty(2-\frac{m_R^2}{H^2})\qty(1-\psi^{(0)}\qty(3/2-\nu_R)-\psi^{(0)}\qty(3/2+\nu_R)+\ln\qty(\frac{M^2}{a(t)^2H^2}))\Bigg]\\&+\mathcal{O}(\lambda^2)
         \end{split}\\
        \label{one-loop_sigma_qq}
        \begin{split}        \sigma_{qq}^{2}=&\frac{2H^3\Gamma(1+\nu_R)\Gamma\qty(\frac{5}{2}-\nu_R)}{\pi^{5/2}\qty(3+2\nu_R)}+\frac{\lambda H^3\Gamma\qty(3/2-\nu_R)\Gamma\qty(\nu_R)}{16\pi^{7/2}\qty(3+2\nu_R)^2}\\&\times\Bigg[3(-3+2\nu_R)\Gamma\qty(3/2-\nu_R)\Gamma\qty(2\nu_R)+\frac{3\times4^{-3+\nu_R}}{\nu_R}\qty(-1+4\nu_R^2)\\&\times\Gamma\qty(1/2+\nu_R)\Bigg(4\frac{m_R^2}{H^2}-12\nu_R-4\frac{m^2_R}{H^2}\nu_R\ln4-4\frac{m_R^2}{H^2}\nu_R\psi^{(0)}\qty(3/2-\nu_R)\\&+8\frac{m_R^2}{H^2}\nu_R\psi^{(0)}(2\nu_R)-4\frac{m_R^2}{H^2}\nu_R\psi^{(0)}\qty(1/2+\nu_R)\Bigg)\\&\qty(-1-\ln\qty(\frac{M^2}{a(t)^2H^2})+\psi^{(0)}\qty(3/2-\nu_R)+\psi^{(0)}\qty(3/2+\nu_R))\Bigg]+\mathcal{O}(\lambda^2),
        \end{split}\\
        \label{one-loop_sigma_qp}
        \begin{split}
        \sigma_{qp}^{2}=&\frac{3\lambda H^4(-3+2\nu_R)\Gamma\qty(3/2-\nu_R)^2\Gamma\qty(\nu_R)^2}{32\nu_R\pi^5}+\mathcal{O}(\lambda^2),
        \end{split}\\
        \label{one-loop_sigma_pp}
        \begin{split}
        \sigma_{pp}^2=&\mathcal{O}(\lambda^2).
        \end{split}
    \end{align}
\end{subequations}
Note that the mass parameter is now dependent on the renormalisation scale $M$. This is an important difference between the second-order and OD stochastic theories, as will be discussed in more detail in Chapter \ref{ch:comparison_models}. Using these parameters in our second-order stochastic equations (\ref{2d_stochastic_eq}) gives us a second-order stochastic theory of quartic self-interacting scalar QFT in de Sitter. However, we note that these expressions still have an IR problem, but it is milder than that of perturbative QFT. Expanding the $\mathcal{O}(\lambda)$ terms to leading order in $m^2/H^2$, we have
\begin{subequations}
    \label{O(lambdaH2m2)_stochastic_parameters}
    \begin{align}
        m_S^{2}=&m_R^2+\frac{3\lambda H^2}{8\pi^2}\qty(2\gamma_E-\ln4+\ln\qty(\frac{M^2}{a(t)^2H^2}))+\mathcal{O}\qty(\lambda m_R^2),\\
        \sigma_{qq}^2=&\frac{2H^3\Gamma(1+\nu_R)\Gamma\qty(\frac{5}{2}-\nu_R)}{\pi^{5/2}\qty(3+2\nu_R)}+\frac{\lambda H^5\qty(-8+3\ln4)}{32\pi^4 m_R^2}+\mathcal{O}\qty(\lambda H^3),\\
        \sigma_{qp}^2=&-\frac{3\lambda H^6}{32\pi^4m_R^2}+\mathcal{O}\qty(\lambda H^4),\\
        \sigma_{pp}^2=&0+\mathcal{O}\qty(\lambda^2).
    \end{align}
\end{subequations}
We see that the sum will converge when $\lambda\ll m^2/H^2$; however, since we have corrected at this order, the error associated with the stochastic parameters is actually $\mathcal{O}\lb\frac{\lambda^2H^4}{m^4}\rb$. Thus, the second-order stochastic theory is limited to $\lambda^2\ll m^4/H^4$. This is a limitation of the matching procedure required to obtain the stochastic parameters, since we rely on the results of perturbative QFT. Crucially, the IR problem is less serious in our stochastic approach compared with perturbative QFT: $\mathcal{O}(\lambda^2 H^4/m^4)$ as opposed to $\mathcal{O}(\lambda H^4/m^4)$. 

Converting our stochastic noise back to using the $(\phi,\pi)$ variables, our stochastic mass is given by
\begin{equation}
    \label{stochastic_mass_O(lambdaH2m2)}
    \keybox{
    m_S^{2}=m_R^2+\frac{3\lambda H^2}{8\pi^2}\qty(2\gamma_E-\ln4+\ln\qty(\frac{M^2}{a(t)^2H^2}))+\mathcal{O}\qty(\lambda m_R^2)
    }
\end{equation}
and the noise matrix is given by

\begin{equation}
    \label{O(lambdaH2m2)_phi-pi_noise_matrix}
    \keybox{
    \begin{split}
    \sigma^{2}=\frac{H^3\Gamma\qty(\nu_R)\Gamma\qty(\frac{5}{2}-\nu_R)}{2\pi^{5/2}}&\begin{pmatrix}
        1&-\frac{2m_R^2}{H(3+2\nu_R)}\\-\frac{2m_R^2}{H(3+2\nu_R)}&\frac{4m_R^4}{(3+2\nu_R)^2H^2}
    \end{pmatrix}
    \\&+\lambda
    \begin{pmatrix}
    \frac{3H^5\qty(-2+\ln4)}{32\pi^4m_R^2}+\mathcal{O}\qty(H^3) & -\frac{3H^6}{32\pi^4m_R^2}+\mathcal{O}\qty(H^4) \\ -\frac{3H^6}{32\pi^4m_R^2}+\mathcal{O}\qty(H^4) & \mathcal{O}\qty(H^5)
    \end{pmatrix}.
    \end{split}
    }
\end{equation}

Thus,
\begin{equation*}
\keybox{
\begin{split}
\text{Using th}& \text{e stochastic parameters (\ref{stochastic_mass_O(lambdaH2m2)}) and (\ref{O(lambdaH2m2)_phi-pi_noise_matrix}) elevates the stochastic theory (\ref{2d_stochastic_eq}) to }\\& \text{an IR effective theory of quartic self-interacting scalar QFT in de Sitter.}
\end{split}
}
\end{equation*}

\subsection{Numerical solutions to the second-order eigenspectrum}
\label{subsec:numerical_solutions_stochastic}

So far, we have a second-order stochastic effective theory that has only been solved perturbatively, and is thus no better than perturbative QFT. We will now elevate our results by solving the stochastic equations numerically to obtain non-perturbative results. We continue to use the $(q,p)$ coordinates so that we can use the free eigenstates (either Eq. (\ref{free_eigenstates_spp=0}) or (\ref{free_eigenstates_sqp=0}), depending on our choice of $\sigma_{pp}^2$). Making the ansatz that our eigensolutions to the eigenequations (\ref{FP_eigenequations}) can be written as
\begin{subequations}
    \label{expansion_free_eigensolutions}
    \begin{align}
        \Psi_{N}(q,p)&=\sum_{rs}c^{(N)}_{rs}\Psi_{rs}^{(0)}(q,p),\\
        \Psi_{N}^*(q,p)&=\sum_{rs}c^{*(N)}_{rs}\Psi_{rs}^{(0)*}(q,p),
    \end{align}
\end{subequations}
where $c_{rs}^{(*)(N)}$ are two sets of coefficients to be determined numerically. Substituting Eq. (\ref{expansion_free_eigensolutions}) into (\ref{FP_eigenequations}) gives
\begin{subequations}
    \label{expansion_expression_FP_eigenequation}
    \begin{align}
        \sum_{rs}c_{rs}^{(N)}\mathcal{L}_{FP}\Psi_{rs}^{(0)}(q,p)&=-\Lambda_N\sum_{rs}c_{rs}^{(N)}\Psi_{rs}^{(0)}(q,p),\\
        \sum_{rs}c_{rs}^{(N)}\mathcal{L}^*_{FP}\Psi_{rs}^{(0)*}(q,p)&=-\Lambda_N\sum_{rs}c_{rs}^{*(N)}\Psi_{rs}^{(0)*}(q,p).
    \end{align}
\end{subequations}
Applying the Fokker-Planck operator to the free eigenstates will give us
\begin{equation}
    \label{FP_operator-->matrix}
    \mathcal{L}_{FP}^{(*)}\Psi_{rs}^{(0)(*)}(q,p)=\sum_{r's'}\mathcal{M}^{(*)}_{rsr's'}\Psi_{r's'}^{(0)(*)},
\end{equation}
where the matrices $\mathcal{M}^{(*)}$ are given by
\begin{equation}
    \label{matrices_M}
    \mathcal{M}_{rsr's'}=\qty(\Psi_{r's'}^{(0)*},\mathcal{L}_{FP}\Psi_{rs}^{(0)})=\mathcal{M}_{r's'rs}^*.
\end{equation}
Explicit expressions for these matrices can be found but they are complicated. Applying Eq. (\ref{FP_operator-->matrix}) to (\ref{expansion_expression_FP_eigenequation}) and making use of the completeness of the free eigenstates (\ref{biorthogonal&completeness_relations}), one can write
\begin{subequations}
    \label{matrix_eigenequation}
    \begin{align}
        \sum_{r's'}\mathcal{M}^T_{rsr's'}c_{r's'}^{(N)}&=-\Lambda_{N}c_{rs}^{(N)},\\
        \sum_{r's'}(\mathcal{M}^*)^T_{rsr's'}c_{r's'}^{*(N)}&=-\Lambda_{N}c_{rs}^{*(N)}.
    \end{align}
\end{subequations}
Thus, by diagonalising the matrices $(\mathcal{M}^{(*)})^T$, we can obtain the eigenvalues $\Lambda_N$ and the coefficients $c_{rs}^{(*)(N)}$ and hence the full solution to the Fokker-Planck equation.

In theory, this sum is infinite and our matrices are infinite-dimensional. Therefore, we have to choose a value of $r$ and $s$ ($r_{max}$ and $s_{max}$ respectively) at which we truncate the series so that we can practically diagonalise the matrices. This approximation only works if the expansion in our chosen eigenstates (\ref{expansion_free_eigensolutions}) converges as $r_{max}$ and $s_{max}$ become large. Indeed, we can use this fact to improve the accuracy of the spectral expansion by evaluating the eigenvalues for a range of $r$ and $s$ and then fitting an appropriate curve that converges at infinity. Here, we will use an exponential fit. This essentially gives us the eigenvalue at infinity. There will naturally be some error associated with this fit but, as we will see, it is exceedingly small. The convergence speeds up as $m^2/H^2$ increases with constant $\lambda$ and as $\lambda$ decreases with constant $m^2/H^2$. This is the case where the free solution is the dominant one.

\begin{figure}[ht]
    \centering
    \includegraphics[width=150mm]{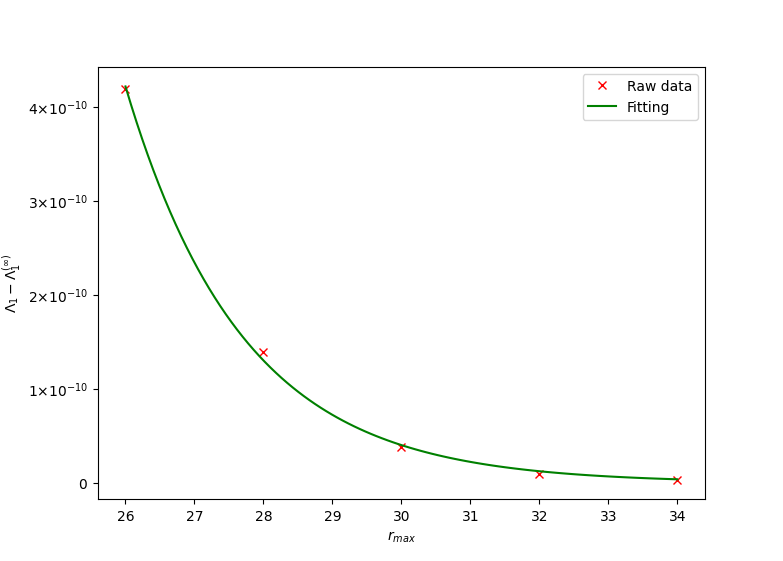}
    \caption{The difference between the first non-zero eigenvalue as found at the highest order of truncation $\Lambda_1$ and the fit $\Lambda_1^{(\infty)}=0.004530997449(4)$ for a range of $(r_{max},s_{max})$ at $m^2/H^2=0.01$ and $\lambda=0.0005$. Red crosses are the data found by numerically diagonalising the matrix $\mathcal{M}^T$ up to the truncation $(r_{max},s_{max})$ and the green line is the fit, which is exponential. Note that we always take $r_{max}=s_{max}$ and hence use a single number, $r_{max}$, to label the $x$-axis.}
    \label{fig:fit_evals_v_trunc}
\end{figure}

For the purposes of this thesis, we will focus on calculating the first non-zero eigenvalue in our spectral expansion, $\Lambda_{1}$. To make the idea of truncation and fitting more concrete, we consider the specific example in Fig. \ref{fig:fit_evals_v_trunc} where we calculate $\Lambda_{1}$ at $m^2/H^2=0.01$ and $\lambda=0.0005$ with the level of truncation $(r_{max},s_{max})$ ranging from (26,26) to (34,34). The fit in Fig. \ref{fig:fit_evals_v_trunc} gives the eigenvalue at infinity $\Lambda_1^{(\infty)}=0.004530997449(4)$. The error is exceedingly small, of order $10^{-12}$. Even if one just studies Fig. \ref{fig:fit_evals_v_trunc} roughly, one can see that the value of $\Lambda_1$ changes on the scale of $10^{-10}$ when going from a truncation at (26,26) to (34,34), 7 orders of magnitude below the leading significant figure of the eigenvalue. This is so small that we can consider our numerical approach to have negligible error. This is the scale of errors for all data taken in this work and therefore we can ignore numerical errors and drop the superscript $(\infty)$ when dealing with numerical results henceforth.

\section{Summary of key results}
\label{sec:key_results}

This chapter has all been original work, so it seems sensible to summarise and highlight the key results. We have developed a \textit{second-order stochastic effective field theory} for the long-distance behaviour of light scalar fields in de Sitter spacetime with a quartic self-interaction. The stochastic equations for this effective theory are
\begin{equation}
\keybox{
    \label{2d_stochastic_eq_again}
    \begin{pmatrix}\Dot{\phi}\\\Dot{\pi}\end{pmatrix}=\begin{pmatrix}\pi\\-3H\pi-V'(\phi)\end{pmatrix}+\begin{pmatrix}\xi_{\phi}(t,\mathbf{x})\\\xi_{\pi}(t,\mathbf{x})\end{pmatrix},
    }
\end{equation}
where $V(\phi)=\frac{1}{2}m^2_S\phi^2+\frac{1}{4}\lambda_S\phi_S^4$ with a white noise contribution
\begin{equation}
    \label{white_noise_general_again}
    \keybox{
    \expval{\xi_i(t)\xi_j(t')}=\sigma_{ij}^2\delta(t-t').
    }
\end{equation}
The stochastic parameters are:
\begin{equation}
    \label{stochastic_mass_O(lambdaH2m2)_again}
    \keybox{
    m_S^{2}=m_R^2+\frac{3\lambda H^2}{8\pi^2}\qty(2\gamma_E-\ln4+\ln\qty(\frac{M^2}{a(t)^2H^2}))+\mathcal{O}\qty(\lambda m_R^2),
    }
\end{equation}
\begin{equation}
    \label{stochastic_lambda_again}
    \keybox{
    \lambda_S=\lambda+\mathcal{O}(\lambda^2),
    }
\end{equation}
\begin{equation}
    \label{O(lambdaH2m2)_phi-pi_noise_matrix_again}
    \keybox{
    \begin{split}
    \sigma^{2}=\frac{H^3\Gamma\qty(\nu_R)\Gamma\qty(\frac{5}{2}-\nu_R)}{2\pi^{5/2}}&\begin{pmatrix}
        1&-\frac{2m_R^2}{H(3+2\nu_R)}\\-\frac{2m_R^2}{H(3+2\nu_R)}&\frac{4m_R^4}{(3+2\nu_R)^2H^2}
    \end{pmatrix}
    \\&+\lambda
    \begin{pmatrix}
    \frac{3H^5\qty(-2+\ln4)}{32\pi^4m_R^2}+\mathcal{O}\qty(H^3) & -\frac{3H^6}{32\pi^4m_R^2}+\mathcal{O}\qty(H^4) \\ -\frac{3H^6}{32\pi^4m_R^2}+\mathcal{O}\qty(H^4) & \mathcal{O}\qty(H^5)
    \end{pmatrix}.
    \end{split}
    }
\end{equation}

Non-perturbative results can be obtained numerically, pushing the second-order stochastic effective theory beyond the regime of validity for perturbative QFT. Its regime of validity in the parameter space is
\begin{equation}
    \label{regime_of_validity_second-order_stochastic}
    m\lesssim H\qquad,\qquad\lambda^2\ll m^4/H^4.
\end{equation}
This is the solid, orange region in Fig. \ref{fig:model_limitations_comparison}.
\chapter{A Comparison of the Three Approximations}
\label{ch:comparison_models}

\section{Introduction to Chapter \ref{ch:comparison_models}}

The previous three chapters have outlined three different approximations for computing the long-distance behaviour of scalar correlation functions in de Sitter spacetime. In particular, the previous chapter introduces a novel approach: the second-order stochastic effective theory. Here, we perform a detailed analysis on how these three approximations compare, with the second-order theory front and centre. The goal of this chapter is to convince ourselves that this novel theory gives sensible results when compared with the established approximations so that we can really say that it is an effective theory of scalar fields in de Sitter. The results given in this chapter are in my [K2] and [K3].

The chapter will go as follows. Sec. \ref{sec:regimes_of_validity} will summarise the regimes in the parameter space where we expect our three approximations to work. We will then launch into the detailed analysis in Sec. \ref{sec:comparing_approx}, where we will see how the three approximations compare with each other throughout the parameter space. In particular, we will see that the second-order stochastic effective theory agrees and disagrees where one would expect: a good sign! I will conclude with some final remarks in Sec. \ref{sec:comparison_conclusion}. 

\section{Regimes of validity}
\label{sec:regimes_of_validity}

\begin{figure}[ht]
    \centering
    \includegraphics[width=150mm]{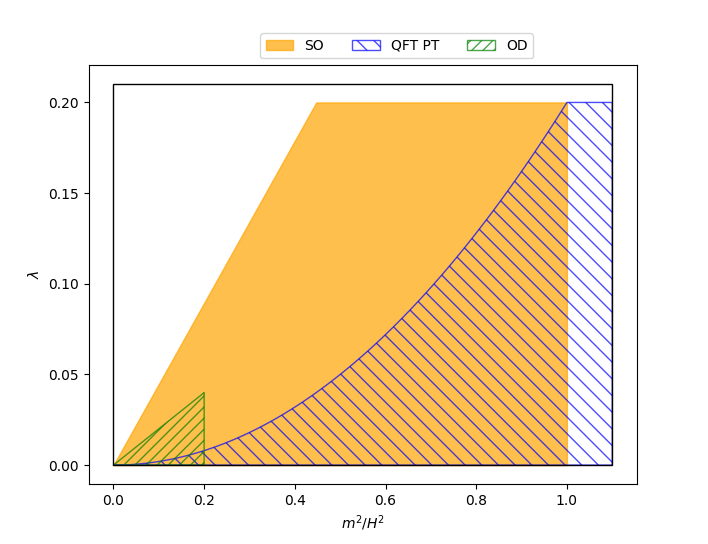}
    \caption{This shows the regimes in which we expect our approximations to work. Perturbative QFT, OD stochastic and second-order (SO) stochastic are expected to work in the blue left hashed, green right hashed and orange regions respectively. Note that there is some overlap. The pure white space is where none of these approximations work.}
    \label{fig:model_limitations_comparison}
\end{figure}

The three approximations considered in this thesis are the established perturbative QFT and overdamped stochastic approach, and the novel second-order stochastic effective theory. For the massive, self-interacting theory considered, each approximation covers a different regime in the parameter space. These are
\begin{subequations}
    \label{model_limitations}
    \begin{align}
        \text{Perturbative QFT: }\qquad\lambda\ll &\frac{m^4}{H^4}, \qquad \lambda\ll1,\\
        \text{OD stochastic: }\qquad\lambda\ll&\frac{m^2}{H^2}, \qquad m\ll H, \\
        \text{Second-order stochastic: }\qquad\lambda^2\ll &\frac{m^4}{H^4}, \qquad m\lesssim H.
    \end{align}
\end{subequations}
We make a graphical comparison of these regimes in Fig. \ref{fig:model_limitations_comparison}. For the purposes of  making the boundaries obvious, we choose ``$\ll1$'' to mean ``$<0.2$'', though in reality we wouldn't expect these boundaries to be so clear cut. 

The blue left hashed region represents the parameter space described by perturbative QFT. We can see that for light fields $m\lesssim H$, this region is entirely covered by the second-order stochastic theory. This is unsurprising given that the stochastic correlators were found directly from the 2-point functions of perturbative QFT. Beyond the light field limit, perturbative QFT continues to extend (though it is still limited to $\lambda\ll1$ - it is after all a perturbative theory!). This extension is not covered by either stochastic approaches as they both require light fields. 

The overdamped stochastic approach - the green, right hashed region - is limited to near-massless $m\ll H$ fields, but does go beyond perturbative QFT due to the non-perturbative methods available to it. Further, it is far simpler to compute stochastic correlation functions than its QFT counterparts, hence its popularity within its regime of validity. 

The OD stochastic approach is encompassed by the second-order stochastic effective theory, as represented by the orange region in Fig. \ref{fig:model_limitations_comparison}.  However, the second-order stochastic effective theory goes further, also encompassing perturbative QFT entirely in the light field limit. We can also see that there is a large chunk of the parameter space, even for near-massless fields, that is only covered by the second-order stochastic theory. The introduction of $\mathcal{O}(\lambda)$ corrections to the stochastic parameters means it goes beyond the OD approach, even in the limit $m\ll H$, while the non-perturbative methods available mean that it can extend beyond perturbative QFT\footnote{Note that, due to the matching procedure, it is still limited to the region $\lambda\ll 1$ as the stochastic parameters are found perturbatively.}. This suggests that the second-order stochastic effective theory can be used to probe hitherto untapped regions of the parameter space and is therefore a useful tool in the toolbox. However, since the theory is not directly derived from the underlying QFT, we cannot state with absolute certainty that it is correct beyond the regime of perturbative QFT. We will consider more detailed comparisons in the next section, which should go some way to assuage our doubts; however, we require further analysis with other approximations to be more certain. 

\section{Comparing approximations}
\label{sec:comparing_approx}

We will now give a more detailed comparison of the three approximations via the computation of the exponent for the leading term in the long-distance behaviour of the scalar 2-point functions. For QFT, this corresponds to the quantity given in Eq. (\ref{QFT_exponent}) to $\mathcal{O}\lb\frac{\lambda H^4}{m^4}\rb$; for clarity it is repeated here as
\begin{equation}
    \label{QFT_exponent_O(lambdaH4m4)}
    \Lambda_1^{(QFT)}=\lb\frac{3}{2}-\nu_R\rb H+\frac{3\lambda H^3}{8\pi^2m_R^2}.
\end{equation}
For the two stochastic approximations - OD and second-order - the quantities in question are the first-excited eigenvalues, $\Lambda_1^{(OD)}$ and $\Lambda_1^{(SO)}$, of their respective spectral expansions. They are computed numerically, using the methods outlined in Sec. \ref{subsec:numerical_sol_OD} and \ref{subsec:numerical_solutions_stochastic}.

The comparisons will generally be made with respect to the second-order stochastic theory. We will show that it agrees with the two established approximations in the regime where they should, and that it disagrees in the regime where they breakdown. This gives us an indication that the second-order stochastic theory is behaving as it should though, as stated, further analysis is required to make this statement more concrete. 

\subsection{Some examples}

We will start the comparison by considering 2 examples. We will plot $\Lambda_1$ for all three approximations as a function of the coupling $\lambda$ for fixed $m_R^2/H^2$ and scale $M=a(t)H$ \footnote{For simplicity, I will drop the subscript `R' for the remainder of the chapter. Further, I will consider $M=a(t)H$ in all subsequent calculations unless otherwise stated.}. The first example will be for $m^2/H^2=0.1$ (Fig. \ref{fig:m20.1_match_v_OD_v_QFT}), where we expect the OD stochastic approach to work, while the second will be for $m^2/H^2=0.3$ (Fig. \ref{fig:m20.3_match_v_OD_v_QFT}), where we expect it to fail. In both examples, we expect perturbative QFT to hold for small $\lambda$ and fail as $\lambda$ increases, since it is in this regime that $\lambda\rightarrow m^4/H^4$. These two approximations are given by the yellow dotted (OD stochastic) and blue dashed (perturbative QFT). 

Similarly to perturbative QFT, the second-order stochastic theory should be expected to hold for small $\lambda$ and begin to fail as we increase $\lambda$ in both plots. This is a less severe failure as the breakdown is now for $\lambda\rightarrow m^2/H^2$. We will also consider how the choice of the $\sigma_{pp}^2$ noise amplitude affects our results. The red and green lines represent the choices $\sigma_{pp}^2=0$ and $\sigma_{pp}^2=\sigma_{pp}^{2(NLO)}$ respectively, as computed in Chapter \ref{ch:second-order_stochastic}. For both cases, we also consider the free (dot-dashed) and $\mathcal{O}(\lambda)$ (solid) stochastic parameters, given in Eq. (\ref{free_matched_stochastic_parameters}) and (\ref{one-loop_qp_stochastic_parameters})/(\ref{O(lambdaH2m2_stochastic_parameters_NLO_match}) respectively, so we can ascertain the effect that interacting stochastic parameters has on the results. Indeed, we will see that interactions are crucial so that physical results such as $\Lambda_1$ are not dependent on our choice of $\sigma_{pp}^2$.

\subsubsection{Example 1: $m^2/H^2=0.1$}

\begin{figure}[ht]
    \centering
    \includegraphics[width=150mm]{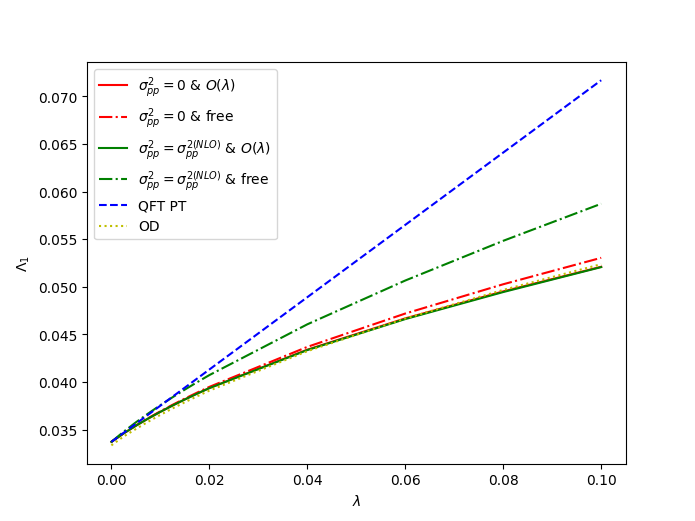}
    \caption{A plot of the first excited eigenvalue $\Lambda_1$ as a function of $\lambda$ for $m^2/H^2=0.1$ using perturbative QFT (blue, dashed), OD stochastic (yellow, dotted) and second-order stochastic approaches. Dot-dashed and solid lines indicate the second-order stochastic parameters are free and interacting respectively, with the noise choice $\sigma_{pp}^2=\sigma_{pp}^{2(NLO)}$ (green)  and $\sigma_{pp}^2=0$ (red).}
    \label{fig:m20.1_match_v_OD_v_QFT}
\end{figure}

The first example is for $m^2/H^2=0.1$. This is chosen because the mass is sufficiently small such that the OD stochastic approach will be valid beyond perturbative QFT. Consider Fig. \ref{fig:m20.1_match_v_OD_v_QFT}. One can see that for small $\lambda$, all three approximations converge. This is as expected because it is in this limit that all three approximations are valid and should thus agree. As we move towards higher $\lambda$, $\frac{\lambda H^4}{m^4}$ quickly becomes comparable to 1 and therefore the perturbative QFT eigenvalue diverges from the other curves. This divergence is large, which is no surprise because even at $\lambda=0.01$, $\frac{\lambda H^4}{m^4}=1$ so we are already out of the regime of validity for perturbative QFT.

For interacting stochastic parameters, one can see that there is near perfect agreement between the second-order and OD stochastic approaches. Further, it is clear that the choice of $\sigma_{pp}^2$ is inconsequential. However, for free stochastic parameters, this choice is important; one can see that for $\sigma_{pp}^2=0$, we have very good agreement with the OD stochastic approach (though not as good as when we use interacting stochastic parameters!) whereas the choice $\sigma_{pp}^2=\sigma_{pp}^{2(NLO)}$ has poor agreement as we move to high $\lambda$. This suggests the $\mathcal{O}\lb\frac{\lambda H^2}{m^2}\rb$ correction to the stochastic parameters is very important.

\begin{figure}[ht]
    \centering
    \includegraphics[width=150mm]{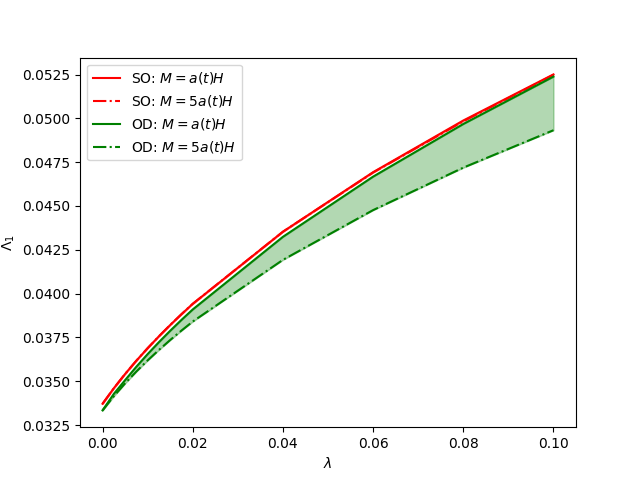}
    \caption{The first excited eigenvalue $\Lambda_1$ as a function of $\lambda$ for $m_R(a(t)H)^2/H^2=0.1$. The red and green lines show results from second-order and OD stochastic theories respectively. The solid and dot-dashed lines are for renormalisation scale choices $M=a(t)H$ and $M=5a(t)H$ respectively. The green shaded region indicates the size of the error that choosing the scale has on the OD stochastic approach. The equivalent red region is negligible because the second-order theory accounts for it via renormalisation.}
    \label{fig:m20.1_match_renorm}
\end{figure}

It is worth noting that this excellent agreement between the two stochastic approximations is due to the choice of renormalisation scale $M=a(t)H$. One can see from Eq. (\ref{stochastic_mass_O(lambdaH2m2)}) that the stochastic mass $m_S$ depends on the renomalisation scale as $\sim \ln\lb\frac{M^2}{a(t)^2H^2}\rb$, which vanishes for the choice $M=a(t)H$. Thus, it is not surprising that the second-order and OD stochastic approaches agree. However, if one were to choose the renormalisation scale differently, the agreement would not be so good. For example, if one chooses $M=5 a(t)H$, the renormalised mass parameter (\ref{renormalised_mass_dim_reg_deSitter}) will no longer equal 0.1; it will have some shift of $\mathcal{O}(\lambda)$. The second-order stochastic theory accounts for this shift via the $M$-dependence in the stochastic mass $m_S$ parameter (\ref{stochastic_mass_O(lambdaH2m2)}), whereas the OD theory does not because it doesn't incorporate any UV renormalisation. Thus, the two will be different for such a choice. Fig. \ref{fig:m20.1_match_renorm} shows the effect of a different choice up to $M=5a(t)H$. One can see that there is very little change to $\Lambda_1$ for the second-order theory and that the change is much larger for the OD case.

\subsubsection{Example 2: $m^2/H^2=0.3$}

\begin{figure}[ht]
    \centering
    \includegraphics[width=150mm]{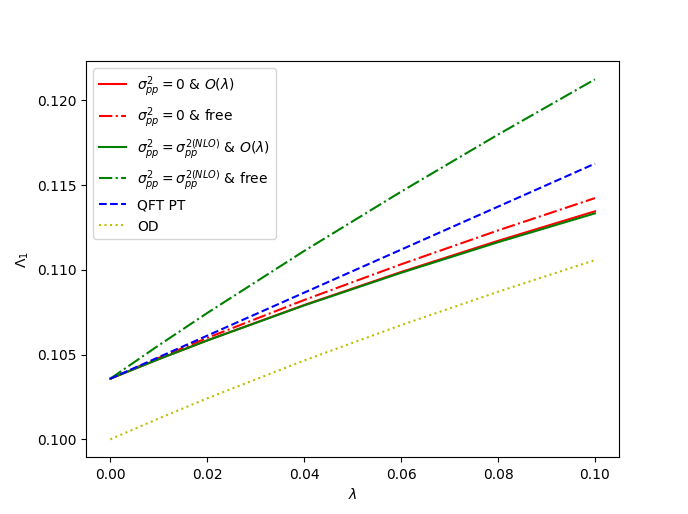}
    \caption{A plot of the first excited eigenvalue $\Lambda_1$ as a function of $\lambda$ for $m^2/H^2=0.3$ using perturbative QFT (blue, dashed), OD stochastic (yellow, dotted) and second-order stochastic approaches. Dot-dashed and solid lines indicate the second-order stochastic parameters are free and interacting respectively, with the noise choice $\sigma_{pp}^2=\sigma_{pp}^{2(NLO)}$ (green)  and $\sigma_{pp}^2=0$ (red).}
    \label{fig:m20.3_match_v_OD_v_QFT} 
\end{figure}

For our second example, we will consider a larger mass $m^2/H^2=0.3$ such that the OD stochastic results become less reliable. Consider Fig. \ref{fig:m20.3_match_v_OD_v_QFT}. One can see that for small $\lambda$ the second-order stochastic and perturbative QFT results agree well but as one increases $\lambda$ the two results diverge from each other. This is once again because increasing $\lambda$ results in an increase of $\frac{\lambda H^4}{m^4}$. Conversely, even at small $\lambda$, the OD stochastic approach gives a different value for the eigenvalue compared to the other two approaches, suggesting that even at $m^2/H^2=0.3$ we are at too high a mass for the OD stochastic approach to be trustworthy.

Note also that a similar argument to the previous example holds for the choice of $\sigma_{pp}^2$. For free stochastic parameters, there is a dependence on this choice whereas when $\mathcal{O}\lb\frac{\lambda H^2}{m^2}\rb$ corrections are included, this dependence becomes much weaker. This underlines the importance of including these corrections.

\subsection{Direct comparisons of the approximations}
\label{subsec:direct_comparisons}

From these two examples, we have an idea of how these three approximations are related. Specifically, we have seen that the second-order stochastic theory behaves as expected in so far as there is agreement and disagreement in the regimes where one would expect to find them. To make this more quantitative, we will now consider more carefully the difference between the three approximations by taking the relative difference between eigenvalues. These will be plotted with an appropriate $x$-axis as follows:
\begin{subequations}
    \label{comparison_plots}
    \begin{align}
    \text{SO v OD:}&\qquad \frac{\Lambda_1^{(SO)}-\Lambda_1^{(OD)}}{\Lambda_1^{(SO)}}\qquad \text{v} \qquad m^2/H^2,\\
    \text{QFT v SO:}&\qquad \frac{\Lambda_1^{(QFT)}-\Lambda_1^{(SO)}}{\Lambda_1^{(SO)}}\qquad \text{v} \qquad \lambda H^4/m^4,\\
    \text{QFT v OD:}&\qquad \frac{\Lambda_1^{(QFT)}-\Lambda_1^{(OD)}}{\Lambda_1^{(OD)}}\qquad \text{v} \qquad \lambda H^4/m^4.
    \end{align}
\end{subequations}
The choice of the $x$-axis is made so that the established approximations - OD stochastic and perturbative QFT - breakdown in the large limit. Note also that, following from our above examples, we will just focus on the case where the stochastic parameters are given by Eq. (\ref{one-loop_qp_stochastic_parameters}). 

\subsubsection{OD v second-order stochastic approaches}
\label{subsubsec:OD_v_matching_comparison}

\begin{figure}[ht]
    \centering
    \includegraphics[width=150mm]{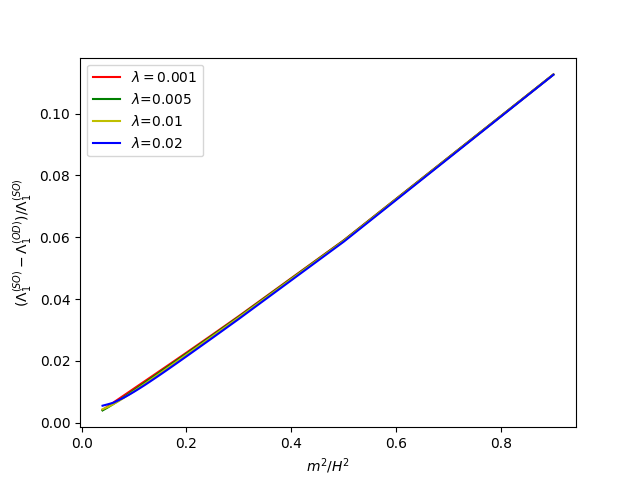}
    \caption{$\frac{\Lambda_1^{(SO)}-\Lambda_1^{(OD)}}{\Lambda_1^{(SO)}}$ against $m^2/H^2$ for $\lambda=0.001$ (red), $0.005$ (green), $0.01$ (yellow), $0.02$ (blue).}
    \label{fig:eval_v_m2_OD_v_match_diff}
\end{figure}

First, we will consider the difference between the second-order and OD stochastic results. We take the relative difference between the second-order and OD eigenvalues $\frac{\Lambda_1^{(SO)}-\Lambda_1^{(OD)}}{\Lambda_1^{(SO)}}$ as a function of $m^2/H^2$. The use of this scale is so that as one increases $m^2/H^2$ the OD stochastic approach becomes less reliable so we expect to see a difference between the two results. In Fig. \ref{fig:eval_v_m2_OD_v_match_diff}, we have plotted the relative difference for different values of $\lambda$ for the case when $\sigma_{pp}^2=0$. We immediately see that all the curves follow the same linearly increasing behaviour. As we increase $m^2/H^2$ to the right of the figure, we see that the relative difference increases as expected.

\subsubsection{Perturbative QFT v second-order stochastic approaches}
\label{subsubsec:QFT_v_matching_comparison}

We will now do the same analysis with a comparison of the second-order stochastic and perturbative QFT eigenvalues where we plot the eigenvalue difference $\frac{\Lambda_1^{(QFT)}-\Lambda_1^{(SO)}}{\Lambda_1^{(SO)}}$ for several values of $\lambda$ (solid lines in Fig. \ref{fig:eval_v_m2_QFT_v_OD_diff}). The difference is that we will now use $\frac{\lambda H^4}{m^4}$ on the $x$-axis since this is the parameter where we will see a breakdown of perturbative QFT. We see the expected behaviour; for small $\frac{\lambda H^4}{m^4}$, all the curves converge to 0. As one increases $\frac{\lambda H^4}{m^4}$, we see an increasing relative difference between the two eigenvalues due to a breakdown of the perturbative QFT.

\subsubsection{OD stochastic v perturbative QFT approaches}
\label{subsubsec:OD_v_QFT_comparison}

\begin{figure}[ht]
    \centering
    \includegraphics[width=150mm]{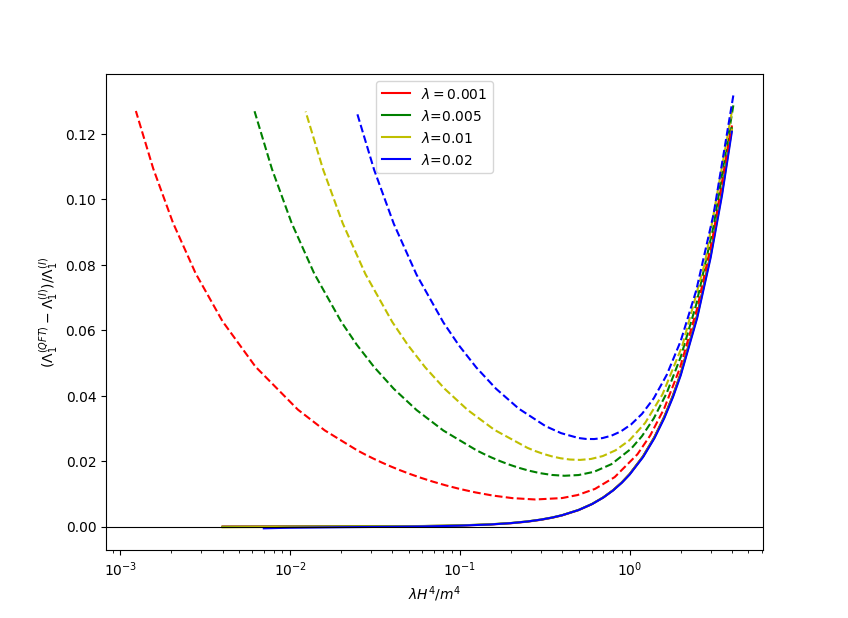}
    \caption{$\frac{\Lambda_1^{(QFT)}-\Lambda_1^{(I)}}{\Lambda_1^{(I)}}$, where $I\in\{\text{SO},\text{OD}\}$ as a function of $\frac{\lambda H^4}{m^4}$ for $\lambda=0.001$ (red), $\lambda=0.005$ (green), $\lambda=0.01$ (yellow) and $\lambda=0.02$ (blue). The solid lines show the relative difference between the SO and QFT eigenvalues, which lie on directly on top of each other for most $\frac{\lambda H^4}{m^4}$ values, while the dashed lines show the relative difference between the OD and QFT eigenvalues.}
    \label{fig:eval_v_m2_QFT_v_OD_diff}
\end{figure}

The final comparison we will make is between perturbative QFT and the OD stochastic approach. The dotted lines in Fig. \ref{fig:eval_v_m2_QFT_v_OD_diff} plots the eigenvalue difference between the two approximations, $\frac{\Lambda_1^{(QFT)}-\Lambda_1^{(OD)}}{\Lambda_1^{(OD)}}$ as a function of $\frac{\lambda H^4}{m^4}$ for the four $\lambda$ values. 

On the right hand side of the plot, we can see that the deviation from the QFT result follows the same pattern as that of the second-order stochastic approach. This is unsurprising because, as we move to higher $\frac{\lambda H^4}{m^4}$, we are moving to smaller $m^2/H^2$, the limit where the OD and second-order stochastic approaches agree. In this regime, perturbative QFT is breaking down so we see a high relative difference between it and the stochastic approaches. As we move to smaller values of $\frac{\lambda H^4}{m^4}$, the dotted curves dip to some minimum before turning upward. As one moves left, there is an increasing relative difference between the two; this is now due to the breakdown of the OD stochastic approach since we are getting to high $m^2/H^2$ values. One can see that the second-order stochastic approach continues towards a zero relative difference, indicating the region where the OD approach breaks down but the other two approximations are still valid.

\section{Concluding remarks}
\label{sec:comparison_conclusion}

Chapters \ref{ch:qft_dS}-\ref{ch:comparison_models} are the core of my thesis, with the development of the second-order stochastic effective theory forming the meat of my PhD work. I have outlined three approximations to scalar QFT in de Sitter spacetime, all of which can be used to compute the long-distance behaviour of scalar correlation functions. The two established approximations - the OD stochastic approach and perturbative QFT - are included here largely so that we can compare the novel second-order stochastic effective theory with concrete, well-studied results. We have shown in the previous section that, within the relevant regimes of validity, the second-order stochastic theory agrees with the other approximations. This is a crucial first check; if they disagreed, questions would be raised about the validity of this approach. A further tick in the box is that the second-order theory appears to disagree with the results from the other approximations when they are outwith their regimes of validity. Since we suspect the second-order theory has scope beyond the established approximations, this indicates novel results. We would like to be more thorough and consider other established methods that extend to different regimes in the parameter space, such as the $1/N$ expansion\cite{Beneke:2013,Nacir:2016} or Monte-Carlo simulations\cite{montvay:1994}, but this goes beyond the scope of this thesis. For now, we can content ourselves with the knowledge that the second-order stochastic effective theory is behaving well and producing sensible results.
\chapter{Conclusion}

\label{ch:conclusions}

\section{Future Work}

We have done a huge amount of work to get the second-order stochastic theory to be a reliable and useful model for scalar fields in de Sitter. The final questions I will address in this thesis are: can this be improved further and where is this applicable? The short answer to these questions is `yes' and `inflationary cosmology'! I will now flesh out these statements.

Thus far, we have considered two other approximations for comparison with the second-order stochastic theory: perturbative QFT and overdamped stochastic theory. An immediate question follows: are there other approximations that can be used to test the second-order stochastic theory further? At present, we are assuming that the matching procedure introduced to relate the stochastic parameters with their QFT counterparts holds firm beyond the regime of perturbative QFT. There is no reason to indicate that it won't at this stage, and all our results thus far suggest that it will, but it would be good to check this against other approximations. Some examples would be the $1/N$ approximation\cite{Beneke:2013,Nacir:2016} or, turning to more numerical methods, Monte Carlo simulations\cite{montvay:1994}. From the other angle, one could ask whether the stochastic formalism set out can be extended to incorporate other effects. The work of Cohen and Green et. al. \cite{Cohen:2020,Cohen:2021,Green:2022,Cohen:2023} considers multiplicative and coloured noise, which could be useful in relaxing the condition of the cut-off method that the subhorizon modes are free. Our approach incorporates UV effects but it is not clear how they enter into the long-distance behaviour; we simply say that it must if we are to have an effective stochastic theory of QFT. A deeper comparison of these two stochastic procedures could give an interesting insight into what UV behaviour can be captured by such a theory.

Consideration of these models could bring further benefits. At present, the limitation of the second-order theory to $\lambda\ll m^2/H^2$ only enters into the calculation of the stochastic parameters; beyond that, the stochastic equations can be solved non-perturbatively. In turn, these parameters are only perturbative because the QFT approximation we are using for their computation is perturbative. It is therefore possible that other approximations to QFT could allow us to update our stochastic parameters, thus extending the regime of validity in which the second-order stochastic theory is valid. An alternative method that would meet these requirements is the use of Monte Carlo simulations, which can be used to perform non-perturbative calculations but is computationally intensive. If one could use it to compute the stochatic parameters, then one could harness its power through the far simpler stochastic theory. One drawback, of course, is that the stochastic parameters would now have to be numerical in nature, but this is not a problem, especially given the non-perturbative computations of the stochastic equations are also numerical: it would simply mean we couldn't write nice, analytic expressions!

These extensions would still result in a reliance of the second-order stochastic theory on other approximations of QFT in order to obtain physical results. Ideally, we would like a method whereby we can derive the second-order stochastic equations from the underlying microscopic picture, much as was done for the overdamped theory. This not only would give us a more reliable backstory for the theory, but would also generalise it so that we could consider other types of potentials, making it far more versatile. This seems very hard to achieve as it appears one would need a much better understanding of the underlying QFT than we presently have. One area where we could make some headway would be to consider how the stochastic theory arises out of an open quantum system (OQS) approach to QFT in de Sitter\cite{Burgess:2014,Kaplanek:2020,Burgess:2022}. In an OQS, one separates the system from the environment, integrating out the environment modes such that one is left with an effective theory of the system. This is very similar to how the stochastic approach is treated in most instances in the literature and indeed how the overdamped theory was introduced in this thesis. The main difference at present is that, in OQS, the separation between system and environment is time-independent, unlike the cutoff procedure used for the stochastic approach. This is not \textit{a priori} a requirement of OQS; it is more a case that it has not yet been discussed in the literature. 

A further benefit of considering the stochastic theory through the OQS lens would be to pinpoint the key behaviours that the stochastic theory captures. Since it is ultimately a semi-classical approximation, there is a limit to how much quantum behaviour it will actually capture. We have seen one example of this already; the complex phase that exists in the analytic continuation from timelike to spacelike 2-point functions doesn't feature in the stochastic results. These questions coincide with more generic questions about the ``quantumness'' of the early Universe and how such a deeply-quantum, microscopic system became the classical Universe that we observe today\footnote{For a deeper discussion of this, see Ref. \cite{Colas:2022,Martin:2023} and the references within}.

This brings us to our final few remarks on the future, which is to bring this thesis back to the original motivation: can the second-order stochastic theory be used to compute inflationary observables? The answer, we sincerely hope, is yes! I have outlined the computation of scalar correlation functions, which are at the core of such observables, so this tool is now readily available for use in a `real' cosmological setting. One immediate extension is to consider observational bounds on spectator dark matter through the computation of isocurvature perturbations, as was done in Ref. \cite{Markkanen:2018_dm,Jukko:2021}. Additionally, it has recently been shown, through the use of the second-order stochastic theory, that spectator fields can play a pivotal role in primordial black hole formation without ruining cosmological models\cite{Cable:2023_PBH}. These are just two examples of many areas where the second-order stochastic theory could be useful. Moreover, if one were to extend this theory to consider a time-dependent $H$ (which is highly non-trivial!), then it would also be an applicable model for the inflaton as well.

\section{Summary of Thesis Achievements}

There has been a lot of material covered in this thesis and so, as it draws to a close, it is worth discussing the key results and overarching achievements that lie within. On that note, the aim of this thesis was:
\begin{equation*}
    \keybox{
    \begin{split}
    \text{To introduce the}& \text{ second-order stochastic theory as an effective theory describing the long}\\&\text{-distance behaviour of scalar fields in de Sitter spacetime.} 
    \end{split}
    }
\end{equation*}
To do this, we have taken inspiration from two other established approximations of scalar QFT in de Sitter: perturbation theory of QFT and the overdamped stochastic approach.

In Chapter \ref{ch:qft_dS}, I introduced scalar QFT in de Sitter from the ground up, beginning with the geometry of de Sitter spacetime and continuing to the quantisation of scalar fields and the introduction of correlation functions. All of this is well-established in the literature and forms the foundation for the work to follow. Indeed, much of this work, and the work of many, would be obsolete if it were straightforward to compute these correlation functions in this general framework. Fortunately for me and my job security, they are not and so we must turn to approximations. It is with this in mind that I introduced to this thesis the first of three: perturbative QFT. Considering a massive scalar field with a quartic self-interaction $\lambda$, I performed the perturbative expansion to compute the scalar 2-point and connected 4-point functions to $\mathcal{O}(\lambda)$. The bulk of this section was a literature review, compiling information from many sources in a novel way. However, to my knowledge at the time of writing, there are two key results which don't appear explicitly in the literature:
\begin{itemize}
    \item An explicit computation and subsequent expression for the renormalised mass (\ref{renormalised_mass_dim_reg_deSitter}) and the correction one must make to the mass (\ref{renormalised_mass_correction}) in the $\overline{\text{MS}}$ scheme in order to write a finite expression for the 2-point function to $\mathcal{O}(\lambda)$. 
    \item An expression for the IR behaviour of the connected 4-point function beyond the limit $m\ll H$ (\ref{spacelike_quantum_4-pt_func}) and a discussion of additional contributing terms.
\end{itemize}

In Chapter \ref{ch:od_stochastic}, I introduced the second of our three approximations - the overdamped stochastic approach - which is the inspiration behind the extension to second-order. I outlined the physical motivation behind the stochastic approach and, starting from the underlying QFT, derived the stochastic equations. I then introduced a spectral expansion method to solve these equations in order to compute the overdamped correlation functions perturbatively, and outlined the overshoot/undershoot numerical method to do non-perturbative calculations. Again, this is predominantly a literature review, with results quoted in a novel way. One result of particular note is:
\begin{itemize}
    \item A general calculation of overdamped 4-point functions and an explicit expression for the connected timelike and spacelike field 4-point functions, Eq. (\ref{connected_OD_4pt_function}) and (\ref{spacelike_connected_OD_4pt_function}) respectively.
\end{itemize}

Chapter \ref{ch:second-order_stochastic} is all new, with the bulk of it making up my three papers [K1], [K2] and [K3]. This is where I introduce the second-order stochastic theory and promote it to an IR effective theory of scalar fields in de Sitter spacetime. The key results are as follows:
\begin{itemize}
    \item The introduction of second-order stochastic equations (\ref{2d_stochastic_eq}) and the computation of stochastic correlation functions via a spectral expansion. 
    \item The computation of the stochastic parameters for free fields (\ref{free_matched_stochastic_parameters}), promoting the second-order stochastic approach to an effective theory for free scalar fields in de Sitter. This was achieved by choosing the stochastic parameters in such a way that the free field stochastic and quantum 2-point functions agreed. 
    \item The introduction of a perturbative calculation to obtain the stochastic correlation functions to leading order in a small $\lambda_S$ expansion, and hence the computation of the stochastic parameters to $\mathcal{O}(\lambda)$ (\ref{stochastic_mass_O(lambdaH2m2)_again})-(\ref{O(lambdaH2m2)_phi-pi_noise_matrix_again}). This was achieved by choosing the parameters in such a way that both the $\mathcal{O}(\lambda)$ stochastic and quantum 2- and 4-point functions agreed. This promoted the second-order stochastic approach to an effective theory of quartic self-interacting scalar fields in de Sitter.
    \item Finally, the outline of a numerical method for solving the stochastic equations non-perturbatively.
\end{itemize}

The final chapter of the bulk thesis, Chapter \ref{ch:comparison_models}, is dedicated to a rigorous comparison between the three approximations, in order to show that the second-order stochastic theory gives sensible results. The essential point is that it does: it agrees with the other approximations when it should, and disagrees where we would expect the others to break down.

Thus, to draw the curtain on this thesis, I state once more the key point: we have developed a second-order stochastic theory that can be used to calculate the IR behaviour of correlation functions for a light scalar field with a quartic self-coupling in de Sitter spacetime.

\appendix

\chapter{QFT Computations}

\section{The (unsuccessful) numerical calculation of timelike correlation functions at $\mathcal{O}(\lambda)$}
\label{app:numerical_calculation_correlators}

In this appendix, I will outline the numerical method we used to compute timelike quantum 2-point correlation functions. Unfortunately, there are some problems with this method in the form of divergences, which are difficult to deal with numerically and thus our calculations aren't particularly reliable, as I will discuss. Results in this section aren't used in the bulk text; specifically in Sec. \ref{subsec:2-pt_function_one-loop} and \ref{subsec:4-pt_functions}, we discuss alternative methods to compute quantum 2-pt and 4-pt functions that can be used to compare with our stochastic theories. I include this appendix so that the reader has a better understanding of some problems facing a ``brute force'' approach to computing quantum correlators.

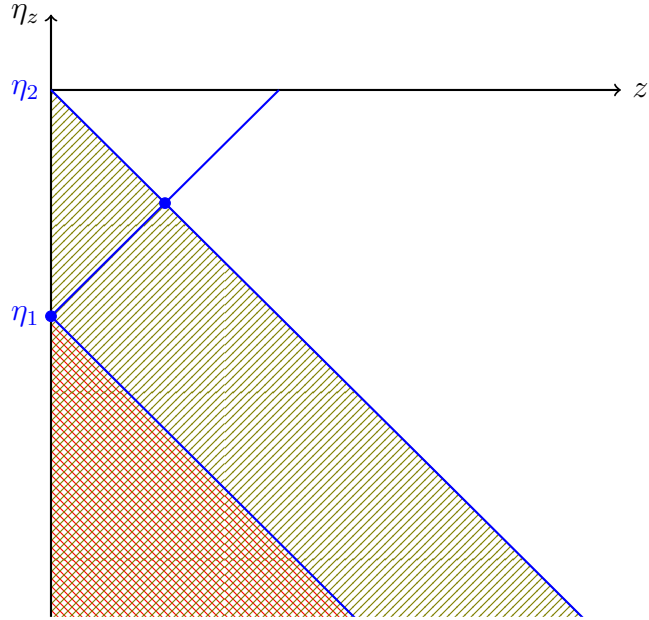
\begin{figure}[ht]
    \centering
    \begin{tikzpicture}
        \filldraw[pattern=north west lines,pattern color = red] (4,0) -- (0,4) -- (0,0) -- cycle;
        \filldraw[pattern=north east lines,pattern color = olive] (0,0) -- (7,0) -- (0,7) -- cycle;
        \draw[black,thick,->] (0,0) -- (0,8) node[left]{$\eta_z$};
        \draw[black,thick,->] (0,7) -- (7.5,7) node[right]{$z$};
        \draw[blue,thick] (7,0) -- (0,7) node[left]{$\eta_2$};
        \draw[blue,thick] (4,0) -- (0,4) node[left]{$\eta_1$};
        \draw[blue,thick] (0,4) -- (3,7);
        \draw[white,thick] (-1,0) -- (8,0);
        \filldraw[blue] (1.5,5.5) circle (2pt);
        \filldraw[blue] (0,4) circle (2pt);
    \end{tikzpicture}
    \caption{The light cone structure of the 2-point functions. Blue lines indicate the regions where $y_{iz}=0$ for $i\in\{1,2\}$. The red, right-hashed and green, left-hashed regions indicate the limits over which the integrals on the first and second lines of Eq. (\ref{O(lambda)_2-pt_func_integral_A&B_etatildez}) respectively are performed.}
    \label{fig:2-pt_light_cones}
\end{figure}

The connected timelike 2-point function at $\mathcal{O}(\lambda)$ is given by the second line of Eq. (\ref{one-loop_UVdiv_2-pt_function}); explicitly
\begin{equation}
    \label{O(lambda)_2-pt_func_integral}
    \begin{split}
    \bra{0}T\hat{\phi}(x_1)\hat{\phi}(x_2)\ket{0}^{(1)}=3i\lambda\expval{\hat{\phi}^2}\int d^4z a(\eta_z)^4&\Bigg[ i\Delta^F(z,x_1)i\Delta^F(z,x_2)\\&-\Delta^+(z,x_1)\Delta^+(z,x_2)\Bigg],
    \end{split}
\end{equation}
where the UV-finite field variance $\expval{\hat{\phi}^2}$ is given in Eq. (\ref{finite_field_variance_dim_reg_deSitter_ana}). We will take $x_i=(\eta_i,\mathbf{0})$ and $z=(\eta_z,\mathbf{z})$, and we switch to conformal time for convenience. We take $\eta_2\gg\eta_1$ to obtain the IR behaviour of the correlator. Splitting the Feynman and Wightman functions using the functions $A(y)$ and $B(y)$, as introduced in Eq. (\ref{A&B_2-pt_amplitudes}), the integral becomes
\begin{equation}
    \label{O(lambda)_2-pt_func_integral_A&B}
    \begin{split}
        \bra{0}T\hat{\phi}(x_1)\hat{\phi}(x_2)\ket{0}^{(1)}=&12\pi\lambda\expval{\hat{\phi}}^2\int_{-\infty}^0 d\eta_z\frac{1}{(H\eta_z)^4}\int_0^{\infty}dz z^2\\&\Bigg[\lb A(y_{2z})B(y_{1z})+iB(y_{1z})B(y_{2z})\rb\theta(\eta_1-\eta_z)\theta(y_{1z})\\&+A(y_{1z})B(y_{2z})\theta(\eta_2-\eta_z)\theta(y_{2z})\Bigg],
    \end{split}
\end{equation}
where we have used spherical symmetry to write $\int d^4z=4\pi\int d\eta_z\int dz z^2$, where we will use $z=\abs{\mathbf{z}}$ henceforth. As will be explained in a moment, it is convenient to define the coordinate $\Tilde{\eta}_z=\eta_z+z$, such that the integral (\ref{O(lambda)_2-pt_func_integral_A&B}) becomes
\begin{equation}
    \label{O(lambda)_2-pt_func_integral_A&B_etatildez}
    \begin{split}
        \bra{0}T\hat{\phi}(x_1)\hat{\phi}(x_2)\ket{0}^{(1)}=&\frac{12\pi\lambda\expval{\hat{\phi}}^2}{H^4}\\&\times\Bigg[\int_{-\infty}^{\eta_1} d\Tilde{\eta}_z\int_0^{\infty}dz \frac{z^2}{(\Tilde{\eta}_z-z)^4}\lb A(y_{2z})B(y_{1z})+iB(y_{1z})B(y_{2z})\rb
        \\&+\int_{-\infty}^{\eta_2} d\Tilde{\eta}_z\int_0^{\infty}dz \frac{z^2}{(\Tilde{\eta}_z-z)^4}A(y_{1z})B(y_{2z})\Bigg],
    \end{split}
\end{equation}
where we have applied the $\theta$-functions, as per Eq. (\ref{heaviside_function}), and we use $y_{iz}=y(x_i,z)$. Naively, one could now attempt to compute this integral numerically; however, the integrand contains poles. To see this, consider the de Sitter invariant (\ref{de_Sitter_invariant}) in terms of the coordinates $(\Tilde{\eta}_z,z)$:
\begin{equation}
    \label{de_sitter_invariant_tildeeta}
    y(\eta_i,\Tilde{\eta}_z,z)=\frac{(\eta_i-\Tilde{\eta}_z+z)^2-z^2}{2\eta_i(\Tilde{\eta}_z-z)}.
\end{equation}
The hypergeometric function appearing in $A(y_{iz})$ (c.f. Eq. (\ref{A&B_2-pt_amplitudes}) is divergent for $y_{iz}=0$\footnote{Note that, similarly, $B(y_{iz})$ is also divergent when $y_{iz}=-2$ but, thanks to the $\theta$-functions in the integrand (\ref{O(lambda)_2-pt_func_integral_A&B}), these never arise in our integral.}. Thus, there are poles at
\begin{subequations}
    \label{eta_poles}
    \begin{align}
        \Tilde{\eta}_z&=\eta_i,\\
        \Tilde{\eta_z}&=\eta_i+2z.
    \end{align}
\end{subequations}
These correspond to the boundaries of the future and past light cone boundaries from $\eta_i$, as depicted in Fig. \ref{fig:2-pt_light_cones}. Note that we have chosen the coordinate $\Tilde{\eta}_z$ such that it runs parallel to the past light cone boundaries. 

\begin{figure}
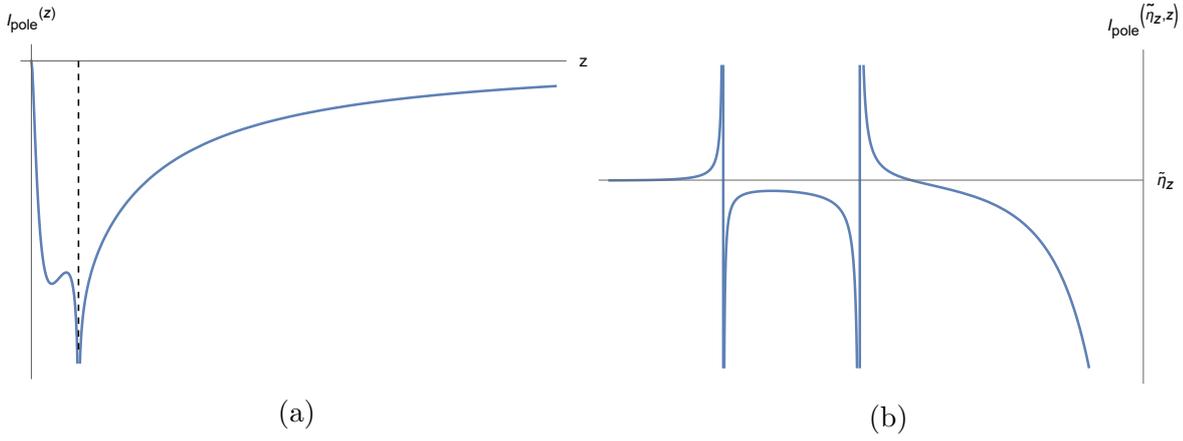

    \begin{subfigure}[ht]{0.45\textwidth}
        \includegraphics[width=\textwidth]{appendices/2-pt_func_poles_z-integral.jpg}
        \caption{}
        \label{fig:2-pt_func_poles_z-integral}
    \end{subfigure}
    \begin{subfigure}[ht]{0.45\textwidth}
        \includegraphics[width=\textwidth]{appendices/2-pt_func_poles.jpg}
        \caption{}
        \label{fig:2-pt_func_poles_integrand}
    \end{subfigure}
    \caption{A plot outlining the poles in the (a) integrand $I_{pole}^{(z)}$ as a function of $z$ prior to the $z$ integral, and (b) integrand $I_{pole}^{(\Tilde{\eta}_z,z)}$ as a function of $\Tilde{\eta}_z$ prior to the $\Tilde{\eta}_z$ integral. The vertical, dashed line in (a) indicates the pole at $z=\frac{\eta_2-\eta_1}{2}$, while the left and right poles in (b) are where $\Tilde{\eta}=\eta_1$ and $\Tilde{\eta}_z=\eta_1+2z$ respectively.}
    \label{fig:2-pt_func_poles}
\end{figure}

The divergences don't feature in the integrals of the first line of Eq. (\ref{O(lambda)_2-pt_func_integral_A&B_etatildez}) because they only integrate to $\eta_1$, and otherwise never see a light cone boundary. The problems arise when we consider the second line; the integral in question is
\begin{equation}
    \label{divergent_integral}
    I_{pole}=\int_{-\infty}^{\eta_2} d\Tilde{\eta}_z\int_0^{\infty}dz \frac{z^2}{(\Tilde{\eta}_z-z)^4}A(y_{1z})B(y_{2z}).
\end{equation}
We will first compute the $\Tilde{\eta}_z$ integral, dealing with the poles that arise here, before turning to the $z$-integral. For convenience, we define 
\begin{subequations}
    \label{divergent_integrals}
    \begin{align}
    \label{divergent_integrand}
        I_{pole}^{(\Tilde{\eta}_z,z)}&=\frac{z^2}{(\Tilde{\eta}_z-z)^4}A(y_{1z})B(y_{2z}),\\
    \label{divergent_z-integrand}    
        I_{pole}^{(z)}&=\int_{-\infty}^{\eta_2} d\Tilde{\eta}_z \frac{z^2}{(\Tilde{\eta}_z-z)^4}A(y_{1z})B(y_{2z}).
    \end{align}
\end{subequations}
An example of the pole structure of these two quantities is given in Fig. \ref{fig:2-pt_func_poles}\footnote{Note that numerical values have been used to obtain these plots. The specific values are not important; the position (as a function of $(\Tilde{\eta_z},z)$) and behaviours of the poles will be the same regardless of the choice of numbers.}.

One can see from Fig. \ref{fig:2-pt_func_poles_integrand} that the structure around the poles is opposing; on the left side it diverges to $+\infty$ whereas on the right it diverges to $-\infty$. This suggests that the divergent behaviour should cancel. To see this, we separate our integral as follows:
\begin{equation}
    \label{eta_integral_split}
    \begin{split}
    I_{pole}^{(z)}=&\int_{-\infty}^{\eta_1-\epsilon}d\Tilde{\eta}_z I_{pole}^{(\Tilde{\eta}_z,z)}
    +\int_{\eta_1-\epsilon}^{\eta_1+\epsilon}d\epsilon' I_{pole}^{(\Tilde{\eta}_z,z)}\eval_{\Tilde{\eta}_z\simeq\eta_1+\epsilon'}
    +\int_{\eta_1+\epsilon}^{\eta_1+2z-\epsilon}d\Tilde{\eta}_z I_{pole}^{(\Tilde{\eta}_z,z)}
    \\&+\int_{\eta_1+2z-\epsilon}^{\eta_1+2z+\epsilon}d\epsilon'    I_{pole}^{(\Tilde{\eta}_z,z)}\eval_{\Tilde{\eta}_z\simeq\eta_1+2z+\epsilon'}
    +\int_{\eta_1+2z+\epsilon}^{\eta_2}d\Tilde{\eta}_z I_{pole}^{(\Tilde{\eta}_z,z)},
    \end{split}
\end{equation}
where $\epsilon$ is some small, positive number used to offset the numerical integrals from the poles\footnote{When $\eta_1+2z>\eta_2$, the second pole no longer contributes and we just have to deal with the one pole.}. The first, third and fifth integrals can now be computed numerically without any trouble. Meanwhile, the second and fourth integrals can be computed analytically by making a small $\epsilon$ expansion. They both work in the same way. Using the notation $\eta_*$ to indicate one of the two poles (\ref{eta_poles}), the integrand (\ref{divergent_integrand} can be computed using Eq. (\ref{A&B_2-pt_amplitudes}) to give
\begin{subequations}
    \label{epsilon-expanded_integrand}
    \begin{align}
    I_{pole}^{(\Tilde{\eta}_z,z)}\eval_{\Tilde{\eta}_z=\eta_*+\epsilon}&\simeq a+\frac{b}{\epsilon}+c\ln\epsilon+\mathcal{O}(\epsilon),\\
    I_{pole}^{(\Tilde{\eta}_z,z)}\eval_{\Tilde{\eta}_z=\eta_*-\epsilon}&\simeq a-\frac{b}{\epsilon}+c\ln\epsilon+\mathcal{O}(\epsilon),
    \end{align}
\end{subequations}
where $a$, $b$ and $c$ are some mass-dependent quantities that are known. Then we can write
\begin{equation}
    \label{cancel_divergences}
    \begin{split}
    \int_{\eta_*-\epsilon}^{\eta_*+\epsilon}d\epsilon' I_{pole}^{(\Tilde{\eta}_z,z)}\eval_{\Tilde{\eta}_z\simeq\eta_*+\epsilon'}&=\int_0^\epsilon d\epsilon'\lb I_{pole}^{(\Tilde{\eta}_z,z)}\eval_{\Tilde{\eta}_z\simeq\eta_*+\epsilon'}+I_{pole}^{(\Tilde{\eta}_z,z)}\eval_{\Tilde{\eta}_z\simeq\eta_*-\epsilon'}\rb\\
    &=\int_{\eta_*-\epsilon}^{\eta_*+\epsilon}d\epsilon'\lb2a+2c\ln\epsilon+\mathcal{O}(\epsilon)\rb\\
    &=4\epsilon(a+c\ln\epsilon)+\mathcal{O}(\epsilon^2).
    \end{split}
\end{equation}
One can see from this that the contribution from the pole is small, vanishing as $\epsilon\rightarrow0$. Thus, we can compute the $\Tilde{\eta}_z$ integral without too much difficulty.

There are two limits that arise in this integral that could cause problems:
\begin{enumerate}[(i)]
    \item $z\rightarrow0$ (left dot in Fig. \ref{fig:2-pt_func_poles}).
    \item $\eta_1+2z\rightarrow\eta_2$ (right dot in Fig. \ref{fig:2-pt_func_poles}).
\end{enumerate}

The first limit (i) is where the two poles merge into one. This is not problematic because we can always make $\epsilon$ small enough such that they never exactly equal. When performing the $z$-integral, we then just need to cut at some value that is $\gtrsim\epsilon$, rather than exactly 0. As we know the pole contribution isn't large, it will be a negligible effect we miss.

The second limit (ii) does cause problems. In this limit, the method of cancelling the divergences, as outlined in Eq. (\ref{cancel_divergences}), fails because the upper bound, $\Tilde{\eta_z}=\eta_1+2z+\epsilon$, will go beyond the limits of our integration. We could make a similar argument to (i) where we just reduce the range of integration to $\eta_2-\epsilon$ and go from there; however, it is no longer clear that the pole will have a negligible contribution at this point. Indeed, one can see from Fig. \ref{fig:2-pt_func_poles_z-integral} that there exists a pole at $z=\frac{\eta_2-\eta_1}{2}$, which precisely corresponds to this limit. The divergent behaviours will not cancel in this pole, as they did for those in the $\Tilde{\eta}_z$ integral. It is not clear to us how to deal with this problem.

\begin{figure}[ht]
    \centering
    \includegraphics[width=150mm]{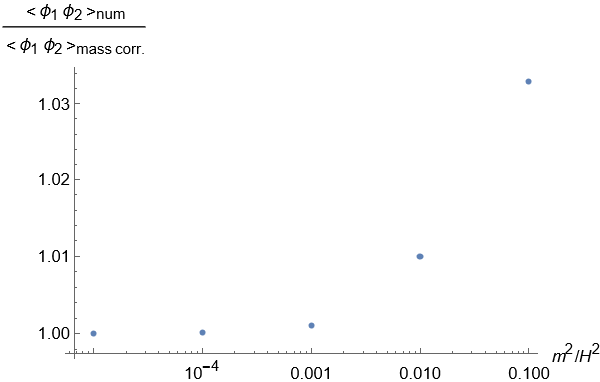}
    \caption{The ratio of the 2-pt functions computed using the numerical method outlined here and the mass redefinition method of Sec. \ref{subsec:2-pt_function_one-loop} as a function of $m^2/H^2$ for $\eta_1=-20$ and $\eta_2=2$.}
    \label{fig:numerical_v_mass-redef_2-pt_func}
\end{figure}

We have attempted to just remove the problematic pole by hand, to give us an idea of how important this contribution is. The 2-pt function (\ref{O(lambda)_2-pt_func_integral_A&B_etatildez}) can then be computed numerically and compared with the results of Sec. \ref{subsec:2-pt_function_one-loop}, where we used the mass redefinition. The results are given in Fig. \ref{fig:numerical_v_mass-redef_2-pt_func}.

One can see that for light fields, there is very good agreement between the two methods. As one increases the mass, the agreement becomes significantly worse. This suggests that the contribution from the problematic pole becomes significant for heavier fields.

Note that attempts were also made to perform this computation for the connected 4-point function. The same types of issues arise in this case; indeed, they are more severe due to the more complex pole structure evident in 4-point functions.

\section{Renormalisation in Minkowski spacetime}
\label{app:renormalisation_minkowski}

In this appendix, I will consider the UV renormalisation of the scalar 2-point function in Minkowski, for comparison with similar results in de Sitter (c.f. Sec. \ref{subsec:2-pt_function_one-loop}). Our equation of motion is
\begin{equation}
    \label{minkowski_scalar_eom}
    \Ddot{\phi}-\nabla^2\phi+m_B^2\phi+\lambda\phi^3=0,
\end{equation}
where $m_B$ and $\lambda$ are the bare mass and quartic self-coupling respectively. For free fields ($\lambda=0$), the equal-time Feynman propagator is given by \cite{Huang_book:1998}
\begin{equation}
    \label{minkowski_free_propagator}
    i\Delta_M^F(\mathbf{x})=\frac{m_B}{4\pi^2\abs{\mathbf{x}}}K_1(m_B\abs{\mathbf{x}}),
\end{equation}
where $K_1(z)$ is the modified Bessel function of the second kind and the subscript `M' indicates we're in Minkowski spacetime. Performing a Fourier transform gives the well-known Feynman propagator in $k$-space as
\begin{equation}
    \label{minkowski_free_propagator_k-space}
    i\Tilde{\Delta}^F_M(k)=\frac{i}{k^2+m_B^2}
\end{equation}\\
where the tilde indicates this is a quantity in $k$-space. If we now compute the Feynman propagator to $\mathcal{O}(\lambda)$, the result is simply to add a contribution to the mass term in the free Feynman propagator, namely
\begin{equation}
    \label{minkowski_one-loop_propagator_k-space}
    \bra{0_M}T\hat{\phi}(x)\hat{\phi}(x')\ket{0_M}=\int\dbar^4k\frac{i}{k^2+m_B^2+3\lambda\expval{\hat{\phi}^2}_M}e^{ik\cdot(x-x')}.
\end{equation}
However, the scalar field variance $\expval{\hat{\phi}^2}_M$ contains quadratic and logarithmic divergences in the UV regime; explicitly,
\begin{equation}
    \label{field_variance_small_x_minkowski}
    \expval{\hat{\phi}^2}_{M,PS}=i\Delta_M(\mathbf{x})\eval_{\abs{\mathbf{x}}\rightarrow0}=\frac{1}{4\pi^2\abs{\mathbf{x}}^2}-\frac{m_B^2}{16\pi^2}\qty(-2\ln\qty(m_B\abs{\mathbf{x}})+1-2\gamma_E+\ln4).
\end{equation}
In \textit{point-splitting regularisation}, one absorbs the divergent terms in Eq. (\ref{field_variance_small_x_minkowski}) into the mass parameter. Then, the 2-point function to $\mathcal{O}(\lambda)$ is simply given by
\begin{equation}
    \label{2-pt_func_Minkowski_point-split}
    \bra{0}\hat{T}\hat{\phi}(t,\mathbf{0})\hat{\phi}(t,\mathbf{x})\ket{0}=i\Delta_M^F(\mathbf{x})\eval_{m_B^2\rightarrow m_R^2+3\lambda\expval{\hat{\phi}^2}_{M,PS}^{(fin)}},
\end{equation}
where the finite part of the field variance is given by
\begin{equation}
    \label{finite_field_variance_minkowski_point-split}
    \expval{\hat{\phi}^2}_{M,PS}^{(fin)}=-\frac{m_B^2}{16\pi^2}\qty(1-2\gamma_E+\ln4),
\end{equation}
where the subscript `PS' indicates that it is computed via point-splitting regularisation.

To align with particle experiments, one tends to be more interested in renormalisation via \textit{dimensional regularisation}. The premise behind this is to shift the spacetime dimensions to non-integer values in order to isolate and remove divergent terms. Consider the scalar field variance in $D=4-\epsilon$ dimensions
\begin{equation}
    \label{scalar_field_variance_k_minkowski}
    \expval{\hat{\phi}^2}_{M,DR}=\mu^{\epsilon}\int\dbar^Dk\frac{i}{k^2+m_B^2},
\end{equation}
where $\mu$ is the regularisation scale and subscript `DR' indicates we are using dimensional regularisation. This integral can be computed as \cite{renormalisation_book}
\begin{equation}
    \label{field_variance_D-dim_int_minkowski}
    \expval{\hat{\phi}^2}_M=\frac{\mu^{\epsilon}\Gamma\qty(1-D/2)m_B^{D-2}}{(4\pi)^{D/2}},
\end{equation}
which, when expanded about small $\epsilon$ reads
\begin{equation}
    \label{field_variance_O(epsilon)_minkowski}
    \expval{\hat{\phi}^2}_{M,DR}=-\frac{m_B^2}{16\pi^2}\qty(\frac{2}{\epsilon}-\gamma_E+\ln(4\pi)+1+\ln\qty(\frac{\mu^2}{m_B^2}))+\mathcal{O}\qty(\epsilon),
\end{equation}
where the $\mathcal{O}(\epsilon)$ terms are negligible. In the $\overline{\text{MS}}$ scheme, the term of order $\qty(\frac{2}{\epsilon}-\gamma_E+\ln\qty(4\pi))$ is removed by redefining the mass parameter to be
\begin{equation}
    \label{renormalised_mass_dim_reg_minkowski}
    m_R^2=m_B^2-\frac{3\lambda m_B^2}{16\pi^2}\qty(\frac{2}{\epsilon}-\gamma_E+\ln(\frac{4\pi\mu^2}{M^2}))+\mathcal{O}\qty(\lambda^2),
\end{equation}
where $M$ is the renormalisation scale. We define the finite part of the field variance to be
\begin{equation}
    \label{finite_field_variance_mink_dim_reg}
    \expval{\hat{\phi}^2}_{M,\overline{{MS}}}^{(fin)}=-\frac{m_R^2}{16\pi^2}\qty(1+\ln\qty(\frac{M^2}{m_R^2}))+\mathcal{O}(\lambda),
\end{equation}
where the subscript $\overline{MS}$ denotes that this is the finite field variance using the dimensional regularisation $\overline{\text{MS}}$ scheme. Note that the explicit $M$-dependence in Eq. (\ref{finite_field_variance_mink_dim_reg}) is cancelled by the implicit dependence in $m_R^2$. Comparing the two schemes, we see that the difference in the finite term between the two is
\begin{equation}
    \label{finite_field_variance_difference_minkowski}
    \expval{\hat{\phi}^2}_{M,\overline{\text{MS}}}^{(fin)}-\expval{\hat{\phi}^2}_{M,PS}^{(fin)}=-\frac{m_R^2}{16\pi^2}\qty(2\gamma_E-\ln4+\ln\qty(\frac{M^2}{m_R^2}))+\mathcal{O}(\lambda),
\end{equation}
where we assume the renormalised mass parameters computed using point splitting and dimensional regularisation are equivalent.

\section{The IR behaviour of the connected four-point function}
\label{app:IR_lim_4-pt_func}

In this appendix, I will consider more carefully the IR limit of the connected quantum 4-pt function of Sec. \ref{subsec:4-pt_functions}. We start with the $\mathbf{k}$-space 4-pt function, given in Eq. (\ref{4-pt_func_k-space_Wightman}) as
\begin{equation}
    \label{4-pt_func_k-space_Wightman_again}
    \begin{split}
    \Tilde{G}^{(4)}_C(\eta,\{\mathbf{k}_i\})=&-6i\lambda\int^\eta_{-\infty} d\eta_z \frac{1}{(H\eta_z)^4}\\&
    \begin{split}
    \times\Bigg(&\Tilde{\Delta}^-(\eta_z,\eta,\mathbf{k}_1)\Tilde{\Delta}^-(\eta_z,\eta,\mathbf{k}_2)\Tilde{\Delta}^-(\eta_z,\eta,\mathbf{k}_3)\Tilde{\Delta}^-(\eta_z,\eta,\mathbf{k}_4)\\&-\Tilde{\Delta}^+(\eta_z,\eta,\mathbf{k}_1)\Tilde{\Delta}^+(\eta_z,\eta,\mathbf{k}_2) \Tilde{\Delta}^+(\eta_z,\eta,\mathbf{k}_3) \Tilde{\Delta}^+(\eta_z,\eta,\mathbf{k}_4)\Bigg).
    \end{split}
    \end{split}
\end{equation}
For the purposes of studying the general features of the IR limit, we will take $\mathbf{k}_i=\mathbf{k}$ $\forall i$. Using the $\mathbf{k}$-space Wightman function (\ref{k-space_2-pt_func}), the 4-pt function becomes\footnote{I will just use $\nu$ instead of $\nu_R$ here. As this is an $\mathcal{O}(\lambda)$ quantity, the non-trivial part of the renormalised mass won't feature.}
\begin{equation}
    \label{4-pt_func_Hankels}
    \begin{split}
        \Tilde{G}^{(4)}_C(\eta,\{\mathbf{k}_i\})=-6i\lambda\int^\eta_{-\infty} d\eta_z& \frac{1}{(H\eta_z)^4}\frac{\pi^4}{256H^4}(-H\eta)^{6}(-H\eta_z)^{6}\\&\times2i\Im\lsb\mathcal{H}_\nu^{(1)}(-k\eta)^4\mathcal{H}_\nu^{(2)}(-k\eta_z)^4\rsb.
    \end{split}
\end{equation}
Defining the quantities $K=-k\eta$ and $K_z=-k\eta_z$, the integral can be written as
\begin{equation}
    \label{4-pt_func_Hankels_K}
    \begin{split}
        \Tilde{G}^{(4)}_C(\eta,\{\mathbf{k}_i\})=\frac{3\lambda\pi^4 H^4(-\eta)^6}{64k^3}\int^\infty_K dK_z K_z^2\Im\lsb\mathcal{H}_\nu^{(1)}(K)^4\mathcal{H}_\nu^{(2)}(K_z)^4\rsb.
    \end{split}
\end{equation}
Since we are interested in the IR behaviour, we take the limit $K\ll1$ such that we can use the asymptotic behaviour of the Hankel functions
\begin{equation}
    \label{asymptotic_Hankel_func}
    \mathcal{H}_\nu^{(1)}(K)\simeq \frac{2^{-\nu}}{\Gamma(1+\nu)}K^{\nu}-i\frac{2^\nu\Gamma(\nu)}{\pi}K^{-\nu}
\end{equation}
such that
\begin{equation}
    \label{4-pt_func_Hankels_K_asymp}
    \begin{split}
        \Tilde{G}^{(4)}_C(\eta,\{\mathbf{k}_i\})=\frac{3\lambda\pi^4 H^4(-\eta)^6}{64k^3}\int^\infty_K dK_z K_z^2\Bigg[&\frac{2^{4\nu}\Gamma(\nu)^4}{\pi^4}K^{-4\nu}\Im\lsb\mathcal{H}_\nu^{(2)}(K_z)^4\rsb\\&+\frac{2^{2+2\nu}\Gamma(\nu)^3}{\pi^3\Gamma(1+\nu)}K^{-2\nu}\Re\lsb\mathcal{H}_\nu^{(2)}(K_z)^4\rsb\Bigg].
    \end{split}
\end{equation}
While this integral can't be computed in general, we can get some information about the IR behaviour of the 4-pt function. Consider the split of the integral
\begin{equation}
    \label{4-pt_func_Hankels_K_asymp_split}
    \begin{split}
        \Tilde{G}^{(4)}_C(\eta,\{\mathbf{k}_i\})=\frac{3\lambda\pi^4 H^4(-\eta)^6}{64k^3}\lb\int^\Lambda_K+\int_\Lambda^\infty\rb dK_z &K_z^2\Bigg[\frac{2^{4\nu}\Gamma(\nu)^4}{\pi^4}K^{-4\nu}\Im\lsb\mathcal{H}_\nu^{(2)}(K_z)^4\rsb\\&+\frac{2^{2+2\nu}\Gamma(\nu)^3}{\pi^3\Gamma(1+\nu)}K^{-2\nu}\Re\lsb\mathcal{H}_\nu^{(2)}(K_z)^4\rsb\Bigg],
    \end{split}
\end{equation}
for some parameter $\Lambda<1$. Focussing on the IR limit of the integral, $K<K_z<\Lambda$, we can take the limit $K_z<1$ such that we can use the approximate form of the Wightman functions (\ref{k_Wightman_IR_lim})
\begin{equation}
    \label{k_Wightman_IR_lim_again}
    \begin{split}
    \Tilde{\Delta}^\pm(\eta_z,\eta,\mathbf{k})\simeq \frac{\pi}{4Ha(\eta)^{3/2}a(\eta_z)^{3/2}}\Bigg[&\frac{4^{\nu}\Gamma(\nu)^2}{\pi^2}\lb\eta_z\eta\rb^{-\nu}k^{-2\nu}\\&\pm i\frac{1}{\pi\nu}\lb\lb\frac{\eta}{\eta_z}\rb^{\nu}-\lb\frac{\eta_z}{\eta}\rb^{\nu}\rb\Bigg].
    \end{split}
\end{equation}
Then, we can compute the IR region of the integral (\ref{4-pt_func_Hankels_K_asymp_split}) to find that the 4-pt function will have the following behaviour:
\begin{equation}
    \label{IR_behaviour_4-pt_k}
    \Tilde{G}^{(4)}_C(\eta,\{\mathbf{k}_i\})\sim K^{-6\nu}+K^{-3-4\nu}+K^{-3-2\nu}.
\end{equation}
The $K^{-6\nu}$ is just the term that we found in Eq. (\ref{k-space_4-pt_func}) and comes from the $K_z\rightarrow K$ limit. The other two contributions come from the limit $K_z\rightarrow\Lambda$. Converting these to coordinate space via a Fourier transform, one finds that
\begin{equation}
    \label{IR_behaviour_4-pt_x}
    G^{(4)}_C(\eta,\{\mathbf{x}_i\})\sim \abs{Ha(\eta)\mathbf{x}}^{-9+6\nu}+\abs{Ha(\eta)\mathbf{x}}^{-6+4\nu}+\abs{Ha(\eta)\mathbf{x}}^{-6+2\nu}.
\end{equation}
Since $\nu\leq 3/2$, it is immediately clear that the final term is subleading. However, for near-massless fields, $\nu\sim3/2$, the first and second terms give a similar contribution. As one increases the mass of the field, the second term is in fact the leading contribution over the first term. So, it appears that the contribution computed in Sec. \ref{subsec:4-pt_functions} is subleading. 

At this point, it is worth noting why we are doing the 4-pt function computation within this thesis. The aim, as outlined in Chapter \ref{ch:second-order_stochastic}, is to compute an object in perturbative QFT that can be compared with an equivalent stochastic quantity to find the form of the stochastic parameters necessary to reproduce the QFT result. Specifically, the connected 4-pt function is used to match the coupling parameter $\lambda$ because its first contribution is at $\mathcal{O}(\lambda)$ in both approximations. It appears that the piece of the quantum connected 4-pt function required for comparison with the stochastic connected 4-pt function is precisely the contribution coming from the limit $K_z\rightarrow K$. 

Since the stochastic theory is supposed to be an effective theory of the IR regime, it is a valid question to ask whether it can reproduces the term $\sim\abs{Ha(\eta)\mathbf{x}}^{-6+4\nu}$ as this is an important contribution in such a limit. While we have not done detailed calculations on this, mainly due to the challenging of computing the coefficient of such a term from the QFT perspective, we have found that the stochastic approach does naturally compute a term that is of the same order. If one considers the stochastic 4-pt function (\ref{spacelike_4-pt_correlator_f}), one can compute a non-zero contribution at $\mathcal{O}(\lambda)$ for $N=(0,1)$ $\forall i$, which will be $\sim\abs{Ha(\eta)\mathbf{x}}^{-\frac{4\Lambda_{01}}{H}}$. Using the eigenvalues (\ref{free_eigenvalue}), one can see that this is precisely the spacetime behaviour in question.  

One additional comment is that this contribution also has the same behaviour as the disconnected pieces of the quantum 4-pt function (\ref{4-pt_function_Scwhinger-Keldysh}) since the 2-pt function $\bra{0}\hat{\phi}(t,\mathbf{0})\hat{\phi}(t,\mathbf{x})\ket{0}\sim\abs{Ha(t)\mathbf{x}}^{-3+2\nu}$. This suggests that there could be additional ``disconnected'' pieces hiding amongst a seemingly connected piece of the 4-pt function. Currently, it is not clear that this is the case; one would need to do a more rigorous computation of the 4-pt function in order to confirm this hypothesis.

\chapter{Additional Comments on the Stochastic Approach}
\label{app:stochastic}

\section{Second-order stochastic parameters for matching with NLO term}
\label{app:stochastic_parameters_NLO}

For completeness, we will also include the matched stochastic parameters if we choose to reproduce the NLO term in the asymptotic expansion of the 2-point function in perturbative QFT. This choice doesn't make a difference to physical results. Repeating the procedure outlined in Sec. \ref{subsec:perturbative_stochastic_parameters}, one obtains the stochastic parameters to $\mathcal{O}(\lambda H^2/m^2)$ as
\begin{subequations}
    \label{O(lambdaH2m2_stochastic_parameters_NLO_match}
    \begin{align}
    m_S^2&=m^2_R+\frac{\lambda H^2}{8\pi^2}\lb-14+6\gamma_E-3\ln4+3\ln\lb\frac{M^2}{a(t)H^2}\rb\rb+\mathcal{O}\lb\lambda m_R^2\rb\\
    \sigma_{qq}^{2}&=\frac{H^3\alpha_R\nu_R}{4\pi^2\beta_R}\frac{\Gamma(2\nu_R)\Gamma(\frac{3}{2}-\nu_R)4^{\frac{3}{2}-\nu}}{\Gamma(\frac{1}{2}+\nu_R)}+\frac{\lambda H^5(-8+3\ln4)}{32\pi^4 m_R^2}+\mathcal{O}\lb\lambda H^3\rb,\\ \sigma_{qp}^{2}&=-\frac{51\lambda H^6}{32\pi^4m_R^2}+\mathcal{O}\lb\lambda H^4\rb,\\
    \sigma_{pp}^2&=\frac{H^5\beta_R^2\nu_R}{4\pi^2}\frac{\Gamma(-2\nu_R)\Gamma(\frac{3}{2}+\nu_R)4^{\frac{3}{2}+\nu_R}}{\Gamma(\frac{1}{2}-\nu_R)}-\frac{9\lambda H^7(-1+6\ln4)}{4\pi^4m^2_R}+\mathcal{O}(\lambda H^5).
    \end{align}
\end{subequations}

\section{The massless limit of the second-order stochastic equations}
\label{app:so_stochastic_light_field}

In this appendix, I will consider the massless limit of the second-order stochastic theory, focussing on the stochastic noise amplitudes. This is for comparison with the overdamped stochastic theory. For this reason, we will just consider the free noise amplitudes.

\subsection{Second-order noise amplitudes using the cutoff procedure}

Our first focus will be to compare the overdamped stochastic noise with the second-order noise obtained via the cutoff procedure (see Sec. \ref{sec:so_stochastic_equations}). Taking the massless limit of Eq. (\ref{noise_amplitudes_modes}), we obtain
\begin{subequations}
    \label{near-massless_second-order_noise_cutoff}
    \begin{align}
        \sigma_{cut,\phi\phi}^2\eval_{m\ll H}&=\frac{H^3}{4\pi^2}(1+\epsilon^2),\\
        \sigma_{cut,\phi\pi}^2\eval_{m\ll H}&=-\frac{H^4}{4\pi^2}\epsilon^2,\\
        \sigma_{cut,\pi\pi}^2\eval_{m\ll H}&=\frac{H^5}{4\pi^2}\epsilon^4.
    \end{align}
\end{subequations}
We see that, for $\mathcal{O}(\epsilon^2)=0$, the only non-zero noise is $\sigma_{cut,\phi\phi}^2$. In this limit, the second-order stochastic equations become
\begin{equation}
    \label{second-order_eq_massless}
    \begin{pmatrix}\Dot{\phi}\\\Dot{\pi}\end{pmatrix}=\begin{pmatrix}\pi\\-3H\pi-V'(\phi)\end{pmatrix}+\begin{pmatrix}\xi_{\phi}\\0\end{pmatrix}.
\end{equation}
Making the addition constraint $\Dot{\pi}\ll 3H\pi$, we recover the OD stochastic equation (\ref{od_stochastic_eq}) as required. Thus, (unsurprisingly) the cutoff procedure still works in the massless limit, even when we consider second-order stochastic equations.

\subsection{Second-order noise amplitudes using the matching procedure}

One would expect the second-order noise amplitudes, obtained using the matching procedure in Sec. \ref{subsec:free_stochastic_parameters}, to behave in a similar way in the massless limit. The $(q,p)$ noise amplitudes in the massless limit are given by
\begin{subequations}
    \label{q,p_massless_noise}
    \begin{align}
        \sigma_{Q,qq}^{2(0)}\eval_{m\ll H}&=\frac{H^3}{4\pi^2},\\
        \sigma_{Q,qp}^{2(0)}\eval_{m\ll H}&=0,\\
        \sigma_{Q,pp}^{2(0)}\eval_{m\ll H}&=\begin{cases}\sigma_{pp}^{2(NLO)}\eval_{m\ll H}=\frac{36H^5}{\pi^2}\\0.\end{cases}
    \end{align}
\end{subequations}
Converting this to $(\phi,\pi)$ noise gives
\begin{subequations}
    \label{phi,pi_massless_noise}
    \begin{align}
        \sigma_{Q,\phi\phi}^{2(0)}\eval_{m\ll H}&=\begin{cases}\frac{17H^3}{4\pi^2}\\\frac{H^3}{4\pi^2}\end{cases},\\
        \sigma_{Q,\phi\pi}^{2(0)}\eval_{m\ll H}&=\begin{cases}-\frac{12 H^4}{\pi^2}\\0\end{cases},\\
        \sigma_{Q,\pi\pi}^{2(0)}\eval_{m\ll H}&=\begin{cases}\frac{36H^5}{\pi^2}\\0\end{cases}.
    \end{align}
\end{subequations}
The first and second cases represent we have chosen to set $\sigma_{Q,pp}^{2(0)}=\sigma_{pp}^{2(NLO)}$ and $\sigma_{Q,pp}^2=0$ respectively. One can see that with the latter choice, we have agreement with the OD and second-order cutoff noise amplitudes. With the former, we have extra contributions that enter to give us the subleading term in asymptotic expansion of the scalar 2-point functions, a feature that does not appear when considering a cutoff. Hence, they don't give the same noise.

\addcontentsline{toc}{chapter}{Bibliography}
\bibliographystyle{bibliography/hep}
\bibliography{bibliography/bibliography}

\end{document}